\documentclass[aps,prb,eqsecnum,nofootinbib,notitlepage,superscriptaddress]{revtex4-1}
\usepackage[colorlinks=true,linktoc=page,citecolor=red,linkcolor=blue]{hyperref}
\usepackage{comment}
\usepackage{amsmath,amssymb,amsfonts,bbold}
\usepackage{bm,braket,array}
\usepackage[dvipdfmx]{graphicx}
\usepackage[usenames]{color}

\usepackage{comment}
\usepackage{multirow}
\usepackage{accents}
\usepackage[all]{xy}
\usepackage{amscd}

\def\im{\mbox{Im }}
\def\mod{{\rm mod }}
\def\dim{{\rm dim }}
\def\ker{{\rm ker }}
\def\coker{{\rm coker }}

\def\tr{{\rm tr\,}}

\def\p{\partial}
\def\ot{\leftarrow}

\def\wt{\widetilde}

\def\Hom{{\rm Hom}}

\def\ua{\uparrow}
\def\da{\downarrow}

\newcommand{\C}{\mathbb{C}}

\newcommand{\R}{\mathbb{R}}
\newcommand{\Z}{\mathbb{Z}}

\newcommand{\bk}{\bm{k}}

\def\widebar{\accentset{{\cc@style\underline{\mskip10mu}}}} 
\def\wideubar{\underaccent{{\cc@style\underline{\mskip10mu}}}} 

\begin{document}
\title{Topological Crystalline Materials
\\ -General Formulation, Module Structure, and Wallpaper Groups -
}
\author{Ken Shiozaki}
\email{ken.shiozaki@riken.jp}
\affiliation{Department of Physics, University of Illinois at Urbana Champaign, Urbana, IL 61801, USA}
\author{Masatoshi Sato}
\email{msato@yukawa.kyoto-u.ac.jp}
\affiliation{Yukawa Institute for Theoretical Physics, Kyoto University, Kyoto 606-8502, Japan}
\author{Kiyonori Gomi}
\email{kgomi@math.shinshu-u.ac.jp}
\affiliation{Department of Mathematical Sciences, Shinshu University, Nagano, 390-8621, Japan}

\date{\today}
\begin{abstract}
We formulate topological crystalline materials on the basis of the twisted equivariant $K$-theory. Basic ideas of the twisted equivariant $K$-theory are explained with application to topological phases protected by crystalline symmetries in mind, and systematic methods of topological classification for crystalline materials are presented. Our formulation is applicable to bulk gapful topological crystalline insulators/superconductors and their gapless boundary and defect states, as well as bulk gapless topological materials such as Weyl and Dirac semimetals, and nodal superconductors. As an application of our formulation, we present a complete classification of topological crystalline surface states, in the absence of time-reversal invariance. The classification works for gapless surface states of three-dimensional insulators, as well as full gapped two-dimensional insulators. Such surface states and two-dimensional insulators are classified in a unified way by 17 wallpaper groups, together with the presence or the absence of (sublattice) chiral symmetry. We identify the topological numbers and their representations under the wallpaper group operation. We also exemplify the usefulness of our formulation in the classification of bulk gapless phases. We present a new class of Weyl semimetals and Weyl superconductors that are topologically protected by inversion symmetry.
\end{abstract}
\maketitle
\tableofcontents

\section{Introduction}
Since the discovery of topological insulators and topological
superconductors,
much effort has been devoted
to exploring new topological phases of matters.\cite{Hasan2010, Qi2011,
Tanaka-Sato-Nagaosa2012, AndoReview, Sato-Fujimoto2016,SatoAndo2016}  
Whereas only fully gapped systems had been regarded as topological phases
in the early stage of the study, recent developments have clarified 
that bulk gapless materials like Weyl semimetals also exhibit  
non-trivial topological phenomena. 
The existence of surface Fermi arcs and related anomalous transports are
typical topological phenomena in the latter case.

In the exploration of such topological materials, symmetry plays an
important role:
In the absence of symmetry, a fully gapped non-interacting system
may realize only an integer quantum Hall state in up to three
dimensions.\cite{Avron1983} 
Indeed, for realization of topological insulators and topological
superconductors, time-reversal and charge conjugation
(i.e. particle-hole symmetry (PHS)) are essential.\cite{Kane-Mele2,
QiHughesRaghuZhang2009, Schnyder2008}
Furthermore,
systems often have other symmetries specific to their structures.
In particular, materials in condensed matter physics support crystalline
symmetries of space groups or magnetic space groups.
Such crystalline symmetries also stabilize distinct topological structures in
gapful materials as well as gapless ones.~\cite{Teo2008, Mong2010, Fu2010,
Hsieh2012, Fang2012, Mizushima2012, Slanger2013, 
Ueno2013,
Chiu2013, Zhang-Kane-Mele2013, Morimoto2013,  Benalcazar2013, Fulga2014, 
Alexandradinata2014,
ShiozakiSato2014, ShiozakiSatoGomi2015, ShiozakiSatoGomi2016, FangFu2015,
Varjas-deJuan-Lu2015,
Watanabe2016,Young2012, Wang2012, Wang2013,Yang2014, Kobayashi2014,
ChiuSchnyder2014, Watanabe2016, Kobayashi2016, Agterberg2016,
Micklitz2016,Nomoto2016,Mathai-Thiang2017Global}

In this paper, we formulate such topological crystalline materials on
the basis of the $K$-theory.
The $K$-theory approach has successfully revealed
all possible topological phases protected by  general symmetries of
time-reversal and charge conjugation.\cite{Horava2005, Kitaev2009, Ryu2010}  
Depending on the presence or absence of the general symmetries, systems are
classified into Altland-Zirnbauer (AZ) ten fold symmetry
classes.~\cite{Schnyder2008, Altland1997}  
All
possible topological numbers in the AZ classes are identified in any
dimensions.\cite{Kitaev2009, Ryu2010, Teo2010, Stone2011,
Abramovici2012}. 
One of the main purposes of the present paper is to generalize the $K$-theory
approach in the presence of crystalline symmetries.

Partial generalization of the $K$-theory approach have been attempted
previously:
Motivated by the discovery of topological mirror insulator
SnTe,~\cite{Hsieh2012, Tanaka2012a, Dziawa2012, Xu2012}
mirror-reflection symmetric insulators and superconductors have
been classified topologically.\cite{Chiu2013,Morimoto2013}
Furthermore, a complete topological classification of crystalline
insulators/superconductors with
order-two space groups has been accomplished by means of the
$K$-theory.\cite{ShiozakiSato2014, ShiozakiSatoGomi2015, ShiozakiSatoGomi2016}
The order-two space groups include reflection, two-fold rotation, inversion
and their magnetic versions, and 
many proposed topological crystalline insulators and superconductors
have been understood systematically in the latter classification.
The order-two space group classification also has revealed that 
nonsymmorphic glide symmetry provides 
novel $\Z_2$~\cite{FangFu2015, ShiozakiSatoGomi2015} and $\Z_4$ phases~\cite{ShiozakiSatoGomi2016} with
M\"{o}bius twisted surface states.
Material realization of such a glide protected topological phase
has been proposed theoretically~\cite{wang2016hourglass} and 
confirmed experimentally.~\cite{ma2016experimental}
There is also a different proposal for material realization of
the M\"{o}bius twisted surface states in heavy fermion systems.~\cite{chang2016m}

Our present formulation is applicable to any bulk gapful topological crystalline
 insulators/superconductors (TCIs/TCSCs) and their gapless boundary and
 defect states, as well as bulk gapless topological crystalline materials.
On the basis of the twisted equivariant $K$-theory,~\cite{Freed2013,
Thiang2016} we illustrate how space groups and
magnetic space groups are
incorporated into topological classification in a unified manner: 
Following the idea by Freed and Moore~\cite{Freed2013},  
the space group action on Hamiltonians is introduced as a ``twist''
 $(\tau,c)$ of that on the base space, and anti-unitary symmetries are
 specified by a $\Z_2$-valued function $\phi$ for group elements.
Then,
the $K$-group  ${}^{\phi} K_{\cal G}^{(\tau, c)-n}(X)$ on the base space
 $X$ is introduced in terms of
the Karoubi's formulation of the $K$-theory.~\cite{Karoubi2008}  
The $K$-group ${}^{\phi} K_{\cal G}^{(\tau, c)-n}(T^d)$ for the Brillouin
 zone (BZ) torus $T^d$ provides
 topological classification of $d$-dimensional crystalline insulators and
 superconductors subject to symmetry ${\cal G}$.

Bearing in mind applications in condensed matter physics, we clarify
connections between the
$K$-theory and the traditional band theory. We also explain
practical methods
to compute $K$-groups.
In particular, we show the following: 
\begin{itemize}
\item  The crystal data of
Wyckoff positions 
are
naturally taken into account in our formulation.
The $K$-group for space group ${\cal G}$ has elements corresponding to
       Wyckoff positions for ${\cal G}$.

\item Not only crystal structures determine properties of
      materials. Atomic orbital characters of band electrons also
      strongly affect their properties. 
For instance, if we change the physical degrees of freedom from
      $s$-orbital electrons to $p$-orbital ones, the
      topological nature of the material may change.
This remarkable aspect of crystalline materials is involved in our
      formulation as 
the $R(P)$-module structure of the $K$-group,  where 
$R(P)$ is the representation ring of a point group $P$. 
An element $V \in R(P)$ acts on the $K$-group 
as the tensor product for the symmetry operator, which induces the 
change of the representations of 
physical degrees of freedom.

\item TCIs and TCSCs support stable gapless boundary excitations
      associated with bulk topological numbers if the boundary is
      compatible with symmetry responsible for the topological numbers.
This so-called bulk-boundary correspondence is explained by using
      dimension-raising maps, of which the existence is ensured by the Gysin
      exact sequence in the $K$-theory.

\item Defect gapless modes in TCIs and TCSCs are understood as boundary
      gapless states in lower dimensional TCIs and TCSCs.

\item Bulk gapless topological crystalline materials are formulated in
      terms of the $K$-theory. This formulation provides a novel systematic
      method to explore gapless topological crystalline materials. 

%
%
%

\item 
We present the topological table for topological crystalline surface
      states protected by wallpaper groups, in the absence of
      time-reversal symmetry (TRS).
The additive structures of the relevant
      $K$-groups were previously calculated in the
      literature for the spinless case with and without chiral
      symmetry~\cite{Yang1997, LuckStamm2000} 
      and for the spinful case without chiral symmetry.~\cite{DongLiu2015}
We complete the topological classification 
by determining their $R(P)$-module structures and considering the
      spinful case with chiral symmetry.

\item The Mayer-Vietoris exact sequence and the Gysin exact sequence play
      central roles in computing $K$-groups. 
We illustrate the calculation of $K$-groups in various examples.

\end{itemize}

The organization of the paper is as follows. 
In Sec.~\ref{Sec:SpaceGroup}, 
we explain how space group symmetries are incorporated in the
Hamiltonian formalism. 
Nonsymmorphic space groups can be thought of as unavoidable $U(1)$ phase
factors in the projective representations of point
groups. 
Section \ref{Sec:K-theory} is devoted to introducing the twisted
equivariant $K$-theory.
Two alternative but equivalent constructions of $K$-groups are explained.
It is shown that
$K$-groups are not just additive groups, 
but have module structures induced by the tensor product of 
representations of point groups.
The treatment of anti-unitary symmetries in the twisted equivariant
$K$-theory is explained in Sec.~\ref{With TRS and/or PHS}. 
Not only TRS and PHS, but also magnetic
space group symmetries are taken into account in a unified manner. 
%
Using chiral symmetries, we also introduce the integer grading of the
$K$-groups.
In Sec.~\ref{sec:Topological crystalline insulators and superconductors}, 
we formulate TCIs and TCSCs on the basis of the twisted equivariant
$K$-theory. 
Characteristic physical properties of TCIs and TCSCs are disucssed here.  
In Sec.~\ref{sec:Topological nodal semimetals and superconductors}, 
we propose a systematic method to classify bulk gapless topological
crystalline materials.
Weyl and Dirac semimetals and nodal superconductors are treated in a
unified manner. 
As an application of the twisted equivariant $K$-theory, 
in Sec.~\ref{Wallpaper_summary}, we summarize the topological classification of 
crystalline insulators with wallpaper groups in the absence of TRS. 
We illustrate computations of $K$-groups 
in various examples in Sec.~\ref{sec:Example of K-theory classification}. 
Finally, we conclude the paper in Sec.~\ref{sec:Conclusion}.
We explain some useful mathematical details of the twisted equivariant
$K$-theories in Appendices.

\section{Hamiltonian and Space group}
\label{Sec:SpaceGroup}


\subsection{Periodic Bloch Hamiltonian}
\label{Sec:Pre}

\begin{figure}[tb]
\begin{center}
  \includegraphics[width=\linewidth, trim=0cm 2cm 0cm 0cm]{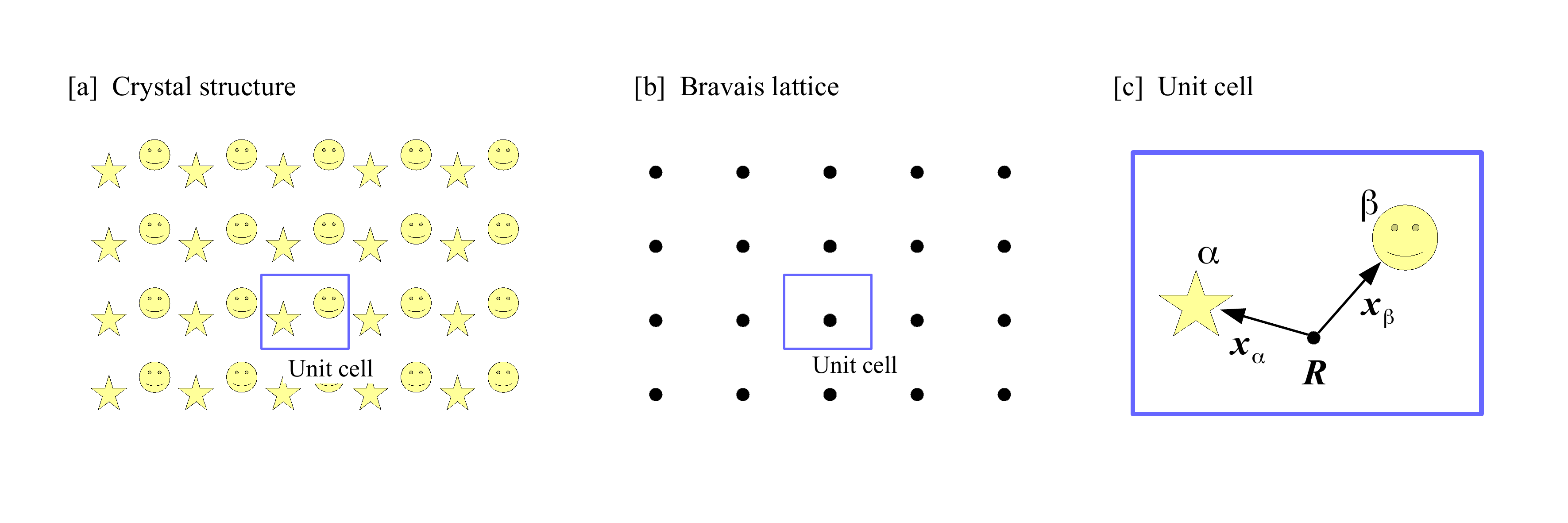}
\end{center}
\caption{A crystal structure [a]. The Bravais lattice [b]. The unit cell [c].}
\label{Fig:Crystal}
\end{figure}

In this paper, we consider one-particle Hamiltonians $\hat H$ with lattice translational symmetry.
Take a proper localized basis, say L\"{o}wdin orbitals $|{\bm R}, \alpha, i\rangle$,
where ${\bm R}$ is a vector of the Bravais lattice $\Pi \cong \Z^d$ for
a given crystal structure in $d$-space dimensions, $\alpha$ is a
label for the $\alpha$-th atom in
the unit cell, and $i$ represents internal degrees of freedom such as
orbital and spin (see Fig.\ref{Fig:Crystal}).
Then the system is well described by the tight-binding Hamiltonian 
\begin{equation}
H= \sum_{\bm{R} , \bm{R}' \in \Pi} \psi_{\alpha i}^{\dag}(\bm{R})
H_{\alpha i,\beta j}(\bm{R}-\bm{R}') 
\psi_{\beta j}(\bm{R}'),
\end{equation}
with
\begin{eqnarray}
H_{\alpha i,\beta j}(\bm{R}-\bm{R}')=\langle {\bm R}, \alpha, i|\hat
H|{\bm R}', \beta, j\rangle. 
\end{eqnarray}
%
%
Because the topological phase of the one-particle Hamiltonian is
examined in the momentum space,
we perform the Fourier transformation of $|{\bm R}, \alpha, i\rangle$
by taking a Bloch basis.
The standard Bloch basis is given by
\begin{align}
&\ket{\bk,\alpha,i}' := \frac{1}{\sqrt{N}}\sum_{\bm{R} \in \Pi} \ket{\bm{R},\alpha,i} e^{i \bm{k} \cdot (\bm{R} + \bm{x}_{\alpha})}, 
\label{DefFourierTr1}
\end{align}
where $\bm{x}_{\alpha}$ is the localized position of the $\alpha$-th atom 
measured from the center of the unit cell specified by $\bm{R}$, and $N$
is the number of unit cells in the crystal.   
This basis, however, is somewhat inconvenient in
topological classification: 
The basis $\ket{\bk,\alpha,i}'$ is not periodic in the
Brillouin zone (BZ) torus $T^d$, 
obeying the twisted periodic boundary condition
\begin{eqnarray}
\ket{\bk +\bm{G},\alpha,i}' = \ket{\bk,\alpha,i}' e^{i \bm{G} \cdot
\bm{x}_{\alpha}} 
\end{eqnarray} 
with $\bm{G}$ a reciprocal vector, so is not the resultant Bloch
Hamiltonian,
\begin{eqnarray}
H'_{\alpha i, \beta j}({\bm k})=\langle {\bm k}, \alpha, i|'\hat{H}|{\bm
 k},\beta, j\rangle'. 
\end{eqnarray}
The non-periodicity of the Hamiltonian gives 
an undesirable complication in topological classificaition.
To avoid this problem, we take here an alternative Bloch basis 
which makes the Hamiltonian $H(\bk)$ 
periodic, 
\begin{align}
&\ket{\bk,\alpha,i} := \frac{1}{\sqrt{N}}
\sum_{\bm{R} \in \Pi} \ket{\bm{R},\alpha,i} e^{i \bm{k} \cdot \bm{R}}. 
\label{DefFourierTr2}
\end{align}
Obviously, the Bloch basis (\ref{DefFourierTr2}) is periodic in the BZ
torus, 
\begin{eqnarray}
|{\bm k}+{\bm G}, \alpha, i\rangle=|{\bm k}, \alpha, i\rangle, 
\end{eqnarray}
and so is the Bloch Hamiltonian $H_{\alpha i, \beta j}(\bk)$,
\begin{equation}
H_{\alpha i, \beta j}({\bm k})=
\langle {\bm k}, \alpha, i|\hat H|{\bm
 k}, \beta, j\rangle.
\end{equation}
We call this basis $(\ref{DefFourierTr2})$ the periodic Bloch basis. 
Here we note that the periodic Bolch basis (\ref{DefFourierTr2}) loses
the information on the localized position $\bm{x}_{\alpha}$ of the
$\alpha$-th atom in the unit
cell, so it may
cause complication in relations between the Berry 
connections and observables. 
Bearing this remark in mind, 
we employ the periodic basis (\ref{DefFourierTr2}) throughout the
present paper.
For simplicity, we often omit the matrix indices $(\alpha, i)$ below,
and we simply denote the Bloch Hamiltonian $H_{\alpha i, \beta
j}({\bm k})$ as $H({\bm k})$. 

\subsection{Space group and unavoidable $U(1)$ factor}
\label{sec3:space}

The Bloch Hamiltonian $H({\bm k})$ has
space group symmetry $G$ for a given crystal structure. 
An element of $G$ is denoted as $\{p|{\bm a}\}\in G$, under which
${\bm x}$ transforms as ${\bm x}\to p{\bm x}+{\bm a}$.
Here $p\in P$ is an element of the point group $P$.  
In this notation, the lattice translation is denoted as $\{1|{\bm t}\}$
with a lattice vector ${\bm t}\in \Pi$. ($\Pi$ is the Bravais lattice.) 
The multiplication in $G$ is given as 
\begin{eqnarray}
\{p|{\bm a}\}\cdot \{p'|{\bm
a}'\}=\{pp'| p{\bm a}'+{\bm a}\},
\end{eqnarray}
and the inverse is 
\begin{eqnarray}
\{p|{\bm
a}\}^{-1}=\{p^{-1}|-p^{-1}{\bm a}\}.   
\end{eqnarray}
For each $p\in P$, one can choose a representative $\{p|{\bm a}_p\}\in
G$, so that
any element $\{p|{\bm
a}\}\in G$ can be written as a product of $\{p|{\bm a}_p\}$ and
a lattice translation $\{1|{\bm t}\}$.
Since the lattice translation trivially acts on the Bloch Hamiltonian,
it is enough to consider a set of representatives $\left\{\{p|{\bm
a}_p\}\in G: p\in P\right\}$ in the topological classification of the
Bloch Hamiltonian. 

For $\{p|{\bm a}_p\}\in G$,
the Bloch Hamiltonian $H({\bm k})$ obeys
\begin{align}
U_{p}({\bm k}) H({\bm k})U_{p}({\bm k})^{-1}= H(p{\bm k}), 
\label{eq:UHU}
\end{align}
with a unitary matrix $U_p({\bm k})$, which is periodic in the BZ, 
$U_p({\bm k}+{\bm G})=U_p({\bm k})$. 
The multiplication in $G$ implies
\begin{align}
U_p(p'{\bm k})U_{p'}({\bm k})=e^{i\tau_{p,p'}(pp'{\bm k})}U_{pp'}({\bm k}),
\label{eq:UU}
\end{align}
where $U_p({\bm k})$, $U_{p'}({\bm k})$ and $U_{pp'}({\bm k})$ are the
unitary matrices for $\{p|{\bm a}_p\}$, $\{p'|{\bm
a}_{p'}\}$ and $\{pp'|{\bm a}_{pp'}\}$, respectively. 
The U(1) factor $e^{i\tau_{p,p'}({\bm k})}$ above arises
because $\{p|{\bm a}_p\}\cdot\{p'|{\bm a}_{p'}\}$ is not equal to
$\{pp'|{\bm a}_{pp'}\}$, in general.
Actually it holds that
\begin{eqnarray}
\{p|{\bm a}_p\}\cdot\{p'|{\bm a}_{p'}\}=\{1|{\bm
\nu}_{p,p'}\}\cdot \{pp'|{\bm a}_{pp'}\}
\end{eqnarray}
with a lattice vector ${\bm \nu}_{p,p'}\equiv p{\bm a}_{p'}+{\bm
a}_p-{\bm a}_{pp'}\in \Pi$. 
Due to the Bloch factor $e^{i{\bm k} \cdot {\bm
R}}$ of $|{\bm k},\alpha, i\rangle$ in Eq.(\ref{DefFourierTr2}),
 the lattice
translation $\{1|{\bm \nu}_{p,p'}\}$ gives the U(1) factor  
\begin{eqnarray}
e^{i\tau_{p,p'}({\bm k})}=e^{-i{\bm k}\cdot{\bm \nu}_{p,p'}}. 
\label{eq:twist_lattice}
\end{eqnarray}

Here note that if ${\bm a}$ for any element of $G$ is given by a
lattice vector ${\bm t}$, then the U(1) factor in
Eq.(\ref{eq:twist_lattice}) can
be $1$ by choosing ${\bm a}_p=0$ for any $p\in P$.
Such a space group is called symmorphic.
On the other hand, if $G$ contains an element $\{p|{\bm a}\}$ with
a non-lattice vector ${\bm a}$, such as glide or screw, a non-trivial
U(1) factor is unavoidable.
The latter space group is called nonsymmorphic.

For spinful fermions, there exists a different source of
the U(1) factor $e^{i\tau_{p.p'}({\bm k})}$ in Eq.(\ref{eq:UU}).
This is because rotation in the spin space is not given as an
original O(3) rotation, but given as its projective U(2)
rotation.
Different from the U(1) factor in Eq.(\ref{eq:twist_lattice}), the
resultant U(1) factor is ${\bm k}$-independent.

\begin{figure}[!]
 \begin{center}
  \includegraphics[width=0.6\linewidth]{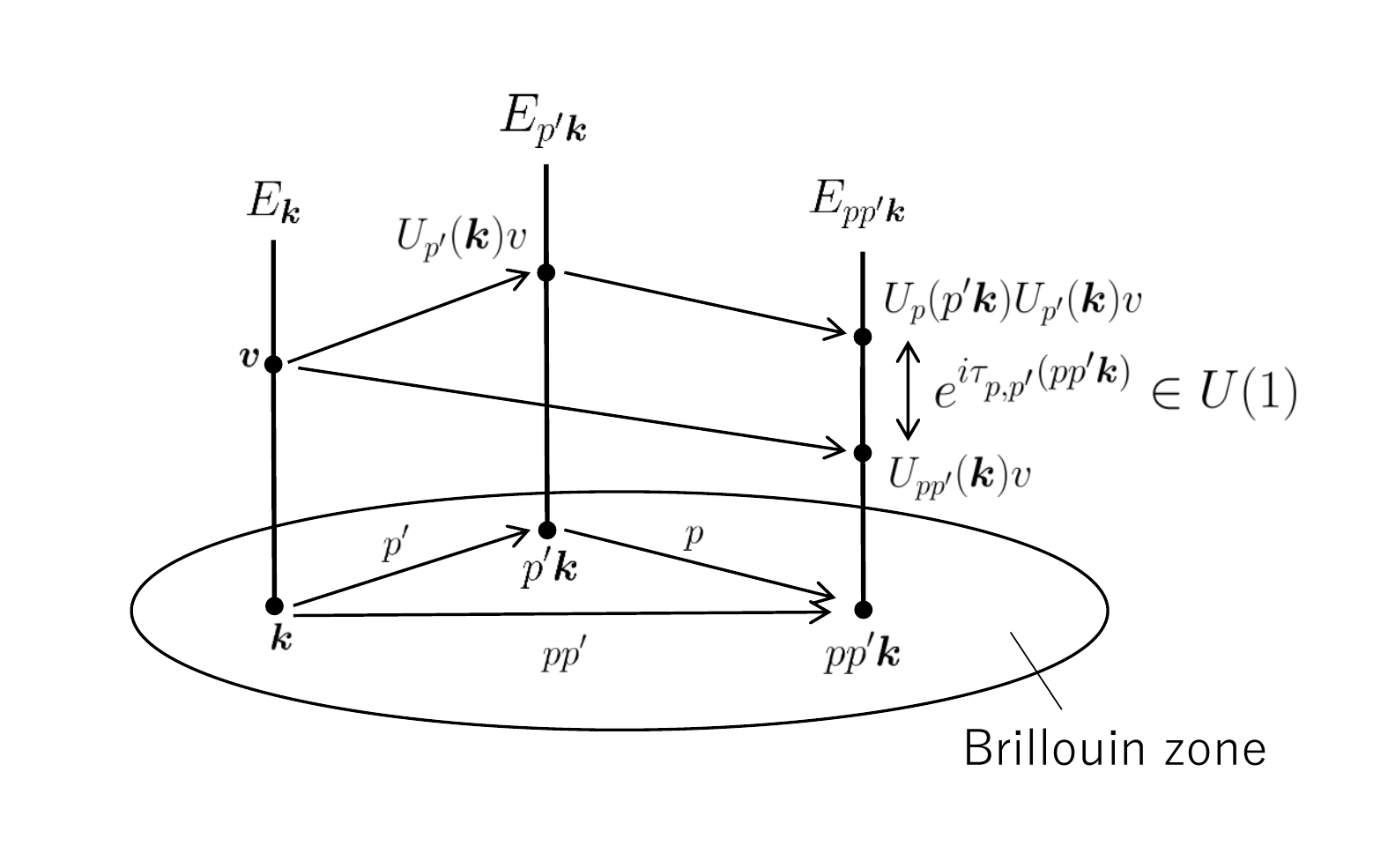}
 \end{center}
 \caption{A $U(1)$ factor associated with group action on a vector bundle on which the Hamiltonian is defined.}
 \label{fig:twist}
\end{figure}

As illustrated in Fig,\ref{fig:twist}, these non-trivial U(1) factors
in Eq.(\ref{eq:UU}) provide a twist in a vector (or Hilbert) space on which the
Bloch Hamiltonian is defined.
In the following, we denote the twist $\tau$ caused by nonsymmorphic
space group $G$ (the projective representation of rotation) as $\tau=\tau_G$
($\tau=\omega$), and if both twists coexist, we denote it as $\tau=\tau_G+\omega$.

\subsubsection{More on space group: group cohomology perspective}

More general treatment of the twist is as follows.
(The reader can skip this section on a first reading.)
Mathematically, space groups and their projective representations are
characterized by inequivalent $U(1)$ phases $\{e^{i
\tau_{p,p'}(\bk)}\}$, which are
classified by the group cohomology. 
The $U(1)$ phases $\{e^{i \tau_{p,p'}(\bk)}\}$
can be considered as an obstruction of the group structure of the group action 
on the (trivial) vector bundle on which the Hamiltonian $H(\bk)$ is defined. 
See Fig.~\ref{fig:twist}. 
To apply the group cohomology classification, 
we introduce the Abelian group $C(T^d,U(1))$ of the $U(1)$-valued functions on the BZ torus $T^d$. 
The Abelian structure of $C(T^d,U(1))$ is given by the usual product of $U(1)$ phases: 
$e^{i \alpha_1(\bk)} \cdot e^{i \alpha_2(\bk)} = e^{i ( \alpha_1(\bk) + \alpha_2(\bk))}$. 
The point group $P$ acts on $C(T^d,U(1))$ by $e^{i (p \cdot \alpha)(\bk)} = e^{i \alpha(p^{-1} \bk)}$, 
where we have denoted the point group action on the BZ by $p \bk$ for $p
\in P$. 
We also introduce the group cochain $C^*(P,C(T^d,U(1)))$. 
The $U(1)$ factor in (\ref{eq:UU}) is a two-cochain $\{ e^{i \tau_{p,p'}(\bk)} \}_{p,p' \in P} \in C^2(P,C(T^d,U(1)))$. 
The associativity $(\hat U_{p_1} \hat U_{p_2}) \hat U_{p_3} = \hat U_{p_1} (\hat U_{p_2} \hat U_{p_3})$ 
implies the two-cocycle condition 
\begin{align}
\delta \tau = 0 \quad \Leftrightarrow \quad 
\tau_{p_2,p_3}(p_1^{-1}\bk)-\tau_{p_1p_2,p_3}(\bk)+\tau_{p_1,p_2p_3}(\bk)-\tau_{p_1,p_2}(\bk) \equiv 0 \qquad  \mod \ 2 \pi, 
\end{align}
and the redefinition of the $U(1)$ factor
$U_p(\bk) \mapsto e^{i \theta_p(\bk)} U_p(\bk)$  
induces the equivalence relation from the two-coboundary 
\begin{align}
\tau \sim \tau+\delta \theta \quad \Leftrightarrow \quad 
\tau_{p_1,p_2}(\bk)\sim \tau_{p_1,p_2}(\bk)+\theta_{p_2}(p_1^{-1}\bk)-\theta_{p_1p_2}(\bk)+\theta_{p_1}(\bk) \qquad  \mod\ 2 \pi. 
\end{align}
(See Appendix \ref{Group_cohomology} for the definition of $\delta$ and
the group cohomology. ) 
Then, we can conclude that 
\begin{itemize}
\item For a given Bravais lattice $\Pi$ and point group $P$, 
the set of inequivalent $U(1)$ phase factors $\{ e^{i \tau_{p,p'}(\bk)} \}$ 
is given by the group cohomology $H^2(P,C(T^d,U(1)))$. 
\end{itemize}
The group cohomology can be divided into two parts~\cite{Gomi2015}
\begin{align}
H^2(P,C(T^d,U(1))) &\cong  H^2(P,H^1(T^d,\Z)) \oplus H^2(P,U(1)), \\
[\tau] &= [\tau_G] + [\omega], 
\end{align}
The latter part $H^2(P,U(1))$ represents 
the classification of the projective representations of the 
point group $P$. 
Moreover, it holds that $H^1(T^d,\Z) \cong \Hom (T^d,U(1)) \cong \Pi$.
(Notice that the BZ torus $T^d$ is the Pontryagin dual 
$\hat \Pi = \Hom(\Pi, U(1))$ of the Bravais lattice $\Pi$.) 
Therefore, the former part coincides with the group cohomology $H^2(P,\Pi)$,
which is known to provide the classification of space groups
for a given Bravais lattice $\Pi$ and a point group $P$.~\cite{Hiller1986}
The two-cocycle $\{\bm{\nu}_{p,p'} \in \Pi\}$ introduced in the 
previous subsection represents an element of the group 
cohomology $H^2(P,\Pi)$.

\subsubsection{Anti space group}
\label{sec:Anti space group}
In addition to ordinary space group operations, 
one may consider a space group operation $U_p({\bm k})$
that changes the sign of the Bloch Hamiltonian,
\begin{eqnarray}
U_p({\bm k})H({\bm k})U_p({\bm k})^{-1}=-H(p{\bm k}) 
\end{eqnarray}
Such an operation is called antisymmetry.~\footnote{
The antisymmetry is equivalent to the antiunitary PHS 
in the many-body Hilbert space.}
The anti space group
symmetry also affects topological nature of the system.
To treat ordinary symmetries and antisymmetries in a unified manner, 
we introduce a function $c(p)=\pm 1$ 
that specifies the symmetry or antisymmetry relations, 
\begin{eqnarray}
U_p({\bm k})H({\bm k})U_p({\bm k})^{-1}=c(p)H(p{\bm k}). 
\end{eqnarray}
It is found that $c(p)$ is a homomorphism on $G$,
i.e. $c(pp')=c(p)c(p')$.

\subsection{Chiral symmetry}
For topological classification based on the $K$-theory, so-called chiral
symmetry plays a special role: As we shall show later, one can change the
dimension of the system keeping the topological structure by
imposing or breaking chiral symmetry. 
Chiral symmetry is defined as
\begin{eqnarray}
\{H({\bm k}), \Gamma\}=0, \quad \Gamma^2=1, 
\end{eqnarray}
where $\Gamma$ is a unitary operator. 
In the presence of space group symmetry, 
\begin{eqnarray}
U_p({\bm k})H({\bm k}) U^{-1}_p({\bm k})=c(p)H(p{\bm k}),
\quad
U_p(p'{\bm k})U_p({\bm k})=e^{i\tau_{p,p'}({pp'\bm k})}U_{pp'}({\bm k}),
\end{eqnarray}
we introduce a compatible chiral symmetry as
\begin{align}
\{H({\bm k}), \Gamma\}=0,
\quad 
U_p({\bm k})\Gamma U_p^{-1}(\bk)
=c(p)\Gamma, \quad 
\Gamma^2=1.
\end{align}

\section{Twisted equivariant $K$-theory}
\label{Sec:K-theory}
%
%


\subsection{Occupied states and $K$-group}
\label{Topological classification and the $K$-group}
Suppose that a Bloch Hamiltonian $H({\bm k})$ is gapped on a compact
momentum space $X$. We
consider the vector bundle $E$ that is
spanned by the occupied states on $X$:
In other words, $E$ is spanned by the states $|\phi({\bm k})\rangle$, ${\bm k}\in X$
in the form of 
\begin{eqnarray}
|\phi({\bm k})\rangle = \sum_{{\cal E}_n({\bm k})<{\cal E}_{\rm
 F}}c_n({\bm k})|u_n({\bm k})\rangle, 
\end{eqnarray}
where $|u_n({\bm k})\rangle$ is an eigenstate of $H({\bm k})$,
\begin{eqnarray}
H({\bm k})|u_n({\bm k})\rangle={\cal E}_n({\bm k})|u_n({\bm k})\rangle,
\quad
\langle u_n({\bm k})|u_m({\bm k})\rangle=\delta_{n,m}. 
\end{eqnarray}
Here ${\cal E}_{\rm F}$ is the Fermi energy, and 
$c_n({\bm k})$ is an arbitrary complex function with the
normalization condition
$\sum_n|c_n({\bm k})|^2=1$.
%
We use the notation $[E]$ to represent a set of vector
bundles that are deformable to $E$.
Vector bundles $[E]$ and $[F]$ can be added as their direct sum,
$[E]+[F]:=[E\oplus F]$.
Namely $|\phi_i({\bm k})\rangle\in [E_i]$ ($i=1,2$) can be added as 
\begin{eqnarray}
\left(
\begin{array}{c}
|\phi_1({\bm k})\rangle\\
|\phi_2({\bm k})\rangle
\end{array}
\right)\in [E_1]+[E_2].
\end{eqnarray}
The zero element $0$ in this summation can be introduced as
      the vector bundle of rank zero. 
Physically, such a rank zero vector is obtained when $H({\bm k})$ does
      not have an occupied state that satisfies ${\cal E}_n({\bm k})<{\cal E}_{\rm F}$.

To compare vector bundles $[E_1]$ and $[E_2]$, 
we consider the pair $([E_1],[E_2])$, where the addition is given by
\begin{eqnarray}
([E_1],[E_2])+([E_1'], [E_2'])=([E_1]+[E_1'], [E_2]+[E_2']). 
\label{eq3:addition}
\end{eqnarray}
Since the ``difference'' between $[E_1]$ and $[E_2]$ does not change
even when a common vector bundle $[F]$ is added to both $[E_1]$ and
$[E_2]$, the pair $([E_1], [E_2])$ can be identified with
$([E_1]+[F], [E_2]+[F])$. 
This motivates us to introduce the following equivalence
relation $\sim$,
\begin{align}
( [E_1] , [E_2] ) \sim ( [E_1'] , [E_2'] ) 
&\overset{\mathrm{def}}{\Longleftrightarrow}
\mbox{${}^{\exists} [F], {}^{\exists}[G]$ such that 
$([E_1], [E_2]) + ([F],[F]) = ([E_1'], [E_2']) +([G],[G])$}. 
\label{eq3:equivalence}
\end{align}
The following properties follow in the equivalence class.
\begin{enumerate}
\renewcommand{\labelenumi}{(\roman{enumi})}
 \item The elements of the form $([E], [E])$ represent the zero for
       the addition in Eq.(\ref{eq3:addition}).

\vspace{1ex}
$\because$ 
From Eq.(\ref{eq3:addition}), we have
       $([E_1],[E_2])+([E],[E])=([E_1]+[E],[E_2]+[E])$, which 
       implies that $([E_1],[E_2])\sim ([E_1]+[E], [E_2]+[E])$. So in
       the equivalence class, the same equation gives 
       $([E_1],[E_2])+([E],[E])=([E_1],[E_2])$, which leads to $([E],[E])=0$.

\item The additive inverse of $([E_1], [E_2])$ is $([E_2], [E_1])$. 

\vspace{1ex}
$\because$ 
From (i), one can show that
      $([E_1],[E_2])+([E_2],[E_1])=([E_1]+[E_2], [E_1]+[E_2])=0$, since
      $E_1\oplus E_2$ is continuously deformed into $E_2\oplus E_1$, so
      $[E_1]+[E_2]=[E_2]+[E_1]$. 
\end{enumerate}
The equivalence classes define an Abelian group, which is
known as the $K$-group or the $K$-theory $K(X)$.
The above properties (i) and (ii) also justify the `formal difference'
notation $[E_1] - [E_2] \in K(X)$ for the pair $([E_1], [E_2])$.  
Accordingly, we often mean by $[E] \in K(X)$ the element $[E] - 0 \in
K(X)$ or equivalently $([E], 0) \in K(X)$. 

The formal difference $[E_1]-[E_2]$ naturally measures the topological difference
between $E_1$ and $E_2$: Indeed, from (i), one finds that if $E_1$ and
$E_2$ are smoothly deformable to each other, then $[E_1]-[E_2]=0 \in K(X)$.
Therefore, we use it to define topological phase on $X$:
When $[E_1]-[E_2]=0$, we say that $E_1$ and $E_2$
belong to the same topological phase on $X$. 
To the contrary, when $[E_1]-[E_2]\neq 0$, we say that $E_1$ and $E_2$
belong to different topological phases on $X$. 
In this definition of topological phases, $[E]-0 \in K(X)$ gives a
topological number of $E$ through the calculation of $K(X)$, 
since a state with no occupied state and the
corresponding vector bundle 0 should be
topologically trivial.

It should be noted here that $[E_1]-[E_2]$ (namely $([E_1], [E_2])$ in the
equivalence relation (\ref{eq3:equivalence})) can
be zero even when
$[E_1]\neq [E_2]$: 
Actually, even if $E_1$ and $E_2$ are not smoothly deformable to each other, 
$E_1\oplus E$ and $E_2\oplus E$ could be by choosing a proper vector
bundle $E$. If this happen, we have $[E_1]+[E]=[E_2]+[E]$, and thus the
above (i) and (ii) lead to 
$([E_1], [E_2])=([E_1], [E_2])+([E],[E])=([E_1]+[E],[E_2]+[E])=0$.
Physically, this result means that a common occupied state can be added
without changing topological difference between $E_1$ and $E_2$.
See Appendix \ref{app:An example of mismatch} for a simple example of a 
mismatch between the $K$-theory and 
the monoid of isomorphism classes of vector bundles.

In topological (crystalline) insulators and superconductors, the vector
bundles $[E]$ of occupied states are
subject to constraints from symmetries. 
The original $K$-theory presented here is not convenient in
order for the symmetry constraints to be taken into account .
In the next section, we introduce a different formulation of $K$-theory,
which is much more suitable for the application in topological
(crystalline) insulators and superconductors.


\subsection{Flattened Hamiltonian and Karoubi's formulation of $K$-theory.}
\label{sec3:Kroubi}

Since $E_i$ $(i=1,2)$ is defined as a vector bundle that is spanned by
occupied states of $H_i({\bm k})$ ($i=1,2$), one may use the
triple $(E, H_1, H_2)$ with $E$ the vector bundle on which $H_i({\bm k})$ acts,
instead of the pair $([E_1], [E_2])$.
In the triple, we impose the additional constraint $H_i^2({\bm k})=1$.
Indeed, any gapped Hamiltonian can satisfy this
constraint by a
smooth deformation without gap closing: Any Bloch Hamiltonian $H({\bm k})$
is diagonalized as
\begin{eqnarray}
H({\bm k})=U({\bm k})
\left(
\begin{array}{ccc}
{\cal E}_1({\bm k}) &  &\\
 & \ddots &\\
&& {\cal E}_n({\bm k})\\
\end{array}
\right)U^{\dagger}({\bm k}), 
\end{eqnarray}
with a unitary matrix $U({\bm k})$, and if $H({\bm k})$ is gapped, then
there is a clear distinction between the empty levels ${\cal E}_{i\le p}({\bm
k})$ and the occupied ones ${\cal E}_{i\ge p+1}({\bm k})$,
\begin{eqnarray}
{\cal E}_{i\le p}({\bm k}) >{\cal E}_{\rm F}> {\cal E}_{i\ge p+1}({\bm k}). 
\end{eqnarray}
Therefore, one may adiabatically deform these levels so that 
${\cal E}_{i\le p}({\bm k})\rightarrow 1$, ${\cal E}_{i\ge p+1}({\bm k})\rightarrow -1$ 
without gap closing. After this deformation, one obtains
\begin{eqnarray}
H({\bm k})=U({\bm k})
\left(
\begin{array}{cc}
{\bm 1}_{p\times p} &  \\
 & -{\bm 1}_{(n-p)\times (n-p)}
\end{array}
\right)U^{\dagger}({\bm k}), 
\end{eqnarray}
which satisfies $H^2({\bm k})=1$.
The flattened Hamiltonian retains the same topological property as the
original one, because the vector bundle spanned by the occupied states
remains the same.
We also regard $H_i$ in the triple as a set of Hamiltonians that are deformable
to $H_i({\bm k})$ keeping the flattened condition $H_i^2({\bm k})=1$. 

In a manner similar to Eq.(\ref{eq3:addition}), 
the addition for the triples is given by
\begin{eqnarray}
(E, H_1, H_2)+(E', H_1', H_2')=(E \oplus E', H_1 \oplus H_1', H_2 \oplus H_2'). 
\label{eq3:addition2}
\end{eqnarray} 
We can also impose
the equivalence relation $\sim$
\begin{align}
&(E, H_1, H_2) \sim (E', H'_1, H'_2)
\nonumber\\
&\overset{\mathrm{def}}{\Longleftrightarrow}
\mbox{${}^\exists (E'', H'', H'')$, ${}^\exists (E''', H''', H''')$
 such that 
$(E, H_1, H_2)+(E'', H'', H'')=(E', H_1', H_2')+(E''', H''', H''')$
}.
\label{eq3:equivalence2}
\end{align}
We denote the equivalence class for the triple $(E, H_1, H_2)$ as $[E,
H_1, H_2]$.
Then, correspondingly to (i) and (ii), the following properties are
obtained, 
\begin{enumerate}
\renewcommand{\labelenumi}{(\roman{enumi}')} 
\item The elements of the form $[E, H, H]$ represent zero in the addition.
\item The additive inverse of $[E, H_1, H_2]$ is 
$[E, H_2, H_1]$, i.e.$-[E, H_1, H_2]=[E, H_2, H_1]$.
\end{enumerate}
The equivalence classes provide an alternative definition of
the $K$-group $K(X)$, which is known as the Karoubi's
formulation of the $K$-theory.
(Karoubi calls the Hamiltonians $H_i$ ($i=1,2$) as gradations.~\cite{Karoubi2008})


In the presence of chiral symmetry $\Gamma$
\begin{eqnarray}
\{\Gamma, H({\bm k}) \}=0, \quad \Gamma^2=1,
\end{eqnarray}
we use the quadruple $(E, \Gamma, H_1, H_2)$ with $E$ the vector
bundle on which $\Gamma$ and $H_i({\bm k})$ act.
Here $H_i({\bm k})$ is flattened, and $H_i$ in the quadruple represents
a set of Hamiltonians that are deformable to $H_i({\bm k})$. 
We can generalize
the notion of equivalence to that on the quadruples $(E, \Gamma, H_1,
H_2)$, and the equivalence classes constitute an Abelian group $K^{-1}(X)$. 


\subsection{Space group and twisted equivariant K-theory}

The Karoubi's formulation can be generalized to insulators subject to
space groups. 
In a crystalline insulator, $H({\bm k})$ is subject to a
constraint from the (anti) space group $G$. 
As mentioned in Sec.\ref{sec3:space}, the space group $G$ acts on
$H({\bm k})$ through the point group $P$ with twist $\tau=\tau_G, \omega,
\tau_G+\omega$. 
The symmetries can be expressed as the following constraint on the Hamiltonian 
\begin{eqnarray}
U_p({\bm k})H({\bm k}) U_p({\bm k})^{-1}=c(p)H(p{\bm k}),
\quad
U_p(p'{\bm k})U_p({\bm k})=e^{i\tau_{p,p'}(pp'{\bm k})}U_{pp'}({\bm k}),
\label{eq3:Up}
\end{eqnarray}
where $p \in P$ is the point group part of an element $\{p|{\bm a}_p\}$ of
$G$, and $U_p({\bm k})$ is a unitary representation matrix of $p$.
The index $c(p)=\pm 1$ specifies
symmetry or antisymmetry. In a manner similar to Sec.\ref{sec3:Kroubi}, a
triple $(E, H_1, H_2)$ with flattened Hamiltonian $H_i$ ($i=1,2$) subject to the
constraint (\ref{eq3:Up}) defines a 
twisted $K$-class $[E, H_1, H_2]\in K_P^{(\tau, c)-0}(X)$, in the
twisted equivariant $K$-theory.
It should be noted here that the direct sum $H({\bm
k})\oplus H'({\bm k})$ satisfies the same constraint (\ref{eq3:Up}) with
the same $c(p)$ and twist $e^{i\tau_{p,p'}({\bm k})}$ if we
consider the corresponding direct sum for $U_p({\bm k})$.
Furthermore, when there exists a compatible chiral symmetry $\Gamma$, 
\begin{align}
&U_p({\bm k})H({\bm k}) U_p({\bm k})^{-1}=c(p)H(p{\bm k}),
\quad
U_p(p'{\bm k})U_p({\bm k})=e^{i\tau_{p,p'}(pp'{\bm k})}U_{pp'}({\bm k})
\nonumber\\
&U_p({\bm k})\Gamma U_p({\bm k})^{-1}=c(p)\Gamma,
\quad \{H({\bm k}), \Gamma\}=0,
\quad \Gamma^2=1,
\end{align}
a quadruple $(E, \Gamma, H_1, H_2)$ subject to this constraint 
defines another twisted $K$-class $[E,\Gamma, H_1, H_2] \in
K_P^{(\tau,c)-1}(X)$.




\subsection{Module structure}
\label{sec:module}
We note that the twisted equivalent $K$-group is not simply an
additive group, but has a more complicated structure.
Indeed, we can multiply an element of the $K$-group by a representation
$R(P)$ of the point group $P$.
To see this, consider a unitary matrix $R(p)$ for an element $p\in P$ in
the representation $R(P)$. 
Then, we can multiply $U_p({\bm k})$ by $R(P)$ taking the tensor product of 
$R(p)$ and $U_p({\bm
k})$, i.e. 
\begin{eqnarray}
R(P)\cdot U_p({\bm k}):=R(p)\otimes U_p({\bm k})
\end{eqnarray}
From the multiplication law in $R(P)$,
$
R(p)R(p')=R(pp'),  
$
we find that the obtained unitary matrix has the same twist as $U_p({\bm k})$
\begin{eqnarray}
\left[R(P)\cdot U_p(p'{\bm k})\right]
\left[R(P)\cdot U_{p'}({\bm k})\right]
=e^{i\tau_{p,p'}(pp'{\bm k})} R(P)\cdot U_{pp'}({\bm k})
\end{eqnarray}
which defines an action of the point group $P$ on the representation space of the tensor product. 
Furthermore, the multiplication of the Hamiltonian $H$ 
by $R(P)$ can be defined as 
\begin{eqnarray}
R(P)\cdot H
({\bm k}):={\bm 1} \otimes H({\bm k}),
\label{eq:RcdotH}
\end{eqnarray}
with the identity matrix ${\bm 1}$ in the representation space of $R(P)$.
Equation (\ref{eq:RcdotH}) gives a Hamiltonian the space group symmetry $G$
\begin{eqnarray}
\left[R(P)\cdot U_p({\bm k})\right]
\left[R(P)\cdot H({\bm k})\right]
\left[R(P)\cdot U_p({\bm k})\right]^{-1}
=\left[R(P)\cdot H(p{\bm k})\right], 
\end{eqnarray}
where $\left[R(P)\cdot U_p({\bm k})\right]^{-1}=[R(p)^{-1} \otimes U_{p}({\bm k})^{-1}]$.
Correspondingly, for the vector space $E$ on which $H$ is
defined, $R(P)\cdot E$ is defined as the tensor product of the representation
space of $R(P)$ and $E$.
Using these definitions,  we can eventually introduce the
multiplication of the triple $(E, H_1, H_2)$ by $R$ as 
\begin{eqnarray}
R(P)\cdot (E, H_1, H_2):=(R(P)\cdot E, R(P)\cdot H_1, R(P)\cdot H_2), 
\end{eqnarray}
which defines the multiplication of the element $[E, H_1, H_2]\in
K_P^{(\tau,c)-0}(X)$ by $R(P)$.
The multiplication by $R(P)$ is compatible with the Abelian group structure
of the $K$-group,
\begin{eqnarray}
R(P)\cdot(E, H_1, H_2)+ R(P)\cdot(E', H'_1, H'_2)
\nonumber\\
=R(P)\cdot (E \oplus E', H_1 \oplus H_1', H_2 \oplus H_2'),
\end{eqnarray} 
and thus the $K$-group is an $R(P)$-module. 
In a similar manner, we can show that $K_P^{(\tau,c)-1}(X)$ is also an
$R(P)$-module.

Remembering that $[E]$ is the space spanned by occupied states
of $H$, one finds that $R\cdot H$ naturally gives the
tensor product of the representation space of $R(P)$ and $[E]$, which we
denote as $R(P)\cdot [E]$.  
Therefore, from the correspondence between $(E, H_1, H_2)$ and
$([E_1], [E_2])$, we can equivalently define the product of $R(P)$ and the element $([E_1], [E_2])$ in the $K$-group as
\begin{eqnarray}
R(P)\cdot ([E_1], [E_2]):=(R(P)\cdot[E_1], R(P)\cdot[E_2]).
\end{eqnarray}
This definition is also useful to identify the $R(P)$-module structure
of the $K$-group.

\section{Coexistence of Anti-unitary symmetry}
\label{With TRS and/or PHS}

So far, we have considered only unitary symmetries.
In this section, we describe how to take into account antiunitary	
symmetries such as TRS, PHS, 
and magnetic space groups. \cite{Freed2013}
Hamiltonians considered here include Bogoliubov-de Gennes Hamiltonians as
well as Bloch Hamiltonians.
We take a suitable basis in which the Hamiltonians are periodic in the BZ
torus, $H({\bm k}+{\bm G})=H({\bm k})$.

  
Suppose that the Hamiltonians $H({\bm k})$ is subject to a symmetry
group ${\cal G}$. 
The symmetry group ${\cal G}$ may include any symmetry operations including
anti-unitary ones.
For $g\in {\cal G}$, we have
\begin{eqnarray}
U_g({\bm k}) H({\bm k}) U_g({\bm k})^{-1} = c(g) H(g {\bm k}),
\label{eq4:c}
\end{eqnarray}
where $g{\bm k}$ denotes the group action on the momentum space for
$g\in {\cal G}$. 
Here $c(g)=\pm 1$ is a function on ${\cal G}$ which specifies symmetry ($c(g)=1$) or
anti-symmetry ($c(g)=-1$). It is a homomorphism on ${\cal G}$,
i.e. $c(gg')=c(g)c(g')$. 
We also introduce a function $\phi(g)=\pm 1$ 
\begin{eqnarray}
U_g({\bm k}) i= \phi(g)i U_g({\bm k}),
\label{eq4:phi}
\end{eqnarray}
with the imaginary unit $i$, in order
to specify unitarity
($\phi(g)=1$) or anti-unitarity ($\phi(g)=-1$) of $U_g({\bm k})$.
Again, it is a homomorphism on ${\cal G}$, i.e. $\phi(gg')=\phi(g)\phi(g')$.
The multiplication in ${\cal G}$ implies that
\begin{eqnarray}
U_g(g'{\bm k})U_{g'}({\bm k})=
e^{i\tau_{g,g'}(gg'{\bm k})}U_{gg'}({\bm k}), 
\label{eq4:tau}
\end{eqnarray}
with a U(1) factor $e^{i\tau_{g,g'}({\bm k})}$.
From the associativity 
\begin{eqnarray}
(U_{g_1}(g_2g_3{\bm k})U_{g_2}(g_3{\bm k}))U_{g_3}({\bm k})
= U_{g_1}(g_2g_3{\bm k})(U_{g_2}(g_3{\bm k})U_{g_3}({\bm k})),
\quad g_1,g_2,g_3\in {\cal G},
\label{eq4:two-cocycle}
\end{eqnarray}
the U(1) factor obeys
\begin{align}
\delta \tau = 0 \quad \Leftrightarrow \quad 
\phi(g_1)\tau_{g_2,g_3}(g_1^{-1}\bk)-\tau_{g_1g_2,g_3}(\bk)+\tau_{g_1,g_2g_3}(\bk)-\tau_{g_1,g_2}(\bk) \equiv 0 \qquad  \mod \ 2 \pi, 
\label{eq4:one-coboundary}
\end{align}
The U(1) gauge ambiguity of $U_p({\bm k})$
\begin{eqnarray}
U_g({\bm k})\rightarrow e^{i\theta_g({\bm k})} U_g({\bm k})
\end{eqnarray}
also induces the equivalence relation 
\begin{align}
\tau \sim \tau+\delta \theta \quad \Leftrightarrow \quad 
\tau_{g_1,g_2}(\bk)\sim \tau_{g_1,g_2}(\bk)+\phi(g_1)\theta_{g_2}(g_1^{-1}\bk)-\theta_{g_1g_2}(\bk)+\theta_{g_1}(\bk) \qquad  \mod\ 2 \pi. 
\end{align}
Equations (\ref{eq4:two-cocycle}) and (\ref{eq4:one-coboundary}) imply
that a set of inequivalent U(1) 
phase factors $\{e^{i\tau_{g,g'}({\bm k})}\}_{g,g'\in {\cal G}}$ gives
an element of the group 
cohomology $H^2({\cal G},C(T^d, U(1)_\phi))$. Here $C(T^d, U(1)_\phi)$ is the
set of U(1)-valued functions on the BZ torus $T^d$, where the Abelian group
structure is given by the usual product of U(1) phases,
$e^{i\alpha_1({\bm k})}\cdot e^{i\alpha_2({\bm k})}
=e^{i(\alpha_1({\bm k})+\alpha_2({\bm k}))}, e^{i\alpha_i({\bm k})}\in
C(T^d, U(1)_{\phi})$, 
and the group ${\cal G}$ acts on $C(T^d, U(1)_\phi)$ by 
$e^{i(g \cdot \alpha)({\bm k})}=e^{i \phi(g) \alpha(g^{-1}{\bm k})}$ 
from the left.
As explained in Appendix \ref{Group_cohomology},
Eq.(\ref{eq4:two-cocycle}) gives the two-cocycle condition, and
Eq.(\ref{eq4:one-coboundary}) is the equivalence relation from the
two-coboundary in the cohomology.
The above three data $(c, \phi, \tau)$ in Eqs.(\ref{eq4:c}),
(\ref{eq4:phi}) and (\ref{eq4:tau})
specify the exact action of ${\cal G}$ on $H({\bm k})$ and the momentum space.

In a manner similar to Sec.\ref{sec3:Kroubi}, we can introduce a $K$-group by
using the Karoubi's formulation. 
For flattened Hamiltonians $H_i({\bm k})$ ($i=1,2$) subject to the
symmetry group ${\cal G}$, we consider a triple 
$
(E, H_1, H_2), 
$
where $E$ is a vector bundle on a compact momentum space $X$, and the
Hamiltonians $H_i$ ($i=1,2$) act on the common vector bundle $E$.
The addition is defined by Eq.(\ref{eq3:addition2}), and the equivalence
relation is imposed by Eq.(\ref{eq3:equivalence2}).
As a result, we obtain the twisted equivariant $K$-group consisting of sets of the
equivalence classes  $[E,H_1,H_2]$, which we denote by
${}^\phi K_{\cal G}^{(\tau,c)}(X)$. 

We introduce the integer grading of the $K$-group, 
${}^\phi K_{\cal G}^{(\tau,c)-n}(X)$, $(n=1,2,3,\dots)$
by imposing $n$ additional chiral symmetries which are compatible with
${\cal G}$, 
\begin{align}
&\Gamma_i H({\bm k}) \Gamma_i^{-1} = - H({\bm k}), \qquad \{ \Gamma_i, \Gamma_j\} = 2 \delta_{ij}, \qquad i = 1 , \dots, n,  \\
&U_g(\bk)\Gamma_i U^{-1}_g({\bm k}) = c(g)\Gamma_i,
\end{align}
together with Eq.(\ref{eq4:c}).
For $n\ge 2$, we also impose the subsector condition
$i\Gamma_{2i-1}\Gamma_{2i}=1$
$(i=1,\dots,[n/2])$:
By dressing antiunitary operators with chiral operators as shown in
Table \ref{tab:n_and_AUS}, the operator $i\Gamma_{2i-1}\Gamma_{2i}$
commutes with
all symmetry operators in ${\cal G}$ as well as with Hamiltonians in the
triple. Thus, we have consistently impose the above condition. 
It is also found that for an odd $n$, there remains a chiral symmetry $\Gamma$
that is compatible with the subsector condition. See Table \ref{tab:n_and_AUS}.
In general, the twist $(\tau, c)$ for the dressed antiunitary operators 
is different from the original one.
However, as summarized in Table.\ref{tab:n_and_AUS}, the twist in each
grading is uniquely determined by the
original twist, so
we use the same notation $(\tau, c)$ to denote the twist in each
grading. 
It is also noted that the chiral operator $\Gamma$ for an odd $n$ obeys
the same symmetry constraints as the Hamiltonian: When $U_g({\bm k})$
acts on the Hamiltonian as symmetry (anti-symmetry), $U_g({\bm k})$
commutes (anticommutes) with $\Gamma$.

The graded twist $(\tau, c)$ has a modulo 8 periodicity (Bott
periodicity)
for the grading
integer $n$. For instance, the dressed antiunitary operator
$\Gamma_7\Gamma_5\Gamma_3\Gamma_1 U_g({\bm k})$ for $n=8$ has the same
$e^{i\tau_{g,g'}({\bm k})}$ and
$c(g)$ as $U_g({\bm k})$.
Therefore, the same modulo 8 periodicity appears in the
$K$-groups, ${}^\phi K_{\cal G}^{(\tau,c)-n-8}(X)
={}^\phi K_{\cal G}^{(\tau,c)-n}(X)$. 
One can introduce ${}^\phi K_{\cal G}^{(\tau,c)+n}(X)$ 
so as to keep the modulo 8 periodicity.
Namely, ${}^\phi K_{\cal G}^{(\tau,c)+n}(X)\equiv {}^\phi K_{\cal
G}^{(\tau,c)-(8m-n)}(X)$ with $8m-n\ge 0$ ($m,n\in \Z$).\footnote{In the
absence of anti-unitary symmetry, the Bott periodicity becomes 2. Thus,
it holds that $K_G^{(\tau,c)+n}(X)=K_G^{(\tau,c)-n}(X)$. }

\begin{table}
\caption{Symmetry operators and twist for each grading.}
\label{tab:n_and_AUS}
\begin{tabular}{c|ccc|ccccc}
Grade & \multicolumn{3}{c|}{Symmetry operators}& \multicolumn{5}{c}{Twist $(\phi(g),\phi(g'))$}\\ 
\hline
$n$ 
&CS 
& $\phi(g)=1$ 
&  $\phi(g)=-1$ 
&$c$
& $(1,1)$
& $(1,-1)$
& $(-1,1)$
& $(-1,-1)$
\\
\hline
$n=0$ 
& 0 
& \multirow{2}{*}{$U_g({\bm k})$}
& \multirow{2}{*}{$U_g({\bm k})$}
& \multirow{2}{*}{$c(g)$}
& \multirow{2}{*}{$e^{i\tau_{g,g'}(\bk)}$} 
& \multirow{2}{*}{$e^{i\tau_{g,g'}(\bk)}$} 
& \multirow{2}{*}{$e^{i\tau_{g,g'}(\bk)}$} 
& \multirow{2}{*}{$e^{i\tau_{g,g'}(\bk)}$} 
\\
$n=1$ 
& $\Gamma_1$ 
&&
\\
$n=2$ 
& 0 
& \multirow{2}{*}{$U_g({\bm k})$} 
 &\multirow{2}{*}{$\Gamma_1U_g({\bm k})$} 
& \multirow{2}{*}{$\phi(g)c(g)$}
& \multirow{2}{*}{$e^{i\tau_{g,g'}(\bk)}$} 
& \multirow{2}{*}{$c(g)e^{i\tau_{g,g'}(\bk)}$} 
& \multirow{2}{*}{$e^{i\tau_{g,g'}(\bk)}$} 
& \multirow{2}{*}{$c(g)e^{i\tau_{g,g'}(\bk)}$} 
\\
$n=3$ 
& $\Gamma_3$ 
&&
\\
$n=4$ 
& 0 
& \multirow{2}{*}{$U_g({\bm k})$} 
 &\multirow{2}{*}{$\Gamma_3\Gamma_1U_g({\bm k})$} 
& \multirow{2}{*}{$c(g)$} 
& \multirow{2}{*}{$e^{i\tau_{g,g'}(\bk)}$} 
& \multirow{2}{*}{$e^{i\tau_{g,g'}(\bk)}$} 
& \multirow{2}{*}{$e^{i\tau_{g,g'}(\bk)}$} 
& \multirow{2}{*}{$-e^{i\tau_{g,g'}(\bk)}$} 
\\
$n=5$ 
& $\Gamma_5$ 
&&
\\
$n=6$ 
& 0 
& \multirow{2}{*}{$U_g({\bm k})$} 
 &\multirow{2}{*}{$\Gamma_5\Gamma_3\Gamma_1U_g({\bm k})$} 
& \multirow{2}{*}{$c(g)\phi(g)$} 
& \multirow{2}{*}{$e^{i\tau_{g,g'}(\bk)}$} 
& \multirow{2}{*}{$c(g)e^{i\tau_{g,g'}(\bk)}$} 
& \multirow{2}{*}{$e^{i\tau_{g,g'}(\bk)}$} 
& \multirow{2}{*}{$-c(g)e^{i\tau_{g,g'}(\bk)}$} 
\\
$n=7$ 
& $\Gamma_7$ 
& & 
\end{tabular}
\end{table}



An important class of symmetries in this category are unitary space
groups with real AZ symmetries (TRS and/or PHS).
They can be treated in a unified way by considering
symmetry group $\Z_2 \times G$ with integer grading.
Here $G$ is a unitary space group, and $\Z_2=\{1, -1\}$ is an order-two
cyclic group that commutes with
all elements of $G$, i.e. $(-1) \cdot g=g\cdot (-1)$, $g\in G$.
To include real AZ symmetries, we take the operators for $\Z_2$ as
$U_{-1}({\bk})={\cal T}$ and $U_{1}({\bm k})=1$, where 
${\cal T}$ is the time-reversal operator with ${\cal T}^2=1$.
We also define 
$U_{(-1) \cdot g}({\bm k})$ as $U_{(-1) \cdot g}({\bm k})=U_g(-{\bk}){\cal T}$.
The presence of such TRS is referred as class AI in
the AZ symmetry classes.
The data $(\phi,c, \tau)$ are summarized as 
\begin{eqnarray}
&&\phi(-1) = -1, \quad c(-1)=1, \quad {\cal T}^2=1, 
\quad
\phi(g)=1, \quad c(g)= \pm 1, 
\nonumber\\
&&U_{g}(g'{\bm k}) U_{g'}({\bm  k}) 
=e^{i\tau_{g,g'}(gg'{\bm k})}U_{gg'}({\bm k}), 
\quad
{\cal T}U_{g}({\bm k}) = e^{i\tau_{-1,g}(-g{\bm k})} U_{g}(-{\bm k}) {\cal T}, 
\end{eqnarray}
Imposing the chiral symmetries $\Gamma_i$ $(i=1,\dots, n)$, one can shift AZ
classes.~\cite{ShiozakiSatoGomi2016} 
The AZ class for the $n$-th grading $K$-group
${}^\phi K_{\Z_2\times G}^{(\tau,c)-n}(X)$
is summarized in Table ~\ref{tab:n_and_AZ}.

\begin{table}
\caption{The relation between the integer grading $n \ ({\rm mod}\  8)$, 
AZ classes, and additional symmetries. }
\label{tab:n_and_AZ}
\begin{tabular}{cccccccccccccc}
$n$ & AZ class & TRS & PHS & $\tau_{T,g}$ & $\tau_{C,g}$ \\
\hline
$n=0$ & AI & $T={\cal T}$ & & 
$T U_g({\bm k}) = e^{i\tau_{-1,g}(-g\bk)} U_g(-\bk) T$ &  \\
$n=1$ & BDI & $T={\cal T}$ & $C=\Gamma_1 {\cal T}$ & 
$T U_g({\bm k}) = e^{i\tau_{-1,g}(-g\bk)} U_g(-\bk) T$ & $C U_g(\bk) = c(g) e^{i\tau_{-1,g}(-g\bk)} U_g(-\bk) C$ \\
$n=2$ & D & & $C = \Gamma_1 {\cal T}$ & & 
$C U_g(\bk) = c(g) e^{i\tau_{-1,g}(-g\bk)} U_g(-\bk) C$ \\
$n=3$ & DIII & $T=\Gamma_3\Gamma_1{\cal T}$ & $C=\Gamma_1 {\cal T}$ & 
$T U_g({\bm k}) = e^{i\tau_{-1,g}(-g\bk)} U_g(-\bk) T$ & $C U_g(\bk) = c(g) e^{i\tau_{-1,g}(-g\bk)} U_g(-\bk) C$ \\
$n=4$ & AII & $T=\Gamma_3\Gamma_1{\cal T}$ & & 
$T U_g({\bm k}) = e^{i\tau_{-1,g}(-g\bk)} U_g(-\bk) T$ &  \\
$n=5$ & CII & $T=\Gamma_3\Gamma_1{\cal T}$ & $C=\Gamma_5\Gamma_3\Gamma_1{\cal T}$ & 
$T U_g({\bm k}) = e^{i\tau_{-1,g}(-g\bk)} U_g(-\bk) T$ & $C U_g(\bk) = c(g) e^{i\tau_{-1,g}(-g\bk)} U_g(-\bk) C$ \\
$n=6$ & C & & $C=\Gamma_5\Gamma_3\Gamma_1{\cal T}$ & & 
$C U_g(\bk) = c(g) e^{i\tau_{-1,g}(-g\bk)} U_g(-\bk) C$ \\
$n=7$ & CI & $T=\Gamma_7\Gamma_5\Gamma_3\Gamma_1{\cal T}$ & $C=\Gamma_5\Gamma_3\Gamma_1{\cal T}$ & 
$T U_g({\bm k}) = e^{i\tau_{-1,g}(-g\bk)} U_g(-\bk) T$ & $C U_g(\bk) = c(g) e^{i\tau_{-1,g}(-g\bk)} U_g(-\bk) C$ \\
\end{tabular}
\end{table}

\section{Topological crystalline insulators and superconductors}
\label{sec:Topological crystalline insulators and superconductors}

In this section, we consider insulators or superconductors that are
gapped in the whole BZ $T^d$.
%
%
Deforming Hamiltonians of the systems, one can
obtain flattened Hamiltonians in the whole BZ without gap closing. 
Using the Karoubi's formulation, these flattened Hamiltonians 
define $K$-groups on $T^d$.
Under the constraint of a symmetry group ${\cal G}$ with the data
$(c,\tau, \phi)$, the obtained $K$-group is the twisted equivariant
$K$-group ${}^{\phi}K_{\cal  G}^{(\tau,c)-n}(T^d)$.
We formulate below TCIs and TCSCs in terms of the $K$-group ${}^{\phi}K_{\cal
G}^{(\tau,c)-n}(T^d)$.


\subsection{$K$-theory classification}
\label{sec:Stable classification of bulk insulators}

First, we define TCIs and TCSCs
on the basis of the $K$-theory:
For this purpose, 
consider two different flattened Hamiltonians, $H_1$
and $H_2$, which are defined on the same vector bundle $E$ and are
subject to the same symmetry constraints for ${}^{\phi}K_{\cal
G}^{(\tau,c)-n}(T^d)$. 
As shown in Sec.\ref{Sec:K-theory}, $[E, H_1, H_2] 
\in {}^{\phi}K_{\cal G}^{(\tau,c)-n}(T^d)$ measures 
a topological difference between $H_1$ and $H_2$, 
so we can define that $H_1$ and $H_2$ are the same (different) 
TCIs or TCSCs if 
$[E, H_1, H_2]=0 \in {}^{\phi}K_{\cal G}^{(\tau,c)-n}(T^d)$ ($[E, H_1,
H_2]\neq 0\in {}^{\phi}K_{\cal G}^{(\tau,c)-n}(T^d)$). 
Some remarks are in order.
\begin{enumerate}
 \item We call $H_1$ and $H_2$ are stably equivalent to each other when
       $[E, H_1, H_2]=0$. $H_1$ and $H_2$ are stably equivalent, if they
       are continuously deformable to each other, but the inverse is not
       true: Indeed, as mentioned in Sec.\ref{Topological classification
       and the $K$-group}, 
$[E, H_1, H_2]$=0 does not necessarily mean that $H_1$ and $H_2$
       are smoothly deformable to each other. Even when they are
       not deformable to each other, $H_1\oplus H'$ and $H_2\oplus H'$
       could be by choosing a proper flattened Hamiltonian $H'$ on $E'$,
       and if this happens, one finds $[E, H_1, H_2]=0$. 
This means that  even if $H_1$ and $H_2$ are not smoothly deformable to
       each other, they could represent the same TCI 
       or TCSC. In this sense, the $K$-theory
       approach presents a loose classification of TCIs and TCSCs.

\item When ${\cal G}$ does not include any anti-symmetry, the identity
      operator $1$ on $E$ is regarded as a flattened Hamiltonian $H_0=1$
      which satisfies all 
      the constraints from ${\cal G}$. Since $H_0=1$ does not have an
      occupied state, the vector bundle spanned by its occupied state
      is of rank zero (i.e.\ empty), and so $H_0=1$ obviously
      describes a topologically trivial state.
      Therefore, for this particular class of ${\cal G}$, one can use
      the identity Hamiltonian as a reference, by which the topological
      index of $H$ is defined as $[E, H, 1]$. When $[E,H,1]$ is
      nonzero, one can say that $H$ is a TCI.

\item Each triple $[E, H_1, H_2]$ has its own symmetry operators
      $U_g({\bm k})$ for $g\in{\cal G}$ defined on $E$. 
      For $H_1$ and $H_2$ in the same triple, the symmetry operators
      commonly act on these Hamiltonians. On the other hand, 
      explicit forms of symmetry operators can be different for
      different triples, as long as the symmetry operators have the same
      data $(\phi, \tau,c)$.

\end{enumerate}

\subsection{Symmetry protected topologically distinct atomic insulators}
\label{sec:Dependence on unit cell and Wyckoff position}

\subsubsection{Wyckoff position}
In the presence of symmetry, short-range entangled states can be
topologically distinct due to symmetry constraints.
TCIs and TCSCs may illustrate
such symmetry protected topological phases in an extreme manner: Atomic
insulators can be topologically different to each other due to space
group symmetry.

An atomic insulator is an insulator where all electrons are tightly
bound to atoms, so its electric properties are local and insensitive to
the boundary condition.
In particular, it does not support topological gapless boundary states.
Nevertheless, in the presence of crystalline space group symmetry, there
arises topological distinction between atomic insulators.

This is because crystalline symmetry restricts possible positions of
atoms in the unit cell.
Each space group (or magnetic space group) has a finite number of
different Wyckoff positions,
according to which atoms are placed in the unit cell, and 
the different Wyckoff
positions remain different under any adiabatic deformation keeping
the space group symmetry.    
This means that atomic insulators with different Wyckoff positions
should be topologically different. 


For example, let us consider atomic insulators with the spatial reflection
symmetry $m$, $x\rightarrow -x$ in one dimension. Spacial reflection in one
dimension has three different Wyckoff positions: (a) 0 (b) 1/2 (c) $x$, $-x$,
which are invariant under reflection up to the lattice
translation $x\rightarrow x+1$.
We illustrate below atomic insulators with Wyckoff positions (a) 0 and
(b) 1/2, respectively:
\begin{align}
&{\rm (a)} \qquad 
\xygraph{
!{<0cm,0cm>;<1cm,0cm>:<0cm,1cm>::}
!{(0,0)}*{\bigcirc},
!{(2,0)}*{\bigcirc},
!{(-2,0)}*{\bigcirc},
!{(4,0)}*{\bigcirc},
!{(-4,0)}*{\bigcirc},
!{(0,-0.4)}*{0},
!{(-5.5,0)}-@{->}!{(5.5,0)},
!{(0,-0.8)}-@{.}!{(0,0.8)},
!{(0,0.6)}*{m},
!{(0.4,0.4)}-@{<->}!{(-0.4,0.4)},
!{(-1,-0.8)}-@{|-|}!{(1,-0.8)},
!{(0,-1)}*{\rm unit\ cell}, 
!{(6,0)}*{x}
}
\\
&{\rm (b)} \qquad 
\xygraph{
!{<0cm,0cm>;<1cm,0cm>:<0cm,1cm>::}
!{(1,0)}*{\bigcirc},
!{(3,0)}*{\bigcirc},
!{(-1,0)}*{\bigcirc},
!{(-3,0)}*{\bigcirc},
!{(-5,0)}*{\bigcirc},
!{(5,0)}*{\bigcirc},
!{(1,-0.4)}*{1/2},
!{(-1,-0.4)}*{-1/2},
!{(-5.5,0)}-@{->}!{(5.5,0)},
!{(0,-0.8)}-@{.}!{(0,0.8)},
!{(0,0.6)}*{m},
!{(0.4,0.4)}-@{<->}!{(-0.4,0.4)},
!{(-1,-0.8)}-@{|-|}!{(1,-0.8)},
!{(0,-1)}*{\rm unit\ cell}, 
!{(6,0)}*{x}
}
\end{align}
Here, ``$\bigcirc$'' represents an atom,   
and the dashed line is the center of the reflection. 
Although the difference between (a) and (b) is
just a difference in choice of the unit cell, the crystal (a) cannot
adiabatically deform 
into (b) keeping the reflection symmetry.
Therefore, they are topologically distinguished from each other.

In the Karoubi's formulation of the $K$-theory, 
the difference between Wyckoff positions 
is manifest in the reflection operator.
Consider the one-dimensional reflection symmetric insulators (a) and (b)
again. 
The reflection
operator $U^{\rm (a)}_{m}(k_x)$ for the atomic
insulator (a) does not coincide with the reflection operator $U^{\rm
(b)}_{m}(k_x)$ for (b), even when atoms in both crystals are identical:
In the crystal (b), after reflection, an additional lattice translation
is needed for an atom in the unit cell to go back to the original position.
As a result, an additional Bloch factor $e^{-ik_x}$ appears in
$U_{m}^{\rm (b)}(k_x)$ as $U_{m}^{\rm (b)}(k_x)=U_{m}^{\rm
(a)}(k_x)e^{-ik_x}$.  
Here it should be noted that the twist in $U_{m}^{\rm (b)}(k_x)$ is
the same as
that in $U_{m}^{\rm (a)}(k_x)$ because
$U_{m}^{\rm (b)}(-k_x)U_{m}^{\rm (b)}(k_x)=U_{m}^{\rm (a)}(-k_x)U_{m}^{\rm (a)}(k_x)$.
Thus, both $U_{m}^{\rm (a)}(k_x)$ and $U_{m}^{\rm (b)}(k_x)$ are
allowed in the same twisted $K$-theory.

\subsubsection{Representation dependence and $R(P)$-module structure}

Let us consider a set of all unitary symmetry operations $g
 \in {\cal G}$, which are characterized by $c(g)=\phi(g)=1$.
The set forms a subgroup of ${\cal G}$ because of the relations
 $c(gg')=c(g)c(g')$ and $\phi(gg')=\phi(g)\phi(g')$.
This unitary symmetry subgroup is given by a space group $G$.
The space group $G$ also provides topologically nontrivial structures. 

To see this, consider the symmetry constraint in Eq.(\ref{eq4:c}). From
Eq.(\ref{eq4:c}),
$H({\bm k})$ at ${\bm k}=0$ commutes with any unitary operator in
the above mentioned space group $G$.
Since the space group $G$ reduces to the point group $P$ at ${\bm k}=0$, 
the constraint implies that
any energy
eigenstate of $H({\bm k})$ at ${\bm k}=0$ should belong to a
representation of $P$.
In particular, occupied states of $H({\bm k})$ at ${\bm k}=0$ constitute
a set of representations of $P$.
It is evident that if occupied states  of $H_1({\bm k})$ and those
of $H_2({\bm k})$  constitute different sets of representations of $P$
at ${\bm k}=0$,   $H_1({\bm k})$ and $H_2({\bm k})$ are not deformable
to each other as long as they keep symmetry $P$ and gaps of the systems. 
In this sense, the representation of $P$ provides topological
differences in insulators and superconductors.

The above arguments also work for atomic insulators.
For illustration, consider again reflection
symmetric atomic insulators in one-dimension.
Below, we show atomic insulators (a1) and (a2) which share the same
Wyckoff position.
\begin{align}
{\rm (a1)} \qquad 
&\xygraph{
!{<0cm,0cm>;<1cm,0cm>:<0cm,1cm>::}
!{(0,0)}*{\bigcirc},
!{(2,0)}*{\bigcirc},
!{(-2,0)}*{\bigcirc},
!{(4,0)}*{\bigcirc},
!{(-4,0)}*{\bigcirc},
!{(0,-0.4)}*{\ket{s}},
!{(2,-0.4)}*{\ket{s}},
!{(-2,-0.4)}*{\ket{s}},
!{(4,-0.4)}*{\ket{s}},
!{(-4,-0.4)}*{\ket{s}},
!{(-5.5,0)}-@{->}!{(5.5,0)},
!{(0,-0.8)}-@{.}!{(0,0.8)},
!{(0,0.6)}*{m},
!{(0.4,0.4)}-@{<->}!{(-0.4,0.4)},
!{(6,0)}*{x}
}
\\
{\rm (a2)} \qquad 
&\xygraph{
!{<0cm,0cm>;<1cm,0cm>:<0cm,1cm>::}
!{(0,0)}*{\bigcirc},
!{(2,0)}*{\bigcirc},
!{(-2,0)}*{\bigcirc},
!{(4,0)}*{\bigcirc},
!{(-4,0)}*{\bigcirc},
!{(0,-0.4)}*{\ket{p}},
!{(2,-0.4)}*{\ket{p}},
!{(-2,-0.4)}*{\ket{p}},
!{(4,-0.4)}*{\ket{p}},
!{(-4,-0.4)}*{\ket{p}},
!{(-5.5,0)}-@{->}!{(5.5,0)},
!{(0,-0.8)}-@{.}!{(0,0.8)},
!{(0,0.6)}*{m},
!{(0.4,0.4)}-@{<->}!{(-0.4,0.4)},
!{(6,0)}*{x}
}
\end{align}
In the atomic insulator (a1), electrons in $s$-orbitals are tightly bound to
atoms, while in (a2), electrons in $p$-orbitals are bound to atoms.
Correspondingly, an occupied state in (a1) is even under reflection,
\begin{eqnarray}
U_{m}^{\rm (a1)}(k_x)|k_x\rangle_{\rm (a1)}=|-k_x\rangle_{\rm (a1)}, 
\end{eqnarray}   
but that in (a2) is odd under reflection,
\begin{eqnarray}
U_{m}^{\rm (a2)}(k_x)|k_x\rangle_{\rm (a2)}=-|-k_x\rangle_{\rm (a2)}. 
\end{eqnarray}   
$U^{\rm (a1)}_{m}(k_x)$ 
and $U^{\rm (a2)}_{m}(k_x)$ have the same twist since we have 
\begin{eqnarray}
U^{\rm (a1)}_{m}(-k_x)
U_{m}^{\rm (a1)}(k_x)|k_x\rangle_{\rm (a1)}
=U^{\rm (a1)}_{m}(-k_x)
|-k_x\rangle_{\rm (a1)}  
=|k_x\rangle_{\rm (a1)},
\nonumber\\
U^{\rm (a2)}_{m}(-k_x)
U_{m}^{\rm (a2)}(k_x)|k_x\rangle_{\rm (a2)}
=-U^{\rm (a2)}_{m}(-k_x)
|-k_x\rangle_{\rm (a2)}  
=|k_x\rangle_{\rm (a2)}.
\end{eqnarray}
Thus, these two insulators can be compared in the same twisted
$K$-theory. 
Obviously, these two insulators are not topologically the same in the
presence of the reflection symmetry.

In the $K$-theory, the representation dependence is properly treated as
the $R(P)$-module structure in Sec.\ref{sec:module}.
In terms of the Karoubi's formulation,  
the atomic insulators (a1) and (a2) are described as the triples with the
same form
\begin{eqnarray}
[E, -1, 1],
\end{eqnarray}
where $E$ is given by $|k_x\rangle_{(\rm a1)}$ and
$|k_x\rangle_{(\rm a2)}$, respectively. 
Indeed, since $E$ is the occupied state for $H=-1$ and no
occupied state exists for $H=-1$, the triple corresponds to
$([E],0)=[E]-0$, which is
naturally identified with $|k_x\rangle_{(a1)}$ and $|k_x\rangle_{(a2)}$,
respectively.
Since $|k_x\rangle_{(a1)}$ and $|k_x\rangle_{(a2)}$ belong to different
representations under the reflection, they correspond to different elements
of the $R(P)$-module in the $K$-theory.


\subsection{Dimensional hierarchy}

A remarkable feature of TCIs and TCSCs is that those in different
dimensions can be related to
each other. Such a hierarchy in spatial dimension  has been useful
for a systematic classification of topological insulators and
superconductors:\cite{Qi2008,Kitaev2009, Ryu2010, Teo2010} 
Furthermore, from this property, topological classification of a class of
crystalline insulators and superconductors protected by order-two point
groups (order-two nonsymmorphic space groups) in any dimensions
reduces to that in zero dimension
(one-dimension),
which makes
it possible to complete topological classification of those classes of
systems in any dimensions.\cite{ShiozakiSato2014, ShiozakiSatoGomi2016} 
In this section, we discuss dimensional hierarchy for
generic TCIs and TCSCs.

\subsubsection{Dimension-raising maps}
\label{sec5:DRM}

The dimensional hierarchy is given by dimension-raising maps 
in the Karoubi's formulation:
Consider a triple $[E, H_1({\bm k}), H_0({\bm k})]\in {}^\phi K_{\cal
G}^{(\tau, c)-n}(X)$ for an even $n$, or a quadruple $[E, \Gamma, H_1({\bm
k}), H_0({\bm k})]\in {}^\phi K_{\cal G}^{(\tau, c)-n}(X)$  
for an odd $n$, which describes a relative topological difference
of crystalline insulators or superconductors in $d$-dimensions.
We assume that $[E, H_1({\bm k}), H_0({\bm k})]\neq 0$ or
$[E, \Gamma, H_1({\bm k}), H_0({\bm k})]\neq 0$, which
implies that $H_1(\bk)$ has a ``nonzero topological charge'' relative to
$H_0(\bk)$ on $X$. 

\begin{figure}[!]
 \begin{center}
  \includegraphics[width=\linewidth, trim=0cm 1cm 0cm 0cm]{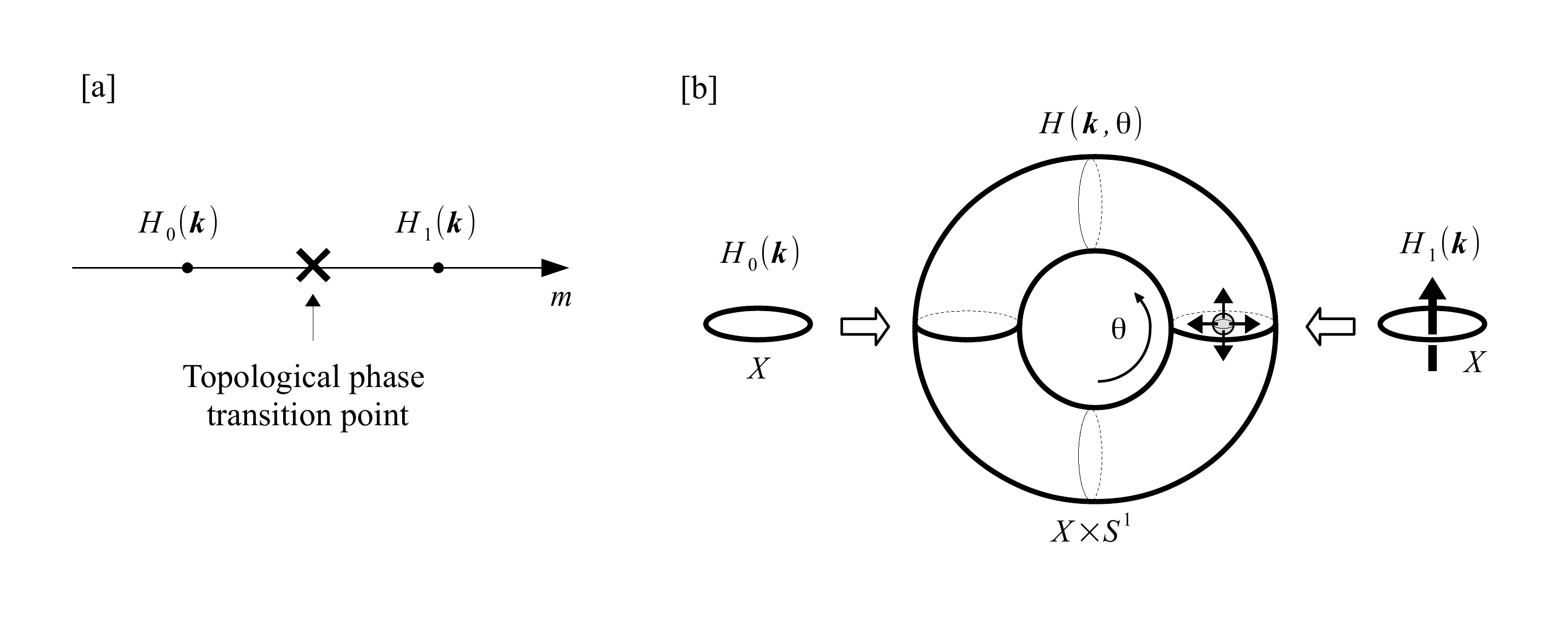}
 \end{center}
 \caption{
 [a] A parameter $m$ connecting two different topological phases. 
 [b] The dimensional raising map from $X$ to $X \times S^1$. 
 }
 \label{Fig:HamiltonianMap}
\end{figure}

To construct dimension-raising maps, we consider a one-parameter
Hamiltonian $H_{10}({\bm k}, m)$, 
where $m\in [-1,1]$ is a parameter 
connecting $H_{10}({\bm k},-1)=H_0({\bm k})$ and $H_{10}({\bm k},1)=H_1({\bm
k})$, and $H_{10}({\bm k}, m)$ keeps the same symmetry constraint as
$H_1({\bm k})$ and $H_0(\bk)$. 
For example, the following one-parameter Hamiltonian satisfies this
requirement,
\begin{eqnarray}
H_{10}({\bm k},m)=\left\{
\begin{array}{ll}
m H_0({\bm k}),& \mbox{for $m\in [-1,0]$}\\
m H_{1}({\bm k}), &\mbox{for $m\in (0,1]$}\\
\end{array}
\right. .
\label{eq5:interpolation}
\end{eqnarray}
Note that $H_{10}({\bm k}, m)$ should have a gap-closing topological phase
transition point 
in the middle region of $m \in [-1,1]$, 
since $H_1(\bk)$ and $H_0(\bk)$ have different topological charges.
See Fig.~\ref{Fig:HamiltonianMap} [a]. 
In $H_{10}({\bm k}, m)$ of Eq.(\ref{eq5:interpolation}),
the gap-closing point is given at $m=0$. 
Depending on the absence (for an even $n$) or presence (for an odd $n$) of
chiral symmetry, we have a map from the Hamiltonians on $X$ to 
a new Hamiltonian $H(\bk,\hat n)$ on $X\times S^d$,
which has the same topological charge as $H_1(\bk)$,
in the following manner.

%

{\it $\gamma$ matrices}---
For preparation, we introduce the following $\gamma$ matrices,
\begin{align}
&\gamma_1^{(k)}=\sigma_y\otimes \underbrace{\sigma_z\otimes \cdots
 \otimes \sigma_z}_{k-1}, 
&\gamma_2^{(k)}&=-\sigma_x\otimes \underbrace{\sigma_z\otimes \cdots
 \otimes \sigma_z}_{k-1},
\nonumber\\
&\gamma_3^{(k)}=\sigma_0\otimes\sigma_y\otimes 
\underbrace{\sigma_z\otimes \cdots
\otimes \sigma_z}_{k-2},
&\gamma_4^{(k)}&=\sigma_0\otimes(-\sigma_x)\otimes
\underbrace{\sigma_z\otimes \cdots \otimes \sigma_z}_{k-2},
\nonumber\\
&\quad\vdots & \vdots
\nonumber\\
&\gamma_{2k-1}^{(k)}=\underbrace{\sigma_0\otimes\cdots\otimes\sigma_0}_{k-1}
\otimes\sigma_y,
&
\gamma_{2k}^{(k)}&=\underbrace{\sigma_0\otimes\cdots\otimes\sigma_0}_{k-1}
\otimes(-\sigma_x),
\end{align}
and 
$\gamma_{2k+1}^{(k)}=\sigma_z\otimes\cdots\otimes\sigma_z$,
which obey $\{\gamma^{(k)}_i,\gamma^{(k)}_j\}=2\delta_{i,j}$.
They also satisfy 
\begin{eqnarray}
\gamma_i^{(k)}\otimes \gamma_{2l+1}^{(l)}=\gamma_i^{(k+l)},
\quad  
\underbrace{\sigma_0\otimes \cdots\otimes \sigma_0}_{k}
\otimes \gamma_{j}^{(l)}=\gamma_{2k+j}^{(k+l)},
\quad
\gamma_{2k+1}^{(k)}\otimes \gamma_{2l+1}^{(l)}=\gamma_{2(k+l)+1}^{(k+l)},
\end{eqnarray}
for $i=1,\dots, 2k$ and $j=1,\dots, 2l$.
We also define $\gamma_1^{(0)}$ as $\gamma_1^{(0)}=1$.
The $\gamma$ matrices are useful to construct
dimension-raising maps.

{\it Maps from nonchiral class}---
For an even $n$, $H_{10}({\bm k}, m)$ does not have
chiral symmetry. 
Here we construct the dimension-raising map that changes the base
space $X$ into $X\times S^{2r-1}$ or $X\times S^{2r}$ $(r=1,2,\dots)$
in this nonchiral case. 
For this purpose, we first formaly increase the rank of the Hamiltonian
\begin{eqnarray}
\mathbb{H}_{10}({\bm k}, m)=H_{10}({\bm k},m)\otimes \gamma_{2r+1}^{(r)},
\label{eq5:Hext}
\end{eqnarray}
and that of symmetry operators $\mathbb{U}_g(\bk)$,  
\begin{eqnarray}
\mathbb{U}_g({\bm k})=\left\{
\begin{array}{ll}
U_g({\bm k})\otimes\underbrace{\sigma_0\otimes\cdots\otimes\sigma_0}_{r}, 
& \mbox{for $c(g)=1$},\\
U_g({\bm k})\otimes\gamma_{2r+1}^{(r)}, 
&\mbox{for $c(g)=-1$},
\end{array}
\right. 
\label{eq5:Uext}
\end{eqnarray}
by using the $\gamma$ matrices.
$\mathbb{H}_{10}({\bm k},m)$ and $\mathbb{U}_g({\bm k})$ 
keep the same symmetry relations as $H_{10}({\bm k}, m)$ and $U_g({\bm
k})$, 
\begin{eqnarray}
\mathbb{U}_g(\bk)
\mathbb{H}_{10}({\bm k}, m)  
\mathbb{U}^{-1}_g(\bk)=c(g)\mathbb{H}_{10}(g\bk),
\quad
\mathbb{U}_g(g'\bk)\mathbb{U}_{g'}(\bk)=e^{i\tau_{g,g'}(\bk)}\mathbb{U}_{gg'}(\bk), 
\quad 
\mathbb{U}_g(\bk )i =\phi(g) i\mathbb{U}_g(\bk ),
\end{eqnarray}
but there appear additional chiral symmetries
\begin{eqnarray}
\{\mathbb{H}_{10}(\bk, m), \mathbb{\Gamma}_i^{(\pm)}\}=0, 
\quad 
(i=1,\dots, r)
\end{eqnarray}
with 
\begin{eqnarray}
\mathbb{U}_g(\bk)
\mathbb{\Gamma}_i^{(\pm)}
\mathbb{U}^{-1}_g(\bk)=c(g)\mathbb{\Gamma}^{(\pm)}_i,
\quad
\{\mathbb{\Gamma}_i^{(+)}, \mathbb{\Gamma}_j^{(+)}\}=2\delta_{i,j},
\quad
\{\mathbb{\Gamma}_i^{(-)}, \mathbb{\Gamma}_j^{(-)}\}=-2\delta_{i,j},
\quad
\{\mathbb{\Gamma}_i^{(+)}, \mathbb{\Gamma}_j^{(-)}\}=0,
\end{eqnarray}
where the chiral operators $\mathbb{\Gamma}_i^{(\pm)}$ $(i=1,\dots,r)$
are defined as
\begin{eqnarray}
\mathbb{\Gamma}_i^{(+)}=1\otimes \gamma_{2i}^{(r)},
\quad 
\mathbb{\Gamma}_i^{(-)}=1\otimes i\gamma_{2i-1}^{(r)},
\label{eq5:Gext}
\end{eqnarray}
Note that
$\mathbb{\Gamma}_i^{(+)}\mathbb{\Gamma}_i^{(-)}$
($i=1,\dots, r$)
commute with $\mathbb{H}_{10}({\bm k}, m)$, $\mathbb{U}_g(\bk)$, and
each other.
Since $\mathbb{H}_{10}({\bm k}, m)$ and $\mathbb{U}_g(\bk)$ reduce to
$H_{10}({\bm k}, m)$ and $U_g({\bm k})$ in the diagonal basis of
$\mathbb{\Gamma}_i^{(+)}\mathbb{\Gamma}_i^{(-)}=\pm 1$, $\mathbb{H}_{10}({\bm
k}, m)$ retains the same topological properties as $H_{10}({\bm k}, m)$.

The following equation defines the dimension-raising map from $H({\bm k}, m)$ on $X$ 
to the Hamiltonian $H(\bk,\hat n)$ on $X\times S^{2r-1}$,
\begin{align}
H(\bk,\hat n)
=\mathbb{H}_{10}(\bk,n_0)
+in_{1} \mathbb{\Gamma}_1^{(-)}
+\cdots
+in_{r} \mathbb{\Gamma}_r^{(-)}
+n_{r+1}\mathbb{\Gamma}_1^{(+)}
+\cdots
+n_{2r-1}\mathbb{\Gamma}_{r-1}^{(+)},
\label{eq:dimensional_raise_nonchiral}
\end{align}
where we introduced the spherical coordinate 
$\hat n=(n_0,\bm{n})=(n_0,n_1,\dots,n_{2r-1})$ with $n_0^2+\bm{n}^2=1$. 
The obtained Hamiltonian is fully gapped and can be flattened because 
$H({\bm k},\hat n)^2=H_{10}({\bm k},n_0)^2+\bm{n}^2$ is
positive definite.
In particular, for $H_{10}({\bm k},m)$ in Eq.(\ref{eq5:interpolation}), 
one can show directly that $H({\bm k},\hat n)^2=1$.

We can also extend symmetry ${\cal G}$ on $X$ into that on $X\times S^{2r-1}$:
The simplest extension is that $g\in {\cal G}$ acts on
$S^{2r-1}$ trivially. For anti-unitary operators, however,
the momentum and the coordinate behave in a different manner under the
trivial action.
While the momentum changes the sign under the
trivial action of anti-unitary operators, the coordinate does not. 
Correspondingly, there exist two different trivial extensions:   
For the momentum sphere $S^{2r-1}$, the trivial extension is given by 
\begin{eqnarray}
U_g(\bk,\hat n)=\left\{
\begin{array}{ll}
\mathbb{U}_g(\bk), & \mbox{for $\phi(g)=1$}\\
\mathbb{\Gamma}_{r-1}^{(+)}\cdots\mathbb{\Gamma}_1^{(+)}\mathbb{U}_g(\bk),
& \mbox{for $\phi(g)=-1$} \\
\end{array}
\right., 
\label{eq5:AUSM}
\end{eqnarray}
which yeilds 
\begin{eqnarray}
U_g(\bk, \hat{n})H(\bk,\hat{n})U_g^{-1}(\bk,
 \hat{n})=c(g)[\phi(g)]^{r-1}H(\bk, n_0,\phi(g){\bm n}), 
\label{eq5:HM}
\end{eqnarray}
and for $S^{2r-1}$ in the coordinate space, 
\begin{eqnarray}
U_g(\bk,\hat n)=\left\{
\begin{array}{ll}
\mathbb{U}_g(\bk), & \mbox{for $\phi(g)=1$}\\
\mathbb{\Gamma}_{r}^{(-)}\cdots\mathbb{\Gamma}_1^{(-)}\mathbb{U}_g(\bk),
& \mbox{for $\phi(g)=-1$} \\
\end{array}
\right.,
\label{eq5:AUSS}
\end{eqnarray}
which leads to
\begin{eqnarray}
U_g(\bk, \hat{n})H(\bk,\hat{n})U_g^{-1}(\bk,
 \hat{n})=c(g)[\phi(g)]^{r}H(\bk, n_0,{\bm n}). 
\end{eqnarray}
Here note that ${\bm n}$ changes the sign under the action of
anti-unitary operators in the former extension. 
(See also Sec.\ref{sec:momentum_sphere})
The mapped Hamiltonian also has chiral symmetry.
\begin{eqnarray}
\{H(\bk,\hat n), \Gamma\}=0,
\quad
\Gamma=\mathbb{\Gamma}^{(+)}_{r}. 
\end{eqnarray}
From Eqs.(\ref{eq5:AUSM}) and (\ref{eq5:AUSS}), one can calculate directly how the twist
$(\tau, c)$ changes for the momentum sphere extension and the coordinate sphere extension,
respectively. In these cases, the change of the twist results in
the change of the grading.
The grading integer $n$ is increased (decreased) by
$2r-1$ for the momentum (coordinate) sphere case.     

Figure \ref{Fig:HamiltonianMap}[b] illustrates the map in the $r=1$
case,
\begin{align}
H(\bk,\theta)&=\mathbb{H}_{10}(\bk, \cos\theta)+i\sin\theta\mathbb{\Gamma}_1^{(-)} 
\nonumber\\
&=H_{10}(\bk, \cos\theta)\otimes \sigma_z -\sin \theta \otimes \sigma_y.
\end{align}
When $\theta=0$ and $\theta=\pi$, the mapped Hamiltonian $H(\bk,
\theta)$ is essentially 
the same as $H_1(\bk)$ and $H_0(\bk)$, respectively. 
Then, with keeping the gap, $H_{1}(\bk)$ and
$H_0(\bk)$ are extended in the $\theta$ direction and they are
glued together.
In the above construction, the nonzero topological charge of $H_1({\bm
k})$, which is illustrated as a ``vortex'' in Fig.\ 
\ref{Fig:HamiltonianMap}[b], 
becomes a ``monopole'' inside $X \times S^1$. Therefore,
$H(\bk,\theta)$ has the same topological charge
as $H_1(\bk)$.
The same argument works for any $r$. 
Thus, the mapped Hamiltonian $H({\bm k}, \hat n)$ also has the same topological charge as
the original Hamiltonian $H_1(\bk)$.

For the dimension-raising map from $X$ to $X\times S^{2r}$, 
we consider the following Hamiltonian  
\begin{align}
H(\bk,\hat n)
=\mathbb{H}_{10}(\bk,n_0)+in_1\mathbb{\Gamma}_1^{(-)}
+\cdots+
in_r\mathbb{\Gamma}_r^{(-)}
+n_{r+1}\mathbb{\Gamma}_1^{(+)}
+\cdots
+n_{2r}\mathbb{\Gamma}_{r}^{(+)},
\label{eq:dimensional_raise_nonchiral2}
\end{align}
which is also gapped and has the
same topological charge as $H_1(\bk)$.
We also has the trivial extension of ${\cal G}$, 
\begin{eqnarray}
U_g(\bk,\hat n)=\left\{
\begin{array}{ll}
\mathbb{U}_g(\bk), & \mbox{for $\phi(g)=1$}\\
\mathbb{\Gamma}_{r}^{(+)}\cdots\mathbb{\Gamma}_1^{(+)}\mathbb{U}_g(\bk),
& \mbox{for $\phi(g)=-1$} \\
\end{array}
\right., 
\label{eq5:AUSM2}
\end{eqnarray}
for the momentum sphere $S^{2r}$, and 
\begin{eqnarray}
U_g(\bk,\hat n)=\left\{
\begin{array}{ll}
\mathbb{U}_g(\bk), & \mbox{for $\phi(g)=1$}\\
\mathbb{\Gamma}_{r}^{(-)}\cdots\mathbb{\Gamma}_1^{(-)}\mathbb{U}_g(\bk),
& \mbox{for $\phi(g)=-1$} \\
\end{array}
\right., 
\label{eq5:AUSS2}
\end{eqnarray}
for the coordinate sphere $S^{2r}$.
The mapped Hamiltonian does not have chiral symmetry.
The above extension increases (decreases) the grading integer $n$ by
$2r$ for the momentum (coordinate) extension.







{\it Map from chiral class}---
For an odd $n$,  $H_{10}(\bk,m)$ has chiral symmetry $\Gamma$, 
the dimension-raising map is constructed in a manner parallel to
the even $n$ case, with a minor modification. 
Using $\Gamma$, we first introduce $\mathbb{\Gamma}$ by  
\begin{eqnarray}
\mathbb{\Gamma}=\Gamma\otimes \gamma_{2r+1}^{(r)}, 
\label{eq:introduce_new_chiral}
\end{eqnarray}
as well as $\mathbb{H}_{10}(\bk, m)$, $\mathbb{U}_g(\bk)$, and
$\mathbb{\Gamma}_i^{(\pm)}$ $(i=1,\dots, r)$ defined in Eqs.(\ref{eq5:Hext}),
({\ref{eq5:Uext}}), and ({\ref{eq5:Gext}}), respectively.
Since $\Gamma$ obeys
$U_g(\bk)\Gamma U_g(\bk)^{-1}=c(g)\Gamma$,
we have
\begin{eqnarray}
\mathbb{U}_{g}(\bk)\mathbb{\Gamma}\mathbb{U}^{-1}_g(\bk)=c(g)\mathbb{\Gamma}. 
\end{eqnarray}
For the dimension-raising map from $X$ to $X\times S^{2r+1}$
$(r=0,1,\dots)$, we
consider 
\begin{align}
H(\bk, \hat n)
=\mathbb{H}(\bk,n_0)
+in_1\mathbb{\Gamma}_1^{(-)}
+\cdots+
in_r\mathbb{\Gamma}_r^{(-)}
+n_{r+1}\mathbb{\Gamma}_1^{(+)}
+\cdots
+n_{2r}\mathbb{\Gamma}_{r}^{(+)}
+n_{2r+1}\mathbb{\Gamma}, 
\label{eq:dimensional_raise_chiral}
\end{align}
where the extension of ${\cal G}$ is given by
\begin{eqnarray}
U_g(\bk,\hat n)=\left\{
\begin{array}{ll}
\mathbb{U}_g(\bk), & \mbox{for $\phi(g)=1$}\\
\mathbb{\Gamma}_{r}^{(+)}\cdots\mathbb{\Gamma}_1^{(+)}
\mathbb{\Gamma}
\mathbb{U}_g(\bk),
& \mbox{for $\phi(g)=-1$} \\
\end{array}
\right., 
\end{eqnarray}
for the momentum sphere $S^{2r+1}$,
and 
\begin{eqnarray}
U_g(\bk,\hat n)=\left\{
\begin{array}{ll}
\mathbb{U}_g(\bk), & \mbox{for $\phi(g)=1$}\\
\mathbb{\Gamma}_{r}^{(-)}\cdots\mathbb{\Gamma}_1^{(-)}\mathbb{U}_g(\bk),
& \mbox{for $\phi(g)=-1$} \\
\end{array}
\right., 
\end{eqnarray}
for the coordinate sphere $S^{2r+1}$. 
The mapped Hamiltonian $H(\bk,\hat n)$ does not
have chiral symmetry.
On the other hand, for the map from $X$ to $X\times S^{2r}$
$(r=1,2,\dots)$, we have
\begin{align}
H(\bk, \hat n)
=\mathbb{H}(\bk,n_0)
+in_1\mathbb{\Gamma}_1^{(-)}
+\cdots+
in_r\mathbb{\Gamma}_r^{(-)}
+n_{r+1}\mathbb{\Gamma}_1^{(+)}
+\cdots
+n_{2r-1}\mathbb{\Gamma}_{r-1}^{(+)}
+n_{2r}\mathbb{\Gamma}, 
\label{eq:dimensional_raise_chiral2}
\end{align}
where the extension of ${\cal G}$ is given by
\begin{eqnarray}
U_g(\bk,\hat n)=\left\{
\begin{array}{ll}
\mathbb{U}_g(\bk), & \mbox{for $\phi(g)=1$}\\
\mathbb{\Gamma}_{r-1}^{(+)}\cdots\mathbb{\Gamma}_1^{(+)}
\mathbb{\Gamma}
\mathbb{U}_g(\bk),
& \mbox{for $\phi(g)=-1$} 
\end{array}
\right., 
\end{eqnarray}
for the momentum sphere $S^{2r}$, 
and 
\begin{eqnarray}
U_g(\bk,\hat n)=\left\{
\begin{array}{ll}
\mathbb{U}_g(\bk), & \mbox{for $\phi(g)=1$}\\
\mathbb{\Gamma}_{r}^{(-)}\cdots\mathbb{\Gamma}_1^{(-)}\mathbb{U}_g(\bk),
& \mbox{for $\phi(g)=-1$} 
\end{array}
\right., 
\end{eqnarray}
for the coordinate sphere $S^{2r+1}$.
The Hamiltonian $H(\bk,\hat n)$ has chiral symmetry,
\begin{eqnarray}
\{H(\bk,\hat n), \Gamma'\}=0,
\quad
\Gamma'=\mathbb{\Gamma}_{r}^{(+)}. 
\end{eqnarray}
The maps in Eqs.(\ref{eq:dimensional_raise_chiral}) and
({\ref{eq:dimensional_raise_chiral2}}) increase (decrease) the grading
integer $n$ by $2r+1$ and $2r$, respectively, for the momentum
(coordinate) sphere extension. 
For the same reason as the even $n$ case, the mapped Hamiltonians in
Eqs.(\ref{eq:dimensional_raise_chiral}) and ({\ref{eq:dimensional_raise_chiral2}}) keep the same topological charge as the
starting Hamiltonian $H_1({\bm k})$.

{\it Isomorphism}--- 
The dimension-raising maps keep the topological charge, with shifting
the grading of the Hamiltonian and the dimension of the base manifold.
In terms of the $K$-theory, these results are summarized as the isomorphism
\begin{eqnarray}
{}^\phi K^{\pi^*(\tau,c)-n}_{\cal G}(X \times S^D) \cong 
\underbrace{{}^{\phi} K^{(\tau,c)-(n-D)}_{\cal G}(X)}_{S^D {\rm  \mathchar`- dependent\ contribution}}
\oplus 
\underbrace{{}^\phi K^{(\tau,c)-n}_{\cal G}(X)}_{S^D {\rm  \mathchar`-
independent\ contribution}}, 
\label{Eq:DimShift_general}
\end{eqnarray}
for the momentum sphere $S^D$, and 
\begin{eqnarray}
{}^\phi K^{\pi^*(\tau,c)-n}_{\cal G}(X \times S^D) \cong 
\underbrace{{}^{\phi} K^{(\tau,c)-(n+D)}_{\cal G}(X)}_{S^D {\rm  \mathchar`- dependent\ contribution}}
\oplus 
\underbrace{{}^\phi K^{(\tau,c)-n}_{\cal G}(X)}_{S^D {\rm  \mathchar`-
independent\ contribution}}, 
\label{Eq:DimShift_general2}
\end{eqnarray}
for the coordinate space sphere $S^D$.
Here ${\cal G}$ acts on $S^D$ trivially, 
and $\pi^*$ is the pull back of the obvious projection $\pi$: $X\times
S^{2r-1}\to X$. 
Strictly speaking, the twist for $U_g({\bm k},\hat n)$ is
defined on $X\times S^{2r-1}$, not on $X$, so
to make it clear, we denote the
twist of $U_g({\bm k},\hat n)$ as $\pi^* (\tau, c)$.
The mapped Hamiltonian $H(\bk,\hat n)$ gives an
element of ${}^\phi K^{\pi^*(\tau,c)-n}_{\cal G}(X \times S^D)$
corresponding to the first term of the right hand side in
Eq.(\ref{Eq:DimShift_general}) or  (\ref{Eq:DimShift_general2}). 
The second terms in Eqs. (\ref{Eq:DimShift_general}) and
(\ref{Eq:DimShift_general})  
are trivial contributions
from Hamiltonians independent of $S^D$. 

The exact relation between a mapped Hamiltonian $H(\bk,\hat n)$
and an element
of the K-group is obtained as follows:
Starting the zero element $[E, H_0,
H_0]=0$ or $[E, \Gamma, H_0, H_0]=0$  
in ${}^\phi K_{\cal G}^{(\tau, c)-(n\mp D)}(X)$, 
we first construct a topologically trivial Hamiltonian $H_0(\bk,\hat n)$ using the dimension-raising map.
Then the element of ${}^\phi K_{\cal G}^{\pi*(\tau,
c)-n}(X\times S^D)$ 
is given by the triple 
 $[E, H, H_0]$ on $X\times S^D$ or the quadruple $[E, \Gamma, H, H_0]$
 on $X\times S^D$.

In Appendix \ref{Sec:Gysin}, we outline the proof of the isomorphisms
by using the Gysin sequence.  
As discussed below, the first terms in the isomorphisms
ensure the existence of gapless
boundary and defect states of TCIs and TCSCs.



\subsubsection{Momentum sphere $S^D$}
\label{sec:momentum_sphere}
In the previous section, we have introduced the momentum sphere $S^D$
parameterized by $\hat{n}=(n_0,{\bm n})$ with $n_0^2+{\bm n}^2=1$.
Here we explain its relation to the actual momentum space.
For the simplest case $S^1$, the momentum sphere can be naturally
identified with the one-dimensional BZ, where $\hat{n}$ is given in the
form of $(n_0, n_1)=(\cos k, \sin k)$ with momentum $k$.
Under the action of anti-unitary operators, $k$ goes to $-k$, so only
$n_1$ changes the sign. This behavior is consistent with Eq.(\ref{eq5:HM}).
Moreover, a general $S^D$ can be regarded as a compactified
$D$-dimensional momentum space. 
Using the following map
\begin{eqnarray}
{\bm k}=\frac{\bm n}{1+n_0}, 
\end{eqnarray}
one can obtain the original decompactified $D$-dimensional momentum space.
Thus, the sign change of ${\bm k}$ is induced by the transformation
$(n_0,{\bm n})\rightarrow (n_0, -{\bm n})$. 
This behavior is also consistent with Eq.(\ref{eq5:HM}).
We also note that $O(D+1)$ rotations of $S^D$ that fix the north
($n_0=1$) and south pole $(n_0=-1)$ induce $O(D)$ rotations around the
origin in the decompactified momentum space. 
This property will be used in Sec.\ref{sec5:more}.

\subsubsection{Examples}

{\it $d=0$ class A $\to$ $d=1$ class AIII}---
Let us consider class A insulators in 0-space dimension. 
The $K$-theory is $K^0(pt) = \Z$ and generator of $K^0(pt)$ 
is represented by the triple $[\C,1,-1]$. 
Then, the mapped Hamiltonian (\ref{eq:dimensional_raise_nonchiral}) reads 
\begin{align}
H(k_x) = \cos k_x \sigma_z - \sin k_x \sigma_y, \quad  
\Gamma = -\sigma_x, 
\label{eq:model_1d_class_aiii}
\end{align}
which leads to the $K$-theory isomorphism 
\begin{align}
K^{-1}(S^1) \cong K^{0}(pt) \oplus K^{-1}(pt) = K^{0}(pt)=\Z. 
\end{align}

{\it $d=1$ class AIII $\to d=2$ class A}---
Let us consider the $K$-theory isomorphism 
\begin{equation}
K^{0}(T^2) 
\cong K^{1}(S^1) \oplus K^{0}(S^1) = K^{-1}(S^1) \oplus K^{0}(S^1) =
\Z \oplus \Z. 
\end{equation}
The second term is a weak index. 
The first term is given by the dimensional raising map. 
From Eq.\ (\ref{eq:model_1d_class_aiii}), 
a Hamiltonian $H(k_x,m)$ connecting the topological phase ($1 \in \Z$) 
and the trivial phase ($0 \in \Z$) is given by 
\begin{align}
H_{10}(k_x,m) = ( m -1 + \cos k_x ) \sigma_z -\sin k_x \sigma_y, \quad 
\Gamma = -\sigma_x, \quad 
m \in [-1,1]. 
\end{align}
Then, the mapped Hamiltonian (\ref{eq:dimensional_raise_chiral}) becomes 
\begin{equation}
H(k_x,k_y) = ( -1 + \cos k_x + \cos k_y ) \sigma_z - \sin k_x \sigma_y - \sin k_y \sigma_x. 
\end{equation}

\subsubsection{More on dimension-raising maps}
\label{sec5:more}

To construct the dimension-raising maps in Sec.\ref{sec5:DRM}, 
we have considered the trivial extension of symmetry ${\cal G}$ from $X$
to $X\times S^D$.
Here we present different dimension-raising maps by using a non-trivial
extension of ${\cal G}$.
For simplicity, we only present here maps from non-chiral systems, but the
generalization to the chiral case is straightforward.   
As shown in Sec.\ref{sec5:DRM}, we have the following
set of equations before increasing the dimension of the base manifold,
\begin{eqnarray}
&&\mathbb{U}_g(\bk)
\mathbb{H}_{10}(\bk)
\mathbb{U}^{-1}_g(\bk)=c(g)\mathbb{H}(g\bk),
\quad
\mathbb{U}_g(\bk)
\mathbb{\Gamma}_i^{(\pm)}
\mathbb{U}^{-1}_g(\bk)=c(g)\mathbb{\Gamma}_i^{(\pm)},
\quad
\mathbb{U}_g(g'\bk)\mathbb{U}_{g'}(\bk)=e^{i\tau_{g,g'}(\bk)}\mathbb{U}_{gg'}(\bk),
\nonumber\\
&&\{\mathbb{\Gamma}_i^{(+)}, \mathbb{\Gamma}_j^{(+)}\}=2\delta_{ij},
\quad
\{\mathbb{\Gamma}_i^{(-)}, \mathbb{\Gamma}_j^{(-)}\}=-2\delta_{ij},
\quad
\{\mathbb{\Gamma}_i^{(-)}, \mathbb{\Gamma}_j^{(-)}\}=0.
\end{eqnarray}
For the nontrivial extension, we take into account $SO(D)$ generators
\begin{eqnarray}
\mathbb{M}^{(++)}_{ij}
=\frac{[\mathbb{\Gamma}_i^{(+)}, \mathbb{\Gamma}_j^{(+)}]}{2i},
\quad
\mathbb{M}^{(--)}_{ij}
=\frac{[\mathbb{\Gamma}_i^{(-)}, \mathbb{\Gamma}_j^{(-)}]}{2i},
\quad
\mathbb{M}^{(+-)}_{ij}
=\frac{[\mathbb{\Gamma}_i^{(+)}, \mathbb{\Gamma}_j^{(-)}]}{2i}.
\end{eqnarray}
By using them, a map from $g\in {\cal G}$ to $\mathbb{V}_g\in
Pin(D)$ (projective group of $O(D)$) can
be expressed as
\begin{eqnarray}
\mathbb{V}_g=\left\{
\begin{array}{ll}
\exp\left[
i\sum_{ij\sigma\sigma'}\mathbb{M}_{ij}^{(\sigma
\sigma')}\theta_{\sigma\sigma'}^{ij}(g)
\right],
&
\mbox{for $p_V(g)=0$}
\\
\mathbb{\Gamma}_1^{(+)}
\exp\left[
i\sum_{ij\sigma\sigma'}\mathbb{M}_{ij}^{(\sigma
\sigma')}\theta_{\sigma\sigma'}^{ij}(g)
\right],
&
\mbox{for $p_V(g)=1$}
\end{array}
\right. ,
\end{eqnarray}
where $p_V(g)$ is the index distinguishing two
different forms of $\mathbb{V}_g$.  
The index $p_V(g)$ satisfies 
\begin{eqnarray}
p_V(gg')=p_V(g)+p_V(g')\quad (\mbox{mod. 2}).
\end{eqnarray}
If the map keeps the group structure of ${\cal G}$ as 
\begin{eqnarray}
\mathbb{V}_g \mathbb{V}_{g'}=e^{i\tau_V(g,g')}\mathbb{V}_{gg'},
\label{eq5:conditionVg}
\end{eqnarray}
where the twist $e^{i\tau_V(g,g')}=\pm 1\in \omega$ is allowed from the
projective nature of $Pin(D)$, 
we can use $\mathbb{U}_g^V({\bm k})$ defined by
\begin{eqnarray}
\mathbb{U}^V_g(\bk)=\mathbb{V}_g\mathbb{U}_g(\bk),
\end{eqnarray}
instead of
$\mathbb{U}_g({\bm k})$, to construct the symmetry operator on $X\times
S^D$ in Eqs.(\ref{eq5:AUSM}) and ({\ref{eq5:AUSS}}) (or Eqs.
(\ref{eq5:AUSM2}) and (\ref{eq5:AUSS2})),  
The presence of $\mathbb{V}_g$ induces an $O(D+1)$ rotation of $S^D$ that fixes
the north pole ($n_0=1$) and the south pole ($n_0=-1$) of $S^D$.    
Since $\mathbb{U}_g^V(\bk)$ obeys
\begin{eqnarray}
\mathbb{U}^V_g(\bk)
\mathbb{H}(\bk)
(\mathbb{U}^{V}_g)^{-1}(\bk)=(-)^{p_V(g)}c(g)\mathbb{H}(g\bk),
\quad
\mathbb{U}^V_g(g'\bk)\mathbb{U}^V_{g'}(\bk)
=[c(g)]^{p_V(g')}e^{i[\tau_{g,g'}(gg'\bk)+\tau_V(g,g')]}\mathbb{U}^V_{gg'}(\bk),
\label{eq5:extra_twist}
\end{eqnarray}
the dimension-raising map in the above presents an extra twist, in addition to
that given by the change of the grading integer.


The above dimension-raising map is summarized as the following
isomorphism in the $K$-theory, 
\begin{align}
{}^\phi K^{\pi^*(\tau,c)-n}_{\cal G}(X \times S^D )
\cong {}^\phi K_{\cal G}^{(\tau,c)+(\tau_V,c_V)-(n\mp D)}(X) \oplus 
{}^\phi K_{\cal G}^{(\tau,c)-n}(X),
\label{eq5:dimension_raising_rotation}
\end{align}
where ${\cal G}$ acts on $S^D$ through 
${\cal G}\to O(D+1)$ with the north and south poles fixed, and $-$ $(+)$ in the
double sign corresponds to the momentum (coordinate) $S^D$.
Here $(\tau_V, c_V)$ denotes the extra twist due to
Eq.(\ref{eq5:extra_twist}).

\subsection{Building block}
As shown in the previous subsection, using the dimension-raising maps,
one can construct a sequence of mapped Hamiltonians on the manifolds 
\begin{align}
X \to X \times S^{r_1} \to X \times S^{r_1} \times S^{r_2} \to X \times S^{r_1} \times S^{r_2} \times S^{r_3} \to \cdots. 
\end{align}
For Hamiltonians fitting in any of the mapped Hamiltonians, their 
topological classification reduces to that of the starting lower
dimensional Hamiltonians on $X$. 
Therefore, $X$ is regarding as a ``building block'' of the classification.
Some examples of building blocks with relevant symmetries are summarized
in Table ~\ref{tab:building_block}.

\begin{table}
\caption{Building blocks.}
\label{tab:building_block}
\begin{tabular}{>{\centering\arraybackslash}m{10cm} | >{\centering\arraybackslash}m{3cm} | >{\centering\arraybackslash}m{3cm} }
Symmetry & Building block Brillouin zone & Related refs. \\
\hline 
No symmetry & $\{ pt \}$ & \\
\hline 
TRS and/or PHS & $\{ pt \}$ & \onlinecite{Schnyder2008, Kitaev2009, Ryu2010} \\
\hline 
Onsite symmetry & $\{ pt \}$ & \\
\hline 
\shortstack{Order-two point group symmetry \\
(reflection, $\pi$-rotation, inversion, reflection $\times$ reflection, $\dots$ )} & $\{pt\}$ & \onlinecite{Chiu2013, Morimoto2013, ShiozakiSato2014} \\
\hline 
\shortstack{Order-two nonsymmorphic space group symmetry \\
(half-lattice translation, glide, two-fold screw, glide $\times$ reflection, $\dots$)} & $S^1$ & \onlinecite{ShiozakiSatoGomi2016} \\
\hline 
General wallpaper group & $T^2$ & \onlinecite{Yang1997, LuckStamm2000, Alexandradinata2014, DongLiu2015, Kruthoff2016} \\
\hline 
General space group & $T^3$ &  \\
\end{tabular}
\end{table}

\subsection{Boundary gapless states}
\label{sec:Classification of boundary gapless states}

\begin{figure}[!]
 \begin{center}
  \includegraphics[width=0.5\linewidth, trim=0cm 0cm 0cm 0cm]{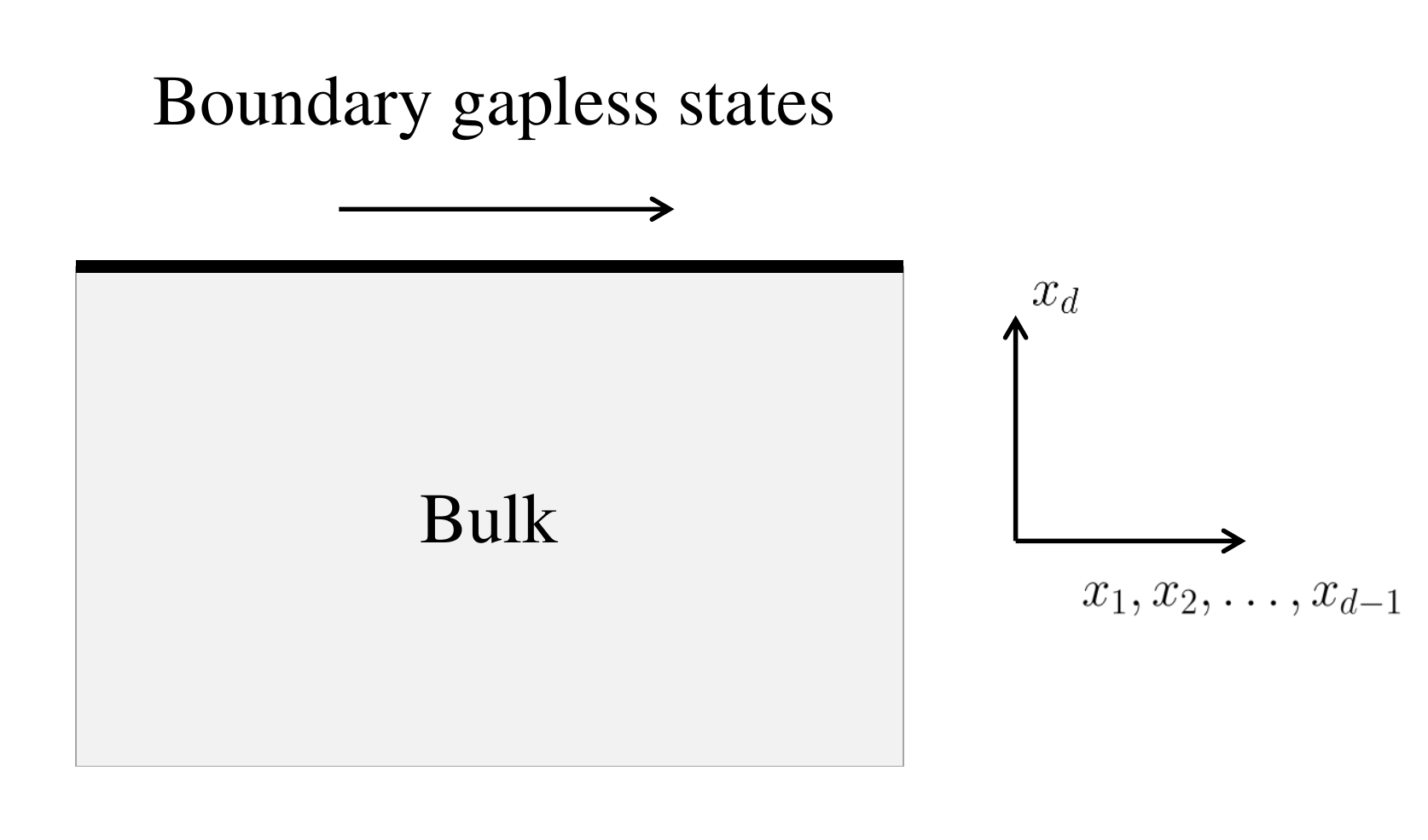}
 \end{center}
 \caption{Bulk-boundary correspondence.}
 \label{fig:bulk_boundary}
\end{figure}

The isomorphism in Eq.(\ref{Eq:DimShift_general}) predicts 
one of the most
important characteristics of TCIs and TCSCs, the existence of gapless boundary states:
Consider a crystalline insulator or superconductor in
$d$-dimensions with the boundary normal to the $x_{d}$-direction
as illustrated in Fig.~\ref{fig:bulk_boundary}. 
Symmetry of the system compatible with the boundary should act trivially
on the $x_{d}$-direction, so it is identical to  
that for ${}^\phi K_{\cal G}^{\pi^*(\tau, c)-n}(X\times S^1)$ in
Eq.(\ref{Eq:DimShift_general}), where $S^1$ is the momentum sphere  
conjugate to $x_{d}$, $X$ is surface BZ conjugate to $x_1,\dots,x_{d-1}$,
and the data of symmetry, $(\phi,\tau, c)$,
$n$, and ${\cal G}$, are properly chosen.
The $K$-group ${}^\phi K_{\cal G}^{\pi^*(\tau, c)-n}(X\times S^1)$
determines topological properties of the system with the boundary. 
In particular, if the system has a non-zero topological number
corresponding to the first term ${}^\phi K_{\cal G}^{(\tau,c)-(n-1)}(X)$ of the right hand side in
Eq.(\ref{Eq:DimShift_general}), 
the TCI or TCSC
hosts topologically protected gapless states on the boundary.
This is a manifestation of the bulk-boundary correspondence:
A non-trivial element of the first term implies the existence of
a topologically twisted structure of the bulk gapped system in the $k_{d}$-direction, which
manifests the existence of gapless boundary states in the presence of
a boundary normal to the $x_{d}$-direction. 
On the other hand, the second term of the right hand side in
Eq.(\ref{Eq:DimShift_general}) merely provides a``weak topological
index'' that can be supported by $d$-dimensional gapped systems
trivially stacked in
the $x_{d}$-dimension.  
Since the stacked system is $k_{d}$-independent, 
the second term does not provide any gapless state on the boundary normal to
the $x_{d}$-direction. 

These important properties of TCIs and TCSCs are summarized as follows.
\begin{itemize}
\item[($\star$)]
Gapless states for crystalline insulators and superconductors in
	     $d$-dimensions  are topologically classified by 
the $K$-group ${}^\phi K_{\cal G}^{(\tau,c)-(n-1)}(X)$, where $X$ is the
	     $(d-1)$-dimensional surface BZ and symmetry of the system is given
	     by $(\phi,\tau,c)$, $n$, and ${\cal G}$.
Note that the grading of the $K$-group is shifted by $-1$ in comparison
	     with that of symmetry of the system: The grading of
	     $K$-group is $n-1$, while that of symmetry is $n$.
\end{itemize}
The dimensional raising maps (\ref{eq:dimensional_raise_nonchiral}) and
(\ref{eq:dimensional_raise_chiral}) present representative Hamiltonians
with non-zero topological numbers of the $K$-group ${}^\phi K_{\cal
G}^{(\tau,c)-n}(X)$, by which one can confirm the existence of gapless
states on the boundary. 

The gapless states on the surface BZ $X$ have their own
effective Hamiltonians given by self-adjoint Fredholm operators acting on
the infinite
dimensional Hilbert space. 
These Fredholm operators also represent elements of the 
$K$-group, which also classifies  
all possible stable gapless states. 
In the present paper, we do not describe the detail of this formulation of 
the $K$-theory 
since it requires an additional mathematical preparation. 
For the outline, see Ref.~\onlinecite{Adem2016} for example. 
It should be noted that 
in contrast to the classification of bulk gapped insulators and
superconductors, where a pair of Hamiltonians $[E, H_1, H_2]$ are needed
in the $K$-theory, the alternative formulation 
requires only
a {\it single} effective Hamiltonian for gapless states to represent an
element of the $K$-group.  






\subsection{Defect gapless modes}

\subsubsection{Semiclassical Hamiltonian}

Here, we consider topological defects of 
band insulators and superconductors.
Away from the topological defects, the systems are
gapped, and they are described by spatially modulated Bloch
and BdG Hamiltonians,~\cite{volovik2003universe,Teo2010}
\begin{eqnarray}
H(\bk,{\bm r}),  
\end{eqnarray}
where the base space of the Hamiltonian is composed of
momentum $\bk$, defined in the $d$-dimensional BZ
$T^d$, and real-space coordinates ${\bm r}$ of a $D$-dimensional sphere
${\wt S}^D$ surrounding a defect. 
We treat $\bk$ and ${\bm r}$ in the Hamiltonian as
classical variables, i.e., momentum operators $\bk$ and coordinate
operators ${\bm r}$ commute with each other. This semiclassical
approach is justified if the characteristic length of the spatial
inhomogeneity is sufficiently longer than that of the quantum
coherence. A realistic Hamiltonian would not satisfy this
semiclassical condition, but if there is no bulk gapless mode,
then the Hamiltonian can be adiabatically deformed so as to
satisfy the condition. Because the adiabatic deformation does
not close the bulk energy gap, it retains the topological nature
of the system.

The defect defines a $(d-D-1)$-dimensional submanifold. 
We assume that the defect keeps the lattice translation symmetry along
the submanifold.
%
Whereas the exact momentum space is $T^d$, we retain the torus
structure only in the directions of the defect submanifold, and thus
consider a simpler space $T^{d-D-1}\times S^1\times S^{D}$, where $S^D$ is
conjugate to $\wt S^D$, in the following:
This simplification keeps any symmetry compatible with the defect
configuration,  
so it does not affect the classification of
symmetry protected topological defect gapless modes.


\subsubsection{Topological classification}

\begin{figure}[!]
 \begin{center}
  \includegraphics[width=0.7\linewidth, trim=0cm 0cm 0cm 0cm]{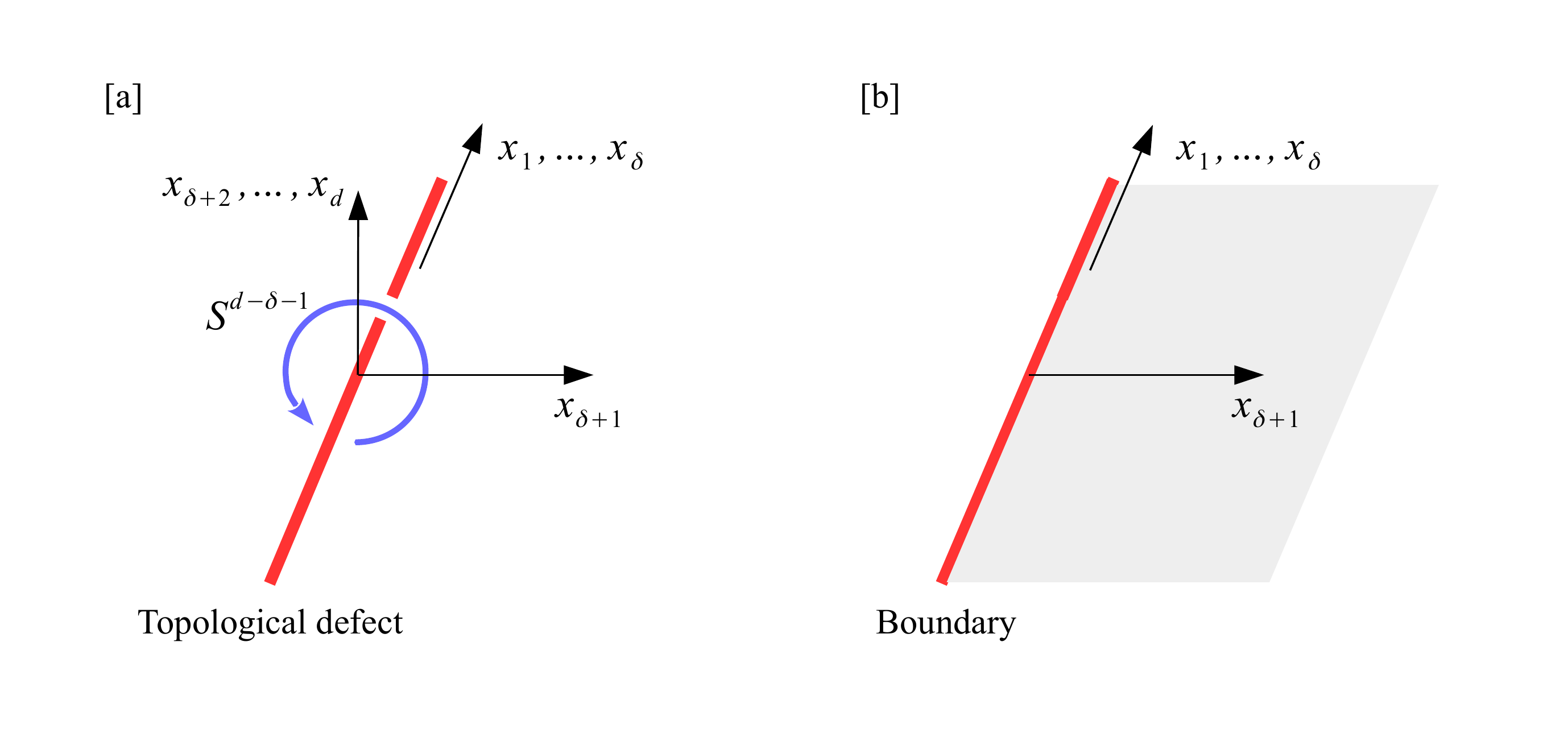}
 \end{center}
 \caption{[a] A topological defect with $\delta$-dimensions in a $d$-dimensional insulator. The blue circle represents a sphere $S^{d-\delta-1}$ surrounding the topological defect. 
[b] A boundary gapless state in $(\delta+1)$-dimensional topological insulators. }
 \label{Fig:Defect}
\end{figure}

Consider a defect described by the
semiclassical Hamiltonian $H(\bk, 
{\bm r})$ on $T^{d-D-1}\times S^1\times S^{D}\times \wt S^D$.
We impose symmetry ${\cal G}$ compatible with the defect
configuration on $H(\bk, {\bm r})$, with the grading integer $n$.
The topological classification of the above system is given by
the $K$-group ${}^\phi K_{\cal G}^{\pi^*(\tau, c)-n}(T^{d-D-1}\times
S^1\times S^D\times \wt S^D)$.
Since $S^D$ and $\wt S^D$ are conjugate to each other, ${\cal G}$ 
acts on them in the same manner. 
The compatibility with the defect
configuration implies that the action of ${\cal G}$ on $S^D$ and $\wt
S^D$ should be $O(D+1)$ rotations with a point fixed. 
Thus one can apply the isomorphism in Eq.(\ref{eq5:dimension_raising_rotation}) to evaluate 
${}^\phi K_{\cal G}^{\pi^*(\tau, c)-n}(T^{d-D-1}\times
S^1\times S^D\times \wt S^D)$:
\begin{align}
&{}^\phi K^{\pi^*(\tau,c)-n}_{\cal G}(T^{d-D-1} \times S^1 \times S^{D} \times
 \tilde S^{D})
\nonumber\\ 
&\cong {}^\phi K_{\cal G}^{(\tau,c)+(\tau_V, c_V)-(n+D)}(T^{d-D-1}\times S^1\times S^D)
\oplus 
\underbrace{
{}^\phi K_{\cal G}^{(\tau,c)-n}(T^{d-D-1}\times S^1\times S^D)}_{\wt
 S^D-{\rm independent}}
\nonumber\\ 
&\cong {}^\phi K_{\cal G}^{(\tau,c)-n}(T^{d-D-1}\times S^1)
\oplus 
\underbrace{
{}^\phi K_{\cal G}^{(\tau,c)+(\tau_V, c_V)-(n+D)}(T^{d-D-1}\times S^1)}_{S^D-{\rm independent}}
\oplus
\underbrace{
{}^\phi K_{\cal G}^{(\tau,c)-n}(T^{d-D-1}\times S^1\times S^D)}_{\wt
 S^D {\rm -independent}}.
\end{align}
Here no extra twist $(\tau_V, c_V)$ appears in the first term of the
right hand side:
The extra twist $(\tau_V, c_V)$ from
the $O(D+1)$ rotation on $\wt S^D$ is canceled by that on $S^D$.
The second and the third terms on the final line in the right hand side
are given by the 
Hamiltonian $H(\bk, {\bm r})$ that are independent of either ${\bm k}$
or ${\bm r}$, so they merely provide a weak topological index and a bulk
topological number irrelevant to the defect, respectively. 
Therefore, only the first term gives a strong topological index for the defect.
We note here that the first term coincides with the $K$-group for
TCIs and TCSCs in
$(d-D)$-dimensions, where the boundary can be identified with the
$(d-D-1)$-dimensional defect submanifold, as illustrated in
Fig.\ref{Fig:Defect}. 
Thus, we obtain the following result.

\begin{itemize}
\item[($\star \star$)]
A defect can be considered as a boundary of a lower dimensional
	     TCI or TCSC.
Defect gapless modes are topologically classified as boundary gapless
	     states of the TCI or TCSC. 
\end{itemize}

\section{Topological nodal semimetals and superconductors}
\label{sec:Topological nodal semimetals and superconductors}

\subsection{Formulation by $K$-theory}
Weyl and Dirac semimetals or nodal superconductors host bulk gapless
excitations as 
band touching points and/or
lines in the BZ.
The gapless excitations have their own topological numbers which ensure 
stability under small perturbations.
There have been a lot of efforts to classify such bulk gapless
topological phases.~\cite{
Kobayashi2014, ChiuSchnyder2014, Watanabe2016, MathaiThiang2016}

Whereas the bulk gapless phases resemble to gapless boundary and defect
modes in TCIs and TCSCs, their theoretical
treatment is different from that of the latter:
While the topological structure of the latter can be examined by 
a bulk Hamiltonian flattened in the entire BZ, that of the
former cannot be, since the information on the band touching structure
is obviously lost by the flattening.
Therefore, one needs a different approach to characterize 
gapless topological phases in the $K$-theory formulation.


A simple way to characterize topological semimetals and nodal
superconductors is to consider subspaces of the
BZ, together with the entire one.\footnote{
We illustrate this view point in terms of the $K$-theory, but 
the same discussion is possible for isomorphism classes of vector bundles.} 
Let $Y \subset T^d$ be a closed subspace in the BZ torus $T^d$. 
The subspace $Y$ may not retain the full symmetry ${\cal
G}$ of the
system, and we denote it as
${\cal G}_Y$, the subgroup of ${\cal G}$ keeping $Y$ invariant.
(Namely, for $g\in {\cal G}_Y$ and $\bk \in Y$, it holds
that $g\bk \in Y$.)
Then, the trivial inclusion $i_Y: Y \to T^d$ induces the following homomorphism 
$i_Y^*$ from the $K$-group on $T^d$ to a $K$-group on $Y$,
\begin{align}
i_Y^*: {}^\phi K_{{\cal G}}^{(\tau,c)-n}(T^d) \to 
{}^{i_Y^*\phi} K^{i_Y^*(\tau,c)-n}_{{\cal G}_Y}(Y).
\label{eq:restriction_sub_skeleton}
\end{align}
Actually, from a triple 
$[E, H_1, H_2]\in {}^\phi K_{{\cal G}}^{(\tau,c)-n}(T^d)$ 
for an even $n$ or a quadruple 
$[E, \Gamma, H_1, H_2]\in {}^\phi K_{{\cal G}}^{(\tau,c)-n}(T^d)$ for an
odd $n$,  one can have an unique element of ${}^{i_Y^*\phi} K_{{\cal
G}_Y}^{i^*_Y(\tau,c)-n}(Y)$, just by restricting the vector bundle $E$
and the Hamiltonians $H_i$ $(i=1,2)$
to the subspace $Y$, and by relaxing the symmetry constraint from
${\cal G}$ to ${\cal G}_Y$. Here we have represented the twist $(\tau, c)$
and $\phi$ for ${\cal G}_Y$ as  $i_Y^*(\tau, c)$ and $i_Y^{*}{\phi}$,
respectively, since they are determined by those data of ${\cal G}$.
Noting that any fully gapped insulator or superconductor subject to the symmetry
${\cal G}$ with the grading integer $n$ is identified with an element of 
${}^\phi K_{{\cal G}}^{(\tau,c)-n}(T^d)$,
we have the following statement:
\begin{itemize}
\item
If one restricts a full gapped crystalline insulator or superconductor
     to a subspace $Y$, the resultant system on $Y$ gives a
     $K$-group element that lies inside
     the image of the 
     homomorphism $i^*_Y$.
\end{itemize}

Now consider a system which is fully gapped on $Y$ but not
necessarily so on the whole BZ $T^d$. 
The restriction on $Y$  also gives an element of ${}^{i_Y^*\phi}
K^{i_Y^*(\tau,c)-n}_{{\cal G}_Y}(Y)$.
Interestingly, the contraposition of the above statement leads to
the following non-trivial statement: 
\begin{itemize}
\item
If the above $K$-group element on $Y$ lies outside the
     image of the homomorphism $i^*_Y$, the original system should
     support a gapless region outside $Y$.
\end{itemize}
Since elements outside the image of $i^*_Y$ is nothing but the cokernel of
$i_Y^{*}$ in mathematics, the second statement can be rephrased as follows.
\begin{itemize}
\item
Non-zero elements of $\coker(i_Y^*)$=${}^{i_Y^*\phi} K_{{\cal
     G}_Y}^{i^*_Y(\tau,c)-n}(Y)/{\rm Im}(i_Y^*)$ provide bulk topological
     gapless phases.
In other words, the $\coker(i_Y^*)$ defines bulk topological gapless
     phases in the $K$-theory formulation. 
\end{itemize}
%
Not all
elements of ${}^{i_Y^*\phi} K_{{\cal G}_Y}^{i^*_Y(\tau,c)-n}(Y)$ can be
obtained from elements of ${}^{\phi} K_{{\cal G}}^{(\tau,c)-n}(T^d)$,
so the cokernel of $i_Y^*$ is not empty in general.
Below, we illustrate this viewpoint in some examples.

\begin{figure}[!]
 \begin{center}
  \includegraphics[width=\linewidth, trim=0cm 0cm 0cm 0cm]{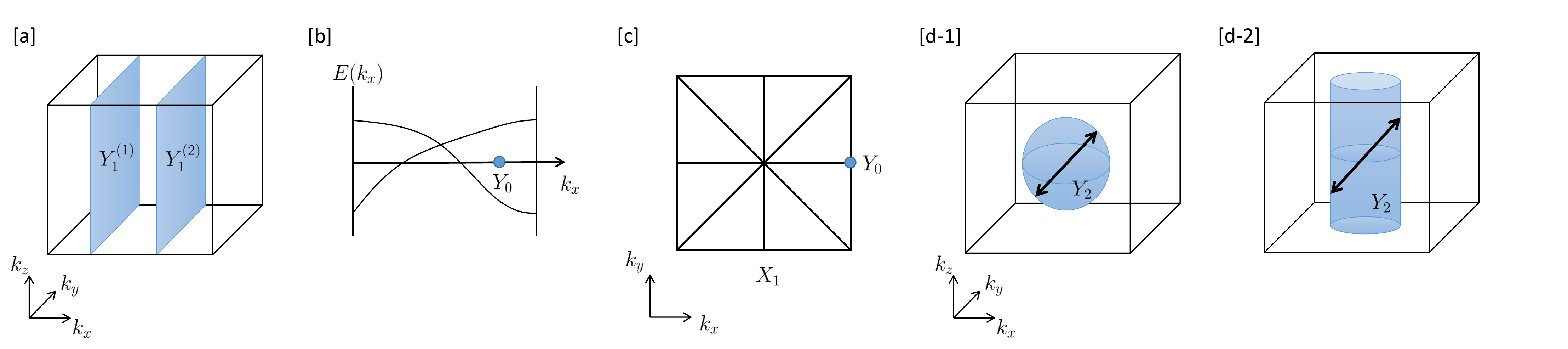}
 \end{center}
 \caption{Subspaces in the BZ torus. 
 [a] Two planes $Y_1$ and $Y_2$ compose the subspace $Y$. 
 [b] The subspace is a single point $Y_0$. 
 [c] The 1-dimensional subspace $X_1$ in the BZ torus $T^2$ and a symmetric point $Y_0$. 
 [d-1] The real projective plane arising from the inversion symmetry acting on the sub sphere $S^2$, 
 [d-2] The Klein bottle from the inversion symmetry acting on the sub torus $T^2$. 
 }
 \label{fig:subskeleton}
\end{figure}

\subsection{Examples}
\subsubsection{Weyl semimetals}
The first example is Weyl semimetals that support bulk band touching points
in the BZ.\cite{Nielsen-Ninomiya1983,Murakami2007, Wan2011, Burkov-Balents2011}
As originally discussed by Nielsen and
Ninomiya~\cite{NielsenNinomiya_homotopy}, the band touching points have  
local monopole charges defined by the Chern number.
The Weyl semimetals are characterized as the cokernel of a
homomorphism between $K$-groups.

Let $Y_1^{(i)}(i=1,2)$ be planes with $k_x = a_i (i=1,2)$ in
Fig.~\ref{fig:subskeleton} [a], and consider 
the disjoint union $Y_1=Y_1^{(1)}\sqcup Y_1^{(2)}$. 
The most general $K$-theory on $Y_1$ is $K(Y_1)=K(Y_1^{(1)})\oplus
K(Y_1^{(2)})$, which does not require 
any symmetry. 
Since the topological index of $K(Y_1^{(i)})$ is the Chern number $ch(a_i)$
on $Y_1^{(i)}$, an element of $K(Y_1)$ is given by $(ch(a_1), ch(a_2))$.

Now consider the trivial inclusion $i_{Y_1}: Y_1\to T^3$,  
which induces the homomorphism
$i_{Y_1}^*$ from ${}^* K_{*}^*(T^3)$ to $K(Y_1)$, 
where ${}^* K_{*}^*(T^3)$ can be any $K$-group for fully gapped
insulators in three dimensions.  
For any fully gapped insulators in three dimensions, 
the Chern number $ch_1(k_x)$ 
at the plane with a constant $k_x$ does not depend on $k_x$, so the
image of $i_{Y_1}^*$ satisfies $ch(a_1)=ch(a_2)$. 
Therefore, 
if the Chern numbers $ch(a_i) (i=1,2)$ of the two planes $Y_1^{(i)} (i=1,2)$
do not match, there should be a stable gapless point 
in the region outside the subspace $Y_1$. 
This means that 
the cokernel of $i_{Y_1}^*$, which is given by $ch(a_1)-ch(a_2)$, 
corresponds to gapless points.

This argument also works for any closed surface $Y$ deformable to a point 
 and its trivial inclusion $i_Y: Y\to T^3$.
 The cokernel of the induced homomorphism $i_Y^{*}$ is nothing but the
 Chern number on $Y$ in this case, which defines the monopole charge of
 Weyl nodes.
%
%
%
%


\subsubsection{Nonsymmorphic gapless materials}
As the second example,  consider the filling constraint from
nonsymmorphic space groups.
In general, a nonsymmorphic space group
gives rise to a constraint on possible filling numbers of band
insulators, as classified by Watanabe et.~al.~\cite{Watanabe2016}   
For example, 
let us consider the glide symmetry $(x,y) \mapsto (x+1/2,-y)$ 
in two dimensions.
The glide operator $G(k_x)$ has the $2\pi$-periodicity
$G(k_x+2\pi)=G(k_x)$ and it also obeys 
$G^2(k_x)=e^{-ik_x}$ 
since two 
consecutive glide operations amount to just a lattice translation, which
results in the Bloch factor $e^{-ik_x}$.
The latter equation implies that eigenvalues of $G(k_x)$ are $\pm
e^{-ik_x/2}$. 
From these equations, it is found that every band forms a pair
on the glide symmetric line $k_y=0$:
For $k_y=0$, the Bloch Hamiltonian commutes with
$G(k_x)$, so any band is an eigenstate of $G(k_x)$.  
Since each eigenvalue of $G(k_x)$ does not have the $2\pi$-periodicity
in $k_x$, bands with opposite eigenvalues  appear in a pair to keep the
$2\pi$-periodicity. 
In particular, any fully gapped glide symmetric insulator should have an even
number of occupied states.

Let $Y_0= \{(a,0)\}$ be a point on the glide symmetric line $k_y=0$. 
At the point $Y_0$, the glide symmetry reduces to
a simple $\Z_2$ symmetry, which defines ${\cal G}_{Y=Y_0}$ in Eq.(\ref{eq:restriction_sub_skeleton}).
Since the $\Z_2$ symmetry only has one-dimensional representations,
the $K$-group on $Y_0$ is different from that obtained by the
restriction of the $K$-group for fully gapped two-dimensional glide
symmetric insulator into $Y_0$.
In particular, the former $K$-group allows an odd number of occupied
states at $Y_0$, while the latter does not as mentioned above.
In other words, the cokernel of $i_{Y_0}^*$ in the present case includes
states with an odd number of occupied states at $Y_0$.
This gives a criterion for glide symmetric gapless materials: 
If the filling number of the occupied states at the point $Y_0$ is odd, 
then there should be a gapless point at the glide plane $k_y=0$, as
illustrated in Fig.~\ref{fig:subskeleton} [b].

\subsubsection{A gapless phase protected by representation at symmetric point for wallpaper group p4g} 
Sometimes a representation of occupied states at a high-symmetric point 
enforces a gapless phase. 
An example is a two-dimensional spinful system with the wallpaper group p4g. 
We will discuss the detail in Sec.~\ref{A stable gapless phase protected
by representation at $X$ point: 2d class A}, and here we only highlight
the consequence. 
The point group for p4g is 
the $D_4$ group, which is generated by a $C_4$-rotation and a reflection. 
In such system, the 
$K$-group is characterized by the one-dimensional subspace $X_1$ in
Fig.~\ref{fig:subskeleton} [c]. 

Let us focus on a high-symmetric point $Y_0 = (\pi,0)$. 
Since the little group at $Y_0$ is $D_2 = \Z_2 \times \Z_2$, 
a state at $Y_0$ obeys a linear representation of $D_2$.
The linear representation is given by a direct sum of 
irreducible representations of $D_2$, i.e.
$A_1, A_2, B_1, B_2$ in the Mulliken notation. 
As shown in Sec.~\ref{A stable gapless phase protected
by representation at $X$ point: 2d class A},
for fully gapped systems,  the occupied states at $Y_0$ should be 
a direct sum of $(A_1 \oplus A_2)$ and $(B_1 \oplus B_2)$ representations. 
The contraposition of this result implies that, 
if a occupied state at $Y_0$ obeys the other 
representations, say  $(A_1 \oplus B_1)$,
the system should have a gapless point on the one-dimensional subspace $X_1$. 
In this case, the other representations correspond to elements of
the cokernel obtained from the trivial inclusion
$i_{Y_0}: Y_0\to X_1$.  

\subsubsection{A $\Z_2$ topological charge induced only by inversion symmetry}

The final example is a bulk three-dimensional $\Z_2$ gapless phase
protected by inversion
symmetry, which has not been discussed before.
The detailed discussion will be presented in
Sec.~\ref{sec:inversion_fermi_pt}. 

As a subspace, we consider a sphere $Y_2=S^2$ of which the center is
an inversion symmetric point. See Fig.~\ref{fig:subskeleton} [d-1]. 
The inversion acts on $S^2$ as the antipodal map, so $S^2$ subject to
inversion is regarded as the quotient $S^2/\Z_2=RP^2$.
The $K$-group on $Y_2$ is $K(RP^2)=\Z_2 \oplus \Z$, where   
the $\Z_2$ index $\nu$ (mod. 2) is associated with the
torsion part of the first Chern class on $RP^2$.
The $\Z$ part is irrelevant to the gapless phase, and thus we focus on
the $\Z_2$ part here.
(The $\Z$ part 
is a trivial contribution counting the number of occupied
states.) 

%
%

When the system is fully gapped, 
the $\Z_2$ invariant $\nu$ should be trivial 
since $S^2$ can shrink to a point preserving inversion symmetry.
This means the following criterion for inversion symmetric
gapless phases:  
if the $\Z_2$ invariant $\nu$ is nontrivial on an inversion
symmetric sub-sphere $S^2$, 
then there should be a gapless region inside $S^2$. 
In this case, the cokernel of the trivial inclusion $i_{Y_2}:
Y_2\to T^3$ is the $\Z_2$ part of $K(RP^2)$. 
We present the model Hamiltonian of the gapless phase in
Sec.~\ref{sec:inversion_fermi_pt}. 

A similar  $\Z_2$ invariant can be defined also for  
a torus with inversion symmetry (See Fig.~\ref{fig:subskeleton} [d-2]). 
In Sec.~\ref{sec:inversion_fermi_pt}, 
we also show that the interplay between inversion symmetry and TRS 
defines a $\Z_2$ invariant associated with the 
Stiefel-Whitney classes on $RP^2$.

\section{The classification of topological insulators with wallpaper group symmetry}
\label{Wallpaper_summary} 

In this section, we summarize the $K$-theories over the BZ torus $T^2$ in the presence of 17 wallpaper groups 
with and without the chiral symmetry. 
Our results do not include TRS or PHS, which is a future problem. 
We present these $K$-groups as $R(P)$-modules, where $P$ is the point group associated with each wallpaper group, 
which can contrast with previous works.~\cite{Yang1997, LuckStamm2000, DongLiu2015, Kruthoff2016}
The detail of calculations of the $K$-groups will appear in the near future.~\cite{GomiShiozakiSato_Wallpaper}
In the next section \ref{sec:Example of K-theory classification}, 
we pick up a few examples of wallpaper groups 
in order to show how to compute the $K$-group and 
apply to the bulk insulators and surface states. 

As explained in Sec.\ \ref{sec:Stable classification of bulk insulators}, 
the $K$-group $K^{\tau-n}_P(T^2)$ ($n=0,1$) on $T^2$ means the stable classification of 2d bulk insulators in class A ($n=0$) and class AIII $(n=1)$. 
At the same time, as explained in Sec.~\ref{sec:Classification of boundary gapless states}, 
the $K$-group $K^{\tau-n}_P(T^2)$ expresses the classification of 2d surface gapless states in class A ($n=1$) and class AIII ($n=0$). 
It is worth summarizing these relations to avoid confusion: 
$$
\renewcommand{\arraystretch}{1.5}
\begin{tabular}{>{\centering\arraybackslash}m{3cm} | >{\centering\arraybackslash}m{3cm} | >{\centering\arraybackslash}m{3cm} }
$K$-group & Stable classification of bulk insulators & Surface gapless states \\
\hline 
$K^{\tau-0}_P(T^2)$ & class A  & class AIII \\
$K^{\tau-1}_P(T^2)$ & class AIII  & class A \\
\end{tabular}
\renewcommand{\arraystretch}{1.0}
$$

\begin{table*}[!]
\label{Tab:BravaisLattice}
\begin{center}
\caption{2d Bravais lattices, unit cells, point groups, and wallpaper groups.}
\renewcommand{\arraystretch}{2.0}
\begin{tabular}[t]{|c|c|c|c|}
\hline
Bravais lattice & Unit cell & Point group & Wallpaper group \\
\hline
\multirow{2}{*}{Oblique} & \multirow{2}{*}{
$$
\xygraph{
!{<0cm,0cm>;<0.6cm,0cm>:<0cm,0.6cm>::}
!{(0,0)}="a"
!{(2,0)}="b"  ([]!{+(0,-0.2)} {a \hat x})
!{(0.6,1.5)}="c" ([]!{+(-1.1,0)} {b \hat x + c \hat y})
!{(2.6,1.5)}="d" 
"a"-@{->}"b"
"a"-@{->}"c"
"b"-"d"
"c"-"d"
}
$$
}
& $C_1$ & p1 \\ \cline{3-4}
& & $C_2$ & p2 \\
\hline 
\multirow{2}{*}{Rectangular} & 
\multirow{2}{*}{
$$
\xygraph{
!{<0cm,0cm>;<0.6cm,0cm>:<0cm,0.6cm>::}
!{(0,0)}="a"
!{(2,0)}="b"  ([]!{+(0,-0.2)} {a \hat x})
!{(0,1.5)}="c" ([]!{+(-0.4,0)} {b \hat y})
!{(2,1.5)}="d" 
"a"-@{->}"b"
"a"-@{->}"c"
"b"-"d"
"c"-"d"
}
$$
}
& $D_1$ & pm, pg \\ \cline{3-4}
& & $D_2$ & pmm, pmg, pgg \\
\hline 
\multirow{2}{*}{Rhombic} & 
\multirow{2}{*}{$$
\xygraph{
!{<0cm,0cm>;<0.6cm,0cm>:<0cm,0.6cm>::}
!{(0,0)}="a"
!{(1.4,1.0)}="b"  ([]!{+(0.6,-0.4)} {a \hat x + b \hat y})
!{(-1.4,1.0)}="c" ([]!{+(-1.2,-0.4)} {-a \hat x + b \hat y})
!{(0,2.0)}="d" 
"a"-@{->}"b"
"a"-@{->}"c"
"b"-"d"
"c"-"d"
}
$$}
& $D_1$ & cm \\ \cline{3-4}
& & $D_2$ & cmm \\
\hline 
\multirow{2}{*}{Square} & 
\multirow{2}{*}{
$$
\xygraph{
!{<0cm,0cm>;<0.6cm,0cm>:<0cm,0.6cm>::}
!{(0,0)}="a"
!{(1.7,0)}="b"  ([]!{+(0.5,0)} {a \hat x})
!{(0,1.7)}="c" ([]!{+(-0.4,0)} {a \hat y})
!{(1.7,1.7)}="d" 
"a"-@{->}"b"
"a"-@{->}"c"
"b"-"d"
"c"-"d"
}
$$
}
& $C_4$ & p4 \\ \cline{3-4}
&& $D_4$ &p4m, p4g \\ \cline{3-4}
\hline 
\multirow{5}{*}{Hexagonal} & 
\multirow{5}{*}{$$
\xygraph{
!{<0cm,0cm>;<0.6cm,0cm>:<0cm,0.6cm>::}
!{(0,0)}="a"
!{(2.0,0)}="b"  ([]!{+(0.2,-0.3)} {a \hat x})
!{(1.0,1.732)}="c" ([]!{+(-0.2,0.8)} { a \Big( \frac{1}{2} \hat x + \frac{\sqrt{3}}{2} \hat y \Big)})
!{(3.0,1.732)}="d" 
"a"-@{->}"b"
"a"-@{->}"c"
"b"-"d"
"c"-"d"
}
$$}
& $C_3$ & p3 \\ \cline{3-4}
&& $C_6$ &p6 \\ \cline{3-4}
&& $D_3$ &p31m \\ \cline{3-4}
&& $D_3$ &p3m1 \\ \cline{3-4}
&& $D_6$ &p6m\\ 
\hline
\end{tabular}
\renewcommand{\arraystretch}{1.0}
\end{center}
\end{table*}

There are five Bravais lattices in 2d crystals, which are 
listed in Table~\ref{Tab:BravaisLattice} with point groups and wallpaper groups. 
In addition to the 17 different wallpaper groups, 
the nontrivial projective representations of the point group are the other sources of symmetry classes. 
Such contributions can be measured by the group cohomology of the point group as explained in Sec.~\ref{sec3:space}. 
For the rotational point group $C_n$, 
the group cohomology is trivial $H^2(\Z_n;U(1)) = 0$. 
For the dihedral group $D_n$, 
there is an even/odd effect: 
$H^2(D_{2n};U(1)) = \Z_2$, 
$H^2(D_{2n-1};U(1)) = 0$. 
Eventually, there are 24 inequivalent symmetry classes. 

Tabs.~\ref{Tab:ClassificationTable_A} and \ref{Tab:ClassificationTable_AIII} 
summarize the $K$-groups for all wallpaper groups. 
We used notations of $R(P)$-modules. 
To connect our notations to crystallography, 
we provide the character tables of 2d point groups 
in Tabs~\ref{Tab:CharacterD_2}, 
\ref{Tab:CharacterD_3}, 
\ref{Tab:CharacterD_4}, 
and 
\ref{Tab:CharacterD_6}, 
where our notations of irreps.\ and 
Mulliken's notations are displayed. 
The representation rings of 2d point groups 
and the module structures of the nontrivial projective representations 
are listed in Table~\ref{Tab:RepRing}, 
which are obtained by the tensor product representations
(see Sec.~\ref{sec:A little bit about representations of D4} for the case of $D_4$).

\begin{table*}[!]
\begin{center}
\caption{Character table of $D_2$.}
\begin{tabular}[t]{cc|cccc}
\hline 
Irrep. & Mulliken & $1$ & $m_x$ & $m_y$ & $m_xm_y$ \\
\hline
$1$ & $A_1$ & $1$ & $1$ & $1$ & $1$ \\
$t_x$ & $B_2$ & $1$ & $-1$ & $1$ & $-1$ \\
$t_y$ & $B_1$ & $1$ & $1$ & $-1$ & $-1$ \\
$t_x t_y$ & $A_2$ & $1$ & $-1$ & $-1$ & $1$ \\
\hline 
\end{tabular}
\label{Tab:CharacterD_2}
\end{center}
\end{table*}
\begin{table*}[!]
\begin{center}
\caption{Character table of $D_3$.}
\begin{tabular}[t]{cc|ccc}
\hline 
irrep. & Mulliken & $1$ & $\{C_3,C_3^{-1}\}$ & $\{\sigma, \sigma C_3, \sigma C_3^2\}$ \\
\hline
$1$ & $A_1$ & $1$ & $1$ & $1$ \\
$A$ & $A_2$ & $1$ & $1$ & $-1$ \\
$E$ & $E$ & $2$ & $-1$ & $0$ \\
\hline 
\end{tabular}
\label{Tab:CharacterD_3}
\end{center}
\end{table*}
\begin{table*}[!]
\begin{center}
\caption{Character table of $D_4$.}
\begin{tabular}[t]{cc|ccccc}
\hline 
irrep. & Mulliken& $1$ & $\{C_4,C_4^{-1}\}$ & $C_2$ & $\{ \sigma, \sigma C_2 \}$ & $\{\sigma C_4, \sigma C_4^3\}$ \\
\hline
$1$ & $A_1$ & $1$ & $1$ & $1$ & $1$ & $1$ \\
$A$ & $A_2$ & $1$ & $1$ & $1$ & $-1$ & $-1$ \\
$B$ & $B_1$ & $1$ & $-1$ & $1$ & $1$ & $-1$ \\
$AB$ & $B_2$ & $1$ & $-1$ & $1$ & $-1$ & $1$ \\
$E  $ & $E$ & $2$ & $0$ & $-2$ & $0$ & $0$ \\
\hline 	
\end{tabular}
\label{Tab:CharacterD_4}
\end{center}
\end{table*}
\begin{table*}[!]
\begin{center}
\caption{Character table of $D_6$.}
\begin{tabular}[t]{cc|ccccccccc}
\hline 
irrep. & Mulliken & $1$ & $\{C_6,C_6^{-1}\}$ & $\{C_3,C_3^{-1}\}$ & $\{C_2\}$ & $\{\sigma, \sigma C_3 , \sigma C_3^2\}$ & $\{\sigma C_6, \sigma C_2, \sigma C_6^5 \}$ \\
\hline
$1$ & $A_1$ & $1$ & $1$ & $1$ & $1$ & $1$ & $1$ \\
$A$ & $A_2$ & $1$ & $1$ & $1$ & $1$ & $-1$ & $-1$ \\
$B$ & $B_1$ & $1$ & $-1$ & $1$ & $-1$ & $1$ & $-1$ \\
$AB$ & $B_2$ & $1$ & $-1$ & $1$ & $-1$ & $-1$ & $1$ \\
$E$ & $E_1$ & $2$ & $1$ & $-1$ & $-2$ & $0$ & $0$ \\
$BE$ & $E_2$ & $2$ & $-1$ & $-1$ & $2$ & $0$ & $0$ \\
\hline 
\end{tabular}
\label{Tab:CharacterD_6}
\end{center}
\end{table*}

\begin{table*}[!]
\begin{center}
\caption{The representation rings of the 2d point groups and the module structure of the nontrivial projective representations of $D_2,D_4$ and $D_6$.}
\begin{tabular}[t]{c|l|c}
\hline 
Point group $P$ & Representation ring $R(P)$ & Abelian group \\ 
\hline 
$C_n$ & $R(\Z_n) = \Z[t]/(1-t^n)$ & $\Z^n$ \\
$D_2$ & $R(D_2) = \Z[t_1,t_2]/(1-t_1^2,1-t_2^2)$ & $\Z^4$ \\
$D_3$ & $R(D_3) = \Z[A,E]/(1-A^2,E-AE,E^2-1-A-E)$ & $\Z^3$ \\
$D_4$ & $R(D_4) = \Z[A,B,E]/(1-A^2,1-B^2,E-AE,E-BE,E^2-1-A-B-AB)$ & $\Z^5$  \\
$D_6$ & $R(D_6) = \Z[A,B,E]/(1-A^2,1-B^2,E-AE,E^2-1-A-BE)$ & $\Z^6$ \\
\hline 
\hline 
Point group $P$ & $R(P)$-module of nontrivial projective representations &  Abelian group \\ 
\hline
$D_2$ & $R^{\omega}(D_2) = (1+t_1+t_2+t_1t_2)$ & $\Z$ \\
$D_4$ & $R^{\omega}(D_4) = (1+A+E)$ & $\Z^2$ \\
$D_6$ & $R^{\omega}(D_6) = (1+A+E,E+BE)$ & $\Z^3$ \\
\hline
\end{tabular}
\label{Tab:RepRing}
\end{center}
\end{table*}

\begin{table*}[!]
\begin{center}
\caption{
The stable classification of 2d class A topological insulators with wallpaper groups/ 
the classification of 2d class AIII surface gapless states with wallpaper groups. 
In the fifth column, the overbraces represent $K$-groups as Abelian groups. 
The red characters mean that these direct summands are generated by 
vector bundles with the first Chern number. 
}
\begin{tabular}[t]{c|c|c|c|c}
\hline 
Wallpaper group & spinless/spinful & Twist & $R(P)$ & $K^{\tau-0}_{P}(T^2)$ \rule{0pt}{12pt}\\
\hline 
p1 & spinless/spinful & $0$ & $\Z$ & $\textcolor{red}{\Z} \oplus \Z$ \rule{0pt}{20pt}\\
p2 & spinless/spinful & $0$ & $R(\Z_2)$& $\textcolor{red}{\overbrace{R(\Z_2)}^{\Z^2}} \oplus \overbrace{R(\Z_2)}^{\Z^2} \oplus \overbrace{(1-t)}^{\Z} \oplus \overbrace{(1-t)}^{\Z}$ \\
p3 & spinless/spinful & $0$ & $R(\Z_3)$& $\textcolor{red}{\overbrace{R(\Z_3)}^{\Z^3}} \oplus \overbrace{R(\Z_3)}^{\Z^3} \oplus \overbrace{(1-t)}^{\Z^2} $ \\
p4 & spinless/spinful & $0$ & $R(\Z_4)$& $\textcolor{red}{\overbrace{R(\Z_4)}^{\Z^4}} \oplus \overbrace{R(\Z_4)}^{\Z^4} \oplus \overbrace{(1-t+t^2-t^3)}^{\Z}$ \\
p6 & spinless/spinful & $0$ & $R(\Z_6)$& $\textcolor{red}{\overbrace{(1-t+t^2)}^{\Z^4}} \oplus \overbrace{R(\Z_6)}^{\Z^6}$ \\
\hline
pm & spinless/spinful & $0$ & $R(\Z_2)$ & $\overbrace{R(\Z_2)}^{\Z^2}\oplus \overbrace{(1-t)}^{\Z}$ \\
cm & spinless/spinful & $0$ & $R(\Z_2)$ & $\overbrace{R(\Z_2)}^{\Z^2}$ \\
pmm & spinless & $0$ & $R(D_2)$ & $\overbrace{R(D_2)}^{\Z^4} \oplus \overbrace{(1-t_1)}^{\Z^2} \oplus \overbrace{(1-t_2)}^{\Z^2} \oplus \overbrace{\big( (1-t_1)(1-t_2) \big)}^{\Z}$ \\
pmm & spinful & $\omega$ & $R(D_2)$ & $\overbrace{R^{\omega}(D_2)}^{\Z}$ \\
cmm & spinless & $0$ & $R(D_2)$ & $\overbrace{R(D_2)}^{\Z^4} \oplus \overbrace{\big( (1-t_1)(1-t_2) \big)}^{\Z} \oplus \overbrace{\big( (1-t_1)(1-t_2) \big)}^{\Z}$ \\
cmm & spinful & $\omega$ & $R(D_2)$ & $\overbrace{R^{\omega}(D_2)}^{\Z} \oplus \overbrace{\big( (1-t_1)(1-t_2) \big)}^{\Z}$ \\
p31m & spinless/spinful & $0$ & $R(D_3)$ & $\overbrace{R(D_3)}^{\Z^3} \oplus \overbrace{(1+A-E)}^{\Z} \oplus \overbrace{(1+A-E)}^{\Z}$ \\
p3m1 & spinless/spinful & $0$ & $R(D_3)$ & $\overbrace{R(D_3)}^{\Z^3} \oplus \overbrace{R(D_3)/(E)}^{\Z} \oplus \overbrace{R(D_3)/(E)}^{\Z}$ \\
p4m & spinless & $0$ & $R(D_4)$ & $\overbrace{R(D_4)}^{\Z^5} \oplus \overbrace{(1+A-E)}^{\Z^2} \oplus \overbrace{(1+B-E)}^{\Z^2}$ \\
p4m & spinful & $\omega$ & $R(D_4)$ & $\overbrace{R^{\omega}(D_4)}^{\Z^2} \oplus \overbrace{\big( (1+A)(1-B) \big)}^{\Z}$ \\
p6m & spinless & $0$ & $R(D_6)$ & $\overbrace{R(D_6)}^{\Z^6} \oplus\overbrace{\big( (1+A)(1-B)(1-E) \big)}^{\Z} \oplus \overbrace{\big( (1+B)(1+A-E) \big)}^{\Z}$ \\
p6m & spinful & $\omega$ & $R(D_6)$ & $\overbrace{R^{\omega}(D_6)}^{\Z^3}\oplus \overbrace{\big( (1+B)(1+A-E) \big)}^{\Z}$ \\
\hline
pg & spinless/spinful & $\tau_{\rm pg}$ & $R(\Z_2)$ & $\overbrace{(1+t)}^{\Z}$ \\
pmg & spinless & $\tau_{\rm pmg}$ & $R(D_2)$ & $\overbrace{(1+t_1,1-t_2)}^{\Z^3}\oplus \overbrace{\big( (1-t_1)(1-t_2) \big)}^{\Z}$ \\
pmg & spinful & $\tau_{\rm pmg}+\omega$ & $R(D_2)$ & $\overbrace{(1+t_1,1-t_2)}^{\Z^3}\oplus \overbrace{\big( (1-t_1)(1-t_2) \big)}^{\Z}$ \\
pgg & spinless & $\tau_{\rm pgg}$ & $R(D_2)$ & $\overbrace{(1+t_1t_2)}^{\Z^2} \oplus \overbrace{((1-t_1)(1-t_2))}^{\Z}$ \\
pgg & spinful & $\tau_{\rm pgg}+\omega$ & $R(D_2)$ & $\overbrace{(1+t_1t_2)}^{\Z^2} \oplus \overbrace{((1-t_1)(1-t_2))}^{\Z}$ \\
p4g & spinless & $\tau_{\rm p4g}$ & $R(D_4)$ & $\overbrace{(1+A-E,1-B)}^{\Z^3} \oplus \overbrace{(1+A-E)}^{\Z^2} \oplus \overbrace{(1+A+B+AB+2E)}^{\Z}$ \\
p4g & spinful & $\tau_{\rm p4g}+\omega$ & $R(D_4)$ & $\overbrace{(1+A+E)}^{\Z^2} \oplus \overbrace{\big( (1+A)(1-B) \big)}^{\Z} \oplus \overbrace{(1+A+B+AB-2E)}^{\Z}$ \\
\hline
\end{tabular}
\label{Tab:ClassificationTable_A}
\end{center}
\end{table*}

\begin{table*}[!]
\begin{center}
\caption{
The stable classification of 2d class AIII topological insulators with wallpaper groups/ 
the classification of 2d class A surface gapless states with wallpaper groups. 
In the fifth column, the overbraces mean $K$-groups as Abelian groups. 
} 
\begin{tabular}[t]{c|c|c|c|c}
\hline 
Wallpaper group & spinless/spinful & Twist & $R(P)$ & $K^{\tau-1}_{P}(T^2)$ \rule{0pt}{12pt}\\
\hline 
p1 & spinless/spinful & $0$ & $\Z$& $\Z \oplus \Z$ \rule{0pt}{20pt}\\
p2 & spinless/spinful & $0$ & $R(\Z_2)$& $0$ \rule{0pt}{20pt}\\
p3 & spinless/spinful & $0$ & $R(\Z_3)$& $0$ \rule{0pt}{20pt}\\
p4 & spinless/spinful & $0$ & $R(\Z_4)$& $0$ \rule{0pt}{20pt}\\
p6 & spinless/spinful & $0$ & $R(\Z_6)$& $0$ \rule{0pt}{20pt}\\
\hline
pm & spinless/spinful & $0$ & $R(\Z_2)$ & $\overbrace{R(\Z_2)}^{\Z^2} \oplus \overbrace{(1-t)}^{\Z}$  \\
cm & spinless/spinful & $0$ & $R(\Z_2)$ & $\overbrace{(1+t)}^{\Z} \oplus \overbrace{(1-t)}^{\Z}$ \\
pmm & spinless & $0$ & $R(D_2)$ & $0$ \rule{0pt}{20pt}\\
pmm & spinful & $\omega$ & $R(D_2)$&$\overbrace{(1-t_1 t_2)}^{\Z^2}\oplus \overbrace{\big( (1+t_1)(1-t_2) \big)}^{\Z} \oplus \overbrace{\big( (1-t_1)(1+t_2) \big)}^{\Z}$ \\
cmm & spinless & $0$ & $R(D_2)$ & $0$ \rule{0pt}{20pt}\\
cmm & spinful & $\omega$ & $R(D_2)$ & $\overbrace{(1-t_1 t_2)}^{\Z^2}$ \\
p31m & spinless/spinful & $0$ & $R(D_3)$ & $\overbrace{(1-A)}^{\Z}$ \\
p3m1 & spinless/spinful & $0$ & $R(D_3)$ & $\overbrace{(1-A)}^{\Z}$ \\
p4m & spinless & $0$ & $R(D_4)$ & $0$ \rule{0pt}{20pt}\\
p4m & spinful & $\omega$ & $R(D_4)$ & $\overbrace{(1-A)}^{\Z^2} \oplus \overbrace{\big( (1-A)(1+B) \big)}^{\Z}$ \\
p6m & spinless & $0$ & $R(D_6)$ & $0$ \rule{0pt}{20pt}\\
p6m & spinful & $\omega$ & $R(D_6)$ & $\overbrace{(1-A)}^{\Z^2}$ \\
\hline
pg & spinless/spinful & $\tau_{\rm pg}$ & $R(\Z_2)$ & $\overbrace{(1+t)}^{\Z} \oplus \overbrace{I}^{\Z_2}$ \\
pmg & spinless & $\tau_{\rm pmg}$ & $R(D_2)$ & $\overbrace{\big( (1-t_1)(1+t_2) \big)}^{\Z}$ \\
pmg & spinful & $\tau_{\rm pmg}+\omega$ & $R(D_2)$ & $\overbrace{\big( (1-t_1)(1+t_2) \big)}^{\Z}$ \\
pgg & spinless & $\tau_{\rm pgg}$ & $R(D_2)$ & $\overbrace{I}^{\Z_2}$ \\
pgg & spinful & $\tau_{\rm pgg}+\omega$ & $R(D_2)$ & $\overbrace{I}^{\Z_2}$ \\
p4g & spinless & $\tau_{\rm p4g}$ & $R(D_4)$ & $0$ \rule{0pt}{20pt}\\
p4g & spinful & $\tau_{\rm p4g}+\omega$ & $R(D_4)$ & $\overbrace{\big( (1-A)(1-B)\big)}^{\Z}$ \\
\hline
\end{tabular}
\label{Tab:ClassificationTable_AIII}
\end{center}
\end{table*}

\makeatletter
\renewcommand{\theequation}{%
\arabic{section}.\arabic{subsection}.\arabic{equation}}
\@addtoreset{equation}{subsection}
\makeatother
\section{Example of $K$-theory classification}
\label{sec:Example of K-theory classification}
In this section, 
we illustrate the $K$-theory calculations in various examples. 
Through concrete problems, 
we introduce basics of the $K$-theory calculations such as 
the module structure, 
the Mayer-Vietoris sequence, 
the exact sequence for the pair $(X,Y)$, 
and the dimensional raising map. 
We also explain the vector bundle representation and Hamiltonian representation of the $K$-groups.

\subsection{$K$-theory on point: representations of symmetry group}
We start with $K$-theories $K^{\omega-n}_P(pt)$ of a point with symmetry group $P$. 
$\omega \in Z^2(P;\R/2 \pi\Z)$ fixes $U(1)$ phase factors associated with projective representations 
\begin{align}
U_p U_{p'} = e^{i \omega_{p,p'}} U_{p p'}. 
\end{align}

For class A ($n=0$), the $K$-theory is nothing but the 
Abelian group generated by the $\omega$-projective representations. 
We denote it by $R^\omega(P)$: 
\begin{align}
R^{\omega}(P) := K^{\omega-0}_P(pt). 
\end{align}
The tensor product of $\omega$- and $\omega'$-projective representations 
has the twist $\omega+\omega' \in Z^2(P;\R/2 \pi \Z)$. 
Especially, $R(P)$, the $K$-group generated by linear representations 
which have the trivial twist $\omega_{p,p'} \equiv 0$, becomes a ring. 

For class AIII ($n=1$), the $K$-group is trivial 
\begin{align}
K^{\omega-1}_P(pt) = 0
\end{align}
because of the chiral symmetry.

\subsubsection{Cyclic group $\Z_3$}
For example, consider the cyclic group $C_3 = \Z_3 = \{1,\sigma,\sigma^2\}$. 
There are three 1-dimensional irreps.\ $\C_0, \C_1, \C_2$ 
characterized by eigenvalues of $U_{\sigma} = 1, \zeta, \zeta^2$ with $\zeta = e^{2 \pi i/3}$, respectively. 
So we have 
\begin{equation}
R(\Z_3) = K^0_{\Z_3}(pt) = \Z \oplus \Z \oplus \Z \ \ {\rm as\ an\ Abelian\ group}. 
\end{equation}
On the vector bundle representation, 
an element $(n_0, n_1, n_2) \in R(\Z_3)$ is represented by the following direct sum
\begin{equation}
[V] \in R(\Z_3), \qquad 
V = [\C_0]^{\oplus n_0} \oplus [\C_1]^{\oplus n_1} \oplus [\C_2]^{\oplus n_2}. 
\end{equation}
In the Karoubi's representation, the same element is represented by two Hamiltonians acting on $V$ as follows
\begin{equation}
[V, H_0, H_1], \qquad 
H_0= 1_{n_0 \times n_0} \oplus 1_{n_1 \times n_1} \oplus 1_{n_2 \times n_2}, \qquad 
H_1 = -1_{n_0 \times n_0} \oplus -1_{n_1 \times n_1} \oplus -1_{n_2 \times n_2}.
\end{equation}
The tensor representation $V \otimes V'$ induces the ring structure in $R(\Z_3)$. 
The irreps.\ $\C_i (i = 0,1,2)$ acts on the element $(n_0, n_1,n_2)$ as 
\begin{equation}
\C_i \otimes ( [\C_0]^{\oplus n_0} \oplus [\C_1]^{\oplus n_1} \oplus [\C_2]^{\oplus n_2} ) = 
[\C_{i}]^{\oplus n_0} \oplus [\C_{i+1}]^{\oplus n_1} \oplus [\C_{i+2}]^{\oplus n_2}, 
\end{equation}
where subscripts $i,i+1, i+2$ are defined modulo $3$. 
In short, $R(\Z_3)$ is isomorphic to the quotient of the polynomial ring
\begin{equation}
R(\Z_3) = \Z[t]/(1-t^3) = \{n_0 + n_1 t + n_2 t^2 | n_0,n_1,n_2 \in \Z \}. 
\end{equation}

\subsubsection{Dihedral group $D_2$}
Consider the dihedral group $D_2 = \{1,m_x, m_y, m_x m_y\}$. 
There are four 1-dimensional linear irreps.\ shown in Table~\ref{Tab:CharacterD_2}. 
Tensor products of these irreps.\ lead to 
the quotient of the polynomial ring: 
\begin{equation}
R(D_2) = K^{0}_{D_2}(pt) = \Z[t_x,t_y]/(1-t_x^2, 1-t_y^2). 
\end{equation}

Because of $H^2(D_2;\R/2 \pi \Z) = \Z_2$, there is a nontrivial twist $[\omega] \in H^2(D_2;\R/2 \pi \Z)$. 
An example of a nontrivial two-cocycle $\omega$ is given by 
\begin{align}
e^{i \omega_{p,p'}} \qquad = \qquad 
\begin{tabular}{c|cccc}
$p \backslash p'$ & 1 & $m_x$ & $m_y$ & $m_x m_y$ \\
\hline 
1 & 1 & 1 & 1 & 1 \\
$m_x$ & 1 & 1 & i & -i \\
$m_y$ & 1 & -i & 1 & i \\
$m_x m_y$ & 1 & i & -i & 1 \\
\end{tabular}
\end{align}
There is one 2-dimensional $c$-projective irrep. 
We denote it by $W$ that is represented by the Pauli matrices 
\begin{align}
U_{1} = \begin{pmatrix}
1 & 0 \\
0 & 1 \\
\end{pmatrix}, \qquad 
U_{m_x} = \begin{pmatrix}
0 & 1 \\
1 & 0 \\
\end{pmatrix}, \qquad 
U_{m_y} = \begin{pmatrix}
0 & -i \\
i & 0 \\
\end{pmatrix}, \qquad 
U_{m_x m_y}
= \begin{pmatrix}
1 & 0 \\
0 & -1
\end{pmatrix}. 
\end{align}
The $K$-group is 
\begin{align}
R^{\omega}(D_2) = K_{D_2}^{\omega-0}(pt) = \Z
\end{align}
as an Abelian group. 
The tensor product $V \otimes W$ by a linear representation $V \in R(D_2)$ 
is just the multiplication $V \otimes W \cong W^{\oplus \dim V}$ by the rank of $V$, which leads to 
the $R(D_2)$-module structure 
\begin{align}
R^{\omega}(D_2) = (1+t_x+t_y+t_x t_y)
= \{(1+t_x+t_y+t_x t_y) f(t_x,t_y) | f(t_x,t_y) \in R(D_2) \}. 
\end{align}

%

\subsection{Onsite symmetry}
Let us consider the $K$-theory associated with the onsite unitary symmetry $G$ 
\begin{align}
&U_g H(\bk) U_g^{-1} = H(\bk), \qquad g \in G, \\
&U_g U_h = e^{i\omega_{g,h}} U_{g h}, \qquad \omega_{g,h} \in Z^2(G,\R/2 \pi \Z). 
\end{align}
For class AIII (n=1), we assume the 
onsite symmetry commutes with the chiral symmetry 
\begin{align}
\Gamma H(\bk) \Gamma^{-1} = - H(\bk), \qquad 
U_g \Gamma = \Gamma U_g. 
\end{align}
In such cases, the Hamiltonian $H(\bk)$ 
is decomposed as a direct sum 
\begin{align}
H(\bk) = \bigoplus_{\rho} H_{\rho}(\bk)
\end{align}
of irreducible $\omega$-projective representations. 
In each sector, the Hamiltonian behaves as a class A or AIII insulator. 
The topological classification is recast as 
\begin{align}
K_G^{\omega-n}(X)
\cong R^{\omega}(G) \otimes_{\Z} K^{n}(X). 
\end{align}

For example, 
we can immediately have the topological 
classification of 2d class A insulators with 
onsite unitary $\Z_n$ symmetry: 
\begin{align}
K_{\Z_n}^{0}(T^2)
\cong  R(\Z_n) \otimes K(T^2)
= R(\Z_n) \otimes_{\Z} (\Z \oplus \Z) 
= R(\Z_n) \oplus R(\Z_n). 
\end{align}
The first direct summand represents 
atomic insulators with representations of $\Z_n$. 
The second direct summand is generated by the 
Chern insulators with irreducible representations of $\Z_n$. 
\color{black}

\subsection{Reflection symmetry}
Let us consider reflection symmetric 1d class A/AIII crystalline insulators. 
The $\Z_2 = \{1,m\}$ group acts on the BZ circle $S^1$ as a reflection: 
\begin{equation}
\tilde S^1 \ \ = \ \ 
\xygraph{
!{<0cm,0cm>;<1cm,0cm>:<0cm,1cm>::}
!{(0,0)}*{\bullet}="a"
!{(2,0)}*{\bullet}="b"
!{(2.4,0.4)}="c"
!{(2.4,-0.4)}="d"
!{(2.6,0)}*{m}="e", 
"a" -@/^1.0cm/ "b", 
"a" -@/_1.0cm/ "b", 
!{(-0.5,0)}-@{.}!{(2.5,0)}
"c" -@{<->} "d"
}
\label{Fig:TildeS1}
\end{equation}
We denoted the circle $S^1$ with the reflection action by $\tilde S^1$. 
There are two fixed points at $k_x = 0, \pi$. 

There is no nontrivial twist: $H^2(\Z_2; C(\tilde S^1, U(1))) = 0$. 
One can fix the $U(1)$ phases associated with the square of 
$\Z_2$ action to 1: 
\begin{align}
U_m(-k_x) U_m(k_x) = {\bf 1},  
\end{align} 
where ${\bf 1}$ is the identity matrix. 

In the Karoubi's representation, 
each $K$-group $K^{n}_{\Z_2}(\tilde S^1)$ means the 
topological classification of the Hamiltonians with the following symmetry 
\begin{align}
&{\rm Class\ A}\  (n=0):  
&&U_m(k_x) H(k_x) U_m(k_x)^{-1} = H(-k_x), \\
&{\rm Class\ AIII}\  (n=1):  
&&\left\{ \begin{array}{l}
\Gamma H(k_x) \Gamma^{-1} = -H(k_x), \\
U_m(k_x) H(k_x) U_m(k_x)^{-1} = H(-k_x), \\
\Gamma U_m(k_x) = U_m(k_x) \Gamma,  \\
\end{array}\right. 
\end{align}

\subsubsection{Calculation of $K$-group by the Mayer-Vietoris sequence}
One way to calculate the $K$-group $K^{-n}_{\Z_2}(\tilde S^1)$ is to use the Mayer-Vietoris sequence.~\cite{bott2013differential}
See Appendix \ref{app:Mayer-Vietoris sequence} for the details of the Mayer-Vietoris sequence. 
We divide $\tilde S^1 = U \cup V$ into two subspaces
\begin{align}
U = \{e^{ik} \in \tilde S^1 | k \in [-\pi/2,\pi/2]\}, &&
V = \{e^{ik} \in \tilde S^1 | k \in [\pi/2,3\pi/2]\}, 
\end{align}
as shown below: 
$$
U \sqcup V =\ \  
\begin{xy}
(0,0)*+!R{V}="A"*{\bullet}, 
"A"+<1cm,0.8cm>="B"*{},
"A"+<1cm,-0.8cm>="C"*{},
"A"+<2.4cm,0cm>*+!L{U}="D"*{\bullet}, 
"D"+<-1cm,0.8cm>="E"*{},
"D"+<-1cm,-0.8cm>="F"*{},
\ar@/^/@{-} "A";"B",
\ar@/_/@{-} "A";"C", 
\ar@/_/@{-} "D";"E",
\ar@/^/@{-} "D";"F" 
\end{xy}
$$
Each of the lines $U$ and $V$ is homotopic to a point preserving the reflection symmetry as: 
\[
U \sqcup V \ \ \sim \ \ \{0\} \sqcup \{\pi\} \ \ = \ \ 
\begin{xy}
(0,0)*{\bullet}="A"*{}, 
"A"+<2cm,0cm>*{\bullet}="D"*{}, 
"A"+<0.4cm,0cm>*{\{\pi\}}, 
"D"+<-0.4cm,0cm>*{\{0\}}, 
"A"+<-1.0cm,0cm>*{\Z_2}, 
"D"+<1.0cm,0cm>*{\Z_2}, 
\ar @(lu,ld) "A";"A"
\ar @(ru,rd) "D";"D"
\end{xy}
\]
The intersection $U \cap V$ is homotopic to two points $\Z_2 \times pt$ that are exchanged by the $\Z_2$ action:  
\[
U \cap V \ \ \sim \ \ \Z_2 \times pt \ \ = \ \ \{\frac{\pi}{2}, -\frac{\pi}{2}\} \ \ = 
\xygraph{
!{<0cm,0cm>;<1cm,0cm>:<0cm,1cm>::}
!{(1,0.6)}="a"*{\bullet}
!{(1,-0.6)}="b"*{\bullet}
!{(1,0.5)}="c"*{}
!{(1,-0.5)}="d"*{}
!{(1.4,0)}*{m}="e"
"c" -@{<->} "d"
}
\]

The Mayer-Vietoris sequence associated to the sequence of the inclusions 
\begin{align}
(\tilde S^1 =) \ U \cup V \ot U \sqcup V \ot U \cap V 
\end{align}
is the six term exact sequence of the $K$-theory 
\begin{equation}
\begin{CD}
K_{\Z_2}^{1}(U \cap V) @<<< K_{\Z_2}^{1}(U) \oplus K_{\Z_2}^{1}(V) @<<< K_{\Z_2}^{1}(\tilde S^1) \\
@VVV @. @AAA \\
K_{\Z_2}^{0}(\tilde S^1) @>>> K_{\Z_2}^{0}(U) \oplus K_{\Z_2}^{0}(V) @>>> K_{\Z_2}^{0}(U \cap V).
\end{CD}
\label{Seq:S1Ref}
\end{equation}
In this sequence, we have 
\begin{align}
K^{n}_{\Z_2}(U) \cong K^{n}_{\Z_2}(\{0\}) \cong 
\left\{ \begin{array}{ll}
R(\Z_2) & (n=0) \\
0 & (n=1) \\
\end{array}\right., && 
K^{n}_{\Z_2}(V) \cong K^{n}_{\Z_2}(\{\pi\}) \cong 
\left\{ \begin{array}{ll}
R(\Z_2) & (n=0) \\
0 & (n=1) \\
\end{array}\right.,
\end{align}
and 
\begin{align}
K^{n}_{\Z_2}(U \cap V) 
\cong K^n_{\Z_2}(\{\frac{\pi}{2}, - \frac{\pi}{2} \}) 
\cong K^{n}(\{\frac{\pi}{2}\}) 
\cong \left\{ \begin{array}{ll}
\Z & (n=0) \\
0 & (n=1) \\
\end{array}\right..
\end{align}
Thus, the sequence (\ref{Seq:S1Ref}) is recast into 
\begin{equation}
\begin{CD}
0 @<<< 0 @<<< K_{\Z_2}^{1}(\tilde S^1) \\
@VVV @. @AAA \\
K_{\Z_2}^{0}(\tilde S^1) @>>> R(\Z_2) \oplus R(\Z_2) @>\Delta>> \Z.
\end{CD}
\label{Seq:S1Ref_2}
\end{equation}
Here, the homomorphism $\Delta$ is given by 
\begin{align}
\Delta : R(\Z_2) \oplus R(\Z_2) \to \Z, && \Delta (f(t),g(t)) = f(1)-g(1), 
\end{align}
under the presentation $R(\Z_2) = \Z[t]/(1-t^2)$. 
We have 
\begin{align}
K_{\Z_2}^{0}(\tilde S^1) \cong \mathrm{Ker} (\Delta), &&
K_{\Z_2}^{1}(\tilde S^1) \cong \mathrm{Coker} (\Delta). 
\end{align}
$\mathrm{Ker}(\Delta)$ is spanned by $\{(1,1),(t,t),(0,1-t)\} \subset R(\Z_2) \oplus R(\Z_2)$, 
so we have $\mathrm{Ker}(\Delta) = \Z^3$ as an Abelian group. 
The base elements $(1,1)$ and $(t,t)$ span the $R(\Z_2)$-module $R(\Z_2)$, and $(0,1-t)$ the ideal $(1-t) = \{(1-t) f(t) | f(t) \in R(\Z_2) \}$ in $R(\Z_2)$.
As a result, we get the following $R(\Z_2)$-modules as $K$-groups 
\begin{align}
{\rm Class\ A}: K_{\Z_2}^{0}(\tilde S^1) \cong  \overbrace{R(\Z_2)}^{\Z^2} \oplus \overbrace{(1-t)}^{\Z}, &&
{\rm Class\ AIII}: K_{\Z_2}^{1}(\tilde S^1) \cong 0. 
\label{Eq:KGroupS1}
\end{align}

\subsubsection{Characterization of $K$-group by fixed points}
Notice the injection in (\ref{Seq:S1Ref_2}), 
\begin{align}
\begin{CD}
0 @>>> K_{\Z_2}^{0}(\tilde S^1) @>>> \underbrace{R(\Z_2)}_{k_x=0} \oplus \underbrace{R(\Z_2)}_{k_x=\pi}, 
\end{CD}
\end{align}
means that the $K$-group $K_{\Z_2}^{0}(\tilde S^1)$ 
can be characterized by the representations at the two fixed points. 
In general, 
representations of the little group at fixed points provide topological invariants 
which enable us to distinguish different elements in a $K$-group.  

Let $\{ e_1, e_2, e_3 \}$ be a basis of the $K$-group $K_{\Z_2}^{0}(\tilde S^1)$  
characterized by the following fixed point representations, 
$$
\begin{tabular}[t]{|c|c|c|}
\hline 
Base & $\underbrace{R(\Z_2)}_{k_x=0}$ & $\underbrace{R(\Z_2)}_{k_x=\pi}$ \\ 
\hline 
$e_1$ & $1$ & $1$ \\ 
$e_2$ & $t$ & $t$ \\ 
$e_3$ & $1$ & $t$ \\ 
\hline 
\end{tabular}
$$
Because of the $R(\Z_2)$-module structures $e_2 = t \cdot e_1$ and $t \cdot (e_1-e_3) = -(e_1-e_3)$, 
two base elements $e_1, e_2$ compose $R(\Z_2)$ and $e_1-e_3$ generates $(1-t)$.

\subsubsection{Vector bundle representation}
We give $\Z_2$ equivariant vector bundle representations for the basis $\{ e_1,e_2,e_3 \}$. 
We will construct $\Z_2$ equivariant vector bundles  $\{ E_1, E_2, E_3 \}$ with the following fixed point data: 
$$
\begin{tabular}[t]{|c|c|c|}
\hline 
Vector bundle & $E|_{k_x=0}$ & $E|_{k_x=\pi}$ \\ 
\hline 
$E_1$ & $\C_0$ & $\C_0$ \\ 
$E_2$ & $\C_1$ & $\C_1$ \\ 
$E_3$ & $\C_0$ & $\C_1$ \\ 
\hline 
\end{tabular}
$$
Here $\C_0$ and $\C_1$ are representations with $U_{m} = 1, -1$, respectively.  

$e_1$ is represented by a $\Z_2$ equivariant complex vector bundle $E_1$ of rank 1 with $\Z_2$ action $\rho_{m} : E_1 \to E_1$ as 
\begin{align}
e_1 = \left[ \left( E_1 = S^1 \times \C, \ \ \rho_{m}(k_x,v) = (-k_x,v) \right) \right] . 
\label{Eq:1DTCIVectE1}
\end{align}
By using the Bloch states, $E_1$ is equivalent to a Bloch state $\ket{k_x}_1$ which satisfies the reflection symmetry as 
\begin{align}
e_1 = \left[ \left( \ket{k_x}_1, \ \ \hat U_{m} \ket{k_x}_1 = \ket{-k_x}_1 \right) \right] . 
\end{align}
(Recall that the (local) Bloch states $\Phi(\bk) = \{ \ket{\bk,n} \}_{n=1, \dots N}$ correspond to 
(local) sections of the frame bundle $F(E)$ associated with a vector bundle $E$.)
The Bloch state $\ket{k_x}_1$ is translated to the real space base $\ket{R_x}_1 = \sum_{k_x \in S^1} \ket{k_x}_1 e^{-i k_x R_x}$ with the 
reflection symmetry 
\begin{align}
e_1 = \left[ \left(\ket{R_x}_1, \ \ \hat U_{m} \ket{R_x}_1 = \ket{-R_x}_1 \right) \right] , 
\end{align}
The base $\ket{R_x}_1$ corresponds to $s$-orbitals 
localized at the center of unit cells  
\begin{align}
e_1 = \left[ \ \ 
\xygraph{
!{<0cm,0cm>;<1cm,0cm>:<0cm,1cm>::}
!{(0,0)}*{\bigcirc},
!{(2,0)}*{\bigcirc},
!{(-2,0)}*{\bigcirc},
!{(4,0)}*{\bigcirc},
!{(-4,0)}*{\bigcirc},
!{(0,-0.4)}*{\ket{s}},
!{(2,-0.4)}*{\ket{s}},
!{(-2,-0.4)}*{\ket{s}},
!{(4,-0.4)}*{\ket{s}},
!{(-4,-0.4)}*{\ket{s}},
!{(-5,0)}-@{->}!{(5,0)},
!{(0,-0.8)}-@{.}!{(0,0.8)},
!{(0,0.6)}*{m},
!{(0.4,0.4)}-@{<->}!{(-0.4,0.4)},
!{(-1,-0.8)}-@{|-|}!{(1,-0.8)},
!{(0,-1)}*{\rm unit\ cell}, 
}
\ \ \right] 
\label{Fig:1DTCIe1}
\end{align}
where the reflection axis is placed at the center of the unit cell. 

The base $e_2 = t \cdot e_1$ is represented by the $\Z_2$ equivariant vector bundle $E_2 = \C_1 \otimes E_1$ as follows
\begin{align}
e_2 = \left[ \left( E_2 = S^1 \times \C, \ \ \rho_{m}(k_x,v) = (-k_x,-v) \right) \right]. 
\label{Eq:1DTCIVectE2}
\end{align}
The Bloch state and localized orbitals representation read 
\begin{align}
e_2 = \left[ \left( \ket{k_x}_2, \ \ \hat U_{m} \ket{k_x}_2 = - \ket{-k_x}_2\right) \right], 
\end{align}
\begin{align}
e_2 = \left[ \left(\ket{R_x}_2, \ \ \hat U_{m} \ket{R_x}_2 = - \ket{-R_x}_2\right) \right]. 
\end{align}
$\ket{R_x}_2$ corresponds to $p$-orbitals localized at the center of unit cells: 
\begin{align}
e_2 = \left[ \ \ 
\xygraph{
!{<0cm,0cm>;<1cm,0cm>:<0cm,1cm>::}
!{(0,0)}*{\bigcirc},
!{(2,0)}*{\bigcirc},
!{(-2,0)}*{\bigcirc},
!{(4,0)}*{\bigcirc},
!{(-4,0)}*{\bigcirc},
!{(0,-0.4)}*{\ket{p}},
!{(2,-0.4)}*{\ket{p}},
!{(-2,-0.4)}*{\ket{p}},
!{(4,-0.4)}*{\ket{p}},
!{(-4,-0.4)}*{\ket{p}},
!{(-5,0)}-@{->}!{(5,0)},
!{(0,-0.8)}-@{.}!{(0,0.8)},
!{(0,0.6)}*{m},
!{(0.4,0.4)}-@{<->}!{(-0.4,0.4)},
!{(-1,-0.8)}-@{|-|}!{(1,-0.8)},
!{(0,-1)}*{\rm unit\ cell}, 
}
\ \ \right] 
\end{align}

The last base $e_3$ is represented by the following $\Z_2$ equivariant vector bundle $E_3$ 
\begin{align}
e_3 = \left[ \left( E_3 = S^1 \times \C, \ \ \rho_{m}(k_x,v) = (-k_x, e^{-i k_x} v) \right) \right].
\end{align}
If one uses the Bloch state $\ket{k_x}_3$, then
\begin{align}
e_3 = \left[ \left( \ket{k_x}_3, \ \ \hat U_{m} \ket{k_x}_3 = e^{-i k_x} \ket{-k_x}_3\right) \right].
\end{align}
If one instead uses the localized orbital $\ket{R_x}_3$, then
\begin{align}
e_3 = \left[ \left(\ket{R_x}_3, \ \ \hat U_{m} \ket{R_x}_3 = \ket{-R_x-1}_3\right) \right], 
\label{Eq:1DTCIe3_RealSpace}
\end{align}
where $\ket{R_x}_3$ corresponds to the localized $s$-orbitals at the boundary of unit cells: 
\begin{align}
e_3 = \left[ \ \ 
\xygraph{
!{<0cm,0cm>;<1cm,0cm>:<0cm,1cm>::}
!{(1,0)}*{\bigcirc},
!{(3,0)}*{\bigcirc},
!{(-1,0)}*{\bigcirc},
!{(-3,0)}*{\bigcirc},
!{(1,-0.4)}*{\ket{s}},
!{(3,-0.4)}*{\ket{s}},
!{(-1,-0.4)}*{\ket{s}},
!{(-3,-0.4)}*{\ket{s}},
!{(-5,0)}-@{->}!{(5,0)},
!{(0,-0.8)}-@{.}!{(0,0.8)},
!{(0,0.6)}*{m},
!{(0.4,0.4)}-@{<->}!{(-0.4,0.4)},
!{(-1,-0.8)}-@{|-|}!{(1,-0.8)},
!{(0,-1)}*{\rm unit\ cell}, 
}
\ \ \right] 
\label{Fig:1DTCIe3}
\end{align}
Here, we assumed that the $s$-orbital belonging to the unit cell $R_x$ is localized at $R_x + \frac{1}{2}$. 
An alternative choice, for example, $R_x -\frac{1}{2}$, 
leads to the $\Z_2$ equivariant vector bundle 
\begin{align}
\left[ \left( E_3' = S^1 \times \C, \ \ \rho_{m}(k_x,v) = (-k_x, e^{i k_x} v) \right) \right]. 
\end{align}
$E_3'$ is isomorphic to $E_3$ as a $\Z_2$ equivariant vector bundle, 
thus, $E_3$ and $E_3'$ give the same $K$-class $e_3$. 

As explained in Sec.~\ref{sec:Dependence on unit cell and Wyckoff position}, 
even if the localized $s$-orbitals described by (\ref{Fig:1DTCIe1}) and (\ref{Fig:1DTCIe3}) are physically the same, 
the corresponding $K$-classes are different. 
The $K$-classes depend on the choice of the unit cell center.

\subsubsection{Karoubi's triple representation}
Here we give an alternative representation of $K$-groups, that is the Karoubi's triple representation. 
First, from the vector bundle representation, 
we can get Karoubi's triple representations $e_i = [(E_i 1,-1)] \ (i=1,2,3)$ for the base elements of the $K$-group 
with the following fixed point data: 
$$
\begin{tabular}[t]{|c|c|c|}
\hline 
Triple & $k_x=0$ & $k_x=\pi$ \\ 
\hline 
$(E_1,1,-1)$ & $(\C_0,1,-1)$ & $(\C_0,1,-1)$ \\ 
$(E_2,1,-1)$ & $(\C_1,1,-1)$ & $(\C_1,1,-1)$ \\ 
$(E_3,1,-1)$ & $(\C_0,1,-1)$ & $(\C_1,1,-1)$ \\ 
\hline 
\end{tabular}
$$

A benefit of using the Karoubi's triple is 
that we can construct representatives of $e_3$ as a Hamiltonian acting on
the vector bundles $E_1$ and $E_2$. 
$E_1 \oplus E_2$ is written by using the Bloch basis as 
\begin{align}
E_1 \oplus E_2 = 
\left( \Phi_{E_1 \oplus E_2}(k_x) = \left( \ket{k_x}_1 , \ket{k_x}_2 \right), \qquad 
\hat U_{m} \Phi_{E_1 \oplus E_2}(k_x) = \Phi_{E_1 \oplus E_2}(-k_x) U_{\sigma}(k_x), \ \ U_{\sigma}(k_x) = \sigma_z \right) , 
\end{align}
where $\sigma_z = \begin{pmatrix}
1 & 0 \\
0 & -1 \\
\end{pmatrix}$ is the $z$ component of the Pauli matrices $\sigma_i (i=x,y,z)$. 
A Hamiltonian on the $E_1 \oplus E_2$ should satisfy the reflection symmetry 
\begin{align}
\sigma_z H(k_x) \sigma_z = H(-k_x). 
\end{align}
We can show that the following triple represents the base $e_3$: 
\begin{align}
e_3 = \left[ (E_1 \oplus E_2 , H_0 = \cos (k_x) \sigma_z + \sin (k_x) \sigma_y, H_1 = -\sigma_0 ) \right] 
\end{align}
Actually, 
the empty and occupied states $\ket{\phi_{\pm}(k_x)}$ of the Hamiltonian $H_0$, 
$H_0 \ket{\phi_{\pm}(k_x)} = \pm \ket{\phi_{\pm}(k_x)}$, are given by the following Bloch states
\begin{align}
&\ket{\phi_+(k_x)} = \frac{1}{2}(1+e^{-i k_x}) \ket{k_x}_1 + \frac{1}{2}(1-e^{-i k_x}) \ket{k_x}_2, \\
&\ket{\phi_-(k_x)} = \frac{1}{2}(1-e^{-i k_x}) \ket{k_x}_1 + \frac{1}{2}(1+e^{-i k_x}) \ket{k_x}_2 
\end{align}
with reflections symmetry 
\begin{align}
\hat U_{m} \ket{\phi_{+}(k_x)} = e^{-i k_x} \ket{\phi_{+}(k_x)}, &&
\hat U_{m} \ket{\phi_{-}(k_x)} = - e^{-i k_x} \ket{\phi_{-}(k_x)}, 
\end{align}
which means the empty state $\ket{\phi_+(k_x)}$ is the $\Z_2$ equivariant bundle $E_3$, 
and $\ket{\phi_-(k_x)}$ is $E_4 = \C_1 \otimes E_3$, i.e.\ $E_1 \oplus E_2$ is isomorphic to $E_3 \oplus E_4$. 
Then, by using the stable equivalence, we have 
\begin{equation}\begin{split}
(E_1 \oplus E_2 , H_0 = \cos (k_x) \sigma_z + \sin (k_x) \sigma_y, H_1 = -\sigma_0 ) 
&\sim (E_3 \oplus E_4 , H_0 = 1 \oplus (-1), H_1 = (-1) \oplus (-1)) \\
&\sim (E_3, H_0 = 1, H_1 = -1).  
\end{split}\label{Eq:EqivalenceE1E2E3}\end{equation}

Note that if we construct the Wannier orbital $\ket{W_+(R_x)}$ from the energy eigenstate $\ket{\phi_+(k_x)}$ 
by $\ket{W_+(R_x)} := \sum_{k_x \in S^1} \ket{\phi_+(k_x)} e^{-i k_x R_x}$, 
we recover the real space orbital picture (\ref{Eq:1DTCIe3_RealSpace}).

\subsubsection{Real space picture of the isomorphism $E_1 \oplus E_2 \cong E_3 \oplus E_4$}
The above equivalence relation (\ref{Eq:EqivalenceE1E2E3}) is based on the isomorphism 
$E_1 \oplus E_2 \cong E_3 \oplus E_4$. 
This can be understood by the continuum deformation of the real space orbitals. 

The $\Z_2$ equivariant vector bundle $E_1 \oplus E_2$ is 
represented by the real space orbitals in which $s$ and $p$-orbitals are placed at the center of unit cell: 
\begin{align}
E_1 \oplus E_2 =  \ \ 
\xygraph{
!{<0cm,0cm>;<1cm,0cm>:<0cm,1cm>::}
!{(0,0.2)}*{\bigcirc},
!{(2,0.2)}*{\bigcirc},
!{(-2,0.2)}*{\bigcirc},
!{(4,0.2)}*{\bigcirc},
!{(-4,0.2)}*{\bigcirc},
!{(0,0.6)}*{\ket{s}},
!{(2,0.6)}*{\ket{s}},
!{(-2,0.6)}*{\ket{s}},
!{(4,0.6)}*{\ket{s}},
!{(-4,0.6)}*{\ket{s}},
!{(0,-0.2)}*{\bigcirc},
!{(2,-0.2)}*{\bigcirc},
!{(-2,-0.2)}*{\bigcirc},
!{(4,-0.2)}*{\bigcirc},
!{(-4,-0.2)}*{\bigcirc},
!{(0,-0.6)}*{\ket{p}},
!{(2,-0.6)}*{\ket{p}},
!{(-2,-0.6)}*{\ket{p}},
!{(4,-0.6)}*{\ket{p}},
!{(-4,-0.6)}*{\ket{p}},
!{(-5,0)}-@{->}!{(5,0)},
!{(0,-1)}-@{.}!{(0,1.1)},
!{(0,1.2)}*{m},
!{(0.4,1)}-@{<->}!{(-0.4,1)},
!{(-1,-1)}-@{|-|}!{(1,-1)},
!{(0,-1.2)}*{\rm unit\ cell}, 
}
\end{align}
To deform the orbital positions, 
first, we mix the $s$ and $p$ orbital as $\ket{s\pm p} : = \frac{\ket{s} \pm \ket{p}}{\sqrt{2}}$. 
Then, we can continuously translate the localized orbital $\ket{s+p}$ to right and $\ket{s-p}$ to left preserving the reflection symmetry as shown below:
\begin{align}
E_1 \oplus E_2 \cong  \ \ 
\xygraph{
!{<0cm,0cm>;<1cm,0cm>:<0cm,1cm>::}
!{(0.4,0.2)}*{\bigcirc},
!{(2.4,0.2)}*{\bigcirc},
!{(-1.6,0.2)}*{\bigcirc},
!{(4.4,0.2)}*{\bigcirc},
!{(-3.6,0.2)}*{\bigcirc},
!{(0.4,0.6)}*{\ket{s+p}},
!{(2.4,0.6)}*{\ket{s+p}},
!{(-1.6,0.6)}*{\ket{s+p}},
!{(4.4,0.6)}*{\ket{s+p}},
!{(-3.6,0.6)}*{\ket{s+p}},
!{(-0.4,-0.2)}*{\bigcirc},
!{(1.6,-0.2)}*{\bigcirc},
!{(-2.4,-0.2)}*{\bigcirc},
!{(3.6,-0.2)}*{\bigcirc},
!{(-4.4,-0.2)}*{\bigcirc},
!{(-0.4,-0.6)}*{\ket{s-p}},
!{(1.6,-0.6)}*{\ket{s-p}},
!{(-2.4,-0.6)}*{\ket{s-p}},
!{(3.6,-0.6)}*{\ket{s-p}},
!{(-4.4,-0.6)}*{\ket{s-p}},
!{(-5,0)}-@{->}!{(5,0)},
!{(0,-1)}-@{.}!{(0,1.1)},
!{(0,1.2)}*{m},
!{(0.4,1)}-@{<->}!{(-0.4,1)},
!{(0,0.2)}-@{->}!{(0.3,0.2)},
!{(0,-0.2)}-@{->}!{(-0.3,-0.2)},
!{(-1,-1)}-@{|-|}!{(1,-1)},
!{(0,-1.2)}*{\rm unit\ cell}, 
}
\end{align}
Note that $\hat U_{m}\ket{s \pm p} = \ket{s \mp p}$. 
After the half translation, and the inverse transformation $(\ket{s+p}, \ket{s-p}) \mapsto (\ket{s},\ket{p})$, 
we get the $\Z_2$ equivariant vector bundle $E_3 \oplus E_4$ : 
\begin{align}
E_1 \oplus E_2 \cong  \ \ 
\xygraph{
!{<0cm,0cm>;<1cm,0cm>:<0cm,1cm>::}
!{(1,0.2)}*{\bigcirc},
!{(3,0.2)}*{\bigcirc},
!{(-1,0.2)}*{\bigcirc},
!{(-3,0.2)}*{\bigcirc},
!{(1,0.6)}*{\ket{s}},
!{(3,0.6)}*{\ket{s}},
!{(-1,0.6)}*{\ket{s}},
!{(-3,0.6)}*{\ket{s}},
!{(-1,-0.2)}*{\bigcirc},
!{(1,-0.2)}*{\bigcirc},
!{(-3,-0.2)}*{\bigcirc},
!{(3,-0.2)}*{\bigcirc},
!{(-1,-0.6)}*{\ket{p}},
!{(1,-0.6)}*{\ket{p}},
!{(-3,-0.6)}*{\ket{p}},
!{(3,-0.6)}*{\ket{p}},
!{(-4,0)}-@{->}!{(4,0)},
!{(0,-1)}-@{.}!{(0,1.1)},
!{(0,1.2)}*{m},
!{(0.4,1)}-@{<->}!{(-0.4,1)},
!{(-1,-1)}-@{|-|}!{(1,-1)},
!{(0,-1.2)}*{\rm unit\ cell}, 
}
\ \  = E_3 \oplus E_4. 
\end{align}

The isomorphism $E_1 \oplus E_2 \cong E_3 \oplus E_4$ is written as the $k_x$-dependent unitary transformation 
in the Bloch basis 
\begin{equation}\begin{split}
\Phi_{E_1 \oplus E_2}(k_x) = (\ket{k_x}_1, \ket{k_x}_2)
&\mapsto \Phi_{E_3 \oplus E_4}(k_x) 
= \Phi_{E_1 \oplus E_2}(k_x) \ V(k_x), \ \ \Phi_{E_3 \oplus E_4}(k_x) = (\ket{\phi_+(k_x)}, \ket{\phi_-(k_x)}), 
\end{split}\end{equation}
where $V(k_x)$ consists of $W \cdot T_{1/2}(k_x) \cdot W^{-1}$ with 
$W = \begin{pmatrix}
\frac{1}{\sqrt{2}} & \frac{1}{\sqrt{2}} \\
\frac{1}{\sqrt{2}} & -\frac{1}{\sqrt{2}} \\
\end{pmatrix}$ the change of the basis as $( \ket{s}, \ket{p} ) \mapsto (\ket{s+p}, \ket{s-p})$ and $T_{1/2}(k_x) = 
\begin{pmatrix}
e^{-i k_x/2} & 0 \\
0 & e^{i k_x/2} \\
\end{pmatrix}$ the half lattice translation for $\ket{s+p}, \ket{s-p}$. 
This unitary transformation connects Hamiltonians on $E_1 \oplus E_2$ and $E_3 \oplus E_4$. 
For example, the Hamiltonian $\hat H$ represented on the basis $\Phi_{E_3 \oplus E_4}$ as 
$\hat H \Phi_{E_3 \oplus E_4}(k_x) = \Phi_{E_3 \oplus E_4}(k_x) \ \sigma_z$ is represented on the basis $\Phi_{E_1 \oplus E_2}(k_x)$ as 
\begin{align}
\hat H \Phi_{E_1 \oplus E_2}(k_x) = \Phi_{E_1 \oplus E_2}(k_x) H_{E_1 \oplus E_2}(k_x), && 
H_{E_1 \oplus E_2}(k_x) = V(k_x) \sigma_z V(k_x)^{\dag} = \cos k_x \sigma_z + \sin k_x \sigma_y. 
\end{align}
This is nothing but the equivalence relation (\ref{Eq:EqivalenceE1E2E3}).

\subsection{Half lattice translation symmetry}
\subsubsection{Preliminarily}
The most simple nonsymmorphic symmetry is half lattice translation symmetry in 1d. 
The symmetry group is $\Z_2 = \{1, \sigma\}$ and the nontrivial $\Z_2$ action is the half lattice translation $\sigma : x \mapsto x+\frac{1}{2}$. 
The twist $\tau_{p,p'}(k_x) \in Z^2(\Z_2;C(S^1;\R/2 \pi Z))$ is fixed as 
\begin{equation}
e^{i \tau_{p,p'}(k_x)} \qquad = \qquad 
\begin{tabular}{c|cc}
$p \backslash p'$ & $1$ & $\sigma$ \\ 
\hline 
$1$ & $1$ & $1$ \\ 
$\sigma$ & $1$ & $ e^{-i k_x}$ \\ 
\end{tabular}
\label{Tab:1DTwist}
\end{equation}
On the Hamiltonian, the half translational symmetry is written as 
\begin{align}
U_{\sigma}(k_x) H(k_x) U^{-1}_{\sigma}(k_x) = H(k_x), &&
[ U_{\sigma}(k_x) ]^2 = e^{- i k_x}. 
\end{align}

\begin{figure}[!]
 \begin{center}
  \includegraphics[width=0.5\linewidth, trim=0cm 0cm 0cm 0cm]{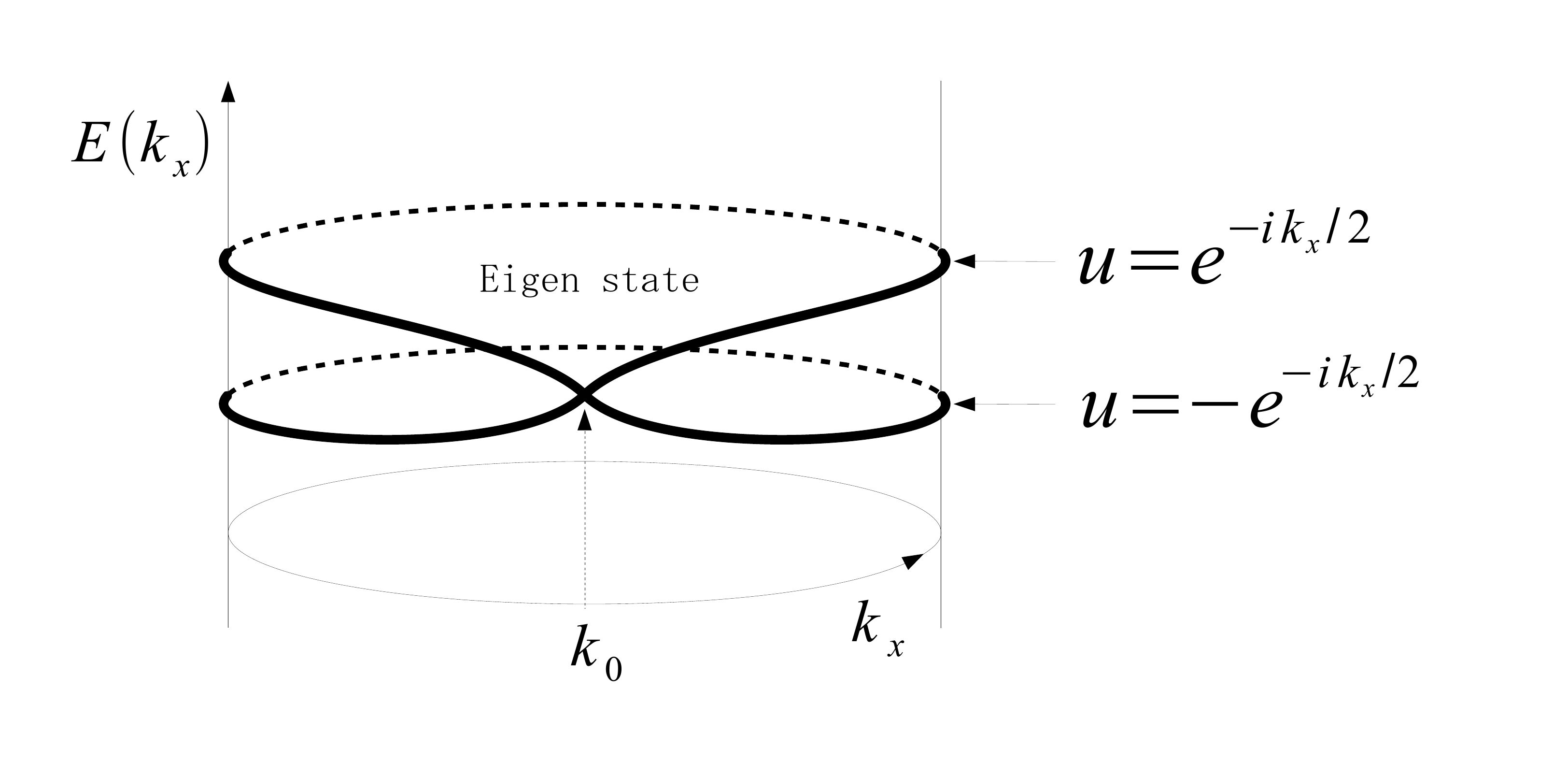}
 \end{center}
 \caption{Structure of the energy eigenstates with the half lattice translation symmetry. }
 \label{Fig:1DTwist}
\end{figure}
A characteristic property of the half lattice translation is 
the crossing of the pair of eigenstates of $U_{\sigma}(k_x)$. 
Because of $[ U_{\sigma}(k_x) ]^2 = e^{- i k_x}$, 
eigenvalues of $U_{\sigma}(k_x)$ can not be globally defined on the BZ $S^1$. 
We have eigenvalues $u(k_x) = \pm e^{-i k_x/2}$ in local region of $S^1$. 
Globally, two eigenstates with eigenvalues $u(k_x) =  \pm e^{-i k_x/2}$ are connected since 
the continuum change of the eigenvalue by $k_x \mapsto k_x + 2 \pi$ leads to the interchange of the eigenvalues 
\begin{equation}
(e^{-ik_x/2},-e^{-ik_x/2}) \underset{k_x \mapsto k_x + 2 \pi}{\mapsto} (-e^{-ik_x/2},e^{-ik_x/2}). 
\label{Eq:EigenvalueTwist}
\end{equation}
See Fig. \ref{Fig:1DTwist}. 
Especially, the pair of eigenstates with $u = \pm e^{-i k_x/2}$ should cross somewhere. 

From the interchange of eigenvalues (\ref{Eq:EigenvalueTwist}), 
we expect that when we use the Mayer-Vietoris sequence the gluing of two lines at 
$k_x = \pi/2$ and $-\pi/2$ should have relative twisting of the eigenstates of $U_{\sigma}(k_x)$. 
If we take the gluing condition for $k_x = \pi/2$ in a proper way, then that for $k_x = - \pi/2$ is twisted, as shown in (\ref{Eq:1DTwist_MV}) below.

\subsubsection{Topological classification}
We want to calculate the twisted equivariant $K$-theory $K^{\tau+ n}_{\Z_2}(S^1)$, 
where $\Z_2$ trivially acts on $S^1$ as $\sigma : k_x \mapsto k_x$, 
and the twist $\tau$ is given by (\ref{Tab:1DTwist}). 
To apply the Mayer-Vietoris sequence to $S^1 = U \cup V$, 
We divide $S^1$ into two intervals 
\begin{align}
U = \{e^{ik_x} \in \tilde S^1 | k_x \in [-\pi/2,\pi/2]\}, &&
V = \{e^{ik_x} \in \tilde S^1 | k_x \in [\pi/2,3\pi/2]\}. 
\label{Eq:1DTwistS1ToUV}
\end{align}
The intersection is 
\begin{equation}
U \cap V = \{ \frac{\pi}{2} \} \sqcup \{ -\frac{\pi}{2} \}. 
\end{equation}
The sequence of the inclusions 
\begin{equation}
\begin{CD}
\xy*\cir<1cm>{}\endxy 
@<<<
\xygraph{
!{<0cm,0cm>;<1cm,0cm>:<0cm,1cm>::}
!{(-0.3,0)}*{\cir<1cm>{l^r}}
!{(0.3,0)}*{\cir<1cm>{r^l}}
}@<<<
\xygraph{
!{<0cm,0cm>;<1cm,0cm>:<0cm,1cm>::}
!{(0,0.7)}*{\bullet}
!{(0.5,0.7)}*{\{\frac{\pi}{2} \}}
!{(0,-0.7)}*{\bullet}
!{(0.6,-0.7)}*{\{-\frac{\pi}{2} \}}
}\\
S^1 = U \cup V @<<< U \sqcup V @<<< U \cap V \\
\end{CD}
\end{equation}
induces the six term exact sequence of the twisted equivariant $K$-theory 
\begin{equation}
\begin{CD}
K_{\Z_2}^{\tau|_{U \cap V}+1}(U \cap V) @<<< K_{\Z_2}^{\tau|_U+1}(U) \oplus K_{\Z_2}^{\tau|_V+1}(V) @<<< K_{\Z_2}^{\tau+1}(S^1) \\
@VVV @. @AAA \\
K_{\Z_2}^{\tau+0}(S^1) @>>> K_{\Z_2}^{\tau|_U+0}(U) \oplus K_{\Z_2}^{\tau|_V+0}(V) @>\Delta>> K_{\Z_2}^{\tau|_{U \cap V}+0}(U \cap V).
\end{CD}
\label{Seq:1DTwist}
\end{equation}
Here, the twists on $U, V,  U \cap V$ are given by the restrictions of the twist $\tau_{\sigma,\sigma}(k_x) = e^{- ik_x}$ to them, and these twists are trivial. 
In fact, the twists $\tau|_U, \tau|_V, \tau|_{U \cap V}$ are exact 
\begin{align}
\label{Eq:1DTwistTrivU}
&\tau|_U = \delta \beta^U, && \beta^U_{1}(k_x) = 1, \ \ \beta^U_{\sigma}(k_x) = e^{-i k_x/2}, && k_x \in \left[-\frac{\pi}{2},\frac{\pi}{2} \right], \\
&\tau|_V = \delta \beta^V, && \beta^V_{1}(k_x) = 1, \ \ \beta^V_{\sigma}(k_x) = e^{-i k_x/2}, && k_x \in \left[\frac{\pi}{2},\frac{3\pi}{2} \right], \\
\label{Eq:1DTwistTrivUV}
&\tau_{U \cap V} = \delta \beta^{U \cap V}, && \beta^{U \cap V}_1( \pm \frac{\pi}{2} ) = 1, \ \ \beta^{U \cap V}_{\sigma}( \pm \frac{\pi}{2} ) = e^{ \mp i \pi/4 }. && 
\end{align}
Note that $\beta^U_{\sigma}(k_x)$ and $\beta^V_{\sigma}(k_x)$ correspond to local eigenvalues of $U_{\sigma}(k_x)$. 
In these trivializations, two eigenstates are connected at $\{ i \}$ and twisted at $\{ -i \}$. 
By using these trivializations, we have 
\begin{align}
&K_{\Z_2}^{\tau|_U+n}(U)
\overset{\beta^U}{\cong} K_{\Z_2}^{n}(U)
\cong K_{\Z_2}^{n}(pt) 
\cong \left\{ \begin{array}{ll}
R(\Z_2) & (n=0) \\
0 & (n=1) \\
\end{array} \right., \\
&K_{\Z_2}^{\tau|_U+n}(V)
\overset{\beta^V}{\cong} K_{\Z_2}^{n}(V)
\cong K_{\Z_2}^{n}(pt) 
\cong \left\{ \begin{array}{ll}
R(\Z_2) & (n=0) \\
0 & (n=1) \\
\end{array} \right., \\
&K_{\Z_2}^{\tau|_{U \cup V}+n}(U \cap V)
\overset{\beta^{U \cap V}}{\cong} K_{\Z_2}^{n}( U \cap V)
\cong K_{\Z_2}^{n}( \{i \} \sqcup \{-i\}) 
\cong \left\{ \begin{array}{ll}
R(\Z_2) \oplus R(\Z_2) & (n=0) \\
0 & (n=1) \\
\end{array} \right.. 
\end{align}
Then, one may expect that the homomorphism $\Delta : K_{\Z_2}^{\tau|_U+0}(U) \oplus K_{\Z_2}^{\tau|_V+0}(V)\to  K_{\Z_2}^{\tau|_{U \cap V}+0}(U \cap V)$ 
is given by 
\begin{align}
j_U^* - j_V^* : K^{n}_{\Z_2}(pt) \oplus K^{n}_{\Z_2}(pt) \to K^{n}_{\Z_2}(\{ \frac{\pi}{2} \} \cap \{ -\frac{\pi}{2}\}) , && 
(f(t), g(t)) \mapsto (\underbrace{f(t) - g(t)}_{\{\pi/2\}}, \underbrace{f(t) - g(t)}_{\{-\pi/2\}}) \qquad 
({\rm wrong!}). 
\end{align}
This is really wrong because of not respecting the global structure of the twist. 
The correct one is 
\begin{align}
\Delta = \alpha^U j_U^* - \alpha^V j_V^* : K^{n}_{\Z_2}(pt) \oplus K^{n}_{\Z_2}(pt) \to K^{n}_{\Z_2}(\{ \frac{\pi}{2} \} \cap \{ -\frac{\pi}{2}\})
\label{Eq:1DTwistDelta'}
\end{align}
with $\alpha^U , \alpha^V : K^{n}_{\Z_2}(U \cap V) \to K^{n}_{\Z_2}(U \cap V)$  defined by 
\begin{align}
&\alpha^U := \beta_{U \cap V} (\beta^U)^{-1}, && \alpha^U_1(\pm \frac{\pi}{2}) = 1, \ \ \alpha^U_{\sigma}(\pm \frac{\pi}{2}) = 1, \\
&\alpha^V := \beta_{U \cap V} (\beta^V)^{-1}, && \alpha^V_1(\pm \frac{\pi}{2}) = 1, \ \ \alpha^V_{\sigma}(\pm \frac{\pi}{2}) = \pm 1.
\end{align}
Here $\alpha^V_{\sigma}= -1$ corresponds to the change of the eigenvalues as $(1,-1) \mapsto (-1) \cdot (1,-1) = (-1,1)$, which is equivalent to 
the action of $t \in R(\Z_2)$. Thus we have 
\begin{align}
\Delta : (f(t), g(t)) \mapsto (\underbrace{f(t) - g(t)}_{\{\pi/2\}}, \underbrace{f(t) - t g(t)}_{\{-\pi/2\}}). 
\label{Eq:1DTwist_MV}
\end{align}

From the the Mayer-Vietoris sequence (\ref{Seq:1DTwist}), we have 
\begin{align}
K^{\tau+0}_{\Z_2}(S^1) \cong \mathrm{Ker} (\Delta), && 
K^{\tau+1}_{\Z_2}(S^1) \cong \mathrm{Coker} (\Delta). 
\label{Eq:1DTwistKGroup}
\end{align}
From (\ref{Eq:1DTwist_MV}), we find $\mathrm{Ker}(\Delta) = \Z$ as an Abelian group, and 
the generator of $\Z$ is characterized by $(1+t,1+t) \in R(\Z_2) \oplus R(\Z_2)$. 
Thus we have 
\begin{align}
K^{\tau+0}_{\Z_2}(S^1) \cong \overbrace{(1+t)}^{\Z}  \ \ ({\rm class\ A}). 
\label{Eq:1DTwistKGroup_n=0}
\end{align}
Here, $(1+t)$ is the $R(\Z_2)$-ideal $(1+t) = \{(1+t) f(t) | f(t) \in R(\Z_2)\}$. 

Since $\mathrm{Im}(\Delta) \subset R(\Z_2) \oplus R(\Z_2)$ is spanned by $\{(1,1), (t,t), (1,t)\}$, we have 
\begin{align}
\mathrm{Coker} (\Delta)
= (R(\Z_2) \oplus R(\Z_2)) / \mathrm{Im}(\Delta)
= \Z
\end{align}
as an Abelian group. 
The generator of $\mathrm{Coker} (\Delta) = \Z$ is represented by $[(1,0)]$ with $(1,0) \in R(\Z_2) \oplus R(\Z_2)$, in which the $R(\Z_2)$ action is given by 
$t \cdot (1,0) = (t,0) \sim (1,0)$, leading to $\mathrm{Coker} (\Delta) \cong (1+t)$ as an $R(\Z_2)$-module. 
Thus, we have 
\begin{align}
K^{\tau+1}_{\Z_2}(S^1) \cong \overbrace{(1+t)}^{\Z}  \ \ ({\rm class\ AIII}). 
\label{Eq:1DTwistKGroup_n=1}
\end{align}

\subsubsection{Vector bundle representation for $K^{\tau+0}_{\Z_2}(S^1)$}
Here we give the vector bundle representation and the corresponding real space orbital picture. 
The generator of the $K$-group $e \in K^{\tau+0}_{\Z_2}(S^1) = (1+t)$ is represented by the 
following $\Z_2$ twisted equivariant bundle $E$, 
\begin{align}
e= \left[ \left( E = S^1 \times \C^2, \ \ \rho_{\sigma}(k_x,v) = (k_x, U_{\sigma}(k_x) v), \ \ U_{\sigma}(k_x) = \begin{pmatrix}
0 & e^{-i k_x} \\
1 & 0 \\
\end{pmatrix} \right) \right] . 
\label{Eq:1DTwistVectE}
\end{align}
By using the Bloch states, $e$ is written as 
\begin{align}
e = \left[ \left( \Phi(k_x) = ( \ket{k_x,A}, \ket{k_x,B} ), \ \ \hat U_{\sigma} \Phi(k_x) = \Phi(k_x) U_{\sigma}(k_x) \right) \right] . 
\label{Eq:1DTwistBlochE}
\end{align}
By using the real space basis $\ket{R_x,\alpha} =\sum_{k_x \in S^1} \ket{k_x, \alpha} e^{-i k_x R_x}, \ (\alpha = A,B)$, we can write $e$ as 
\begin{align}
e = \left[ \left( \Phi(R_x) = ( \ket{R_x,A}, \ket{R_x,B} ), \ \ \hat U_{\sigma} \Phi(R_x) = ( \ket{R_x,B}, \ket{R_x+1,A}) \right) \right] . 
\end{align}
Thus, $e$ just describes the two atoms $\ket{R_x,A}$ and $\ket{R_x,B}$ exchanged under the half translation $\hat U_{\sigma}$, 
which is figured as: 
\begin{align}
e= \left[ \ \ 
\xygraph{
!{<0cm,0cm>;<1cm,0cm>:<0cm,1cm>::}
!{(0.5,0)}*{\bigcirc},
!{(-0.5,0)}*{\bigcirc},
!{(1.5,0)}*{\bigcirc},
!{(-1.5,0)}*{\bigcirc},
!{(2.5,0)}*{\bigcirc},
!{(-2.5,0)}*{\bigcirc},
!{(3.5,0)}*{\bigcirc},
!{(-3.5,0)}*{\bigcirc},
!{(4.5,0)}*{\bigcirc},
!{(-4.5,0)}*{\bigcirc},
!{(0.5,-0.4)}*{B},
!{(-0.5,-0.4)}*{A},
!{(1.5,-0.4)}*{A},
!{(-1.5,-0.4)}*{B},
!{(2.5,-0.4)}*{B},
!{(-2.5,-0.4)}*{A},
!{(3.5,-0.4)}*{A},
!{(-3.5,-0.4)}*{B},
!{(4.5,-0.4)}*{B},
!{(-4.5,-0.4)}*{A},
!{(-5,0)}-@{->}!{(5,0)},
!{(0,0.8)}*{\sigma},
!{(-0.4,0.4)}-@/^0.2cm/@{->}!{(0.4,0.4)},
!{(1,0.8)}*{\sigma},
!{(0.6,0.4)}-@/^0.2cm/@{->}!{(1.4,0.4)},
!{(-1,-0.8)}-@{|-|}!{(1,-0.8)},
!{(0,-1)}*{\rm unit\ cell}, 
}
\ \ \right]. 
\label{Fig:1DTCI_NS_e1}
\end{align}

\subsubsection{Vector bundle representation for $K^{\tau+1}_{\Z_2}(S^1)$}
Here we give a representation of the generator $1_q \in K^{\tau+1}_{\Z_2}(S^1) = (1+t)$ by an automorphism $q : E \to E$, 
where $E$ is the $\Z_2$ twisted equivariant bundle introduced in $(\ref{Eq:1DTwistVectE})$. 
Because $E = S^1 \times \C^2$ is trivial as a complex vector bundle of rank $2$, 
$q : E \to E$ amounts to a function with values in $2 \times 2$ unitary matrices $q : S^1 \to U(2)$, 
and $q(k_x)$ commutes with the half lattice translation symmetry 
\begin{align}
U_{\sigma}(k_x) q(k_x) U^{-1}_{\sigma}(k_x) = q(k_x). 
\end{align}
We can define the topological invariant $W$ charactering $q(k_x)$ as 
\begin{align}
W := \frac{1}{2 \pi i } \oint_{S^1} \mathrm{tr} [q^{\dag} d q]. 
\end{align}
The generator model $q(k_x)$ is characterized by $W = 1$. 
The simplest model is given by 
\begin{align}
q(k_x) = 
\begin{pmatrix}
0 & 1 \\
e^{i k_x} & 0
\end{pmatrix}. 
\end{align}
Thus, we have a representation of the generator $1_q  \in K^{\tau+1}_{\Z_2}(S^1)$ as 
\begin{align}
1_q= \left[ \left( q : E \to E, \ \ q(k_x) = \begin{pmatrix}
0 & 1 \\
e^{i k_x} & 0
\end{pmatrix}
 \right) \right] . 
\end{align}
By using the Bloch state representation for $E$ in (\ref{Eq:1DTwistBlochE}), 
$q : E \to E$ is written in the second quantized form 
\begin{align}
1_q= \left[ \ \  \hat q  = \sum_{k_x \in S^1} \left( \psi^{\dag}_{f, A}(k_x), \psi^{\dag}_{f, B}(k_x) \right) q(k_x) \begin{pmatrix}
\psi_{i,A}(k_x) \\
\psi_{i,B}(k_x) \\
\end{pmatrix} \ \  \right] , 
\end{align}
where $\{ i,f \}$ are auxiliary indices which distinguish between initial and final states. 
In the real space basis, $1_q$ can be written as the following hopping model 
\begin{align}
1_q= \left[ \ \  \hat q  = \sum_{R_x \in \Z} \left( \psi^{\dag}_{f, A}(R_x) \psi_{i,B}(R_x) + \psi^{\dag}_{f, B}(R_x) \psi_{i,A}(R_x+1) \right) \ \  \right] , 
\end{align}
which is figured out as 
\begin{align}
1_q= \left[ \ \ 
\xygraph{
!{<0cm,-0.5cm>;<1cm,-0.5cm>:<0cm,0.5cm>::}
!{(-5.5,0)}*{E_i},
!{(0.5,0)}*{\bigcirc},
!{(-0.5,0)}*{\bigcirc},
!{(1.5,0)}*{\bigcirc},
!{(-1.5,0)}*{\bigcirc},
!{(2.5,0)}*{\bigcirc},
!{(-2.5,0)}*{\bigcirc},
!{(3.5,0)}*{\bigcirc},
!{(-3.5,0)}*{\bigcirc},
!{(4.5,0)}*{\bigcirc},
!{(-4.5,0)}*{\bigcirc},
!{(0.5,-0.4)}*{B},
!{(-0.5,-0.4)}*{A},
!{(1.5,-0.4)}*{A},
!{(-1.5,-0.4)}*{B},
!{(2.5,-0.4)}*{B},
!{(-2.5,-0.4)}*{A},
!{(3.5,-0.4)}*{A},
!{(-3.5,-0.4)}*{B},
!{(4.5,-0.4)}*{B},
!{(-4.5,-0.4)}*{A},
!{(-5,0)}-@{->}!{(5,0)},
!{(-5.5,1)}*{E_f},
!{(0.5,1)}*{\bigcirc},
!{(-0.5,1)}*{\bigcirc},
!{(1.5,1)}*{\bigcirc},
!{(-1.5,1)}*{\bigcirc},
!{(2.5,1)}*{\bigcirc},
!{(-2.5,1)}*{\bigcirc},
!{(3.5,1)}*{\bigcirc},
!{(-3.5,1)}*{\bigcirc},
!{(4.5,1)}*{\bigcirc},
!{(-4.5,1)}*{\bigcirc},
!{(0.5,1.4)}*{B},
!{(-0.5,1.4)}*{A},
!{(1.5,1.4)}*{A},
!{(-1.5,1.4)}*{B},
!{(2.5,1.4)}*{B},
!{(-2.5,1.4)}*{A},
!{(3.5,1.4)}*{A},
!{(-3.5,1.4)}*{B},
!{(4.5,1.4)}*{B},
!{(-4.5,1.4)}*{A},
!{(-5,1)}-@{->}!{(5,1)},
!{(0.4,0.1)}-@{->}!{(-0.4,0.9)},
!{(1.4,0.1)}-@{->}!{(0.6,0.9)},
!{(2.4,0.1)}-@{->}!{(1.6,0.9)},
!{(3.4,0.1)}-@{->}!{(2.6,0.9)},
!{(4.4,0.1)}-@{->}!{(3.6,0.9)},
!{(-0.6,0.1)}-@{->}!{(-1.4,0.9)},
!{(-1.6,0.1)}-@{->}!{(-2.4,0.9)},
!{(-2.6,0.1)}-@{->}!{(-3.4,0.9)},
!{(-3.6,0.1)}-@{->}!{(-4.4,0.9)},
!{(-0.7,0.5)}*{1},
!{(0.3,0.5)}*{1},
}
\ \ \right]. 
\label{Fig:1DTCI_NS_e1_AIII}
\end{align}

\subsubsection{Hamiltonian representation for $K^{\tau+1}_{\Z_2}(S^1)$}
We give the Hamiltonian representation for $K^{\tau+1}_{\Z_2}(S^1)$. 
If an automorphism representation $q : E \to E$ is obtained, 
the Hamiltonian $H_q$ with the chiral symmetry $\Gamma H_q + H_q \Gamma = 0$ is given by  
\begin{align}
H_q = 
\begin{pmatrix}
0 & q^{\dag} \\
q & 0
\end{pmatrix}
 \ \ {\rm with\ \ }
\Gamma = 
\begin{pmatrix}
1 & 0 \\
0 & -1
\end{pmatrix}. 
\end{align}
In the second quantized form, this means $\hat H_q = \hat q + \hat q^{\dag}$.

\subsection{Glide symmetry}
Let us consider the glide symmetry which is a nonsymmorphic wallpaper group generated by 
$\sigma : (x,y) \mapsto (x+\frac{1}{2},-y)$. 
The point group is $\Z_2 = \{1,\sigma\}$ which acts on the BZ torus as 
\begin{align}
\sigma : (k_x,k_y) \mapsto (k_x,-k_y). 
\end{align}
The twist $(\tau_{\sf pg})_{p,p'}(k_x,k_y) \in Z^2(\Z_2;C(T^2,\R/2 \pi \Z))$ of the glide symmetry is given by 
\begin{equation}
e^{i (\tau_{\sf pg})_{p,p'}(k_x,k_y)} \qquad = \qquad 
\begin{tabular}{c|cc}
$p \backslash p'$ & $1$ & $\sigma$ \\ 
\hline 
$1$ & $1$ & $1$ \\ 
$\sigma$ & $1$ & $ e^{-i k_x}$ \\ 
\end{tabular}
\label{Tab:2DTwist}
\end{equation}
Hamiltonians with the glide symmetry are written as 
\begin{align}	
U_{\sigma}(k_x,k_y) H(k_x,k_y) U^{-1}_{\sigma}(k_x,k_y) = H(k_x,-k_y), &&
U_{\sigma}(k_x,-k_y) U_{\sigma}(k_x,k_y) = e^{- i k_x}. 
\end{align}

\subsubsection{Topological classification}
To apply the Mayer-Vietoris sequence to the BZ torus $T^2$, 
we divide $T^2$ into two cylinders $U$ and $V$ so that 
\begin{align}
U = \left\{(e^{i k_x},e^{i k_y}) \in T^2 | - \frac{\pi}{2} \leq k_x \leq \frac{\pi}{2} \right\}, && 
V = \left\{(e^{i k_x},e^{i k_y}) \in T^2 | \frac{\pi}{2} \leq k_x \leq \frac{3 \pi}{2} \right\}.
\end{align}
The intersection consists of two circles 
\begin{align}
U \cap V = \{\pi/2\} \times \tilde S^1 \sqcup \{-\pi/2\} \times \tilde S^1. 
\end{align}
$U$ and $V$ are $\Z_2$ equivariantly homotopic to $S^1$: 
\begin{align}
U \sim \{0\} \times \tilde S^1, && 
V \sim \{\pi\} \times \tilde S^1.  
\end{align}
Here, we denote the $\Z_2$-space $S^1$ with the reflection symmetry by $\tilde S^1$ as introduced previously in (\ref{Fig:TildeS1}). 
In the same way as (\ref{Eq:1DTwistTrivU}) - (\ref{Eq:1DTwistTrivUV}), 
the twist on $U, V, U \cap V$ can be trivialized as 
\begin{align}
&(\tau_{\sf pg})|_U = \delta \beta^U, && \beta^U_{1}(k_x,k_y) = 1, \ \ \beta^U_{\sigma}(k_x,k_y) = e^{-i k_x/2}, && k_x \in \left[-\frac{\pi}{2},\frac{\pi}{2} \right], \\
&(\tau_{\sf pg})|_V = \delta \beta^V, && \beta^V_{1}(k_x,k_y) = 1, \ \ \beta^V_{\sigma}(k_x,k_y) = e^{-i k_x/2}, && k_x \in \left[\frac{\pi}{2},\frac{3\pi}{2} \right], \\
&(\tau_{\sf pg})|_{U \cap V} = \delta \beta^{U \cap V}, && \beta^{U \cap V}_1( \pm \frac{\pi}{2} ,k_y) = 1, \ \ \beta^{U \cap V}_{\sigma}( \pm \frac{\pi}{2},k_y ) = e^{ \mp i \pi/4 }. && 
\end{align}
By using these trivializations and the $K$-group of $\tilde S^1$ (\ref{Eq:KGroupS1}), we have 
\begin{align}
&K_{\Z_2}^{(\tau_{\sf pg})|_U+n}(U)
\overset{\beta^U}{\cong} K_{\Z_2}^{n}(U)
\cong K_{\Z_2}^{n}(\{0\} \times \tilde S^1) 
\cong \left\{ \begin{array}{ll}
R(\Z_2) \oplus (1-t) & (n=0) \\
0 & (n=1) \\
\end{array} \right., \\
&K_{\Z_2}^{(\tau_{\sf pg})|_U+n}(V)
\overset{\beta^V}{\cong} K_{\Z_2}^{n}(V)
\cong K_{\Z_2}^{n}(\{\pi\} \times \tilde S^1) 
\cong \left\{ \begin{array}{ll}
R(\Z_2) \oplus (1-t) & (n=0) \\
0 & (n=1) \\
\end{array} \right.,
\end{align}
\begin{equation}\begin{split}
K_{\Z_2}^{(\tau_{\sf pg})|_{U \cup V}+n}(U \cap V)
&\overset{\beta^{U \cap V}}{\cong} K_{\Z_2}^{n}( U \cap V)
\cong K_{\Z_2}^{n}( \{\pi/2 \} \times \tilde S^1 \sqcup \{-\pi/2\} \times \tilde S^1) \\
&\cong \left\{ \begin{array}{ll}
( R(\Z_2) \oplus (1-t) ) \oplus ( R(\Z_2) \oplus (1-t) ) & (n=0) \\
0 & (n=1) \\
\end{array} \right.. 
\end{split}\end{equation}
Thus, the Mayer-Vietoris sequence reads 
\begin{equation}
\begin{CD}
0 @<<< 0 @<<< K_{\Z_2}^{\tau_{\sf pg}+1}(T^2) \\
@VVV @. @AAA \\
K_{\Z_2}^{\tau_{\sf pg}+0}(T^2) @>>> K_{\Z_2}^{(\tau_{\sf pg})|_U+0}(U) \oplus K_{\Z_2}^{(\tau_{\sf pg})|_V+0}(V) @>\Delta>> K_{\Z_2}^{(\tau_{\sf pg})|_{U \cap V}+0}(U \cap V).
\end{CD}
\end{equation}
Then, in the same way as (\ref{Eq:1DTwistDelta'}) - (\ref{Eq:1DTwistKGroup}), 
the $K$-group $K_{\Z_2}^{\tau_{\sf pg}+n}(T^2)$ is given by 
\begin{align}
K_{\Z_2}^{\tau_{\sf pg}+0}(T^2) \cong \mathrm{Ker}(\Delta), &&
K_{\Z_2}^{\tau_{\sf pg}+1}(T^2) \cong \mathrm{Coker}(\Delta), 
\end{align}
with 
\begin{align}
\Delta: 
\underbrace{R(\Z_2) \oplus (1-t)}_{\{0\} \times \tilde S^1} \oplus \underbrace{R(\Z_2) \oplus (1-t)}_{\{\pi\} \times \tilde S^1} 
&\to
\underbrace{R(\Z_2) \oplus (1-t)}_{\{\pi/2\} \times \tilde S^1} \oplus \underbrace{R(\Z_2) \oplus (1-t)}_{\{-\pi/2\} \times \tilde S^1}, &&
(x,y) \mapsto (x-y, x-t y), 
\end{align}
where $\Delta$ is $\Delta = \alpha_U j_U^* - \alpha_V j_V^*$ with $\alpha_U := \beta^{U \cap V}(\beta^U)^{-1}$ and $\alpha_V := \beta^{U \cap V}(\beta^V)^{-1}$. 
Note that $x,y \in R(\Z_2) \oplus (1-t)$ are glued with the twist by $t \in R(\Z_2)$ on the circle $\{-\pi/2\} \times \tilde S^1$. 
On the direct summands $R(\Z_2) \oplus R(\Z_2)$ and $(1-t) \oplus (1-t)$, 
the homomorphism $\Delta$ takes the following forms 
\begin{align}
&\Delta|_{R(\Z_2) \oplus R(\Z_2)} : R(\Z_2) \oplus R(\Z_2) \to R(\Z_2) \oplus R(\Z_2), && \big( f(t), g(t) \big) \mapsto \big( f(t) - g(t), f(t) - t g(t) \big), \\
&\Delta|_{(1-t) \oplus (1-t)}: (1-t)\oplus (1-t) \to (1-t) \oplus (1-t), &&  \big( n(1-t),m(1-t) \big) \mapsto \big( (n - m)(1-t) , (n + m)(1-t) \big).
\end{align}
Note that $t (1-t) = - (1-t)$. 
As a result, we get 
\begin{align}
K_{\Z_2}^{\tau_{\sf pg}+0}(T^2) \cong \overbrace{(1+t)}^{\Z}\ \ \ \ \ ({\rm class\ A}), && 
K_{\Z_2}^{\tau_{\sf pg}+1}(T^2) \cong \overbrace{(1+t)}^{\Z} \oplus \overbrace{I}^{\Z_2} \ \ \ \ \ ({\rm class\ AIII}). 
\label{Eq:2DPgKGroup}
\end{align}
We denoted Abelian groups in the overbraces. 
A generator of $I = \Z_2$ is represented by $a = ((1-t),0) \in (1-t) \oplus (1-t)$. 
The $R(\Z_2)$ action on $I$ is trivial because $t \cdot ((1-t),0) = (-(1-t),0) \sim ((1-t),0)$.

\subsubsection{Alternative derivation : Gysin sequence}
In the last subsection, we computed the $K$-group on the $2$-dimensional torus directly. 
There is an alternative derivation of (\ref{Eq:2DPgKGroup}) by using the Gysin sequence as discussed in Ref.~\onlinecite{ShiozakiSatoGomi2015}. 
Here, we will briefly describe this method. 

Let $\pi:S^1 \times \tilde S^1 \to S^1$ be the projection onto the $k_x$-direction. 
The twisting $\tau_{\sf pg}$ defined in (\ref{Tab:2DTwist}) arises only from the $k_x$-direction, 
which means the twisting $\tau_{\sf pg}$ of the glide symmetry is realized as the pull back 
$\tau_{\sf pg}=\pi^* \tau$ 
of the twisting of the half-lattice translation defined in (\ref{Tab:1DTwist}). 
Applying the Gysin sequence associated with the reflection 
(That is explained in Appendix \ref{Sec:Gysin} and the relevant isomorphism is (\ref{eq:app_Gysin_reflection}).) to 
$T^2 = S^1 \times \tilde S^1$, 
we have the isomorphism of $R(\Z_2)$-modules 
\begin{align}
K^{\tau_{\sf pg}+n}_{\Z_2}(S^1 \times \tilde S^1)
= K^{(\tau,0)+n}_{\Z_2}(S^1) \oplus K^{(\tau,w)+n+1}_{\Z_2}(S^1).
\label{Eq:2DPgGysin}
\end{align}
The first direct summand represents just a ``weak" index, say, the contribution from the $k_y$-independent Hamiltonians, 
which is already given in (\ref{Eq:1DTwistKGroup_n=0}) and (\ref{Eq:1DTwistKGroup_n=1}) as 
\begin{equation}
K^{(\tau,0)+n}_{\Z_2}(S^1) = \left\{ \begin{array}{ll}
(1+t) & (n=0) \\
(1+t) & (n=1) \\
\end{array} \right..
\label{Eq:1DTwistKGroup_pg}
\end{equation}
So, the second direct summand $K^{(\tau,w)+n+1}_{\Z_2}(S^1)$ is a contribution specific to 2d. 
The problem is recast into the 1d problem $K^{(\tau,w)+n+1}(S^1)$. 

In the exponent of the $K$-group $K^{(\tau,w)+n}(S^1)$, 
$c=w$ means the ``antisymmetry class" $c(\sigma)=-1$ introduced in Sec.~\ref{sec:Anti space group} 
which is defined for Hamiltonians by 
\begin{align}
&{\rm class\ A} \ \ (n=0) : 
\left\{ \begin{array}{l}
U_{\sigma}(k_x) H(k_x) U^{-1}_{\sigma}(k_x) = - H(k_x), \\
{[U_{\sigma}(k_x)]}^2 = e^{-i k_x}, \\
\end{array} \right. \\
&{\rm class\ AIII} \ \ (n=1) : 
\left\{ \begin{array}{l}
\Gamma H(k_x) \Gamma^{-1} = - H(k_x), \\
U_{\sigma}(k_x) H(k_x) U^{-1}_{\sigma}(k_x) = - H(k_x), \\
{[U_{\sigma}(k_x)]}^2 = e^{-i k_x}, \\
\Gamma U_{\sigma}(k_x) = - U_{\sigma}(k_x) \Gamma. 
\end{array} \right.  
\end{align}

By the same decomposition $S^1 = U \cup V$ as (\ref{Eq:1DTwistS1ToUV}) and 
the same trivialization of the twist $\tau$ on $U, V, U \cap V$ as (\ref{Eq:1DTwistTrivU} - \ref{Eq:1DTwistTrivUV}), 
we can show the following 
\begin{align}
&K_{\Z_2}^{(\tau|_U,w)+n}(U)
\overset{\beta^U}{\cong} K_{\Z_2}^{(0,w)+n}(U)
\cong K_{\Z_2}^{(0,w)+n}(pt) 
\cong \left\{ \begin{array}{ll}
0 & (n=0) \\
(1-t) & (n=1) \\
\end{array} \right., \\
&K_{\Z_2}^{(\tau|_U,w)+n}(V)
\overset{\beta^V}{\cong} K_{\Z_2}^{(0,w)+n}(V)
\cong K_{\Z_2}^{(0,w)+n}(pt) 
\cong \left\{ \begin{array}{ll}
0 & (n=0) \\
(1-t) & (n=1) \\
\end{array} \right., \\
&K_{\Z_2}^{(\tau|_{U \cup V},w)+n}(U \cap V)
\overset{\beta^{U \cap V}}{\cong} K_{\Z_2}^{(0,w)+n}( U \cap V)
\cong K_{\Z_2}^{(0,w)+n}( \{\pi/2 \} \sqcup \{-\pi/2\}) 
\cong \left\{ \begin{array}{ll}
0 & (n=0) \\
(1-t) \oplus (1-t) & (n=1) \\
\end{array} \right.. 
\end{align}
Here, the $K$-group of the point $K^{(0,w)+n}_{\Z_2}(pt)$ is given as follows. 
For $n=0$, the symmetry restricted to the point is the same as the chiral symmetry $U_{\sigma}(pt) H(pt) U_{\sigma}^{-1}(pt) = - H(pt)$, 
leading to $K^{(0,w)+0}_{\Z_2}(pt) = 0$. 
For $n=1$, from the double chiral symmetries by $U_{\sigma}(pt)$ and $\Gamma$ which anti-commute with each other, 
a symmetry-preserving Hamiltonian takes a form $H(pt) = \widetilde H(pt) \otimes (i U_{\sigma}(pt) \Gamma)$ 
with no symmetry for $\widetilde H(pt)$. 
(Here we assume $\Gamma^2 = U^2_{\sigma}(pt) = 1$.) 
Thus, the symmetry class is the same as class A and we find $K^{(0,w)+1}_{\Z_2}(pt) = \Z$ as an Abelian group. 
The $R(\Z_2)$-module structure is given by the Karoubi's quadruplet representation. 
A generator of $K^{(0,w)+1}_{\Z_2}(pt) = \Z$ is represented by 
\begin{align}
e = \left[ \left(\C_0 \oplus \C_1, \Gamma = \sigma_x, H_0(pt) = \sigma_y, H_1(pt) = - \sigma_y \right) \right] 
\end{align}
where $\sigma_i, (i=x,y,z)$ is the Pauli matrix, and 
$\C_0$ and $\C_1$ are 1-dimensional irreps with eigenvalues $U_{\sigma}(pt) = 1$ and $-1$, respectively. 
The $t \in R(\Z_2)$ action is 
\begin{equation}\begin{split}
t \cdot e 
&= \left[ \left(\C_1 \oplus \C_0, \Gamma = \sigma_x, H_0(pt) = \sigma_y, H_1(pt) = - \sigma_y \right) \right] \\
&= \left[ \left(\C_0 \oplus \C_1, \Gamma = \sigma_x, H_0(pt) = -\sigma_y, H_1(pt) = \sigma_y \right) \right] 
= -e, 
\end{split}\end{equation}
which leads to $K^{(0,w)+1}_{\Z_2}(pt) \cong (1-t)$.

The Mayer-Vietoris sequence for $S^1 = U \cup V$ is given by 
\begin{equation}
\begin{CD}
K_{\Z_2}^{(\tau|_{U \cap V},w)+1}(U \cap V) @<\Delta'<< K_{\Z_2}^{(\tau|_U,w)+1}(U) \oplus K_{\Z_2}^{(\tau|_V,w)+1}(V) @<<< K_{\Z_2}^{(\tau,w)+1}(S^1) \\
@VVV @. @AAA \\
K_{\Z_2}^{(\tau,w)+0}(S^1) @>>> 0 @>>> 0.
\end{CD}
\end{equation}
We have 
\begin{align}
K_{\Z_2}^{(\tau,w)+1}(S^1) 
\cong \mathrm{Ker}(\Delta') , &&
K_{\Z_2}^{(\tau,w)+0}(S^1) 
\cong \mathrm{Coker}(\Delta') , 
\end{align}
where $\Delta' = \alpha^U j^*_U - \alpha^V j_V^* : K_{\Z_2}^{(0,w)+1}(U) \oplus K_{\Z_2}^{(0,w)+1}(V) \to K_{\Z_2}^{(0,w)+1}(U \cap V)$ is 
\begin{align}
\Delta' : (1-t) \oplus (1-t) \to (1-t) \oplus (1-t), && 
\left( n(1-t), m(1-t) \right) \mapsto \left( (n-m)(1-t), (n+m)(1-t) \right). 
\end{align}
As a result, we get 
\begin{align}
K_{\Z_2}^{(\tau,w)+1}(S^1) = 0, && 
K_{\Z_2}^{(\tau,w)+0}(S^1) \cong \overbrace{I}^{\Z_2}, 
\label{Eq:1DTwist_w_KGroup}
\end{align}
where $R(\Z_2)$ trivially acts on $I = \Z_2$. 

Combining (\ref{Eq:2DPgGysin}) with (\ref{Eq:1DTwistKGroup_pg}) and (\ref{Eq:1DTwist_w_KGroup}) 
we re-provide the $K$-group (\ref{Eq:2DPgKGroup}) for 2d TCI with the glide symmetry 
\begin{align}
&K^{\tau_{\sf pg}+0}_{\Z_2}(S^1 \times \tilde S^1)
\cong K^{(\tau,0)+0}_{\Z_2}(S^1) \oplus K_{\Z_2}^{(\tau,w)+1}(S^1)
\cong \overbrace{(1+t)}^{\Z}, \label{Eq:GysinGlide2D_A} \\ 
&K^{\tau_{\sf pg}+1}_{\Z_2}(S^1 \times \tilde S^1)
\cong K^{(\tau,0)+1}_{\Z_2}(S^1) \oplus K_{\Z_2}^{(\tau,w)+0}(S^1)
\cong \overbrace{(1+t)}^{\Z} \oplus \overbrace{I}^{\Z_2}. \label{Eq:2DAIIIGlideGysin}
\end{align}

\subsubsection{Model and topological invariant}
Model vector bundles/Hamiltonians representing $K^{\tau_{\sf pg}+n}_{\Z_2}(T^2)$ are as follows. 
Eqs.\ (\ref{Eq:GysinGlide2D_A}) and (\ref{Eq:2DAIIIGlideGysin}) 
imply that the free parts $(1+t)$ of $K$-groups $K^{\tau_{\sf pg}+n}_{\Z_2}(T^2)$, $(n=0,1)$, arise from 1d models 
which were already introduced in (\ref{Fig:1DTCI_NS_e1}) and (\ref{Fig:1DTCI_NS_e1_AIII}). 

The generating Hamiltonian of $\Z_2$ part $I$ in $K^{\tau_{\sf pg}+1}_{\Z_2}(T^2)$ 
is given by the dimensional raising map from the $K$-group $K^{(\tau,w)+0}(S^1)$. 
As shown in Ref.~\onlinecite{ShiozakiSatoGomi2015}, 
the Karoubi's triple for the generator of $K^{(\tau,w)+0}(S^1)$ is given as 
\begin{align}
\Big[ E = S^1 \times \C^2, U(k_x) = \begin{pmatrix}
0 & e^{-i k_x} \\
1 & 0 \\
\end{pmatrix}_{\mu}, H_1 = 
\begin{pmatrix}
1 & 0 \\
0 & -1 \\
\end{pmatrix}_{\mu}, H_{0} = 
\begin{pmatrix}
-1 & 0 \\
0 & 1 \\
\end{pmatrix}_{\mu} \Big], 
\end{align}
where $E$ is the $\Z_2$ twisted equivariant bundle defined in (\ref{Eq:1DTwistVectE}) 
and the subscript $\mu$ represents the two localized positions inside the unit cell. 
Then, the dimensional raising map (\ref{eq:dimensional_raise_nonchiral}) leads us to 
the Hamiltonian in class AIII with glide symmetry 
\begin{align}
\wt H(k_x,k_y)
= \cos k_y \mu_z \otimes \sigma_z + \sin k_y \mu_0 \otimes \sigma_x, \qquad 
\wt \Gamma = \mu_0 \otimes \sigma_y, \qquad 
\wt U(k_x)= U(k_x) \otimes \sigma_y.  
\label{eq:mode_2d_aiii_glide_z2}
\end{align}
Here, we introduced the Pauli matrices $\mu_{a} (a=0,x,y,z)$ for the $\mu$ space. 
Notice that the $\wt U(k_x)$ acts on the Hamiltonian $\wt H(k_x,k_y)$ as a glide symmetry 
which commutes with the chiral symmetry 
\begin{align}
\wt U(k_x) \wt H(k_x,k_y) \wt U(k_x)^{-1} 
= H(k_x,-k_y), \qquad 
\wt U(k_x)^2 = e^{-i k_x}, \qquad 
[\wt \Gamma, \wt U(k_x)]=0. 
\end{align}

The $\Z_2$ invariant is defined as follows. 
This is the 2d analog of $\Z_2$ invariant in 3d class A insulator with glide symmetry.~\cite{FangFu2015, ShiozakiSatoGomi2015}
Due to the chiral symmetry, 
the flattened Hamiltonian takes the form of ${\rm sign}[H(k_x,k_y)] = \begin{pmatrix}
0 & q(k_x,k_y) \\
q^{\dag}(k_x,k_y) & 0 \\
\end{pmatrix}$, where $q(k_x,k_y)$ is a unitary matrix in the basis producing the expression $\Gamma = \begin{pmatrix}
1 & 0 \\
0 & -1
\end{pmatrix}$. 
On the glide lines $k_y = \Gamma_y, \Gamma_y = 0$ and $\pi$, 
the Hamiltonian is divided by the glide sectors $U(k_x) = \pm e^{-i k_x/2}$. 
Let $q_{\pm}(k_x,\Gamma)$ be the Hamiltonian of the glide sectors with $U(k_x) = \pm e^{-i k_x/2}$. 
Since the two glide sectors are glued at the boundary, 
these Hamiltonians are connected at the BZ boundary $q_{\pm}(\pi,\Gamma_y) = q_{\mp}(-\pi,\Gamma_y)$. 
We define the $\Z_2$ invariant $\nu \in \{0,1/2\}$ by 
\begin{equation}\begin{split}
\nu
:= &
\frac{1}{2 \pi i} \Big[ 
\ln \det q_+(-\pi,0) + \frac{1}{2} \int_{-\pi}^{\pi} d_{k_x} \ln \det q_+(k_x,0) 
\Big] \\
&-\frac{1}{2 \pi i} \Big[ 
\ln \det q_+(-\pi,\pi) + \frac{1}{2} \int_{-\pi}^{\pi} d_{k_x} \ln \det q_+(k_x,\pi) 
\Big] \\
&+ \frac{1}{2} \cdot \frac{1}{2 \pi i} \int_0^{\pi} d_{k_y} \ln \det q(-\pi,k_y) \qquad (\mod\ 1). 
\end{split}\end{equation}
By the use of the Stokes' theorem, it is easy to show that $2 \nu = 0 (\mod\ 1)$, i.e.\ 
$\nu$ is quantized to $\Z_2$ values. 
One can check that the Hamiltonian (\ref{eq:mode_2d_aiii_glide_z2}) has $\nu = 1/2$.

\subsubsection{3d TCIs with glide symmetry}
Applying the dimensional isomorphism (\ref{Eq:DimShift_general}) 
to 2d $K$-groups (\ref{Eq:2DPgKGroup}) leads to the 
topological classification of 3d class A and AIII insulators with glide symmetry 
\begin{align}
&{\rm 3d\ class\ A\ bulk}: &&
K_{\Z_2}^{\tau_{\sf pg}+0}(T^3) \cong \underbrace{\overbrace{(1+t)}^{\Z}}_{k_x} \oplus \underbrace{\overbrace{(1+t)}^{\Z}}_{(k_x,k_z)} \oplus \underbrace{\overbrace{I}^{\Z_2}}_{(k_x,k_y,k_z)}, 
\label{Eq:3DPgKGroup_A} \\
&{\rm 3d\ class\ AIII\ bulk}: &&
K_{\Z_2}^{\tau_{\sf pg}+1}(T^3) \cong \underbrace{\overbrace{(1+t)}^{\Z}}_{k_x} \oplus \underbrace{\overbrace{I}^{\Z_2}}_{(k_x,k_y)} \oplus \underbrace{\overbrace{(1+t)}^{\Z}}_{(k_x,k_z)}.
\label{Eq:3DPgKGroup_AIII}
\end{align}
Here, the underbraces indicate the minimum dimensions 
required for realizing generators. 
For example, $(k_x,k_z)$ means that a generator model is 
adiabatically connected to
a stacking model along the $x$ and $z$-directions. 
It is clear that the so-called ``strong index'' appears only in the last $\Z_2$ group in (\ref{Eq:3DPgKGroup_A}). 
This $\Z_2$ phase in 3d class A insulators with glide symmetry 
was already described in Refs.~\onlinecite{FangFu2015, ShiozakiSatoGomi2015}, so we do not repeat it here.

\subsubsection{2d surface states with glide symmetry}
As explained in Sec.~\ref{sec:Classification of boundary gapless states}, 
the topological classification of 
boundary gapless states is given by the 
$K$-group with the shift of the integer grading $-n \mapsto -(n-1)$. 
Hence, the results (\ref{Eq:2DPgKGroup}) imply the 
classification of surface states: 
\begin{align}
&{\rm 2d\ class\ A\ surface\ gapless\ states}: &&
K_{\Z_2}^{\tau_{\sf pg}+1}(T^2) \cong \underbrace{\overbrace{(1+t)}^{\Z}}_{k_x} \oplus \underbrace{\overbrace{I}^{\Z_2}}_{(k_x,k_y)} 
\label{Eq:2DPgKGroup_A_surface} \\
&{\rm 2d\ class\ AIII\ surface\ gapless\ states}: &&
K_{\Z_2}^{\tau_{\sf pg}+0}(T^2) \cong \underbrace{\overbrace{(1+t)}^{\Z}}_{k_x}. 
\label{Eq:2DPgKGroup_AIII_surface}
\end{align}
The meaning of the underbraces are similar to 
(\ref{Eq:3DPgKGroup_A}) and (\ref{Eq:3DPgKGroup_AIII}), 
indicating the momentum dependence of the spectrum. 
Comparing (\ref{Eq:2DPgKGroup_A_surface}) ( (\ref{Eq:2DPgKGroup_AIII_surface}) ) with 
(\ref{Eq:3DPgKGroup_A}) ( (\ref{Eq:3DPgKGroup_AIII}) ), 
one can see that the bulk-boundary correspondence holds.

\subsection{$C_4$ rotation symmetry}
In this section, we present a $K$-theory computation 
of the TCIs with $C_4$ symmetry in two-dimension for class A and AIII. 
The BZ is a square. 
The point group $C_4 = \Z_4 = \{1,c_4,c_2=c_4^2,c_4^3\}$ acts on $T^2$ by $ c_4: (k_x,k_y) \mapsto (-k_y,k_x).$
There are two fixed points: $\Gamma = (0,0)$ and $M = (\pi,\pi)$. 
$X = (\pi,0)$ is a fixed point of the subgroup $C_2 = \Z_2 = \{1,c_2\} \subset \Z_4$.
We present the representation rings of $\Z_4$ and $\Z_2$ groups as follows
\begin{align}
R(\Z_4) = \Z[t]/(1-t^4), \qquad 
R(\Z_2) = \Z[s]/(1-s^2). 
\end{align}
$R(\Z_4)$ acts on $R(\Z_2)$ by the restriction of representations of $\Z_4$: $t |_{\Z_2} = s$, 
which means $R(\Z_2)$ is $(1+t^2) = \{f(t) (1+t^2) | f(t) \in R(\Z_4) \}$ as an $R(\Z_4)$-module.

\subsubsection{Topological classification}
To compute the $K$-group $K^{n}_{\Z_4}(T^2)$, 
we introduce a subspace $Y \subset T^2$ as follows: 
$$
Y \ \ = \ \ 
\xygraph{
!{<0cm,0cm>;<1cm,0cm>:<0cm,1cm>::}
!{(1,1)}="a",
!{(-1,1)}*+{\bullet}="b" ,
!{(-1,0)}*+{\bullet}="e", 
!{(-1,-1)}*+{\bullet}="c"([]!{+(0,-0.4)} {(0,0)}),
!{(0,-1)}*+{\bullet}="f" ,
!{(1,-1)}*+{\bullet}="d" ,
"a"-@{.}"b",
"d"-@{.}"a",
"b"-"e",
"e"-"c",
"c"-"f",
"f"-"d",
}
\ \ = \ \ 
\xygraph{
!{<0cm,0cm>;<1cm,0cm>:<0cm,1cm>::}
!{(0,0)}*+{\bullet}="a"([]!{+(0,-1.0)} {(0,0)}),
!{(1.5,0)}*+{\bullet}="b"([]!{+(0.5,0)} {(\pi,0)}),
!{(-1.5,0)}*+{\bullet}="c"([]!{+(-0.5,0)} {(0,\pi)}),
!{(0,-0.7)}-@{->}!{(0,-0.3)}, 
"a" -@/^1cm/ "b",
"a" -@/_1cm/ "b",
"a" -@/^1cm/ "c",
"a" -@/_1cm/ "c",
} 
$$
Let us compute the $K$-group on $Y$. 
We can decompose $Y = U \cup V$ to two parts which are $\Z_4$-equivariantly homotopic to points
\begin{equation}
U \sim (\Z_4/\Z_4) \times pt = \{ \Gamma \}, \qquad 
V \sim (\Z_4/\Z_2) \times pt = \{ X, c_4 X \}.
\end{equation}
The intersection is 
\begin{equation}
U \cap V \sim \Z_4 \times pt. 
\end{equation}
The Mayer-Vietoris sequence for $Y = U \cup V$ is 
\begin{equation}\begin{split}
\begin{CD}
0 @<<< 0 @<<< K_{\Z_4}^{1}(Y) \\
@VVV @. @AAA \\
K_{\Z_4}^{0}(Y) @>>> \underbrace{R(\Z_4)}_{\Gamma} \oplus \underbrace{R(\Z_2)}_{X} @>\Delta>> \Z.
\end{CD} 
\end{split}\end{equation}
where $\Delta$ is given by 
\begin{equation}
\Delta : (f(t), g(s) )  \mapsto f(1)-g(1). 
\end{equation}
A basis of $\mathrm{Ker} (\Delta)$ can be chosen as 
\begin{equation}
\{ \underbrace{(1,1), (t,s), (t^2,1), (t^3,s)}_{R(\Z_4)}, \underbrace{(0,1-s)}_{(1-t+t^2-t^3)} \} \subset R(\Z_4) \oplus R(\Z_2).
\end{equation}
The former four base elements compose $R(\Z_4)$ 
and the last base element generates the $R(\Z_4)$-module $(1-t+t^2-t^3)= \{f(t)(1-t+t^2-t^3)|f(t) \in R(\Z_4) \}$. 
We have  
\begin{align}
K_{\Z_4}^{0}(Y) \cong \mathrm{Ker} ( \Delta ) \cong \overbrace{R(\Z_4)}^{\Z^4} \oplus \overbrace{(1-t+t^2-t^3)}^{\Z}, \qquad 
K_{\Z_4}^{1}(Y) \cong 0.
\end{align}

Next, we ``fill in'' the BZ torus $T^2$ with wave functions from $Y$. 
To this end, we use the exact sequence for the pair $(T^2, Y)$. 
\begin{equation}\begin{split}
\begin{CD}
K^1_{\Z_4}(Y) @<<< K^1_{\Z_4}(T^2) @<<< K^1_{\Z_4}(T^2,Y) \\
@VVV @. @AAA \\
K^0_{\Z_4}(T^2,Y) @>>> K^0_{\Z_4}(T^2) @>>>  K^0_{\Z_4}(Y).
\end{CD}
\end{split}\label{Seq:Pair_p4_T^2Y}\end{equation}
The $K$-group of the pair $(T^2,Y)$ is given as follows: 
The quotient $T^2/Y$ can be identified with the sphere $D(\C_1)/S(\C_1)$ 
obtained by shrinking the boundary circle $S(\C_1)$ of the disc $D(\C_1)$.
Here, $\C_1$ is the 1-dimensional complex representation of $\Z_4$, 
say, the generator $C_4 \in \Z_4$ acts on $\C$ by $C_4 \cdot z = i z$, 
and $\Z_4$ naturally acts on $D(\C_1)$, $S(\C_1)$ and $D(\C_1)/S(\C_1)$. 
Then, the Thom isomorphism for the $\Z_4$-equivariant complex vector bundle $\C_1 \to pt$ states (see Appendix \ref{app:Thom}) 
\begin{align}
K^n_{\Z_4}(T^2,Y) 
\cong \widetilde K^n_{\Z_4}(T^2/Y) 
\cong \widetilde K^n_{\Z_4}(D(\C_1)/S(\C_1)) 
\cong K^n_{\Z_4}(D(\C_1),S(\C_1)) 
\cong K^n_{\Z_4}(pt). 
\end{align}
Then, the sequence (\ref{Seq:Pair_p4_T^2Y}) is recast into 
\begin{align}
\begin{CD}
0 @<<< K^1_{\Z_4}(T^2) @<<< 0 \\
@VVV @. @AAA \\
R(\Z_4) @>>> K^0_{\Z_4}(T^2) @>i^*>> R(\Z_4) \oplus (1-t+t^2-t^3) \\
\end{CD}
\end{align}
Since the contribution $K^0_{\Z_4}(Z) = R(\Z_4) \subset K^0_{\Z_4}(T^2) = K^0_{\Z_4}(Z) \oplus \widetilde{K}^0_{\Z_4}(T^2)$ from the fixed point $\Gamma$ is identically mapped by $i^*$, 
we get the exact sequence for the reduced $K$-theory 
\begin{align}
0 \to R(\Z_4) \to \widetilde{K}^0_{\Z_4}(T^2) \to (1-t+t^2-t^3) \to 0. 
\end{align}
One can show that the extension of $(1-t+t^2-t^3)$ by $R(\Z_4)$ is unique. 
(See Appendix \ref{app:ext} for details.) 
We thus get the reduced $K$-group
\begin{align}
\widetilde{K}^0_{\Z_4}(T^2) \cong R(\Z_4) \oplus (1-t+t^2-t^3)
\label{eq:K0_p4_T2}
\end{align}
and the $K$-group 
\begin{align}
K^0_{\Z_4}(T^2) \cong \overbrace{R(\Z_4)}^{\Z^4} \oplus \overbrace{R(\Z_4)}^{\Z^4} \oplus \overbrace{(1-t+t^2-t^3)}^{\Z}, \qquad 
K^1_{\Z_4}(T^2) = 0. 
\label{Eq:P4_KGroup_Modu} 
\end{align}

\subsubsection{Models of $K^0_{\Z_4}(T^2)$}
In this subsection, we give generating models of the $K$-group $K^0_{\Z_4}(T^2)$, the 2d TCIs with $C_4$ symmetry. 
Through the ``lens'' of topological invariants, one can reconstruct the $R(\Z_4)$-module structure (\ref{Eq:P4_KGroup_Modu}). 

As mentioned, the BZ is a square. 
$\Gamma=(0,0)$ and $M = (\pi,\pi)$ are the fixed points of the $C_4$ group, 
and $X = (\pi,0)$ is fixed under the subgroup $C_2 = \Z_2$: 
\begin{align}
\xygraph{
!{<0cm,0cm>;<1cm,0cm>:<0cm,1cm>::}
!{(0,0)}*+{\bullet} ([]!{+(-0.2,-0.2)} {\Gamma}), 
!{(1,0)}*+{\bullet} ([]!{+(+0.2,-0.2)} {X}), 
!{(1,1)}*+{\bullet} ([]!{+(+0.2,+0.2)} {M}), 
!{(0,1)}, 
!{(1,1)}="a"  ,
!{(-1,1)}="b",
!{(-1,-1)}="c",
!{(1,-1)}="d",
"a"-"b", 
"b"-"c", 
"c"-"d", 
"d"-"a", 
}
\end{align}
In general, parts of the $K$-group of class A can be represented by 
vector bundles realized as atomic insulators. 
Put a representation of site symmetry at the Wyckoff positions inside a unit cell. 
There are two Wyckoff Positions (a) and (b) of which the filling number is one: 
\begin{align}
\left( E_a = T^2 \times \C, \ \ \rho_{c_4}(\bk,v) = (c_4  \bk, v) \right) 
\ \ \leftrightarrow \ \ \xygraph{
!{<0cm,0cm>;<1.5cm,0cm>:<0cm,1.5cm>::}
!{(0,0)}*+{\bigcirc},([]!{+(0,-0.3)}{\ket{s}}), 
!{(0.5,0.5)}-!{(-0.5,0.5)}, 
!{(-0.5,0.5)}-!{(-0.5,-0.5)}, 
!{(-0.5,-0.5)}-!{(0.5,-0.5)}, 
!{(0.5,-0.5)}-!{(0.5,0.5)}, 
}
\end{align}
\begin{align}
\left( E_b = T^2 \times \C, \ \ \rho_{c_4}(\bk,v) = (c_4 \bk, e^{-i k_y} v) \right) 
\ \ \leftrightarrow \ \ \xygraph{
!{<0cm,0cm>;<1.5cm,0cm>:<0cm,1.5cm>::}
!{(0.5,0.5)}*+{\bigcirc}([]!{+(0.3,0)}{\ket{s}}), 
!{(0.5,0.5)}-!{(-0.5,0.5)}, 
!{(-0.5,0.5)}-!{(-0.5,-0.5)}, 
!{(-0.5,-0.5)}-!{(0.5,-0.5)}, 
!{(0.5,-0.5)}-!{(0.5,0.5)}, 
}
\end{align}
In the above, the corresponding $\Z_4$-equivariant line bundles are denoted by 
$E_a$ and $E_b$. 
The solid squares in the figures represent the unit cells. 
The $C_4$ action on $E_b$ is determined by 
the $C_4$ action on the real space basis $\hat U_{c_4} \ket{(R_x,R_y),s} = \ket{(-R_y-1, R_x),s}$. 
Here we put the $s$-orbitals at the Wyckoff positions. 
Other representations of $\Z_4$ are obtained by tensor products of elements of $R(\Z_4)$. 
We have another generator $E_c$ of rank 2 that is realized by 
putting $s$-orbitals at the centers of edges of the square: 
\begin{align}
\left( E_c = T^2 \times \C^2, \ \ \rho_{c_4}(\bk,v) = (c_4 \bk, \begin{pmatrix}
0 & e^{-i k_y} \\
1 & 0 \\
\end{pmatrix} v) \right) 
\ \ \leftrightarrow \ \ \xygraph{
!{<0cm,0cm>;<1.5cm,0cm>:<0cm,1.5cm>::}
!{(0.5,0)}*+{\bigcirc}([]!{+(0.3,0)}{\ket{s}}), 
!{(0,0.5)}*+{\bigcirc}([]!{+(0,0.3)}{\ket{s}}), 
!{(0.5,0.5)}-!{(-0.5,0.5)}, 
!{(-0.5,0.5)}-!{(-0.5,-0.5)}, 
!{(-0.5,-0.5)}-!{(0.5,-0.5)}, 
!{(0.5,-0.5)}-!{(0.5,0.5)}, 
}
\end{align}
All other atomic line bundles 
can be direct sums of $E_a, E_b, E_c$ and 
tensor products by representations of $\Z_4$. 

The $K$-group $K^0_{\Z_4}(T^2)$ includes 
line bundles with finite Chern number. 
To construct a line bundle with a nonzero Chern number, 
we gap out a trivial atomic insulator by 
introducing (one-body) interaction. 
Let $E$ be the atomic $\Z_4$-equivariant bundle 
consisting of $s$ and $p_{x+iy}$ orbitals localized at 
the center of the unit cell: 
\begin{align}
\left( E = T^2 \times \C^2, \ \ \rho_{c_4}(\bk,v) = \left(c_4  \bk, \begin{pmatrix}
1 & 0 \\
0 & i \\
\end{pmatrix} v \right) \right) 
\ \ \leftrightarrow \ \ \xygraph{
!{<0cm,0cm>;<1.5cm,0cm>:<0cm,1.5cm>::}
!{(0,0)}*+{\bigcirc},([]!{+(0.15,-0.3)}{\ket{s} \oplus \ket{p_{x+iy}}}), 
!{(0.5,0.5)}-!{(-0.5,0.5)}, 
!{(-0.5,0.5)}-!{(-0.5,-0.5)}, 
!{(-0.5,-0.5)}-!{(0.5,-0.5)}, 
!{(0.5,-0.5)}-!{(0.5,0.5)}, 
}
\end{align}
We define four $\Z_4$-equivariant line bundles as the 
occupied state of the following $C_4$ symmetric Hamiltonians on the bundle $E$, 
\begin{align}
&F_{\Gamma, \pm} : \qquad H(\bk) = \sin k_x \sigma_x + \sin k_y \sigma_y \pm (m - \cos k_x - \cos k_y) \sigma_z, \qquad 0 < m < 2, \\
&F_{M, \pm} : \qquad H(\bk) = \sin k_x \sigma_x + \sin k_y \sigma_y \pm (m - \cos k_x - \cos k_y) \sigma_z, \qquad -2 < m < 0, 
\end{align}
where $\sigma_i (i = x,y,z)$ is the Pauli matrices, and
the subscript $\Gamma/M$ represents the location of the band inversion. 

There are four topological invariants: 
the Chern number $ch_1$ and representations at $\Gamma, M$ and $X$. 
These topological invariants have $R(\Z_4)$-module structures. 
The above models have the following data of topological invariants: 
$$
\begin{tabular}[t]{l|c|c|c|c}
\hline 
 & $ch_1([E])$ & $[E|_{\Gamma}]$ & $[E|_{M}]$ & $[E|_{X}]$ \\
\hline
Bundle & $(1+t+t^2+t^3)$ & $R(\Z_4)$ & $R(\Z_4)$ & $R(\Z_2 = \{1,c_2\})$ \\
\hline
$E_a$ & $0$ & $1$ & $1$ & $1$ \\
$E_b$ & $0$ & $1$ & $t^2$ & $s$ \\
$E_c$ & $0$ & $1+t^2$ & $t+t^3$ & $1+s$ \\
$F_{\Gamma,+}$ & $1$ & $1$ & $t$ & $s$ \\
$F_{\Gamma,-}$ & $-1$ & $t$ & $1$ & $1$ \\
$F_{M,+}$ & $-1$ & $1$ & $t$ & $1$ \\
$F_{M,-}$ & $1$ & $t$ & $1$ & $s$ \\
\hline 
\end{tabular}
$$
From this table, we can read off three generators $e_1, e_2, e_3$ of the $K$-group $K^0_{\Z_2}(T^2)$:
\begin{align}
\begin{tabular}{c|c|c|c|c|c}
\hline 
& & $ch_1([E])$ & $[E|_{\Gamma}]$ & $[E|_{M}]$ & $[E|_{X}]$ \\
\hline
$R(\Z_4)$-module structure & generator & $(1+t+t^2+t^3)$ & $R(\Z_4)$ & $R(\Z_4)$ & $R(\Z_2 = \{1,c_2\})$ \\
\hline
$R(\Z_4)$ 
& $e_1 = [E_a]$ & $0$ & $1$ & $1$ & $1$ \\
& $t e_1$ & $0$ & $t$ & $t$ & $s$ \\ 
& $t^2 e_1$ & $0$ & $t^2$ & $t^2$ & $1$ \\ 
& $t^3 e_1$ & $0$ & $t^3$ & $t^3$ & $s$ \\ 
\hline 
$R(\Z_4)$ 
&$e_2 = [F_{M,+}]-[E_a]$ & $-1$ & $0$ & $t-1$ & $0$ \\ 
&$t e_2$ & $-1$ & $0$ & $t^2-t$ & $0$ \\ 
&$t^2 e_2$ & $-1$ & $0$ & $t^3-t^2$ & $0$ \\ 
&$t^3 e_2$ & $-1$ & $0$ & $1-t^3$ & $0$ \\ 
\hline 
$(1-t+t^2-t^3)$ & $e_3 = [E_c]-(1+t^2) \cdot [E_a]$ & $0$ & $0$ & $-1+t-t^2+t^3$ & $s-1$ \\ 
\hline 
\end{tabular}
\label{tab:p4_top_inv_K}
\end{align}
An arbitrary formal difference $[E_1]-[E_2]$ of two $\Z_4$-equivariant bundles can be a linear combination of 
these generators. 
For example, 
\begin{align}
&[E_b] = e_1 + (t-t^2) e_2 + e_3, \qquad 
[E_c] = (1+t^2) e_1 - e_3, \qquad 
[F_{\Gamma,+}] = e_1-t^2 e_2 + e_3, \\
&[F_{\Gamma,-}] = t e_1 + t^2 e_2 + e_3, \qquad 
[F_{M,-}] = t e_1 - e_2. 
\end{align}
The $R(\Z_4)$-module structure is consistent with 
the algebraic derivation of the $K$-group (\ref{Eq:P4_KGroup_Modu}).

\subsubsection{Constraint on topological invariants}
There is a constraint on the data of topological invariants 
\begin{align}
(ch_1([E]),[E|_{\Gamma}], [E|_{M}], [E|_X]) \in 
(1+t+t^2+t^3) \oplus 
\underbrace{R(\Z_4)}_{K^0_{\Z_4}(\{\Gamma \})} \oplus 
\underbrace{R(\Z_4)}_{K^0_{\Z_4}(\{M \})} \oplus 
\underbrace{R(\Z_2)}_{K^0_{\Z_4}(\{X\})} 
\label{Eq:TopInvP4}
\end{align}
that arises from the fully gapped condition on the whole BZ torus. 
Let us denote the r.h.s\ of (\ref{Eq:TopInvP4}) by ${\rm Top}^0_{\Z_4}(T^2)$. 
The constraint can be considered as the condition that 
the topological invariant lies in the image of an injective homomorphism 
from the $K$-group $K^{0}_{\Z_4}(T^2)$ to the set of topological invariants 
\begin{align}
f_{\rm top}: K^{0}_{\Z_4}(T^2) \to {\rm Top}^0_{\Z_4}(T^2). 
\end{align}
This homomorphism $f_{\rm top}$ is not surjective in general, 
hence the condition 
\begin{align}
x \equiv 0 \quad \mod \ \ \im (f_{\rm top}), \qquad x \in {\rm Top}^0_{\Z_4}(T^2)
\end{align}
makes sense. 
From the data (\ref{tab:p4_top_inv_K}), $\im(f_{\rm top})$ is spanned by 
\begin{align}
\left\{ \begin{array}{l}
(0,1,1,1) \\
(0,t,t,s) \\
(0,t,t,1) \\
(0,t,t,s) \\
(-1,0,t-1,0) \\
(-1,0,t^2-t,0) \\
(-1,0,t^3-t^2,0) \\
(-1,0,1-t^3,0) \\
(0,0,-1+t-t^2+t^3,s-1) \\
\end{array} \right. 
\sim 
\left\{ \begin{array}{l}
(0,1,1,1) \\
(0,t,t,s) \\
(0,t,t,1) \\
(0,t,t,s) \\
(-1,0,t-1,0) \\
(-2,0,t^2-1,0) \\
(-3,0,t^3-1,0) \\
(2,0,0,s-1) \\
(4,0,0,0) \\
\end{array} \right. 
\end{align}
as an Abelian group. 
Let us denote a general element of ${\rm Top}^0_{\Z_4}(T^2)$ by $(ch_1, \Gamma(t), M(t), X(s))$. 
Solving the equation $(ch_1, \Gamma(t), M(t), X(s)) = 0 \ \mod \ \im(f_{\rm top})$ leads us to the constraints 
\begin{align}
&{\rm Constraint\ 1}: \quad \Gamma(1) = M(1) = X(1), \\
&{\rm Constraint\ 2}: \quad ch_1 = \Gamma'(1) - M'(1) +2 X'(1) \ \ \mod \ 4, 
\end{align}
where $\Gamma'(1)$ is the derivative $\Gamma'(1) := \frac{d}{d t}\Gamma(t) |_{t \to 1}$ and so are $M'(1)$ and $X'(1)$. 
The first constraint means that 
the number of occupied states should be uniform around the whole BZ torus. 
The breaking of the first condition implies the existence of the Fermi surface. 
The latter constraint serves as a criterion 
for nontrivial Chern number.~\cite{Fang2012}

\subsection{Wall paper group p4g with projective representation of $D_4$}
In this section, we calculate the 
$K$-group of $T^2$ with the wallpaper group p4g and a
nontrivial projective representation of its point group $D_4$, 
which corresponds to that the degree of freedom at 
a site is spin half-integer. 

\subsubsection{Space group p4g}
The space group p4g is generated by the following two elements 
\begin{align}
\{ c_4 | \hat y/2 \} : (x,y) \to (-y,x+1/2),  &&
\{ \sigma | \hat x/2 \} : (x,y) \to (x+1/2,-y), 
\end{align}
and the primitive lattice translations. 
This corresponds to the choice of non-primitive 
lattice translations $\bm{a}_{c_4} = (0,1/2)$ and $\bm{a}_{\sigma} = (1/2,0)$. 
We define other non-primitive translations by 
\begin{align}
&\bm{a}_{c_2} := \bm{a}_{c_4} + c_4 \bm{a}_{c_4}, \qquad 
\bm{a}_{c_4^3} := \bm{a}_{c_4} + c_4 \bm{a}_{c_2}, \label{eq:ap_p4g_c_4} \\
&\bm{a}_{\sigma c_4} := \bm{a}_{\sigma} +\sigma \bm{a}_{c_4}, \qquad 
\bm{a}_{\sigma c_2} := \bm{a}_{\sigma} +\sigma \bm{a}_{c_2}, \qquad 
\bm{a}_{\sigma c_4^3} := \bm{a}_{\sigma} +\sigma \bm{a}_{c_4^3}, \label{eq:ap_p4g_sigma}
\end{align}
which are summarized as: 
$$
\begin{tabular}[t]{c|c|c|c|c|c|c|c|c}
\hline 
$p \in D_4$ & $1$ & $c_4$ & $c_2$ & $c_4^3$ & $\sigma$ & $\sigma c_4$ & $\sigma c_2$ & $\sigma c_4^3$ \\
\hline
$\bm{a}_p$ & $(0,0)$ & $(0,1/2)$ & $(1/2,1/2)$ & $(1/2,0)$ & $(1/2,0)$ & $(1/2,1/2)$ & $(0,1/2)$ & $(0,0)$ \\
\hline 
\end{tabular}
$$
Under this choice, the two-cocycle $\bm{\nu}_{p_1,p_2} = \bm{a}_{p_1}+p_1 \bm{a}_{p_2} - \bm{a}_{p_1p_2} \in \Pi$ is given by the following table.
$$
\begin{tabular}[t]{c|ccccccccc}
\hline 
$\bm{\nu}_{p_1,p_2}, p_1 \backslash p_2$ & $1$ & $c_4$ & $c_2$ & $c_4^3$ & $\sigma$ & $\sigma c_4$ & $\sigma c_2$ & $\sigma c_4^3$ \\
\hline
$1$ & $(0,0)$ & $(0,0)$ & $(0,0)$ & $(0,0)$ & $(0,0)$ & $(0,0)$ & $(0,0)$ & $(0,0)$ \\
$c_4$ & $(0,0)$ & $(-1,0)$ & $(-1,1)$ & $(0,1)$ & $(0,1)$ & $(-1,1)$ & $(-1,0)$ & $(0,0)$ \\
$c_2$ & $(0,0)$ & $(0,0)$ & $(0,0)$ & $(0,0)$ & $(0,0)$ & $(0,0)$ & $(0,0)$ & $(0,0)$ \\
$c_4^3$ & $(0,0)$ & $(1,0)$ & $(1,-1)$ & $(0,-1)$ & $(0,-1)$ & $(1,-1)$ & $(1,0)$ & $(0,0)$ \\
$\sigma$ & $(0,0)$ & $(0,-1)$ & $(1,-1)$ & $(1,0)$ & $(1,0)$ & $(1,-1)$ & $(0,-1)$ & $(0,0)$ \\
$\sigma c_4$ & $(0,0)$ & $(0,0)$ & $(0,0)$ & $(0,0)$ & $(0,0)$ & $(0,0)$ & $(0,0)$ & $(0,0)$ \\
$\sigma c_2$ & $(0,0)$ & $(0,1)$ & $(-1,1)$ & $(-1,0)$ & $(-1,0)$ & $(-1,1)$ & $(0,1)$ & $(0,0)$ \\
$\sigma c_4^3$ & $(0,0)$ & $(0,0)$ & $(0,0)$ & $(0,0)$ & $(0,0)$ & $(0,0)$ & $(0,0)$ & $(0,0)$ \\
\hline 
\end{tabular}
$$

Next, we move on to the momentum space. 
The point group $D_4$ acts on the square BZ torus by 
$c_4 \cdot (k_x,k_y) = (-k_y,k_x)$ and $\sigma \cdot (k_x,k_y) = (k_x,-k_y)$. 
All the $D_4$ actions are summarized in the following figure: 
\begin{align}
\xygraph{
!{<0cm,0cm>;<1.5cm,0cm>:<0cm,1.5cm>::}
!{(0,0)}*+{\bullet}="O" ([]!{+(-0.1,-0.2)} {\Gamma}) ,
!{(1,1)}="a"  ,
!{(-1,1)}="b",
!{(-1,-1)}="c",
!{(1,-1)}="d",
!{(0,1)}="e",
"a"-"b", 
"b"-"c", 
"c"-"d", 
"d"-"a", 
!{(1,0)}*+{\bullet}  ([]!{+(-0.1,-0.2)} {X}) , 
!{(1,1)}*+{\bullet}  ([]!{+(-0.1,-0.2)} {M}) , 
!{(-1,0)}, 
!{(-1,1)}, 
!{(-1.3,0)}-@{.}!{(1.3,0)},
!{(0,-1.3)}-@{.}!{(0,1.3)},
!{(-1.2,-1.2)}-@{.}!{(1.2,1.2)},
!{(-1.2,1.2)}-@{.}!{(1.2,-1.2)},
!{(1.2,-0.2)}-@{<->}!{(1.2,0.2)}, 
!{(-0.2,1.2)}-@{<->}!{(0.2,1.2)},
!{(1.0,1.3)}-@{<->}!{(1.3,1.0)},
!{(1.0,-1.3)}-@{<->}!{(1.3,-1.0)},
!{(1.5,0)}*+{\sigma}, 
!{(0,1.5)}*+{\sigma c_2}, 
!{(1.4,1.4)}*+{\sigma c_4^3}, 
!{(1.4,-1.4)}*+{\sigma c_4}, 
!{(0.4,0)}-@/_0.2cm/@{->}!{(0,0.4)}, 
!{(0.6,0.2)}*+{c_4}, 
}
\end{align}
$\Gamma$ and $M$ are fixed points of $D_4$
and $X$ is fixed by the sub-group 
$D_2^{(v)} = \{1,c_2,\sigma,\sigma c_2\}$. 
The choices (\ref{eq:ap_p4g_c_4}) and (\ref{eq:ap_p4g_sigma}) 
correspond to 
\begin{align}
&U_{1}(\bk):= 1, && U_{c_2}(\bk) := U_{c_4}(c_4 \bk) U_{c_4}(\bk), && U_{c_4^3}(\bk) := U_{c_4}(c_2 \bk) U_{c_2}(\bk), \label{eq:u_p4g_1} \\
&U_{\sigma c_4}(\bk):=U_{\sigma}(c_4 \bk) U_{c_4}(\bk), && U_{\sigma c_2}(\bk) := U_{\sigma}(c_2 \bk) U_{c_2}(\bk), && U_{\sigma c_4^3}(\bk) := U_{\sigma}(c_4^3 \bk) U_{c_4}(\bk)  \label{eq:u_p4g_2}
\end{align}
for fixed $U_{c_4}(\bk)$ and $U_{\sigma}(\bk)$. 
The two-cocycle $(\tau_{\sf p4g})_{p,p'}(\bk) = - \bk \cdot \bm{\nu}_{p,p'}$ 
on the momentum space is summarized as: 
$$
e^{i(\tau_{\sf p4g})_{p,p'}(pp'\bk)}
\quad = \quad 
\begin{tabular}{c|ccccccccc}
\hline 
$p \backslash p'$ & $1$ & $c_4$ & $c_2$ & $c_4^3$ & $\sigma$ & $\sigma c_4$ & $\sigma c_2$ & $\sigma c_4^3$ \\
\hline
$1$ & $1$ & $1$ & $1$ & $1$ & $1$ & $1$ & $1$ & $1$ \\
$c_4$ & $1$ & $1$ & $1$ & $1$ & $e^{- ik_x}$ & $e^{i k_y}$ & $e^{i k_x}$ & $e^{-i k_y}$ \\
$c_2$ & $1$ & $1$ & $1$ & $1$ & $e^{- i(k_x+k_y)}$ & $e^{-i (k_x-k_y)}$ & $e^{i (k_x+k_y)}$ & $e^{i (k_x+k_y)}$ \\
$c_4^3$ & $1$ & $1$ & $1$ & $1$ & $e^{- ik_y}$ & $e^{-i k_x}$ & $e^{i k_y}$ & $e^{i k_x}$ \\
$\sigma$ & $1$ & $1$ & $1$ & $1$ & $e^{- ik_x}$ & $e^{i k_y}$ & $e^{i k_x}$ & $e^{-i k_y}$ \\
$\sigma c_4$ & $1$ & $1$ & $1$ & $1$ & $e^{- i(k_x+k_y)}$ & $e^{-i (k_x-k_y)}$ & $e^{i (k_x+k_y)}$ & $e^{i (k_x+k_y)}$ \\
$\sigma c_2$ & $1$ & $1$ & $1$ & $1$ & $e^{- ik_y}$ & $e^{-i k_x}$ & $e^{i k_y}$ & $e^{i k_x}$ \\
$\sigma c_4^3$ & $1$ & $1$ & $1$ & $1$ & $1$ & $1$ & $1$ & $1$ \\
\hline 
\end{tabular}
$$
The two-cocycle at symmetric points can be read off as follows. 
At $\Gamma$ point, the two-cocycle is trivial 
\begin{equation}
(\tau_{\sf p4g}|_{\Gamma})_{p,p'} = 1, \qquad  p,p' \in D_4. 
\end{equation}
The restriction to the $M$ point is summarized as: 
$$
e^{i(\tau_{\sf p4g}|_M)_{p,p'}}
\quad =\quad 
\begin{tabular}{c|ccccccccc}
\hline 
$p \backslash p'$ & $1$ & $c_4$ & $c_2$ & $c_4^3$ & $\sigma$ & $\sigma c_4$ & $\sigma c_2$ & $\sigma c_4^3$ \\
\hline
$1$ & $1$ & $1$ & $1$ & $1$ & $1$ & $1$ & $1$ & $1$ \\
$c_4$ & $1$ & $1$ & $1$ & $1$ & $-1$ & $-1$ & $-1$ & $-1$ \\
$c_2$ & $1$ & $1$ & $1$ & $1$ & $1$ & $1$ & $1$ & $1$ \\
$c_4^3$ & $1$ & $1$ & $1$ & $1$ & $-1$ & $-1$ & $-1$ & $-1$ \\
$\sigma$ & $1$ & $1$ & $1$ & $1$ & $-1$ & $-1$ & $-1$ & $-1$ \\
$\sigma c_4$ & $1$ & $1$ & $1$ & $1$ & $1$ & $1$ & $1$ & $1$ \\
$\sigma c_2$ & $1$ & $1$ & $1$ & $1$ & $-1$ & $-1$ & $-1$ & $-1$ \\
$\sigma c_4^3$ & $1$ & $1$ & $1$ & $1$ & $1$ & $1$ & $1$ & $1$ \\
\hline 
\end{tabular}
$$
This can be trivialized by
the one-cochain $\beta \in C^1(D_4,U(1))$ defined by the following table.
$$
\begin{tabular}[t]{c|ccccccccc}
\hline 
$\beta_p$ & $1$ & $c_4$ & $c_2$ & $c_4^3$ & $\sigma$ & $\sigma c_4$ & $\sigma c_2$ & $\sigma c_4^3$ \\
\hline 
$p \in D_4$ & $1$ & $i$ & $-1$ & $-i$ & $-i$ & $1$ & $i$ & $-1$ \\
\hline 
\end{tabular}
$$
We can see $(\tau_{\sf p4g}|_{M} + \delta \beta)_{p,p'} = 0$ for all $p,p' \in D_4$. 
(Note that $(\delta \beta)_{p,p'} = \beta_p \beta_{p'} \beta_{pp'}^{-1}$.)
On the other hand, the 
restriction to the $X$ point is summarized in the table: 
$$
e^{i(\tau_{\sf p4g}|_X)_{p,p'}}
\quad =\quad 
\begin{tabular}{c|ccccccccc}
\hline 
$p \backslash p'$ & $1$ & $c_2$ & $\sigma$ & $\sigma c_2$ \\
\hline
$1$ & $1$ & $1$ & $1$ & $1$ \\
$c_2$ & $1$ & $1$ & $-1$ & $-1$\\
$\sigma$ & $1$ & $1$ & $-1$ & $-1$ \\
$\sigma c_2$ & $1$ & $1$ & $1$ & $1$ \\
\hline 
\end{tabular}
$$
This two-cocycle $\tau_{\sf p4g}|_X$ cannot be trivialized, 
which implies that $\tau_{\sf p4g}|_X$ generates the 
nontrivial group cohomology $H^2(D_2,U(1)) = \Z_2$.

\subsubsection{Projective representation of $D_4$}
In this section, we will 
consider the spin half integer systems with nonsymmorphic $p4g$ symmetry. 
In addition to the twist from the non-primitive lattice translations $\{\bm{a}_p\}_{p \in D_4}$, 
the point group $D_4$ obeys a projective representation of which the 
factor group represents the nontrivial element of $H^2(D_4,U(1)) = \Z_2$. 

A simple way to fix the two-cocycle $\omega \in Z^2(D_4,\R/2 \pi \Z)$ is 
to consider an explicit form of a projective representation of $D_4$. 
Let us consider the following projective representation of $D_4$, 
\begin{align}
U_{c_4} = e^{-i \frac{\pi}{4} \sigma_z}, \qquad 
U_{\sigma} = e^{-i \frac{\pi}{2} \sigma_y} = -i \sigma_y, 
\end{align}
where $\sigma_{\mu} (\mu = x,y,z)$ are the Pauli matrices. 
Under the same choice of representation matrices as (\ref{eq:u_p4g_1}) and (\ref{eq:u_p4g_2}), 
the two-cocycle $\omega \in Z^2(D_4,\R/2 \pi \Z)$ is fixed as in the following table.
\begin{align}
e^{i\omega_{p,p'}}
\quad =\quad 
\begin{tabular}{c|ccccccccc}
\hline 
$p \backslash p'$ & $1$ & $c_4$ & $c_2$ & $c_4^3$ & $\sigma$ & $\sigma c_4$ & $\sigma c_2$ & $\sigma c_4^3$ \\
\hline
$1$ & $1$ & $1$ & $1$ & $1$ & $1$ & $1$ & $1$ & $1$ \\
$c_4$ & $1$ & $1$ & $1$ & $-1$ & $-1$ & $1$ & $1$ & $1$ \\
$c_2$ & $1$ & $1$ & $-1$ & $-1$ & $-1$ & $-1$ & $1$ & $1$ \\
$c_4^3$ & $1$ & $-1$ & $-1$ & $-1$ & $-1$ & $-1$ & $-1$ & $1$ \\
$\sigma$ & $1$ & $1$ & $1$ & $1$ & $-1$ & $-1$ & $-1$ & $-1$ \\
$\sigma c_4$ & $1$ & $1$ & $1$ & $-1$ & $1$ & $-1$ & $-1$ & $-1$ \\
$\sigma c_2$ & $1$ & $1$ & $-1$ & $-1$ & $1$ & $1$ & $-1$ & $-1$ \\
$\sigma c_4^3$ & $1$ & $-1$ & $-1$ & $-1$ & $1$ & $1$ & $1$ & $-1$ \\
\hline 
\end{tabular}
\label{tab:d4_two_cocycle}
\end{align}
Then, the total two-cocycle $\tau$ for the spin half integer degrees of freedom 
with p4g symmetry is given by $\tau = \tau_{\sf p4g} + \omega$. 

The spin half integer $p4g$ symmetry is summarized in terms of Hamiltonians by 
\begin{equation}
\left\{\begin{array}{l}
U_p(\bk) H(\bk) U_p(\bk)^{-1} = H(p \bk), \\
U_{p_1}(p_2 \bk) U_{p_2}(\bk) = e^{i(\tau_{\sf p4g})_{p_1,p_2}(p_1p_2\bk)} \cdot e^{i\omega_{p_1,p_2}} U_{p_1,p_2}(\bk). 
\end{array}\right.
\end{equation}
The two-cocycle $\tau = \tau_{\sf p4g} + \omega \in Z^2 \big( D_4, C(T^2,\R/2 \pi \Z) \big)$ 
is summarized in the following table. 
\begin{align}
e^{i(\tau_{\sf p4g})_{p,p'}(pp'\bk)+i\omega_{p,p'}}
\quad =\quad 
\begin{tabular}{c|ccccccccc}
\hline 
$p \backslash p'$ & $1$ & $c_4$ & $c_2$ & $c_4^3$ & $\sigma$ & $\sigma c_4$ & $\sigma c_2$ & $\sigma c_4^3$ \\
\hline
$1$ & $1$ & $1$ & $1$ & $1$ & $1$ & $1$ & $1$ & $1$ \\
$c_4$ & $1$ & $1$ & $1$ & $-1$ & $-e^{- ik_x}$ & $e^{i k_y}$ & $e^{i k_x}$ & $e^{-i k_y}$ \\
$c_2$ & $1$ & $1$ & $-1$ & $-1$ & $-e^{- i(k_x+k_y)}$ & $-e^{-i (k_x-k_y)}$ & $e^{i (k_x+k_y)}$ & $e^{i (k_x+k_y)}$ \\
$c_4^3$ & $1$ & $-1$ & $-1$ & $-1$ & $-e^{- ik_y}$ & $-e^{-i k_x}$ & $-e^{i k_y}$ & $e^{i k_x}$ \\
$\sigma$ & $1$ & $1$ & $1$ & $1$ & $-e^{- ik_x}$ & $-e^{i k_y}$ & $-e^{i k_x}$ & $-e^{-i k_y}$ \\
$\sigma c_4$ & $1$ & $1$ & $1$ & $-1$ & $e^{- i(k_x+k_y)}$ & $-e^{-i (k_x-k_y)}$ & $-e^{i (k_x+k_y)}$ & $-e^{i (k_x+k_y)}$ \\
$\sigma c_2$ & $1$ & $1$ & $-1$ & $-1$ & $e^{- ik_y}$ & $e^{-i k_x}$ & $-e^{i k_y}$ & $-e^{i k_x}$ \\
$\sigma c_4^3$ & $1$ & $-1$ & $-1$ & $-1$ & $1$ & $1$ & $1$ & $-1$ \\
\hline 
\end{tabular}
\label{twist_spinful_p4g}
\end{align}
The fixed points $\Gamma$ and $M$ obey 
nontrivial projective representations of $D_4$ with 
two-cocycles $\tau_{\sf p4g}|_{\Gamma} + \omega$ and $\tau_{\sf p4g}|_{M} + \omega$, respectively. 
The $X$ point obeys a trivial projective representation of $D^{(v)}_2 = \{1,c_2, \sigma,\sigma c_2\}$
with two-cocycle $\tau_{\sf p4g}|_X + \omega$.

\subsubsection{A little bit about representations of $D_4$}
\label{sec:A little bit about representations of D4}
To compute the $K$-group $K^{\tau_{\sf p4g}+\omega+n}_{D_4}(T^2)$, 
we need to know the representations at high-symmetric points and their 
restrictions to subgroups of $D_4$ realized at low-symmetric lines in BZ. 
The dihedral group $D_4$ has 
four 1-dimensional linear irreps.\ $\{1, A, B, AB\}$, two 2-dimensional linear irrep.\ $\{E\}$, 
and two $2$-dimensional nontrivial projective irreps.\ $\{W,BW\}$. 
It is useful to introduce the character of a representation, which is defined as 
the trace of representation matrices. 
The character table of linear representations of the dihedral group $D_4$ is summarized as the following table: 
\begin{align}
\begin{tabular}[t]{cc|ccccc}
\hline 
irrep. & Mulliken& $\{1\}$ & $\{c_4,c_4^3\}$ & $\{c_2\}$ & $\{\sigma,\sigma c_2\}$ & $\{\sigma c_4,\sigma c_4^3\}$ \\
\hline
$1$ & $A_1$ & $1$ & $1$ & $1$ & $1$ & $1$ \\
$A$ & $A_2$ & $1$ & $1$ & $1$ & $-1$ & $-1$ \\
$B$ & $B_1$ & $1$ & $-1$ & $1$ & $1$ & $-1$ \\
$AB$ & $B_2$ & $1$ & $-1$ & $1$ & $-1$ & $1$ \\
$E  $ & $E$ & $2$ & $0$ & $-2$ & $0$ & $0$ \\
\hline 	
\end{tabular}
\end{align}

For projective representations, 
we need to specify a two-cocycle $\omega_{p,p'} \in Z^2(D_4,\R/2 \pi \Z)$ 
which appears in 
\begin{align}
U(p) U(p') = e^{i \omega_{p,p'}} U(p p'), \qquad p,p' \in D_4.  
\end{align}
Once we fix a two-cocycle $\omega$, 
projective representations with two-cocycle $\omega$, 
dubbed {\it $\omega$-projective representations}, 
make sense. 
Note that fixing of a two-cocycle 
is needed for a projective representation with 
the trivial group cohomology $[\omega] = 0 \in H^2(D_4,U(1))$. 
In the same way, 
the $\omega$-projective character is defined as the trace of 
the representation matrices 
\begin{align}
\chi(p) := \tr (U(P)). 
\end{align}
Clearly, $\chi(p)$ is invariant under the unitary transformation $U(p) \mapsto V U(p) V^{\dag}$. 
Different choices of two-cocycles with the same cohomology class may change the projective character. 
For example, the following table shows the 
projective characters at the symmetric points $\Gamma, M$, and $X$: 
\begin{align}
\begin{tabular}[t]{c|c|c|cccccccc}
\hline 
Symmetric point & two-cocycle & irrep. & $1$ & $c_4$ & $c_2$ & $c_4^3$ & $\sigma$ & $\sigma c_4$ & $\sigma c_2$ & $\sigma c_4^3$ \\
\hline
$\Gamma$ & $\omega$ 
& $W$ & $2$ & $\sqrt{2}$ & $0$ & $-\sqrt{2}$ & $0$ & $0$ & $0$ & $0$ \\
& (defined in (\ref{tab:d4_two_cocycle})) & $BW$ & $2$ & $-\sqrt{2}$ & $0$ & $\sqrt{2}$ & $0$ & $0$ & $0$ & $0$ \\
\hline 	
$M$ & $\tau_{\sf p4g}|_{M} + \omega$ 
& $W$ & $2$ & $-\sqrt{2} i$ & $0$ & $\sqrt{2} i$ & $0$ & $0$ & $0$ & $0$ \\
& & $BW$ & $2$ & $\sqrt{2} i$ & $0$ & $- \sqrt{2} i$ & $0$ & $0$ & $0$ & $0$ \\
\hline 	
$X$ & $\tau_{\sf p4g}|_{X} + \omega$ 
& $1$ & $1$ & & $-i$ & & $1$ & & $-i$ & \\
& & $t_{\sigma c_2}$ & $1$ & & $i$ & & $1$ & & $i$ & \\
& & $t_{\sigma}$ & $1$ & & $i$ & & $-1$ & & $-i$ & \\
& & $t_{\sigma c_2} t_{\sigma}$ & $1$ & & $-i$ & & $-1$ & & $i$ & \\
\hline 	
\end{tabular}
\label{tab:proj_character_d4}
\end{align}
Examples of representations of $D_4$ are shown in Table~\ref{Tab:rep_of_D4}. 

\begin{table*}[!]
\begin{center}
\caption{Examples of projective representations of $D_4$. 
$\zeta$ means $\zeta = e^{- \pi i/4}$. 
}
\begin{tabular}[t]{c|c|ccccccccc}
\hline 
two-cocycle & $U_{\rho}(p)$ & $1$ & $c_4$ & $c_2$ & $c_4^3$ & $\sigma$ & $\sigma c_4$ & $\sigma c_2$ & $\sigma c_4^3$ \\
\hline
& $1$ & $1$ & $1$ & $1$ & $1$ & $1$ & $1$ & $1$ & $1$ \\
& $A$ & $1$ & $1$ & $1$ & $1$ & $-1$ & $-1$ & $-1$ & $-1$ \\
triv. & $B$ & $1$ & $-1$ & $1$ & $-1$ & $1$ & $-1$ & $1$ & $-1$ \\
(linear reps.) & $AB$ & $1$ & $-1$ & $1$ & $-1$ & $-1$ & $1$ & $-1$ & $1$ \\
& $E$ & $\begin{pmatrix}
1 & 0 \\
0 & 1 \\
\end{pmatrix}$ & $\begin{pmatrix}
0 & -1 \\
1 & 0 \\
\end{pmatrix}$ & $\begin{pmatrix}
-1 & 0 \\
0 & -1 \\
\end{pmatrix}$ & $\begin{pmatrix}
0 & 1 \\
-1 & 0 \\
\end{pmatrix}$ & $\begin{pmatrix}
1 & 0 \\
0 & -1 \\
\end{pmatrix}$ & $\begin{pmatrix}
0 & -1 \\
-1 & 0 \\
\end{pmatrix}$ & $\begin{pmatrix}
-1 & 0 \\
0 & 1 \\
\end{pmatrix}$ & $\begin{pmatrix}
0 & 1 \\
1 & 0 \\
\end{pmatrix}$ \\
\hline
$\omega$ & $W$ & 
$\begin{pmatrix}
1 & 0 \\
0 & 1 \\
\end{pmatrix}$ 
&$\begin{pmatrix}
\zeta & 0 \\
0 & \zeta^{-1} \\
\end{pmatrix}$ 
& $\begin{pmatrix}
-i & 0 \\
0 & i \\
\end{pmatrix}$ 
& $\begin{pmatrix}
-i \zeta & 0 \\
0 & i \zeta^{-1} \\
\end{pmatrix}$ 
& $\begin{pmatrix}
0 & -1 \\
1 & 0 \\
\end{pmatrix}$ 
& $\begin{pmatrix}
0 & -\zeta^{-1} \\
\zeta & 0 \\
\end{pmatrix}$
& $\begin{pmatrix}
0 & -i \\
-i & 0 \\
\end{pmatrix}$
& $\begin{pmatrix}
0 & -i \zeta^{-1} \\
-i \zeta & 0 \\
\end{pmatrix}$ \\
& $BW$ & 
$\begin{pmatrix}
1 & 0 \\
0 & 1 \\
\end{pmatrix}$ 
&$\begin{pmatrix}
-\zeta & 0 \\
0 & -\zeta^{-1} \\
\end{pmatrix}$ 
& $\begin{pmatrix}
-i & 0 \\
0 & i \\
\end{pmatrix}$ 
& $\begin{pmatrix}
i \zeta & 0 \\
0 & -i \zeta^{-1} \\
\end{pmatrix}$ 
& $\begin{pmatrix}
0 & -1 \\
1 & 0 \\
\end{pmatrix}$ 
& $\begin{pmatrix}
0 & \zeta^{-1} \\
-\zeta & 0 \\
\end{pmatrix}$
& $\begin{pmatrix}
0 & -i \\
-i & 0 \\
\end{pmatrix}$
& $\begin{pmatrix}
0 & i \zeta^{-1} \\
i \zeta & 0 \\
\end{pmatrix}$ \\
\hline 
\end{tabular}
\label{Tab:rep_of_D4}
\end{center}
\end{table*}

\begin{table*}[!]
\begin{center}
\caption{The table of tensor product representations of $D_4$. }
\begin{tabular}[t]{c|ccccc|cc}
\hline 
$\rho_1 \otimes \rho_2, \rho_1 \backslash \rho_2$ & $1$ & $A$ & $B$ & $AB$ & $E$ & $W$ & $BW$ \\
\hline
$1$ & $1$ & $A$ & $B$ & $AB$ & $E$ & $W$ & $BW$ \\
$A$ & $A$ & $1$ & $AB$ & $B$ & $E$ & $W$ & $BW$ \\
$B$ & $B$ & $AB$ & $1$ & $A$ & $E$ & $BW$ & $W$ \\
$AB$ & $AB$ & $B$ & $A$ & $1$ & $E$ & $BW$ & $W$ \\
$E  $ & $E  $ & $E  $ & $E  $ & $E  $ & $1+A+B+AB$ & $W+BW$ & $W+BW$ \\
\hline 
\end{tabular}
\label{Tab:ProductR(D_4)}
\end{center}
\end{table*}

The tensor product of two linear representations is defined by 
\begin{align}
U_{\rho_1 \otimes \rho_2}(p) := U_{\rho_1}(p) U_{\rho_2}(p), 
\label{eq:tensor_prod_rep_d4}
\end{align}
which induces the ring structure on $R(D_4)$, the Abelian group generated by 
linear representations of $D_4$. 
If $\rho_2$ is a $\omega$-projective representation, 
eq.(\ref{eq:tensor_prod_rep_d4}) defines 
the $R(D_4)$-module structure of 
$R^{\omega}(D_4)$, the Abelian group generated by 
$\omega$-projective representations. 
Table~\ref{Tab:ProductR(D_4)} summarizes the tensor product representations. 
As the notations suggest, $AB$ and $BW$ means $A \otimes B$ and $B \otimes W$, respectively. 
From Table~\ref{Tab:ProductR(D_4)}, 
the representation ring of $D_4$ reads 
\begin{equation}
R(D_4) \cong \Z[A,B,E]/(1-A^2,1-B^2,E-AE,E-BE,E^2-1-A-B-AB). 
\end{equation}
We can read off the $R(D_4)$-module structure 
of the $\omega$-projective representations $R^{\omega}(D_4)$ 
as 
\begin{align}
R^{\omega}(D_4) \cong (1+A+E). 
\end{align}

The restriction of group elements of $D_4$ to its 
subgroup $H$ leads to 
the restriction of the two-cocycle 
\begin{align}
\omega \to \omega|_H \in Z^2(H,U(1))
\end{align}
and the restriction of $\omega$-projective representations of $D_4$ to 
$(\omega|_H)$-projective representations, 
\begin{align}
\rho \to \rho|_H \in R^{\omega|_H}(H).
\end{align}
We summarize the restriction of irreps.\ of $D_4$ in Table~\ref{Tab:RestrictionD_4}. 
\begin{table*}[!]
\begin{center}
\caption{
Subgroups of $D_4 = \{1,c_4,c_2,c_4^3, \sigma,\sigma c_4,\sigma c_2,\sigma c_4^3\}$ 
and restrictions of representations of $D_4$ to subgroups.
In the restriction of two irreps.\ of 
$\omega$-projective irreps., 
we trivialize the two-cocycle $\omega$ by the 
redefinition $U_{c_4} \mapsto \zeta^{-1} U_{c_4}$. 
}
\begin{tabular}{c|l|l|ccccc|cc}
\hline 
subgroup $H$ & elements & $R(H)$ & $1|_H$ & $A|_H$ & $B|_H$ & $AB|_H$ & $E|_H$ & $W|_H$ & $BW|_H$ \\
\hline
$C_4$ & $\{1,c_4,c_2,c_4^3\}$ & $\Z[t]/(1-t^4)$ & $1$ & $1$ & $t^2$ & $t^2$ & $t+t^3$ & $1+t$ & $t^2+t^3$ \\
$D_2^{(v)}$ & $\{1,c_2,\sigma ,\sigma c_2\}$ & $\Z[t_1,t_2]/(t_1^2,t_2^2)$ & $1$ & $t_1t_2$ & $1$ & $t_1t_2$ & $t_1+t_2$ & $W$ & $W$  \\
$D_2^{(d)}$ & $\{1,c_2,\sigma c_4,\sigma c_4^3\}$ & $\Z[t_1,t_2]/(t_1^2,t_2^2)$ & $1$ & $t_1t_2$ & $t_1t_2$ & $1$ & $t_1+t_2$ & $W$ & $W$  \\
$\Z_2$ & $\{1,c_2\}$ & $\Z[s]/(1-s^2)$ & $1$ & $1$ & $1$ & $1$ & $2 s$ & $1+s$ & $1+s$  \\
$\Z_2^{(v)}$ & $\{1,\sigma \} \sim \{1,\sigma c_2\}$ & $\Z[s]/(1-s^2)$ & $1$ & $s$ & $1$ & $s$ & $1+s$ & $1+s$ & $1+s$  \\
$\Z_2^{(d)}$ & $\{1,\sigma c_4\} \sim \{1,\sigma c_4^3\}$ & $\Z[s]/(1-s^2)$ & $1$ & $s$ & $s$ & $1$ & $1+s$ & $1+s$ & $1+s$  \\
\hline 
\end{tabular}
\label{Tab:RestrictionD_4}
\end{center}
\end{table*}

\subsubsection{$K$-group of 1-dimensional subspace $X_1$}
To compute the $K$-group, we introduce $D_4$-invariant subspaces
$X_1, Y_1$ and $Z$: 
$$
X_1 \ \ = \ \ 
\xygraph{
!{<0cm,0cm>;<1cm,0cm>:<0cm,1cm>::}
!{(1,0)}*+{\bullet}="a",
!{(1,1)}*+{\bullet}="b" ,
!{(0,1)}*+{\bullet}="c" ,
!{(-1,1)}*+{\bullet}="d" ,
!{(-1,0)}*+{\bullet}="e", 
!{(-1,-1)}*+{\bullet}="f", 
!{(0,-1)}*+{\bullet}="g" ,
!{(1,-1)}*+{\bullet}="h" ,
!{(0,0)}*+{\bullet}="i" ,
"a"-@{.}"b",
"b"-@{.}"c",
"c"-@{.}"d",
"d"-"e",
"e"-"f",
"f"-"g",
"g"-"h",
"h"-@{.}"a",
"a"-"i",
"c"-"i",
"e"-"i",
"g"-"i",
"b"-"i",
"d"-"i",
"f"-"i",
"h"-"i",
}
\qquad \qquad 
Y_1 \ \ = \ \ 
\xygraph{
!{<0cm,0cm>;<1cm,0cm>:<0cm,1cm>::}
!{(1,0)}*+{\bullet}="a",
!{(1,1)}="b" ,
!{(0,1)}*+{\bullet}="c" ,
!{(-1,1)}="d" ,
!{(-1,0)}*+{\bullet}="e", 
!{(-1,-1)}="f", 
!{(0,-1)}*+{\bullet}="g" ,
!{(1,-1)}="h" ,
!{(0,0)}*+{\bullet}="i" ([]!{+(0.6,-0.6)} {\Gamma}),
!{(0.3,-0.3)}-@{->}!{(0.1,-0.1)}, 
"a"-@{.}"b",
"b"-@{.}"c",
"c"-@{.}"d",
"d"-@{.}"e",
"e"-@{.}"f",
"f"-@{.}"g",
"g"-@{.}"h",
"h"-@{.}"a",
"a"-"i",
"c"-"i",
"e"-"i",
"g"-"i",
} \qquad \qquad 
Z \ \ = \ \ \xygraph{
!{<0cm,0cm>;<1cm,0cm>:<0cm,1cm>::}
!{(1,1)}*+{\bullet}="a",
!{(-1,1)}*+{\bullet}="b" ,
!{(-1,-1)}*+{\bullet}="c", 
!{(1,-1)}*+{\bullet}="d" ,
!{(0,0)}*+{\bullet}="e", 
"a"-@{.}"b",
"b"-@{.}"c",
"c"-@{.}"d",
"d"-@{.}"a",
"a"-"e",
"b"-"e",
"c"-"e",
"d"-"e",
}$$
In the computation below, 
we focus on the following fundamental region in the BZ torus that 
is surrounded by $\Gamma$, $M$ and $X$ points. 
We mark points on the edges of the fundamental region with $I_1, I_2$ and $I_3$. 
See the following figure: 
$$
\xygraph{
!{<0cm,0cm>;<1cm,0cm>:<0cm,1cm>::}
!{(1,0)}="a"([]!{+(0.2,-0.2)} {X}),
!{(1,1)}="b"([]!{+(0.3,0.3)} {M}),
!{(0,1)}="c" ,
!{(-1,1)}="d" ,
!{(-1,0)}="e", 
!{(-1,-1)}="f", 
!{(0,-1)}="g" ,
!{(1,-1)}="h" ,
!{(0,0)}="i"([]!{+(-0.2,-0.2)} {\Gamma}),
!{(0.5,0)}*+{\bullet}([]!{+(0,-0.3)} {I_1}),
!{(0.5,0.5)}*+{\bullet}([]!{+(-0.3,0.3)} {I_2}),
!{(1,0.5)}*+{\bullet}([]!{+(0.3,0)} {I_3}),
!{(1,0)}*+{\bullet}([]!{+(0.2,-0.2)} {X}),
!{(1,1)}*+{\bullet}([]!{+(0.3,0.3)} {M}),
!{(0,0)}*+{\bullet}([]!{+(-0.2,-0.2)} {\Gamma}),
"a"-"b",
"a"-"i",
"b"-"i",
"b"-@{.}"d",
"d"-@{.}"f",
"f"-@{.}"h",
"h"-@{.}"b",
"a"-@{.}"e",
"c"-@{.}"g",
"b"-@{.}"f",
"d"-@{.}"h",
}
$$

First, we compute the $K$-group of $Y_1$ 
by use of the Mayer-Vietoris sequence. 
$Y_1$ is divided into two parts $Y_1 = U \cup V$, 
where $U$ and $V$ have the following $D_4$-equivariant homotopy equivalences
\begin{equation}
U \sim \{\Gamma \} = (D_4/D_4) \times pt , \qquad 
V \sim \{ X, c_4 \cdot X \} \sim (D_4/D_2^{(v)}) \times pt , 
\end{equation}
where $D_2^{(v)} = \{1 , \sigma , \sigma c_2, c_2\}$ is a subgroup of $D_4$. 
The intersection has the $D_4$-equivariant homotopy equivalence
\begin{equation}
U \cap V 
\sim \{I_1, c_4 I_1, c_2 I_1, c_4^3 I_1 \}
\sim (D_4/\Z_2^{(y)}) \times pt = \big\{ \{ 1, \sigma \}, \{c_2, \sigma c_2\}, \{c_4, \sigma c_4^3\}, \{c_4^3, \sigma c_4 \} \big\}. 
\end{equation}
Here we chose $\Z_2^{(y)} = \{1,\sigma\}$ as a $\Z_2$ subgroup. 
In this choice, the intersection $U \cap V $ can be labeled by the $D_4$-space $(D_4/\Z_2^{(y)}) \times pt$ as follows: 
$$
U \cap V \quad \sim \quad 
\xygraph{
!{<0cm,0cm>;<1.5cm,0cm>:<0cm,1.5cm>::}
!{(1,0)}*+{\circ}="a",
!{(1,1)}="b" ,
!{(0,1)}*+{\circ}="c" ,
!{(-1,1)}="d" ,
!{(-1,0)}*+{\circ}="e", 
!{(-1,-1)}="f", 
!{(0,-1)}*+{\circ}="g" ,
!{(1,-1)}="h" ,
!{(0,0)}*+{\circ}="i",
"a"-@{.}"b",
"b"-@{.}"c",
"c"-@{.}"d",
"d"-@{.}"e",
"e"-@{.}"f",
"f"-@{.}"g",
"g"-@{.}"h",
"h"-@{.}"a",
"a"-"i",
"c"-"i",
"e"-"i",
"g"-"i",
!{(0.5,-0.15)}*+{\{1,\sigma\}}, 
!{(-0.6,-0.15)}*+{\ \{c_2, \sigma c_2\}}, 
!{(0.05,0.7)}*+{\{c_4,\sigma c_4^3\}}, 
!{(0.03,-0.7)}*+{\{c_4^3,\sigma c_4\}}, 
}
$$
The $D_4$ group naturally acts on the set $D_4/\Z_2^{(y)}$. 
An alternative choice is $\Z_2^{(x)} = \{1, \sigma c_2\}$.  
The finial expression for the $K$-group $K_{D_4}^{\tau+ c + n}(Y_1)$ 
does not depend on the choices $\Z_2^{(x)}$ and $\Z_2^{(y)}$. 

The six term Mayer-Vietoris sequence associated with the decomposition 
$Y_1 = U \cup V$ is given by 
\begin{equation}\begin{split}
\begin{CD}
0 @<<< 0 @<<< K_{D_4}^{\tau_{\sf p4g}+\omega+1}(Y_1) \\
@VVV @. @AAA \\
K_{D_4}^{\tau_{\sf p4g}+\omega+0}(Y_1) @>>> \underbrace{R^{\omega}(D_4)}_{\Gamma} \oplus \underbrace{R^{\tau_{\sf p4g}|_X+\omega}(D_2^{(v)})}_{X} @>\Delta_0>> \underbrace{R^{\tau_{\sf p4g}|_{I_1} + \omega}(\Z_2^{(y)})}_{I_1}. \\
\end{CD} 
\end{split}\end{equation}
The homomorphism $\Delta_0$ of $R(D_4)$-modules is given by 
\begin{equation}
\Delta_0 : (\rho, g(t_{\sigma c_2},t_{\sigma}))  \mapsto  
\rho|_{\Z_2^{(y)}} \cdot (1+t_{\sigma}) - g(1,t_{\sigma}). 
\end{equation}
$\mathrm{Ker} (\Delta_0)$ is spanned by the following basis
\begin{equation}
\{ \underbrace{(W,t_{\sigma c_2}+t_{\sigma}), (BW,t_{\sigma c_2}+t_{\sigma})}_{(1+A+E)}, \underbrace{(0,1-t_{\sigma c_2}), (0,t_{\sigma}-t_{\sigma c_2} t_{\sigma})}_{(1+B-E)} \}.
\end{equation}
We have 
\begin{equation}
K_{D_4}^{\tau_{\sf p4g}+\omega+0}(Y_1) \cong \mathrm{Ker}(\Delta_0) \cong \overbrace{(1+A+E)}^{\Z^2} \oplus \overbrace{(1+B-E)}^{\Z^2}, \qquad 
K_{D_4}^{\tau_{\sf p4g}+\omega+1}(Y_1) = 0, 
\end{equation}
where $(1+A+E)$ and $(1+B-E)$ are $R(D_4)$-ideals defined by 
\begin{align}
&(1+A+E) = \{(1+A+E) f(A,B,E) | f(A,B,E) \in R(D_4) \}, \\
&(1+B-E) = \{(1+B-E) f(A,B,E) | f(A,B,E) \in R(D_4) \}. 
\end{align}

Next, we compute the $K$-group of the subspace $X_1$. 
Decompose $X_1$ to $U \cup V$ as follows: 
$$
X_1 = U \cup V \ \ = \ \ 
\xygraph{
!{<0cm,0cm>;<1cm,0cm>:<0cm,1cm>::}
!{(1,0)}*+{\bullet}="a",
!{(1,1)}="b" ,
!{(0,1)}*+{\bullet}="c" ,
!{(-1,1)}="d" ,
!{(-1,0)}*+{\bullet}="e", 
!{(-1,-1)}="f", 
!{(0,-1)}*+{\bullet}="g" ,
!{(1,-1)}="h" ,
!{(0,0)}*+{\bullet}="i" ,
"b"-@{.}"d",
"d"-@{.}"f",
"f"-@{.}"h",
"h"-@{.}"b",
"a"-"i",
"c"-"i",
"e"-"i",
"g"-"i",
"i"-!{(0.5,0.5)},
"i"-!{(-0.5,0.5)},
"i"-!{(-0.5,-0.5)},
"i"-!{(0.5,-0.5)},
"e"-!{(-1,0.6)},
"e"-!{(-1,-0.6)},
"g"-!{(-0.6,-1)},
"g"-!{(0.6,-1)},
}\ \ 
\cup \ \ 
\xygraph{
!{<0cm,0cm>;<1cm,0cm>:<0cm,1cm>::}
!{(1,0)}="a",
!{(1,1)}*+{\bullet}="b" ,
!{(0,1)}="c" ,
!{(-1,1)}*+{\bullet}="d" ,
!{(-1,0)}="e", 
!{(-1,-1)}*+{\bullet}="f", 
!{(0,-1)}="g" ,
!{(1,-1)}*+{\bullet}="h" ,
!{(0,0)}="i" ,
"h"-@{.}"b",
"b"-@{.}"d",
"d"-@{.}"f",
"f"-@{.}"h",
"b"-!{(0.5,0.5)},
"d"-!{(-0.5,0.5)},
"d"-!{(-1,0.4)},
"f"-!{(-0.5,-0.5)},
"f"-!{(-1,-0.4)},
"f"-!{(-0.4,-1)},
"h"-!{(0.5,-0.5)},
"h"-!{(0.4,-1)},
}$$
$U$ and $V$ are $D_4$-equivariantly homotopy equivalent to $Y_1$ and the point $(\pi,\pi) \sim D_4/D_4$, respectively. 
The intersection $U \cap V$ is $D_4$-equivariantly homotopy equivalent to the disjoint union of two $D_4$-spaces. 
\begin{equation}
U \cap V \sim (D_4/\Z_2^{(d)}) \sqcup (D_4/\Z_2^{(x)})
\quad \sim \quad 
\xygraph{
!{<0cm,0cm>;<1cm,0cm>:<0cm,1cm>::}
!{(1,0)}="a",
!{(1,1)}*+{\circ}="b" ,
!{(0,1)}="c" ,
!{(-1,1)}*+{\circ}="d" ,
!{(-1,0)}="e", 
!{(-1,-1)}*+{\circ}="f", 
!{(0,-1)}="g" ,
!{(1,-1)}*+{\circ}="h" ,
!{(0,0)}*+{\circ}="i" ,
"h"-@{.}"b",
"b"-@{.}"d",
"d"-@{.}"f",
"f"-@{.}"h",
"i"-"b",
"i"-"d",
"i"-"f",
"i"-"h",
}
\quad  \sqcup \quad 
\xygraph{
!{<0cm,0cm>;<1cm,0cm>:<0cm,1cm>::}
!{(1,0)}="a",
!{(1,1)}*+{\circ}="b" ,
!{(0,1)}="c" ,
!{(-1,1)}*+{\circ}="d" ,
!{(-1,0)}*+{\circ}="e", 
!{(-1,-1)}*+{\circ}="f", 
!{(0,-1)}*+{\circ}="g" ,
!{(1,-1)}*+{\circ}="h" ,
!{(0,0)}="i" ,
"h"-@{.}"b",
"b"-@{.}"d",
"d"-@{.}"f",
"f"-@{.}"h",
"d"-"e",
"e"-"f",
"f"-"g",
"g"-"h",
}
\end{equation}
The Mayer-Vietoris sequence of $X_1 = U \cup V$ is given by 
\begin{equation}\begin{split}
\begin{CD}
0 @<<< 0 @<<< K_{D_4}^{\tau_{\sf p4g}+\omega+1}(X_1) \\
@VVV @. @AAA \\
K_{D_4}^{\tau_{\sf p4g}+\omega+0}(X_1) @>>> K^{\tau_{\sf p4g}+\omega+0}_{D_4}(Y_1) \oplus \underbrace{R^{\omega}(D_4)}_{M} @>\Delta_0>> \underbrace{R(\Z_2^{(d)})}_{I_2} \oplus \underbrace{R(\Z_2^{(x)})}_{I_3}. \\
\end{CD}
\end{split}\end{equation}
From Table~\ref{Tab:RestrictionD_4}, the restrictions of elements in the $K$-group $K^{\tau_{\sf p4g}+\omega+0}_{D_4}(Y_1)$ and $R^{\omega}(D_4)$ to the 
intersection are given by 
\begin{equation}
j_U^* : K^{\tau_{\sf p4g}+\omega+0}_{D_4}(Y_1) \mapsto R(\Z_2^{(d)}) \oplus R(\Z_2^{(x)}), \qquad 
\left\{ \begin{array}{l}
(W, t_{\sigma c_2}+t_{\sigma}) \mapsto (1+t_{\sigma c_4^3},1+t_{\sigma c_2}), \\
(BW,t_{\sigma c_2}+t_{\sigma}) \mapsto (1+t_{\sigma c_4^3},1+t_{\sigma c_2}), \\ 
(0,1-t_{\sigma c_2}) \mapsto (0,1-t_{\sigma c_2}), \\ 
(0,t_{\sigma}-t_{\sigma c_2} t_{\sigma}) \mapsto (0,1-t_{\sigma c_2}), \\ 
\end{array} \right. \ \ 
\end{equation}
\begin{equation}
j_V^* : R^{\omega}(D_4) \mapsto R(\Z_2^{(d)}) \oplus R(\Z_2^{(x)}), \qquad 
\left\{ \begin{array}{l}
W \mapsto (1+t_{\sigma c_4^3},1+t_{\sigma c_2}), \\
BW \mapsto (1+t_{\sigma c_4^3},1+t_{\sigma c_2}). \\ 
\end{array} \right. 
\end{equation}
Then, the kernel of $\Delta_0 = j_U^* - j_V^*$ is spanned by the following basis 
in terms of representations at symmetric points $(\Gamma,X,M)$ 
\begin{multline}
\{ \underbrace{(W,t_{\sigma c_2}+t_{\sigma},W), (BW,t_{\sigma c_2}+t_{\sigma},BW)}_{(1+A+E)}, 
\underbrace{(0,1-t_{\sigma c_2}-t_{\sigma}+t_{\sigma c_2} t_{\sigma},0)}_{(1+A+B+AB-2E)}, 
\underbrace{(0,0,W-BW)}_{(1+A-B-AB)} \} \\
\subset 
R^{\omega}(D_4) \oplus R(D_2^{(v)}) \oplus R^{\omega}(D_4). 
\label{eq:local_data_X_1}
\end{multline}
${\rm Im}(\Delta_0)$ is spanned by 
\begin{equation}
\{(1+t_{\sigma c_4^3},1+t_{\sigma c_2}), (0,1-t_{\sigma c_2})\} \subset R(\Z_2^{(d)}) \oplus R(\Z_2^{(x)}). 
\end{equation}
Notice that a basis of $R(\Z_2^{(d)}) \oplus R(\Z_2^{(x)})$ can be chosen as 
\begin{equation}
\{(1,0),(0,1),(1+t_{\sigma c_4^3},1+t_{\sigma c_2}),(0,1-t_{\sigma c_2})\}.
\end{equation}
Hence, $\mathrm{Coker}(\Delta_0)$ is generated by 
two elements $\{[1,1],[0,1]\}$. 
The $R(D_4)$-actions on these generators, 
\begin{align}
\left\{ \begin{array}{l}
A\cdot [1,1] = [(t_{\sigma c_4^3},t_{\sigma c_2})(1,1)] = [t_{\sigma c_4^3},t_{\sigma c_2}] = -[1,1], \\
B\cdot [1,1] = [(t_{\sigma c_4^3},1)(1,1)] = [t_{\sigma c_4^3},1] = -[1,1], \\
E\cdot [1,1] = [(1+t_{\sigma c_4^3},1+t_{\sigma c_2})(1,1)] = [1+t_{\sigma c_4^3},1+t_{\sigma c_2}] = 0, \\
\end{array}\right. \\
\left\{ \begin{array}{l}
A\cdot [0,1] = [(t_{\sigma c_4^3},t_{\sigma c_2})(0,1)] = [0,t_{\sigma c_2}] = [0,1], \\
B\cdot [0,1] = [(t_{\sigma c_4^3},1)(0,1)] = [0,1], \\
E\cdot [0,1] = [(1+t_{\sigma c_4^3},1+t_{\sigma c_2})(0,1)] = [0,1+t_{\sigma c_2}]= 2 [0,1], \\
\end{array}\right. 
\end{align}
imply the $R(D_4)$-module structures
$\Z[1,1] \cong (1-A-B+AB)$ and 
$\Z[0,1] \cong (1+A+B+AB+2E)$.  
We consequently get the $K$-group of $X_1$ as follows
\begin{align}
&K^{\tau_{\sf p4g}+\omega+0}_{D_4}(X_1) \cong 
\overbrace{(1+A+E)}^{\Z^2} \oplus 
\overbrace{(1+A+B+AB-2E)}^{\Z} \oplus 
\overbrace{(1+A-B-AB)}^{\Z}, \\
&K^{\tau_{\sf p4g}+\omega+1}_{D_4}(X_1) \cong 
\overbrace{(1-A-B+AB)}^{\Z} \oplus 
\overbrace{(1+A+B+AB+2E)}^{\Z}. 
\end{align}

\subsubsection{$K$-group of $Y_1 \vee Z$}
In the same way, 
the $K$-group of the subspace $Z$ is given by the 
Mayer-Vietoris sequence 
\begin{equation}\begin{split}
\begin{CD}
0 @<<< 0 @<<< K_{D_4}^{\tau_{\sf p4g}+\omega+1}(Z) \\
@VVV @. @AAA \\
K_{D_4}^{\tau_{\sf p4g}+\omega+0}(Z) @>>> \underbrace{R^{\omega}(D_4)}_{\Gamma} \oplus \underbrace{R^{\tau_{\sf p4g}|_M + \omega}(D_4)}_{M} @>\Delta_0>> \underbrace{R^{\tau_{\sf p4g}|_{I_2}+\omega}(\Z_2^{(d)})}_{I_2}. \\
\end{CD} 
\end{split}\end{equation}
Here, $\Delta_0$ is given by 
\begin{align}
\Delta_0 : (f(B),g(B)) \mapsto (f(1)-g(1)) (1+t_{\sigma c_4^3}). 
\end{align}
The kernel of $\Delta_0$ is spanned by 
\begin{align}
&\ker (\Delta_0) : \qquad \{ \underbrace{(W,W), (BW,BW)}_{(1+A+E)}, \underbrace{(0,W-BW)}_{1+A-B-AB} \} \subset R^{\omega}(D_4) \oplus R^{\omega}(D_4). 
\end{align}
The generator of the cokernel of $\Delta_0$ is 
represented by $[1] \in R(\Z_2^{(d)})$, and the $R(D_4)$-module structure 
is summarized as $A \cdot [1] = -[1]$, $B \cdot [1] = - [1]$, and $E \cdot [1] = 0$, 
which implies $\coker(\Delta_0) \cong (1-A-B+AB)$. 
We have 
\begin{align}
&K^{\tau_{\sf p4g}+\omega+0}_{D_4}(Z) \cong 
\overbrace{(1+A+E)}^{\Z^2} \oplus 
\overbrace{(1+A-B-AB)}^{\Z}, \\
&K^{\tau_{\sf p4g}+\omega+1}_{D_4}(Z) \cong 
\overbrace{(1-A-B+AB)}^{\Z}.  
\end{align}

Gluing the $K$-groups of $Y_1$ and $Z$ 
at the single fixed point $\Gamma = (0,0)$ of the $D_4$ action, 
we have the $K$-group of $Y_1 \vee Z$, 
where $Y_1 \vee Z$ is defined as the disjoint union $Y_1 \sqcup Z$ with the $\Gamma$ point of $Y_1$ and that of $Z$ identified, 
\begin{align}
&K^{\tau_{\sf p4g}+\omega+0}_{D_4}(Y_1 \vee Z) \cong 
\overbrace{(1+A+E)}^{\Z^2} \oplus 
\overbrace{(1+A-B-AB)}^{\Z} \oplus 
\overbrace{(1+B-E)}^{\Z^2}, \\
&K^{\tau_{\sf p4g}+\omega+1}_{D_4}(Y_1 \vee Z) \cong 
\overbrace{(1-A-B+AB)}^{\Z}.  
\end{align}

\subsubsection{$K$-group over $T^2$}
Next, we ``extend'' the wave function over the subspaces $X_1$ and $Y_1 \vee Z$ to that over the BZ torus $T^2$. 
In other words, we assume that the existence of a finite energy gap 
persists in the whole region of BZ torus $T^2$, which gives rise to 
a kind of global consistency condition on the wave functions with 
p4g symmetry. 
Mathematically, this global constraint can be expressed by the 
exact sequence for the pair $(T^2,X_1)$ 
\begin{equation}\begin{split}
\begin{CD}
K^{\tau_{\sf p4g}+\omega+1}_{D_4}(X_1) @<<< K^{\tau_{\sf p4g}+\omega+1}_{D_4}(T^2) @<<< K^{\tau_{\sf p4g}+\omega+1}_{D_4}(T^2,X_1) \\
@VVV @. @AAA \\
K^{\tau_{\sf p4g}+\omega+0}_{D_4}(T^2,X_1) @>>> K^{\tau+0}_{D_4}(T^2) @>>> K^{\tau_{\sf p4g}+\omega+0}_{D_4}(X_1), \\
\end{CD}
\end{split}\end{equation}
which is the exact sequence of $R(D_4)$-modules
\begin{equation}\begin{split}
\begin{CD}
(1-A-B+AB) \oplus (1+A+B+AB+2E) @<<< K^{\tau_{\sf p4g}+\omega+1}_{D_4}(T^2) @<<< 0 \\
@VV\delta V @. @AAA \\
(1+A+B+AB+2E) @>>> K^{\tau_{\sf p4g}+\omega+0}_{D_4}(T^2) @>>> \begin{array}{l}
K^{\tau_{\sf p4g}+\omega+0}_{D_4}(X_1).
\end{array} \\
\end{CD}
\end{split}\end{equation}
Here, we used 
\begin{equation}
K^{\tau_{\sf p4g}+\omega+n}_{D_4}(T^2,X_1) \cong \widetilde K^n_{D_4}( D_4 \times e^2) \cong K^n(S^2) 
\cong \left\{ \begin{array}{ll}
(1+A+B+AB+2E) & (n=0), \\
0 & (n=1). \\
\end{array}\right.
\end{equation}
Any $R(D_4)$-homomorphism $f : (1-A-B+AB) \to (1+A+B+AB+2E)$ is trivial, because $f(1) = A \cdot f(1) = f( A \cdot 1) = f(-1) = - f(1) = 0$. Therefore $\delta$ is either: (i) trivial; (ii) non-trivial and surjective; or (iii) non-trivial and non-surjective. To determine which is the case, we employ the exact sequence for the pair $(T^2, Y_1 \vee Z)$: 
\begin{align}
\begin{CD}
K^{\tau_{\sf p4g}+\omega+1}_{D_4}(Y_1 \vee Z)  @<<< K^{\tau_{\sf p4g}+\omega+1}_{D_4}(T^2) @<<< K^{\tau_{\sf p4g}+\omega+1}_{D_4}(T^2,Y_1 \vee Z) @= \underbrace{(1 - A + B - AB)}_{\Z} \\
@VVV @. @AAA \\
0 @>>> K^{\tau_{\sf p4g}+\omega+0}_{D_4}(T^2) @>>> K^{\tau_{\sf p4g}+\omega+0}_{D_4}(Y_1 \vee Z) @= \Z^5 \\
\end{CD}
\label{eq:sequence_Y_1+Z}
\end{align}
Here, we used the excision axiom and the Thom isomorphism to get
\begin{equation}
K^{\tau_{\sf p4g}+\omega+n}_{D_4}(T^2,Y_1 \vee Z) \cong K^n_{\Z_2^{(v)}}(e^2, \partial e^2) \cong \widetilde{K}^n_{\Z_2^{(v)}}(S^2)
= 
\left\{ \begin{array}{ll}
0 & (n=0), \\
(1 - A + B - AB), & (n=1), \\
\end{array}\right.
\label{eq:D_4_spinful_T2_Y1_Z}
\end{equation}
where the $\Z_2^{(v)}$-action on the sphere 
is the reflection $S^2 \ni (n_0, n_1, n_2) \mapsto (n_0, n_1, -n_2)$.
In the exact sequence (\ref{eq:sequence_Y_1+Z}), the Abelian group 
$K^{\tau_{\sf p4g}+\omega+0}_{D_4}(Y_1 \vee Z)$ is free. Hence $K^{\tau_{\sf p4g}+\omega+0}_{D_4}(T^2)$ must be torsion free, and the case (iii) is rejected. Now, let us assume that (i) is the case. Then, the exact sequence for the pair $(T^2, X_1)$ implies $K^{\tau_{\sf p4g}+\omega+1}_{D_4}(T^2) \cong K_{D_4}^{\tau_{\sf p4g}+\omega+1}(X_1)$. Substituting this into the exact sequence (\ref{eq:sequence_Y_1+Z}) for $(T^2, Y_1 \vee Z)$, we find that $K^{\tau_{\sf p4g}+\omega+1}_{D_4}(T^2, Y_1 \vee Z)$ surjects onto $(1 + A + B + AB + 2E)$, because any $R(D_4)$-homomorphism $(1 + A + B + AB + 2E) \to (1 - A - B + AB)$ is trivial. However this is impossible in view of (\ref{eq:D_4_spinful_T2_Y1_Z}). As a result, we conclude that (ii) is the case, and we eventually reached the conclusion 
\begin{align}
&K^{\tau_{\sf p4g}+\omega+0}_{D_4}(T^2)
\cong K^{\tau_{\sf p4g}+\omega+0}_{D_4}(X_1)
\cong \overbrace{(1+A+E)}^{\Z^2} \oplus 
\overbrace{(1+A+B+AB-2E)}^{\Z} \oplus 
\overbrace{(1+A-B-AB)}^{\Z}, 
\label{eq:K_p4g_0_T2} \\
&K^{\tau_{\sf p4g}+\omega+1}_{D_4}(T^2) 
\cong K^{\tau_{\sf p4g}+\omega+1}_{D_4}(Z)
\cong \overbrace{(1-A-B+AB)}^{\Z}.
\label{eq:K_p4g_1_T2}
\end{align}

\subsubsection{Models of $K$-group $K^{\tau_{\sf p4g}+\omega+0}_{D_4}(T^2)$} 
In this subsection, 
we will reconstruct the $R(D_4)$-module structure (\ref{eq:K_p4g_0_T2})
from models with small filling number. 
The minimum number of Wyckoff positions inside a unit cell is two, 
which are realized in the two Wyckoff positions labeled by (a) and (b): 
\begin{align}
& ({\rm a}): \left\{ \begin{array}{l}
\bm{x}_A = (-1/4,1/4) \\
\bm{x}_B=(1/4,-1/4)
\end{array}\right. && \mapsto &&
\xygraph{
!{<0cm,0cm>;<1.5cm,0cm>:<0cm,1.5cm>::}
!{(-0.3,0.3)}*+{\textcircled{\scriptsize A}}="A" ,
!{(0.3,-0.3)}*+{\textcircled{\scriptsize B}}="B" ,
!{(-0.6,0.6)}-!{(0.6,0.6)},
!{(-0.6,-0.6)}-!{(0.6,-0.6)},
!{(0.6,0.6)}-!{(0.6,-0.6)},
!{(-0.6,0.6)}-!{(-0.6,-0.6)},
!{(-0.8,0)}-@{.}!{(0.8,0)},
!{(1,0)}, 
!{(0,-0.8)}-@{.}!{(0,0.8)},
!{(0,1.0)}, 
} \\
& ({\rm b}): \left\{ \begin{array}{l}
\bm{x}_A = (-1/4,-1/4) \\
\bm{x}_B=(1/4,1/4)
\end{array}\right. && \mapsto &&
\xygraph{
!{<0cm,0cm>;<1.5cm,0cm>:<0cm,1.5cm>::}
!{(-0.3,-0.3)}*+{\textcircled{\scriptsize A}}="A" ,
!{(0.3,0.3)}*+{\textcircled{\scriptsize B}}="B" ,
!{(-0.6,0.6)}-!{(0.6,0.6)},
!{(-0.6,-0.6)}-!{(0.6,-0.6)},
!{(0.6,0.6)}-!{(0.6,-0.6)},
!{(-0.6,0.6)}-!{(-0.6,-0.6)},
!{(-0.8,0)}-@{.}!{(0.8,0)},
!{(1,0)}, 
!{(0,-0.8)}-@{.}!{(0,0.8)},
!{(0,1.0)}, 
}
\end{align}
In the right figures the solid lines represent the unit cells,
and $\bm{x}_{A}$ and $\bm{x}_B$ are the localized positions from the center of the unit cell. 
In the Wyckoff position (a), each $A$ and $B$ is invariant under the subgroup $C_4 = \{1, c_4, c_2, c_4^3\}$ modulo the lattice translation, 
hence, local orbitals at $A$ and $B$ obey a representation of $C_4$, 
which implies the minimum number of filling of atomic insulators by putting 
degrees of freedom at the Wyckoff position (a) becomes two. 
On the other hand, in the Wyckoff position (b), each $A$ and $B$ position is 
invariant under the subgroup $D^{(d)}_2 = \{1, c_2, \sigma c_4, \sigma c_4^3\}$ modulo the lattice translation, 
thus the local orbitals at $A$ and $B$ obey a nontrivial projective representation of $D_2^{(d)}$ 
if spin is half-integer. 
This means that the minimum number of filling for the Wyckoff position (b) is four. 

The generating models are given as follows. 
First, we consider the Wyckoff position (a). 
Put an $s$-orbital with spin up (down) polarized state of spin 1/2 degrees of freedom at A (B). 
The $D_4$ group acts on these local states by 
$U_{c_4} \ket{s,\ua/\da} = e^{\mp \pi i/4} \ket{s,\ua/\da}$ and 
$U_{\sigma} \ket{s,\ua/\da} = \pm \ket{s,\da/\ua}$. 
By taking the space group transformation into account, 
the corresponding $D_4$-equivariant vector bundle $E_1$ is given by 
\begin{align}
&\xygraph{
!{<0cm,0cm>;<1.5cm,0cm>:<0cm,1.5cm>::}
!{(-0.3,0.3)}*+{\ket{s,\uparrow}},
!{(0.3,-0.3)}*+{\ket{s,\downarrow}},
!{(-0.6,0.6)}-@{.}!{(0.6,0.6)},
!{(-0.6,-0.6)}-@{.}!{(0.6,-0.6)},
!{(0.6,0.6)}-@{.}!{(0.6,-0.6)},
!{(-0.6,0.6)}-@{.}!{(-0.6,-0.6)},
!{(-0.8,0)}-@{.}!{(0.8,0)},
!{(1,0)}, 
!{(0,-0.8)}-@{.}!{(0,0.8)},
!{(0,1.0)}, 
} \quad = \quad 
\left(
E_1= T^2 \times \C^2, \ \  
U_{c_4}(\bk) = \begin{pmatrix}
\zeta & 0 \\
0 & \zeta^{-1} e^{-i k_x} \\ 
\end{pmatrix}, \ \ 
U_{\sigma}(\bk) = \begin{pmatrix}
0 & -e^{-i k_x} \\
1 & 0 \\ 
\end{pmatrix}
\right), 
\label{eq:p4g_E1}
\end{align}
where the matrix is for the A and B space. 
The orbital part can be replaced by other 
1-dimensional representations $d_{xy}, p_{x+iy}, p_{x-iy}$ of $C_4$, 
and spin part can be exchanged. 
In addition to the atomic ground state $E_1$, 
we have the following three independent atomic ground states: 
\begin{align}
&\xygraph{
!{<0cm,0cm>;<1.5cm,0cm>:<0cm,1.5cm>::}
!{(-0.3,0.3)}*+{\ket{s,\da}},
!{(0.3,-0.3)}*+{\ket{s,\ua}},
!{(-0.6,0.6)}-@{.}!{(0.6,0.6)},
!{(-0.6,-0.6)}-@{.}!{(0.6,-0.6)},
!{(0.6,0.6)}-@{.}!{(0.6,-0.6)},
!{(-0.6,0.6)}-@{.}!{(-0.6,-0.6)},
!{(-0.8,0)}-@{.}!{(0.8,0)},
!{(1,0)}, 
!{(0,-0.8)}-@{.}!{(0,0.8)},
!{(0,1.0)}, 
} \quad = \quad 
\left(
E_2= T^2 \times \C^2, \ \  
U_{c_4}(\bk) = \begin{pmatrix}
\zeta^{-1} & 0 \\
0 & \zeta e^{-i k_x} \\ 
\end{pmatrix}, \ \ 
U_{\sigma}(\bk) = \begin{pmatrix}
0 & e^{-i k_x} \\
-1 & 0 \\ 
\end{pmatrix}
\right), 
\end{align}
\begin{align}
&\xygraph{
!{<0cm,0cm>;<1.5cm,0cm>:<0cm,1.5cm>::}
!{(-0.3,0.3)}*+{\ket{d_{xy},\uparrow}},
!{(0.3,-0.3)}*+{\ket{d_{xy},\downarrow}},
!{(-0.6,0.6)}-@{.}!{(0.6,0.6)},
!{(-0.6,-0.6)}-@{.}!{(0.6,-0.6)},
!{(0.6,0.6)}-@{.}!{(0.6,-0.6)},
!{(-0.6,0.6)}-@{.}!{(-0.6,-0.6)},
!{(-0.8,0)}-@{.}!{(0.8,0)},
!{(1,0)}, 
!{(0,-0.8)}-@{.}!{(0,0.8)},
!{(0,1.0)}, 
} \quad = \quad 
\left(
AB \cdot E_1 = T^2 \times \C^{\oplus 2}, \ \  
U_{c_4}(\bk) = \begin{pmatrix}
-\zeta & 0 \\
0 & - \zeta^{-1} e^{-i k_x} \\ 
\end{pmatrix}, \ \ 
U_{\sigma}(\bk) = \begin{pmatrix}
0 & e^{-i k_x} \\
-1 & 0 \\ 
\end{pmatrix}
\right), \\
&\xygraph{
!{<0cm,0cm>;<1.5cm,0cm>:<0cm,1.5cm>::}
!{(-0.3,0.3)}*+{\ket{d_{xy},\da}},
!{(0.3,-0.3)}*+{\ket{d_{xy},\ua}},
!{(-0.6,0.6)}-@{.}!{(0.6,0.6)},
!{(-0.6,-0.6)}-@{.}!{(0.6,-0.6)},
!{(0.6,0.6)}-@{.}!{(0.6,-0.6)},
!{(-0.6,0.6)}-@{.}!{(-0.6,-0.6)},
!{(-0.8,0)}-@{.}!{(0.8,0)},
!{(1,0)}, 
!{(0,-0.8)}-@{.}!{(0,0.8)},
!{(0,1.0)}, 
} \quad = \quad 
\left(
AB \cdot E_2 = T^2 \times \C^{\oplus 2}, \ \  
U_{c_4}(\bk) = \begin{pmatrix}
- \zeta^{-1} & 0 \\
0 & - \zeta e^{-i k_x} \\ 
\end{pmatrix}, \ \ 
U_{\sigma}(\bk) = \begin{pmatrix}
0 & -e^{-i k_x} \\
1 & 0 \\ 
\end{pmatrix}
\right).  
\end{align}
From the table of projective characters (\ref{tab:proj_character_d4}), 
one can read off the representations at symmetric points of the above 
atomic ground states, which are summarized as the following table: 
\begin{align}
\begin{tabular}{c|ccc}
$E$ & $E|_{\Gamma}$ & $E|_X$ & $E|_M$ \\
\hline 
$E_1$ & $W$ & $1+t_{\sigma c_2} t_{\sigma}$ & $W$ \\
$AB \cdot E_1$ & $BW$ & $t_{\sigma c_2} + t_{\sigma} $ & $BW$ \\
$E_2$ & $W$ & $t_{\sigma c_2} + t_{\sigma} $ & $BW$ \\
$AB \cdot E_2$ & $BW$ & $1+t_{\sigma c_2} t_{\sigma} $ & $W$ \\
\end{tabular}
\label{tab:p4g_atomic_(a)}
\end{align}

Compared these data with the $K$-group (\ref{eq:K_p4g_0_T2}), 
one can recognize that the above table (\ref{tab:p4g_atomic_(a)})
lacks the generator with the data 
$(W,t_{\sigma c_2} + t_{\sigma},W)$, 
$(W,1+t_{\sigma c_2} t_{\sigma},BW)$, 
$(BW,1+t_{\sigma c_2} t_{\sigma},BW)$, or 
$(BW,t_{\sigma c_2}+t_{\sigma},W)$. 
As a formal difference of two vector bundles, 
this deficit can be filled with 
the atomic ground state obtained by the 
Wyckoff position (b). 
Let $E_3$ be the atomic ground state 
defined by putting an $s$-orbital with spin 1/2 degrees of freedom at the two 
positions $A$ and $B$ of the Wyckoff label (b): 
\begin{align}
&\xygraph{
!{<0cm,0cm>;<2cm,0cm>:<0cm,2cm>::}
!{(-0.3,-0.3)}*+{\ket{s}\otimes \C^2_{\rm spin}} ,
!{(0.4,0.3)}*+{\ket{s} \otimes \C^2_{\rm spin}},
!{(-0.6,0.6)}-@{.}!{(0.6,0.6)},
!{(-0.6,-0.6)}-@{.}!{(0.6,-0.6)},
!{(0.6,0.6)}-@{.}!{(0.6,-0.6)},
!{(-0.6,0.6)}-@{.}!{(-0.6,-0.6)},
!{(-0.8,0)}-@{.}!{(0.8,0)},
!{(1,0)}, 
!{(0,-0.8)}-@{.}!{(0,0.8)},
!{(0,1.0)} 
} \quad = \quad 
\left(
E_3= T^2 \times \C^4, \ \  
U_{c_4}(\bk) = \begin{pmatrix}
0 & e^{- \frac{\pi}{4} i \sigma_z} e^{-i k_x} \\
e^{-\frac{\pi}{4} i \sigma_z} & 0 \\ 
\end{pmatrix}, \ \ 
U_{\sigma}(\bk) = \begin{pmatrix}
0 & -i \sigma_y e^{-i k_x} \\
-i \sigma_y & 0 \\ 
\end{pmatrix}
\right).
\end{align}
All the projective characters of $E_3$ at symmetric points are zero, 
which leads to the following data of projective representations of $E_3$: 
\begin{align}
\begin{tabular}{c|ccc}
$E$ & $E|_{\Gamma}$ & $E|_X$ & $E|_M$ \\
\hline 
$E_3$ & $W+BW$ & $1+t_{\sigma c_2} + t_{\sigma} + t_{\sigma c_2} t_{\sigma}$ & $W+BW$ \\
\end{tabular}
\label{tab:p4g_atomic_(b)}
\end{align}
Then, the formal difference $[E_3] - [AB \cdot E_1]$ 
provides the remaining generator of the $K$-group (\ref{eq:K_p4g_0_T2}). 

Interestingly, the vector bundle with the data $(BW,1+t_{\sigma c_2}t_{\sigma},BW)$ 
can be realized as a band insulator. 
Let us consider a Hamiltonian $\hat H$ on the atomic insulator $E_3$, 
\begin{equation}
\hat H
:= \psi^{\dag}_{B}(\bm{R}) t_1 \psi_A(\bm{R}) + \psi^{\dag}_{A}(\bm{R}+\hat x) t_2 \psi_A(\bm{R}) + h.c. 
+ ({\rm space\ group\ symmetrization}), 
\label{eq:p4g_model_on_Wyckoff_b}
\end{equation}
where $t_1$ and $t_2$ are nearest and next nearest hopping terms, respectively. 
The space group transformations are defined by 
\begin{align}
&\hat U_{c_4} \psi^{\dag}_{A}(\bm{R}) \hat U_{c_4}^{-1} = \psi^{\dag}_{B}(c_4 \bm{R}) e^{-\frac{\pi i}{4} \sigma_z}, &&
\hat U_{c_4} \psi^{\dag}_{B}(\bm{R}) \hat U_{c_4}^{-1} = \psi^{\dag}_{A}(c_4 \bm{R}+\hat y) e^{-\frac{\pi i}{4} \sigma_z}, \\
&\hat U_{\sigma} \psi^{\dag}_{A}(\bm{R}) \hat U_{\sigma}^{-1} = \psi^{\dag}_{B}(\sigma \bm{R}) (-i \sigma_y), && 
\hat U_{\sigma} \psi^{\dag}_{B}(\bm{R}) \hat U_{\sigma}^{-1} = \psi^{\dag}_{A}(\sigma \bm{R}+\hat x) (-i \sigma_y), 
\end{align}
which leads to constraints 
\begin{align}
&t_1 = \alpha + \beta \frac{\sigma_x - \sigma_y}{\sqrt{2}}, \qquad \alpha, \beta \in \C, 
\label{eq:p4g_E_3_parameter_1} \\
&t_2 = a + b \sigma_z + i c \sigma_x + i d \sigma_y, \qquad a,b,c,d \in \R. 
\label{eq:p4g_E_3_parameter_2}
\end{align}
Let us consider the following Hamiltonian 
\begin{equation}
\hat H_4
:= \psi^{\dag}_{B}(\bm{R}) \frac{1+i}{4} \psi_A(\bm{R}) + \psi^{\dag}_{A}(\bm{R}+\hat x) \frac{\sigma_z}{4} \psi_A(\bm{R}) + h.c. 
+ ({\rm space\ group\ symmetrization}). 
\end{equation}
The one-particle Hamiltonian $H_4(\bk)$ in the momentum space is written as 
\begin{equation}
H_4(\bk) 
= \begin{pmatrix}
\frac{1}{2} (\cos k_x - \cos k_y) \sigma_z & \frac{1-i}{4}(1+e^{-i (k_x+k_y)}) + \frac{1+i}{4}(e^{-ik_x} + e^{-i k_y}) \\
\frac{1+i}{4}(1+e^{i (k_x+k_y)}) + \frac{1-i}{4}(e^{ik_x} + e^{i k_y}) & -\frac{1}{2} (\cos k_x - \cos k_y) \sigma_z \\
\end{pmatrix}. 
\end{equation}
This model conserves the $z$-component of the spin and is fully gapped with 
the dispersion 
\begin{equation}
\varepsilon(\bk) = \pm \sqrt{\frac{6+\cos(2 k_x) + \cos (2 k_y)}{8}}. 
\end{equation}
At the symmetric points, $H_4(\bk)$ takes the following forms
\begin{align}
H_4(\Gamma) = \begin{pmatrix}
0 & \sigma_0 \\
\sigma_0 & 0 \\	
\end{pmatrix},&&
H_4(M) = \begin{pmatrix}
0 & -i \sigma_0 \\
i \sigma_0 & 0 \\
\end{pmatrix},&&
H_4(X) = \begin{pmatrix}
-\sigma_z & 0 \\
0 & \sigma_z \\
\end{pmatrix}. 
\end{align}
Then, the occupied basis $\Psi_{P} (P=\Gamma, M, X)$ at symmetric points reads 
\begin{align}
&\Psi_{\Gamma} = \{\frac{\ket{A,\uparrow}-\ket{B,\uparrow}}{\sqrt{2}},\frac{\ket{A,\downarrow}-\ket{B,\downarrow}}{\sqrt{2}}\}, &&
\Psi_{M} = \{\frac{\ket{A,\uparrow}-i\ket{B,\uparrow}}{\sqrt{2}},\frac{\ket{A,\downarrow}-i\ket{B,\downarrow}}{\sqrt{2}}\}, \\
&\Psi_{X} = \{\ket{A,\ua},\ket{B,\da}\}. 
\end{align}
Let $E_4$ be the occupied state bundle of the Hamiltonian $H_4(\bk)$. 
The representation matrices $U_p(P) (P=\Gamma,M,X)$ on $E_4$ are given by 
\begin{align}
U_{c_4}(\Gamma) = \begin{pmatrix}
-\zeta & 0 \\
0 & -\zeta^{-1}\\ 
\end{pmatrix}, &&
U_{c_4}(M) = \begin{pmatrix}
i \zeta & 0 \\
0 & i \zeta^{-1} \\ 
\end{pmatrix}, &&
U_{c_2}(X)=
\begin{pmatrix}
-i & 0\\
0 & -i \\
\end{pmatrix}, &&
U_{\sigma}(X)=
\begin{pmatrix}
0 & 1\\
1 & 0 \\
\end{pmatrix}, 
\end{align}
which implies that the occupied state bundle has 
the data $E_4 := (BW,1 + t_{\sigma c_2} t_{\sigma},BW)$. 
In the same way, the unoccupied states of $\hat H$ 
has the data $(W,t_{\sigma c_2} + t_{\sigma},W)$. 
We conjecture: 
\begin{itemize}
\item The vector bundles with the data $(W,t_{\sigma c_2} + t_{\sigma},W)$, 
$(W,1+t_{\sigma c_2} t_{\sigma},BW)$, 
$(BW,1+t_{\sigma c_2} t_{\sigma},BW)$, and 
$(BW,t_{\sigma c_2}+t_{\sigma},W)$ cannot be realized as atomic insulators. 
\end{itemize}
If this is true, the band insulator $E_4$ we constructed is a topologically nontrivial ground state 
in the sense that there is no atomic orbital representation, 
which is similar to filling enforced topological insulators protected by space group symmetry.~\cite{PoWatanabe2016filling}
Our model $E_4$ is not filling enforced since atomic ground states obtained by the 
Wyckoff position (a) have the same filling number as $E_4$.

\subsubsection{Models of $K$-group $K^{\tau_{\sf p4g}+\omega+1}_{D_4}(T^2)$: 2d class AIII insulator} 
Now we consider a generating model of the $K$-group $K^{\tau_{\sf p4g}+\omega+1}_{D_4}(T^2)$, 
which is represented by a class AIII insulator with p4g symmetry in spin half-integer systems. 
From (\ref{eq:K_p4g_1_T2}), 
the the topological invariant detecting 
the $K$-group $K^{\tau_{\sf p4g}+\omega+1}_{D_4}(T^2) \cong \Z$ 
can be understood by the subspace $Z$. 
Under the two-cocycle (\ref{twist_spinful_p4g}), 
the reflection $U_{\sigma c_4^3}(\bk)$ satisfies $U_{\sigma c_4^3}(ky,kx) U_{\sigma c_4^3}(kx,ky) = -1$. 
We can define the mirror winding number $w_{\sigma c_4^3}$ on the 
invariant line of $\sigma c_4^3$-reflection as 
\begin{align}
w_{\sigma c_4^3} := 
\frac{1}{i} \cdot \frac{1}{4 \pi i} \oint_{-\pi}^{\pi} d k \ \tr \Big[ U_{\sigma c_4^3}(k,k) \Gamma H(k,k)^{-1} \p_k H(k,k) \Big] \in 2 \Z, 
\label{eq:mirror_winding_p4g_aiii}
\end{align}
where $\Gamma$ is the chiral operator. 
That the mirror winding number $w_{\sigma c_4^3}$ 
is an even integer is ensured by the 
absence of the total winding number associated with the same line. 

We give an example of a nontrivial model. 
We define a Hamiltonian $H(\bk)$ on the atomic vector bundle $E_1 \otimes \C^2$ 
where $E_1$ is introduced in (\ref{eq:p4g_E1}) and $\C^2$ represents 
internal degrees of freedom on which the point group $D_4$ acts trivially. 
Let $\hat H$ be the following model with 
nearest neighbor and next-nearest neighbor hopping, 
\begin{align}
\hat H 
&:= 
\psi^{\dag}_{B s \da}(\bm{R}) e^{-\pi i/4} \sigma_x \psi_{A s \ua}(\bm{R}) 
+ t \psi^{\dag}_{A s \ua}(\bm{R}+\hat x) \sigma_y \psi_{A s \ua}(\bm{R}) 
+ m \psi^{\dag}_{A s \ua}(\bm{R}) \sigma_y \psi_{A s \ua}(\bm{R}) \\
& \qquad + ({\rm space\ group\ symmetrization}) \\
&= \sum_{\bk} (\psi^{\dag}_{A s \ua}(\bk), \psi^{\dag}_{B s \da}(\bk)) H(\bk)
\begin{pmatrix}
\psi_{A s \ua}(\bk) \\
\psi_{B s \da}(\bk)
\end{pmatrix}, \\
H(\bk) &= \begin{pmatrix}
(m+2t \cos k_x + 2t \cos k_y) \sigma_y & e^{\pi i/4} (1 - i e^{i k_y} - e^{-i k_x+i k_y}+i e^{-i k_x}) \sigma_x \\
e^{-\pi i/4} (1 + i e^{-i k_y} - e^{i k_x-i k_y}-i e^{i k_x}) \sigma_x & (m+2t \cos k_x + 2t \cos k_y) \sigma_y \\
\end{pmatrix}, 
\label{eq:p4g_aiii_bulk_model}
\end{align}
where $\sigma_{\mu} (\mu=x,y,z)$ are the Pauli matrices for the 
internal degrees of freedom. 
The space group transformations are defined by 
\begin{align}
&\hat U_{c_4} \psi^{\dag}_{A s \ua}(\bm{R}) \hat U_{c_4}^{-1} = \psi^{\dag}_{A s \ua}(c_4 \bm{R}) e^{-\pi i/4}, \qquad 
\hat U_{c_4} \psi^{\dag}_{B s \da}(\bm{R}) \hat U_{c_4}^{-1} = \psi^{\dag}_{B s \da}(c_4 \bm{R}+\hat y) e^{\pi i/4}, \\
&\hat U_{\sigma} \psi^{\dag}_{A s \ua}(\bm{R}) \hat U_{\sigma}^{-1} = \psi^{\dag}_{B s \da}(\sigma \bm{R}), \qquad 
\hat U_{\sigma} \psi^{\dag}_{B s \da}(\bm{R}) \hat U_{\sigma}^{-1} = \psi^{\dag}_{A s \ua}(\sigma \bm{R}+\hat x) (-1). 
\end{align}
The chiral operator is $\Gamma = \sigma_z$. 
The one-particle Hamiltonian $H(\bk)$ has the mirror winding number 
\begin{align}
w_{\sigma c_4^3} = \left\{\begin{array}{ll}
2 & (t<-\frac{|m|}{4}) \\
0 & (-\frac{|m|}{4}<t<\frac{|m|}{4}) \\
-2 & (\frac{|m|}{4}<t) \\
\end{array}\right..
\end{align}

The module structure (\ref{eq:K_p4g_1_T2}) of 
the $K$-group can be understood 
from the mirror winding number (\ref{eq:mirror_winding_p4g_aiii}). 
From the character table \ref{Tab:rep_of_D4}, 
the operator $U_{\sigma c_4^3}(\bk)$ 
is changed under the actions of $A$ and $B$ irreps.\ as 
$U_{\sigma c_4^3}(\bk) \mapsto - U_{\sigma c_4^3}(\bk)$, 
which implies that the mirror winding number $w_{\sigma c_4^3}$ 
is the invariant of the $R(D_4)$-module $(1-A-B+AB)$.

\subsubsection{Models of $K$-group $K^{\tau_{\sf p4g}+\omega+1}_{D_4}(T^2)$: 2d class A surface state}
The $K$-group $K^{\tau_{\sf p4g}+\omega+1}_{D_4}(T^2)$ with grading $n=1$ 
classifies gapless states in 2d BZ torus $T^2$ with p4g symmetry 
in spin half-integer systems. 
The corresponding 3d model Hamiltonian and topological 
invariants immediately follow from (\ref{eq:mirror_winding_p4g_aiii}) 
and (\ref{eq:p4g_aiii_bulk_model}). 
The mirror Chern number is defined on the $\sigma c_4^3$-invariant plane~\cite{DongLiu2015}
\begin{align}
ch_{\sigma c_4^3}
:= \frac{1}{i} \cdot \frac{i}{2 \pi} \oint_{-\pi}^{\pi} d k \oint_{-\pi}^{\pi} d k_z \ \tr \Big[ U_{\sigma c_4^3}(k,k,k_z) {\cal F}_{k k_z}(k,k,k_z) \Big] \in 2 \Z, 
\end{align}
where ${\cal F}_{k k_z}$ is the Berry curvature on the $\sigma c_4^3$-invariant plane. 
From the dimensional raising map, 
the 2d class AIII Hamiltonian (\ref{eq:p4g_aiii_bulk_model}) becomes 
\begin{align}
\widetilde H(k_x,k_y,k_z) 
=
\begin{pmatrix}
(m+2t \cos k_x + 2t \cos k_y + 2t \cos k_z) \sigma_y + \sin k_z \sigma_z & e^{\pi i/4} (1 - i e^{-i k_y} - e^{-i k_x+i k_y}+i e^{-i k_x}) \sigma_x \\
e^{-\pi i/4} (1 + i e^{i k_y} - e^{i k_x-i k_y}-i e^{i k_x}) \sigma_x & (m+2t \cos k_x + 2t \cos k_y + 2t \cos k_z) \sigma_y + \sin k_z \sigma_z \\
\end{pmatrix}.
\end{align}

\subsubsection{A stable gapless phase protected by representation at $X$ point: 2d class A}
\label{A stable gapless phase protected by representation at $X$ point: 2d class A}
The $K$-group (\ref{eq:K_p4g_0_T2}) is 
characterized by the representations at 
the symmetric points $\Gamma$, $X$ and $M$. 
From the local data of the $K$-group (\ref{eq:local_data_X_1}) on $X_1$, 
one can find that not every representations at the point $X$ are allowed. 
Only two representations 
\begin{align}
t_{\sigma c_2} + t_{\sigma}, 1 + t_{\sigma c_2} t_{\sigma} \in R^{\tau_{\sf p4g}|_X+\omega}(D_2^{(v)})
\label{eq:p4g_condition_x_a}
\end{align}
survive on the subspace $X_1$. 
The evenness of the rank is due to the nonsymmorphic property of the 
wallpaper group $p4g$. 
In addition to a simple condition on the number of filing, 
(\ref{eq:p4g_condition_x_a}) means there is an additional 
condition: 
\begin{itemize}
\item If a band spectrum is isolated from other bands on the subspace $X_1$, 
then the representation at $X$ point should be a direct sum of $t_{\sigma c_2} + t_{\sigma}, 1 + t_{\sigma c_2} t_{\sigma} \in R^{\tau_{\sf p4g}|_X+\omega}(D_2^{(v)})$. 
\end{itemize}
The contraposition of this condition provides a criterion of stable gapless phases: 
\begin{itemize}
\item If the representation of a valence band at the $X$ point is not a direct sum of 
$t_{\sigma c_2} + t_{\sigma}, 1 + t_{\sigma c_2} t_{\sigma} \in R^{\tau_{\sf p4g}|_X+\omega}(D_2^{(v)})$, 
then there should be a gapless point on the subspace $X_1$, 
unless the valence band at the $X$ point touches the conduction band. 
\end{itemize}

\begin{figure}[!]
 \begin{center}
  \includegraphics[width=\linewidth, trim=0cm 0cm 0cm 0cm]{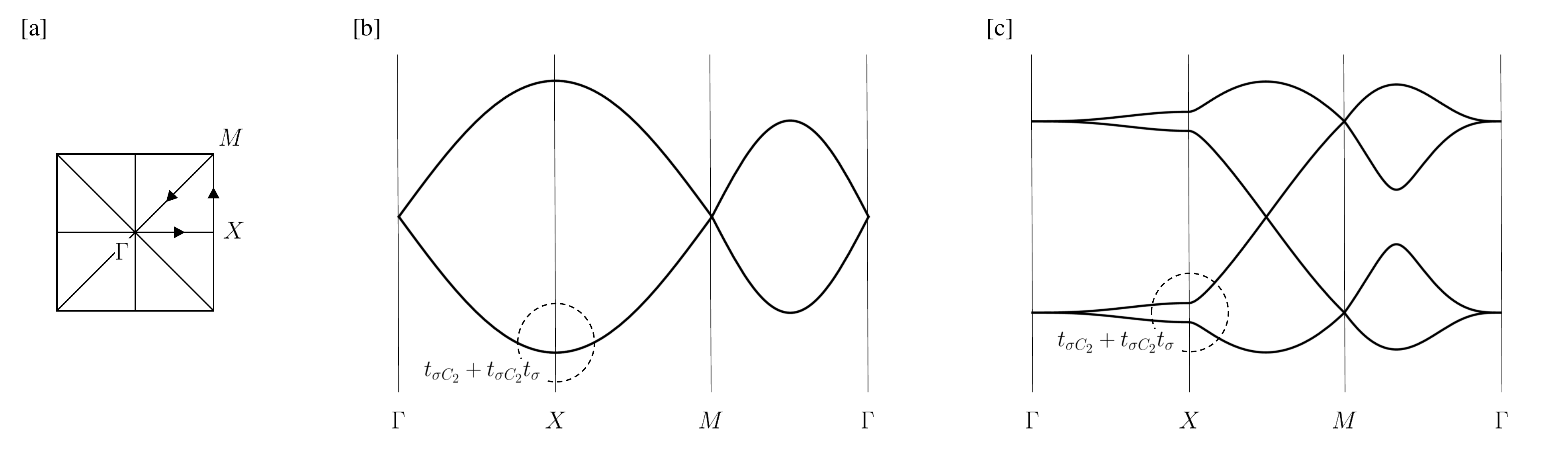}
 \end{center}
 \caption{Band crossings protected by the representation at the $X$ point. 
 [a] subspace $X_1$. 
 [b] The energy spectrum on the subspace $X_1$ of Hamiltonian (\ref{eq:hamiltonian_p4g_gapless}) which corresponds to 
 the parameters $(\alpha,\beta,a,b,c,d) = (0,1,0,0,0,0)$ of eqs.\ (\ref{eq:p4g_E_3_parameter_1}) and (\ref{eq:p4g_E_3_parameter_2}). 
 [c] Another parameter choice $(\alpha,\beta,a,b,c,d) = (1+i,1,0.2,0,0,0)$.}
 \label{fig:p4g_gapless}
\end{figure}

We give a simple model in a form (\ref{eq:p4g_model_on_Wyckoff_b}). 
Let us consider the following Hamiltonian on the atomic insulator $E_3$: 
\begin{align}
\hat H_5
&:= \psi^{\dag}_{B}(\bm{R}) \frac{\sigma_x-\sigma_y}{2} \psi_A(\bm{R}) + h.c. 
+ ({\rm space\ group\ symmetrization}) 
\label{eq:hamiltonian_p4g_gapless} \\
&= \sum_{\bk} (\psi^{\dag}_{A}(\bk), \psi^{\dag}_{B}(\bk)) H_5(\bk)
\begin{pmatrix}
\psi_{A}(\bk) \\
\psi_{B}(\bk)
\end{pmatrix}, \\
H_5(\bk) &= \begin{pmatrix}
0 & \frac{\sigma_x-\sigma_y}{2}(1 - e^{-i k_x-i k_y}) + \frac{\sigma_x+\sigma_y}{2} (e^{-i k_y}-e^{-i k_x}) \\
\frac{\sigma_x-\sigma_y}{2}(1 - e^{i k_x+i k_y}) + \frac{\sigma_x+\sigma_y}{2} (e^{i k_y}-e^{i k_x}) & \\
\end{pmatrix}. 
\end{align}
At the $X$ point 
the Hamiltonian becomes $H_5(X) = \begin{pmatrix}
0 & \sigma_x \\
\sigma_x & 0 \\
\end{pmatrix}$ and the occupied states at $X$ belong to the 
representation $t_{\sigma c_2} + t_{\sigma c_2} t_{\sigma} \in R^{\tau_{\sf p4g}|_X + \omega}(D_2^{(v)})$. 
Then, the above criterion implies that 
there should be a topologically stable gapless point on the subspace $X_1$ 
as long as the mass gap at the $X$ point is preserved. 
Fig.~\ref{fig:p4g_gapless} [b] shows the energy spectrum of (\ref{eq:hamiltonian_p4g_gapless}). 
Fig.~\ref{fig:p4g_gapless} [c] shows the perturbed energy spectrum from (\ref{eq:hamiltonian_p4g_gapless}). 
The band crossing on the subspace $X_1$ is protected by the representation of the $X$ point.

\subsection{Weyl semimetals and nodal superconductors protected by inversion symmetry}
\label{sec:inversion_fermi_pt}
In this section, 
we introduce a $\Z_2$ invariant protecting Weyl semimetals and nodal superconductors 
defined from the inversion symmetry which is not discussed in the literature. 

\subsubsection{$\Z_2$ invariant from unoriented surface}
\label{Z2 invariant from unoriented surface}
We start with a $\Z_2$ invariant arising from unoriented BZ manifold. 
Let $X$ be a 2d unoriented manifold. 
Complex bundles $E$ on $X$ 
can be classified by 
their first Chern classes $c_1(E) \in H^2(X;\Z)$. 
If $X$ is nonorientable, $H^2(X;\Z)$ may have a torsion part. 
For example, the real projective plane $RP^2$ shows $H^2(RP^2;\Z) = \Z_2$, 
which implies that we have a ``$\Z_2$ topological insulator'' on $RP^2$. 

\begin{figure}[!]
 \begin{center}
  \includegraphics[width=0.5\linewidth, trim=0cm 2cm 0cm 0cm]{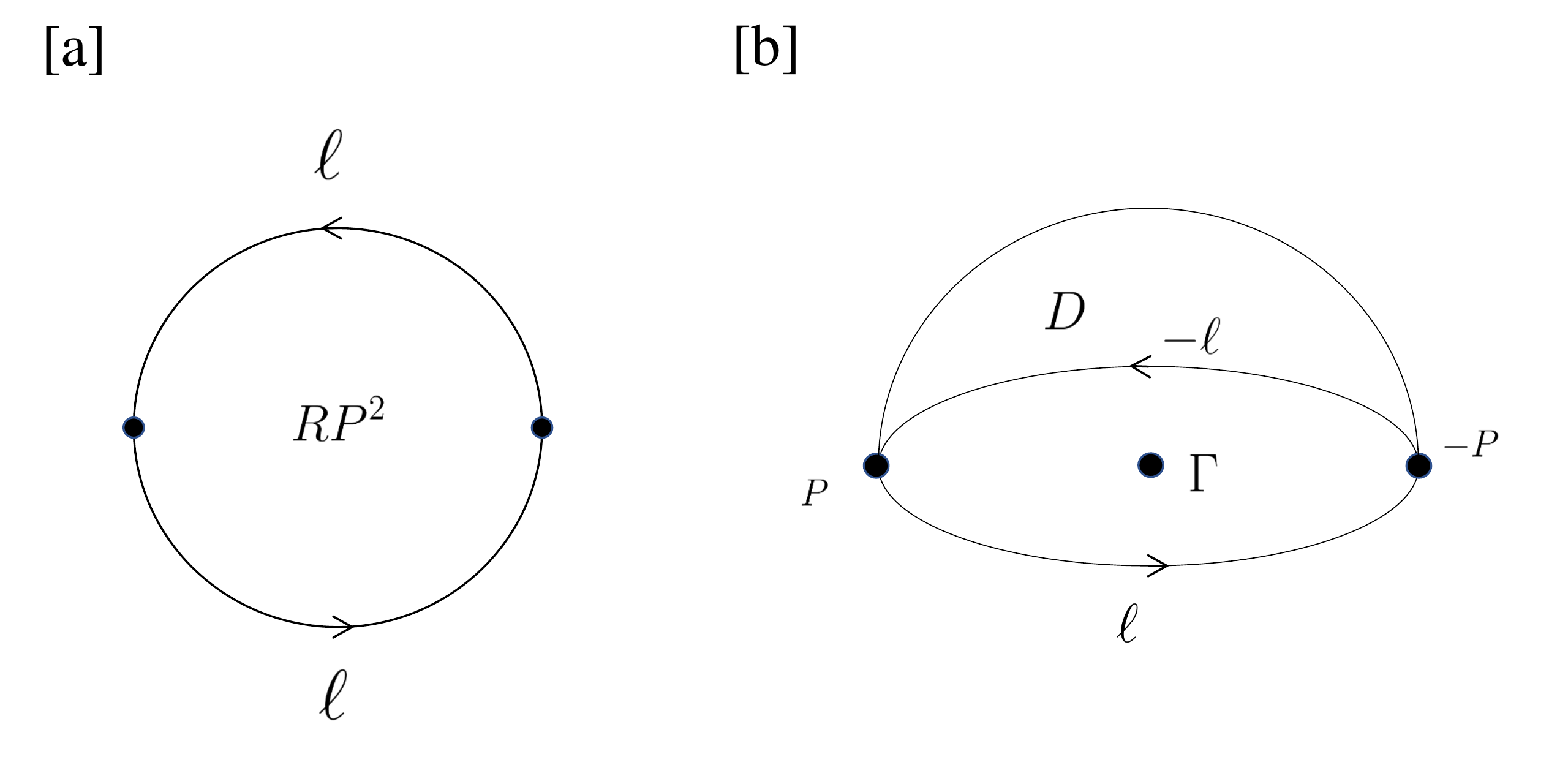}
 \end{center}
 \caption{
 [a] The real projective plane and a noncontractible loop $\ell$. 
 [b] The line $\ell$ connecting the inversion symmetric pair of points $P$ and $-P$. 
 The surface $D$ bounds $\ell$ and its inversion symmetric pair $-\ell$. 
 }
 \label{fig:inversion}
\end{figure}

The torsion part of the first Chern class can be detected as follows.~\cite{Freed1986}
Let ${\cal A}(\bk)$ be the Berry connection of occupied states on $RP^2$. 
Let $\ell$ be a noncontractible loop on $RP^2$. 
Then, $RP^2$ can be considered as a disc $D$ surrounded by the loop $\ell$ and its copy. 
See Fig.~\ref{fig:inversion} [a]. 
Then, the $\Z_2$ invariant $c_1 \in \{0,1/2\}$ is defined by 
\begin{align}
c_1 := \frac{i}{2\pi}\ln {\rm hol}_{\ell}({\cal A})+\frac{1}{2} \frac{i}{2\pi}\int_{D} \tr {\cal F} \ (\mathrm{mod}\ 1), 
\label{eq:c1_RP2}
\end{align}
where ${\rm hol}_{\ell}({\cal A}) \in U(1)$ is the Berry phase ($U(1)$ holonomy) along the loop $\ell$, 
and ${\cal F}$ is the Berry curvature. 
$c_1$ is quantized to $0$ or $1/2$ because of the Stokes' theorem 
\begin{align}
2 c_1 = \frac{i}{2\pi}\ln {\rm hol}_{\p D} ({\cal A})+ \frac{i}{2\pi}\int_{D} \tr {\cal F} = 0 \ (\mod \ 1). 
\end{align}
A nontrivial model Hamiltonian will be presented in Sec.~\ref{sec:z2_inversion}. 

It is worth reminding the definition of the Berry phase in the cases where 
the Berry connection ${\cal A}$ on the loop $\ell$ needs multiple patches.   
In such cases, the Berry phase is defined by integral of parallel transports on patches 
and transition functions. 
Let $\{ U_i \}_{i=1, \dots N}$ be a cover including the loop $\ell$. 
We divide $\ell$ to $N$ components so that $\ell_i \subset U_i$. 
Let $p_i$ be junction points of $\ell_i$, namely, $\p \ell_i = p_{i+1} - p_i$. 
Then the $U(1)$ holonomy is defined by 
\begin{align}
{\rm hol}_{\ell}({\cal A})
= e^{-\int_{\ell_1} \tr {\cal A}_1 } \cdot \det g_{1,2}(p_2) \cdot 
e^{-\int_{\ell_2} \tr {\cal A}_2 } \cdot \det g_{2,3}(p_3) \cdots 
e^{-\int_{\ell_N} \tr {\cal A}_N } \cdot \det g_{N,1}(p_1), 
\end{align}
where ${\cal A}_i$ is the Berry connection on $U_i$ 
and $g_{i,j}$ is the transition function on $U_i \cap U_j$. 

A similar construction is possible for the Klein bottle and also 
the torsion part of higher Chern classes $c_d(E)$, $d > 1$.~\cite{Freed1986}

\subsubsection{$\Z_2$ invariant from the inversion symmetry}
\label{sec:z2_inversion}
Now we discuss an application of 
the $\Z_2$ invariant (\ref{eq:c1_RP2}) to Weyl semimetals and nodal superconductors. 
Let us consider an inversion symmetric 3d Hamiltonian 
\begin{align}
U(\bk) H(\bk) U(\bk)^{-1} = H(-\bk), \qquad 
U(-\bk) U(\bk) = 1. 
\end{align}
The existence of the $\Z_2$ invariant is understood as follows. 
We pick a closed surface $\Sigma$ on which 
the inversion symmetry freely acts. 
We effectively have a Hamiltonian
on the quotient $\Sigma/\Z_2$ which is a nonorientable manifold. 
This implies there is a $\Z_2$ invariant similar to (\ref{eq:c1_RP2}). 

Let us define the $\Z_2$ invariant. 
We pick a pair of inversion symmetric points $P$ and $-P$. 
Let $\ell$ be an oriented line from $P$ to $-P$. 
In the presence of the inversion symmetry, 
even if the line $\ell$ is not closed, 
one can define a well-defined Berry phase associated with the line $\ell$. 
The Bloch states at $P$ and $-P$ are related by a unitary matrix $V(P)$ as 
\begin{align}
U(-P) \Phi(-P) = \Phi(P) V(P), 
\label{eq:def_v(p)_inversion}
\end{align}
where $\Phi(\bk)$ is the frame of occupied states 
$\Phi(\bk) = \big (\ket{\phi_1(\bk)}, \dots, \ket{\phi_m(\bk)} \big)$. 
We define the Berry phase associated with the line $\ell$ by 
\begin{align}
{\rm hol}_{\ell}({\cal A}) 
:= e^{-\int_{P,\ell}^{-P} \tr {\cal A}} \cdot \det [V(P)] \in U(1). 
\end{align}
(Here we have assumed that $\ell$ is covered by a single patch.) 
The phase ${\rm hol}_{\ell}({\cal A})$ is gauge invariant since 
the gauge dependences of the parallel transport and the unitary matrix $V(P)$ are canceled. 

It should be noticed that 
there is ambiguity in ${\rm hol}_{\ell}({\cal A})$ arising from $U(\bk)$. 
The change of sign $U(\bk) \mapsto - U(\bk)$ induces the $\pi$ phase shift 
${\rm hol}_{\ell}({\cal A}) \mapsto -{\rm hol}_{\ell}({\cal A})$. 
This ambiguity cannot be eliminated, however, $\Z_2$ distinction is well-defined 
if $U(\bk)$ is fixed. 

In the same way as (\ref{eq:c1_RP2}), 
we can define the $\Z_2$ invariant. 
The line $\ell$ and its inversion symmetric line $-\ell$
together form a closed loop $\ell \cup (-\ell)$ in the BZ. 
We choose a surface $D$ whose boundary is $\ell \cup (-\ell)$. 
See Fig.~\ref{fig:inversion} [b]. 
Then, the same formula as (\ref{eq:c1_RP2}) defines the $\Z_2$ invariant $c_1 \in \{0,1/2\}$. 
Notice that the $\Z_2$ invariant $c_1$ depends on both the line $\ell$ and the surface $D$. 

Now we give a nontrivial model Hamiltonian. 
Let 
\begin{align}
\ket{\bk} := 
\frac{1}{|\bk|} 
\begin{pmatrix}
k_x+i k_y \\
k_z
\end{pmatrix}, \qquad 
\bk \neq \bm{0}, 
\label{eq:inversion_model_line_bundle}
\end{align}
be a single occupied state with two orbitals near $\bk=\bm{0}$. 
The associated 2 by 2 Hamiltonian is given by 
\begin{align}
H(\bk)
= |\bk|^2 ({\bf 1}_{2 \times 2}- 2 \ket{\bk} \bra{\bk} )
= \begin{pmatrix}
-k_x^2-k_y^2+k_z^2 & -2k_z(k_x-ik_y) \\
-2k_z(k_x+ik_y)& k_x^2+k_y^2-k_z^2
\end{pmatrix}. 
\label{eq:model_inversion}
\end{align}
For example, the BdG Hamiltonian of $(d_{zx}+i d_{zy})$-wave superconductors takes this form. 
$\bk = \bm{0}$ point is the gapless point of this Hamiltonian. 
This model has the symmetry $H(-\bk) = H(\bk)$. 
Let us compute the $\Z_2$ invariant associated with 
the north hemisphere of a $|\bk| = {\rm const.}$\ sphere as shown in 
Fig.~\ref{fig:inversion} [b]. 
Under the choice $U(\bk) = {\bf 1}_{2 \times 2}$, 
the inversion symmetry $\ket{-\bk} = - \ket{\bk}$ means that 
the $V(\bk)$ in (\ref{eq:def_v(p)_inversion}) is $V(\bk) = -1$. 
Introduce the spherical coordinate 
$\bk = |\bk| (\sin \theta \cos \phi, \sin \theta \sin \phi, \cos \theta)$. 
The Berry connection and the curvature of $\ket{\bk}$ are given by 
${\cal A} = \frac{i}{2} (1-\cos 2 \theta) d \phi$ and 
${\cal F} = i \sin 2 \theta d \theta \wedge d \phi$, respectively. 
It is easy to show that the $\Z_2$ invariant (\ref{eq:c1_RP2}) 
becomes $c_1 = 1/2\ (\mod\ 1)$. 
On the other hand, the trivial nonsingular Hamiltonian $H(\bk) = {\rm diag}(1,-1)$ 
shows $c_1 = 0 \ (\mod \ 1)$. 
Thus, $c_1 = 1/2 \ (\mod \ 1)$ protects the gapless point of the Hamiltonian (\ref{eq:model_inversion}). 

Notice that the singular point of the Hamiltonian (\ref{eq:model_inversion}) has no Chern number, 
so the singularity of (\ref{eq:model_inversion}) can be stabilized only after the inversion symmetry is enforced. 

\begin{figure}[!]
 \begin{center}
  \includegraphics[width=\linewidth, trim=0cm 1cm 0cm 0cm]{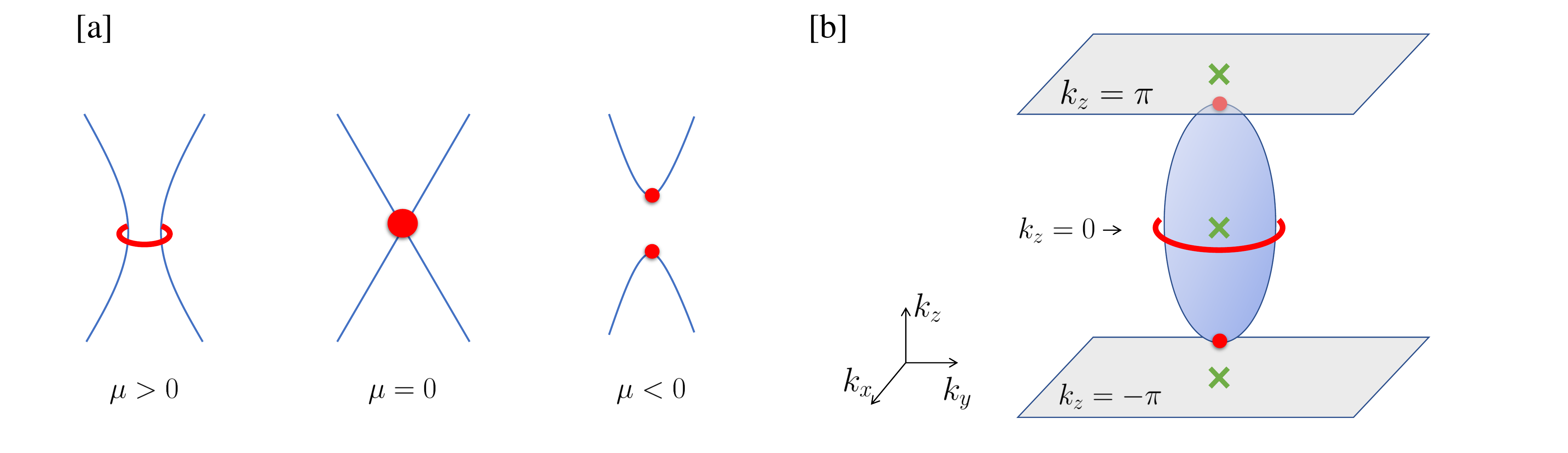}
 \end{center}
 \caption{
 [a] The red curves and points represent singular gapless points. 
 The blue curves represent the ``Fermi surfaces'' of the diagonal part of the Hamiltonian (\ref{eq:model_inversion_dwave}). 
 [b] The pair of two $\Z_2$ charges. The green x marks are inversion symmetric points. 
 }
 \label{fig:inversion_dwave}
\end{figure}

Let us consider more implication of the $\Z_2$ nontriviality. 
To make it easy to understand, 
we use the notation of the 
BdG Hamiltonian of $(d_{zx}+id_{zy})$-wave superconductors with 
a spin $s_z$ conserved system. 
But the following discussion can be applied to any inversion symmetric systems. 
Let us consider a Hamiltonian 
\begin{align}
H_{d}(\bk)
&= \begin{pmatrix}
\frac{k_x^2 + k_y^2}{2m}-\frac{k_z^2}{2m'} -\mu & \Delta k_z(k_x+i k_y) \\
\Delta k_z(k_x-i k_y) & -\frac{k_x^2 + k_y^2}{2m}+\frac{k_z^2}{2m'} + \mu
\end{pmatrix}, \qquad m,m' > 0. 
\label{eq:model_inversion_dwave} 
\end{align}
Depending on the sign of the ``chemical potential'' $\mu$, 
the singular points of the Hamiltonian (\ref{eq:model_inversion_dwave}) 
form a ring ($\mu >0$), 
single point ($\mu = 0$), 
and pair of two points with Chern number ($\mu < 0$) 
as shown in Fig.~\ref{fig:inversion_dwave} [a]. 
An important point is that 
both ring and point like singularities 
have the same $\Z_2$ invariant $c_1=1/2$, 
provided that the inversion symmetric sphere 
surrounds these singular regions. 

The inversion symmetric version of Nielsen-Ninomiya's theorem holds true. 
Let us consider a lattice analog of (\ref{eq:model_inversion_dwave}) along the $z$-direction 
\begin{align}
H_{d,{\rm lattice}}(k_x,k_y,k_z)
&= \begin{pmatrix}
\frac{k_x^2 + k_y^2}{2m} - t \cos k_z -\mu & \Delta k_z(k_x+i k_y) \\
\Delta k_z(k_x-i k_y) & -\frac{k_x^2 + k_y^2}{2m} + t \cos k_z + \mu
\end{pmatrix}, \qquad m,t > 0. 
\label{eq:model_inversion_dwave_lattice} 
\end{align}
For the parameter region $-t < \mu < t$, 
the ``Fermi surface'' of the diagonal part of (\ref{eq:model_inversion_dwave_lattice}) 
forms a spheroid 
as shown in Fig.~\ref{fig:inversion_dwave} [b]. 
There is a ring singularity with $\Z_2$ charge $c_1=1/2$ on the $k_z=0$ plane. 
Moreover, near the $(0,0,\pi)$ point, 
there are two point like singularities which have the 
$\Z_2$ charge $c_1=1/2$ as a pair. 
Nielsen-Ninomiya's theorem is that 
in the closed BZ torus 
the single $\Z_2$ charge $c_1=1/2$ is forbidden. 
Like this example, if there is a ring node 
near an inversion symmetric point $(0,0,0)$, 
there should be another node with $\Z_2$ charge $c_1=1/2$. 

\subsubsection{Generalization to higher dimensions}
It is easy to generalize the discussion so far to higher space dimensions with inversion symmetry. 
Let us consider $d$-dimensional systems with inversion symmetry 
$U(\bk) H(-\bk) U(\bk)^{-1}= H(-\bk), \bk = (k_1, \dots, k_d)$. 
We focus on an inversion symmetric $(d-1)$-dimensional sphere $S^{d-1}$. 
The $K$-theory on the sphere $S^{d-1}$ is given by~\cite{Adams1962}
\begin{align}
K_{\Z_2}(S^{d-1})
\cong K(S^{d-1}/\Z_2)
= K(RP^{d-1})
= \Z_{p} \oplus \Z, \qquad 
p = \left\{\begin{array}{ll}
2^{(d-1)/2} & (d = {\rm odd}) \\
2^{(d-2)/2} & (d = {\rm even}) \\
\end{array}\right. .
\end{align}
Here, $\Z_2$ acts on $S^{d-1}$ as the antipodal map. 
The free part $\Z$ of the $K$-group $K_{\Z_2}(S^{d-1})$ 
is generated by the trivial line bundle $[1]$ on $RP^{d-1}$. 
The torsion part $\Z_p$ is generated by 
the formal difference $[\xi'] - [1]$, 
where $\xi'$ is the complexification $\xi'=\xi\otimes\mathbb{C}$ of 
the tautological real line bundle $\xi$ over $RP^{d-1}$.~\cite{Adams1962}
$\Z_p$ implies that $(\xi')^{\oplus p}$ is stably isomorphic to 
the trivial bundle $1^{\oplus p}$. 
The $\Z_2$-equivariant line bundle on $S^{d-1}$ corresponding to $\xi'$ 
is given by a form similar to (\ref{eq:inversion_model_line_bundle}), 
\begin{align}
\ket{\bm{n}} 
=\left\{\begin{array}{ll}
(n_1 + i n_2, n_3 + i n_4, \dots, n_{d})^T & (d = {\rm odd}) \\
(n_1 + i n_2, n_3 + i n_4, \dots, n_{d-1} + i n_{d})^T & (d = {\rm even}) \\
\end{array}\right., 
\end{align}
where $\bm{n} = (n_1, \dots, n_d)$, $|\bm{n}|=1$ is the coordinate of $S^{d-1}$. 

For $d \leq 6$ (which corresponds $\Z_2$ or $\Z_4$ classifications), 
elements in the $K$-group can be distinguished by the Chern classes. 
Recall that the total Chern class $c(E) = 1 + \sum_{j>0} c_j(E)$ of a given complex bundle $E$ over a space $M$ 
takes values in the cohomology ring $H^*(M;\Z)$. 
The Whitney sum induces the cup product $c(E \oplus F) = c(E) c(F)$ in $H^*(M;\Z)$. 
The cohomology of $RP^{d-1}$ is given by 
\begin{equation}\begin{split}
H^{j}(RP^{d-1};\mathbb{Z})=
\left\{ \begin{array}{ll}
\mathbb{Z} & (\mbox{$j = 0$; and $j = d - 1$ for even $d$}) \\
\mathbb{Z}_2 & (\mbox{even $j$ with $0< j < d-1$; and $j = d - 1$ for odd $d$}) \\
0 & (\mathrm{otherwise}) \\
\end{array} \right. .
\end{split}\end{equation}
The nonzero elements of $H^{2j}(RP^{d-1};\mathbb{Z})=\mathbb{Z}_2, (0 < j \le [d/2])$ are 
given by the cup products $t^j\in{H}^{2j}(RP^{d-1};\mathbb{Z})$ of the 
first Chern class $t=c_1(\xi')$ of the tautological line bundle. 
The generator $[\xi']$ of the torsion part of the $K$-group has 
the Chern class $c(\xi') = 1+t$. 

For example, the torsion part of $K(RP^4) = \Z_4 \oplus \Z$ is generated by 
$[\xi'] = (1,1) \in K(RP^4)$. 
In this case, 
the Chern class can distinguish all elements of $\Z_4$, 
since 
$c(\xi' \oplus \xi') = 1+t^2, c(\xi' \oplus \xi' \oplus \xi') = 1+t+t^2$, and  
$c(\xi' \oplus \xi' \oplus \xi' \oplus \xi') = 1$. 
I.e.\ the first and second Chern classes detect all the $\Z_4$ phases. 

On the other hand, 
the torsion part of $K(RP^6) = \Z_8 \oplus \Z$ cannot be 
detected by the Chern classes. 
This is because the $4 \in \Z_8$ phase 
is trivial in the Chern class $c( (\xi')^{\oplus 4} ) = (1+t)^4 = 1 \in H^*(RP^6;\Z)$.

\begin{table*}[]
\begin{center}
\caption{
A topological charges of Fermi points in inversion symmetric systems. 
$d$ is the space dimension. 
In class AI and AII, the inversion symmetry commutes with the TRS. 
$\wt{K}(RP^{d-1})$, $\wt{KO}(RP^{d-1})$, and $\wt{KSp}(RP^{d-1})$
represent the reduced complex, real, and quaternionic $K$-theories, respectively. 
}
\begin{tabular}[t]{ccccccccccc}
\hline \hline
AZ class & $K$-group & $d=1$ & $d=2$ & $d=3$ & $d=4$ & $d=5$ & $d=6$ & $d=7$ & $d=8$ \\
\hline
A & $\wt{K}(RP^{d-1})$ & $0$ & $0$ & $\mathbb{Z}_2$ & $\mathbb{Z}_2$ & $\mathbb{Z}_4$ & $\mathbb{Z}_4$ &$\mathbb{Z}_8$ & $\mathbb{Z}_8$ \\ 
AI & $\wt{KO}(RP^{d-1})$ & $0$ & $\mathbb{Z}_2$ & $\mathbb{Z}_4$ & $\mathbb{Z}_4$ & $\mathbb{Z}_8$ & $\mathbb{Z}_8$ &$\mathbb{Z}_8$ & $\mathbb{Z}_8$ \\ 
AII & $\wt{KSp}(RP^{d-1})$ & $0$ & $0$ & $0$ & $0$ & $\mathbb{Z}_2$ & $\mathbb{Z}_4$ &$\mathbb{Z}_8$ & $\mathbb{Z}_8$ \\ 
\hline \hline
\end{tabular}
\label{TabISFP}
\end{center}
\end{table*}

\subsubsection{Time-reversal symmetry with inversion symmetry: Stiefel-Whitney class}
The interplay of TRS and inversion symmetry 
gives rise to Fermi points with a nontrivial topological charge 
and some topological charges can be captured by Stiefel-Whitney (SW) classes. 

Let us consider the class AI TRS with inversion symmetry which {\it commutes} with the TRS 
\begin{align}
&T H(\bk) T^{-1} = H(-\bk), \qquad T^2 = 1, \\
&U(\bk) H(\bk) U(\bk)^{-1} = H(-\bk), \qquad U(-\bk) U(\bk) = 1, \\
&T U(\bk) = U(-\bk) T, 
\end{align}
where $T$ is anti-unitary. 
We, here, focus on the class AI which is the TRS for spin integer systems. 
In the cases of class AII TRS $T^2 = -1$, there is no torsion part in lower space dimensions, 
hence, we only show the $K$-group in Table~\ref{TabISFP}.
The combined symmetry $T U(\bk)$ acts on the BZ without changing the momentum as 
\begin{align}
T U(\bk) H(\bk) (T U(\bk))^{-1} = H(\bk), \qquad (TU(\bk))^2 = 1, 
\end{align}
so $T U(\bk)$ induces the real structure on the occupied states. 
Since the inversion symmetry $U(\bk)$ commutes with the combined symmetry $T U(\bk)$, 
the $K$-theory of a sphere $S^{d-1}$ surrounding the symmetric point $\bk=\bm{0}$ is 
recast into that of the quotient $S^{d-1}/\Z_2 = RP^{d-1}$. 
The real $K$-theory $KO(RP^{d-1})$ of the real projective space is known:~\cite{Adams1962} 
\begin{equation}\begin{split}
{}^{\phi} K^{\tau+0}_{\Z_2}(S^{d-1}) = KO(RP^{d-1})= \Z_{2^g} \oplus \Z, 
\end{split}\end{equation}
where $g$ is the number of integers $s$ such that $0<s\leq d-1$ and $s\equiv 0,1,2$, or $4$ mod 8. 
Here, the twisting $\tau$ represents the commutation relation between $T$ and $U(\bk)$. 
See Table~\ref{tab:n_and_AZ} for some examples. 
The torsion part of $KO(RP^{d-1})$ is additively generated by the formal difference 
$([\xi]-[1])$ where $\xi$ is the tautological real line bundle over $RP^{d-1}$. 

A generating $\Z_2$-equivariant real line bundle over $S^{d-1}$ corresponding to $\xi$ is given as follows. 
Let $\ket{\bk}$ be a line bundle with TRS and inversion 
\begin{align}
\ket{\bk} = \frac{1}{|\bk|} (k_1, k_2, \dots, k_d)^T, \qquad 
\ket{\bk} = \ket{\bk}^*, \qquad 
\ket{-\bk} = -\ket{\bk}. 
\end{align}
Notice that $\bk = \bm{0}$ is singular. 
The restriction of the line bundle $\ket{\bk}$ to a sphere $|\bk| = {\rm const.}$ leads 
to the generator of the torsion part. 
A Hamiltonian of which the occupied state is $\ket{\bk}$ is given by 
\begin{align}
H(\bk)
= |\bk|^2 ({\bf 1}_{d \times d} - 2 \ket{\bk} \bra{\bk}). 
\label{eq:model_ai_inversion}
\end{align}

In the same way as the complex $K$-theory of $RP^{d-1}$, 
$\Z_2$ and $\Z_4$ classifications of the $K$-groups $KO(RP^{d-1})$ 
can be characterized by the SW classes. 
A real bundle $E$ over a manifold $M$ defines the total 
SW class $w(E) = 1 + \sum_{j>0} w_j(E) \in H^*(M;\Z_2)$. 
The real projective space $RP^{d-1}$ has the following cohomology with $\Z_2$ coefficients 
\begin{equation}\begin{split}
H^{j}(RP^{d-1};\mathbb{Z}_2)=
\left\{ \begin{array}{ll}
\mathbb{Z}_2 & (0\leq j \leq d-1) \\
0 & (\mathrm{otherwise}) \\
\end{array} \right.. 
\end{split}\end{equation}
As the cohomology ring, $H^*(RP^{d-1};\Z_2)$ is isomorphic to $\Z_2[t]/(1-t^{d})$. 
The tautological real line bundle $\xi$ over $RP^{d-1}$ has the data $w(\xi) = 1 + t$. 
From the structure of the SW classes $w(E \oplus F) = w(E) w(F)$, 
one can show that the $\Z_2$ and $\Z_4$ subgroups in the torsion part of the $K$-group 
$KO(RP^{d-1})$ can be characterized by the SW classes. 

Let us construct the $\Z_2$-equivariant first SW class on $S^{d-1}$. 
A similar invariant defined by TRS and $C_4$-rotation symmetry is discussed in Ref.~\onlinecite{alexandradinata2016berry}. 
Choose a point $P$ and its inversion symmetric pair $-P$ in the BZ. 
Let $\ell$ be an oriented path from $P$ to $-P$. 
Let $\Phi(\bk), (\bk \in \ell)$ be a frame of occupied states which is smoothly defined on the line $\ell$. 
We fix the gauge freedom of $\Phi(\bk)$ so that 
the combined symmetry $TU(\bk)$ is represented by a $\bk$-independent unitary matrix $W$ as 
$T U(\bk) \Phi(\bk) = \Phi(\bk) W$ on the line $\ell$. 
Because of the inversion symmetry, 
$\Phi(P)$ and $\Phi(-P)$ are related as 
$U(-P) \Phi(-P) = \Phi(P) V(P)$ 
with $V(P)$ a unitary matrix. 
From the assumption $T U(\bk) = U(-\bk) T$, one can show that 
$W V(P)^* = V(P) W$, which leads to the
$\Z_2$ quantization of the determinant $\det[V(P)] = \pm 1$. 
This determinant $\det[V(P)]$ is the $\Z_2$-equivariant version of the 
first SW class. 
Notice that the change of sign $U(\bk) \mapsto -U(\bk)$ 
induces $V(P) \mapsto - V(P)$, 
thus, the $\Z_2$ invariant $\det[V(P)]$ is relatively well-defined from the trivial occupied state. 

On the other hand, unfortunately, there is no simple expression of the second SW class $w_2(E)$ 
for a given occupied states bundle $E$ with $T$ and $U(\bk)$ symmetries.  

Here, we give two examples in low dimensions. 

In 2-spatial dimensions, the model Hamiltonian (\ref{eq:model_ai_inversion}) reads 
\begin{align}
H_{2d}(k_x,k_y)
= \begin{pmatrix}
-k_x^2+k_y^2 & -2 k_x k_y \\
-2 k_x k_y & k_x^2 - k_y^2 \\
\end{pmatrix}. 
\label{eq:model_ai_inversion_d=2}
\end{align}
Such a Hamiltonian is realized in a 
$d$-wave superconductor and a $d$-density wave in 2-dimensions. 
The TRS and inversion symmetry are given as $T = K$ and $U(\bk) = {\bf 1}_{2 \times 2}$, 
where $K$ means the complex conjugate. 
The occupied state is $\ket{\bk} = (k_x, k_y)^T/ |\bk|, (\bk \neq \bm{0})$. 
This occupied state satisfies the gauge fixing condition 
$TU(\bk) \ket{\bk} = \ket{\bk}$, that is, $W = 1$.  
Because of $U(\bk) \ket{\bk} = - \ket{-\bk}$, 
the $\Z_2$ invariant is $\det[V(\bk)] = -1$. 
Thus, the singular point $\bk = \bm{0}$ of the Hamiltonian 
(\ref{eq:model_ai_inversion_d=2}) is stable 
unless $T$ or $U(\bk)$ symmetry is broken. 

In 3-spatial dimensions, the model Hamiltonian (\ref{eq:model_ai_inversion}) reads 
\begin{align}
H_{3d}(k_x,k_y,k_z)
= \begin{pmatrix}
-k_x^2+k_y^2+k_z^2 & -2 k_x k_y & -2 k_x k_z \\
-2 k_x k_y & k_x^2 - k_y^2+k_z^2 & -2 k_y k_z \\
-2 k_x k_z & -2 k_y k_z & k_x^2 + k_y^2-k_z^2 \\
\end{pmatrix}. 
\label{eq:model_ai_inversion_d=3}
\end{align}
The occupied states of this Hamiltonian 
have the $\Z_4$ charge of the 
$KO$-theory $KO(RP^2) = \Z_4 \oplus \Z$. 
Actually, in the same way as 2-dimensions, 
the occupied state $\ket{\bk} = (k_x,k_y,k_z)^T/|\bk|$ 
has the $\Z_2$ charge $\det[V(\bk)] = -1$. 
From the property of $w$,
the first SW class of the direct sum $\ket{\bk} \oplus \ket{\bk}$ is trivial,
but the second SW class is non-trivial.
The first and second SW classes of the direct sum $\ket{\bk} \oplus \ket{\bk} \oplus \ket{\bk}$ are both non-trivial.

\section{Conclusion}
\label{sec:Conclusion}
In this paper, we formulate topological crystalline materials on the
basis of the twisted equivariant $K$-theory.
We illustrate how space and magnetic space groups are
incorporated into topological classification of both gapful and gapless
crystalline materials in a unified manner. 
The twisted equivariant $K$-theory ${}^{\phi} K_{\cal G}^{(\tau, c)-n}(T^d)$ 
on the BZ torus $T^d$ serves the stable classification of bulk TCIs and
TCSCs and their boundary and defect gapless states. 
$K$-theories are not just additive groups, but are equipped with the module
structures for point groups so that the
classification naturally includes the
information on crystals such as point group representations and Wyckoff
positions.
Using isomorphisms between $K$-theories, we also discuss
bulk-boundary and bulk-defect correspondences in the presence of
crystalline symmetry. 
In Sec.~\ref{sec:Topological nodal semimetals and superconductors},
we propose a systematic method to classify bulk gapless topological
crystalline materials in terms of $K$-theory.
We show that the cokernel of the map $i^*_Y$ between
$K$-theories, which is induced by the 
inclusion $i_Y$ of a subspace $Y$ into the BZ torus $T^d$, defines 
bulk gapless topological materials.
In Sec.~\ref{Wallpaper_summary}, we present topological
table with wallpaper groups in the absence of TRS
and PHS.
In particular, the module structures for point groups are identified
in the wallpaper classification, of which information is important to
understand crystalline materials. 
Furthermore, we illustrate computations of $K$-groups 
for various systems in Sec.~\ref{sec:Example of K-theory classification}. 

More computations of
$K$-groups are necessary to fully explore topological crystalline materials.
Even for relatively simple wallpaper groups, the full computation is
missing in the presence of
TRS and/or PHS, although a part of
computations have been done by the present authors.\cite{ShiozakiSatoGomi2016}
In three  dimensions, most of $K$-groups with (magnetic) space groups
have not been known yet. 
Our present formulation provides a precise and systematic framework to step
into the unexplored field of topological crystalline materials.

{\it Note Added.}---
While this manuscript was being prepared, we became aware of a recent independent work by Kruthoff {\it et al.},~\cite{Kruthoff2016} which discussed the topological classification of bulk insulators and stable nodal structures in the presence of space groups, mainly focusing on class A spinless systems. 
They also gave the classification of class A spinless topological crystalline insulators in two dimensions with wallpaper groups, which is consistent with us and Refs.~\onlinecite{Yang1997, LuckStamm2000}.

\acknowledgements

K.S.\ thanks useful discussions with Aris Alexandradinata and Takahiro Morimoto. 
K.S.\ is supported by JSPS Postdoctoral Fellowship for Research Abroad.
M.S.\ is supported by the "Topological Materials Science''
(No.JP15H05855) KAKENHI on Innovative Areas from JSPS of Japan. 
K.G.\ is supported by JSPS KAKENHI Grant Number JP15K04871.

\appendix

\makeatletter
\renewcommand{\theequation}{%
\Alph{section}.\arabic{equation}}
\@addtoreset{equation}{section}
\makeatother

\section{An example of mismatch between $K$-theory and isomorphism classes of vector bundles}
\label{app:An example of mismatch}
A simple example of the mismatch between 
the $K$-theory and the set of isomorphic classes of 
vector bundles is real vector bundle over $S^2$. 
The tangent bundle $T S^2$ is not isomorphic to 
the trivial rank two vector bundle $\underline{\R} \oplus \underline{\R}$, 
since $TS^2$ does not have any nowhere vanishing sections. 
On the other hand, in the sense of stable equivalence, 
$T S^2$ is trivialized by 
adding a trivial line bundle $\underline{\R}$ on $S^2$ 
because $\underline{\R}$ is isomorphic to the normal bundle $N S^2$. 
So we found 
\begin{align}
T S^2 \oplus \underline{\R}
\cong 
T S^2 \oplus N S^2 
\cong 
\underline{\R} \oplus \underline{\R} \oplus \underline{\R}, 
\end{align}
which implies $T S^2$ and $\underline{\R} \oplus \underline{\R}$ 
give the same element $[T S^2] = [\underline{\R} \oplus \underline{\R}] \in KO(S^2)$ 
in the $K$-theory.

\section{Group cohomology}
\label{Group_cohomology}

Let $G$ be a finite group. A $G$-bimodule is by definition an Abelian group $M$ with a left action 
$m \mapsto g_L \cdot m$ of $g_L \in G$ 
and a right action 
$m \mapsto m \cdot g_R$ of $g_R \in G$ 
which are compatible $(g_L \cdot m) \cdot g_R = g_L \cdot (m \cdot g_R)$.
An example is the trivial $G$-module $M$, which is an Abelian group $M$ with the left and right actions of $G$ by the identity $m \mapsto m$. Another example relevant to the body of this paper is $M = C(X, \R/2\pi \Z)_\phi$. This is the group $C(X, \R/2\pi \Z)$ of $\R/2\pi\Z$-valued functions on $X$ endowed with the left action $\alpha({\bm k}) \mapsto \phi(g) \alpha(g^{-1}{\bm k})$ of $g_L \in G$ and the trivial right action $\alpha({\bm k}) \mapsto \alpha({\bm k})$, where $\phi: G \to \{1,-1\}$ is a homomorphism indicating that the symmetry $g$ is unitary ($\phi(g)=1$) or antiunitary ($\phi(g)=-1$).

Given a $G$-bimodule $M$, we write $C^n(G; M) = C(G^n, M)$ for the set of maps $\tau : G^n \to M$ for $n = 1, 2, \ldots$. 
In the case of $n = 0$, we put $C^0(G; M) = M$. With the addition $(\tau + \tau')(g_1, \ldots, g_n) = \tau(g_1, \ldots, g_n) + \tau'(g_1, \ldots, g_n)$, the set $C^n(G; M)$ gives rise to an Abelian group. We define a homomorphism $\delta : C^n(G; M) \to C^{n+1}(G; M)$ to be
\begin{equation*}
(\delta \tau)(g_1, \ldots, g_{n+1})
= g_1 \cdot \tau(g_2, \ldots, g_{n+1}) 
+ \sum_{i = 1}^n (-1)^i \tau(g_1, \ldots, g_ig_{i+1}, \ldots, g_{n+1})
+ (-1)^{n+1} \tau(g_1, \ldots, g_n) \cdot g_{n+1},
\end{equation*}
by using the left action of $G$ in the first term and the right action in the last. We can directly verify $\delta \delta = 0$, so that $(C^*(G; M), \delta)$ is a cochain complex. As usual, we write $Z^n(G; M) = \mathrm{Ker}(\delta) \cap C^n(G; M)$ for the subgroup of $n$-cocycles and $B^n(G; M) = \mathrm{Im}(\delta) \cap C^n(G; M)$ for the subgroup of $n$-coboundaries. Then the group cohomology of $G$ with coefficients in the $G$-bimodule $M$ is defined by 
$$
H^n(G; M) = Z^n(G; M)/B^n(G; M).
$$

As a matter of fact, the group cohomology $H^n(G; M)$ with coefficients in the trivial $G$-module $M$ is isomorphic to the Borel equivariant cohomology $H^n_G(pt; M)$ of the point with coefficients in $M$. In particular, $H^2(G; \R/2\pi\Z) \cong H^2_G(pt; \R/2\pi\Z) \cong H^3_G(pt; \Z)$ by the exponential exact sequence.

\section{More on vector bundle formulation}
\label{Seq:Vect_Bndl_Formulation}

As in \ref{Topological classification and the $K$-group}, the $K$-group $K(X) = K^0(X)$ of a space $X$ can be defined as the group of pairs $([E], [F])$ or formal differences $[E] - [F]$. This is a standard formulation of the $K$-theory \cite{Atiyah1966}, and we can generalize this to formulate some twisted equivariant $K$-theory as well [\onlinecite{Freed2013}]: Suppose that a finite group $G$ acts on $X$ and a two-cocycle $\tau = \{ \tau_{g, g'}({\bm k}) \} \in Z^2(G; C(X, \R/2\pi\Z))$ is given. A complex vector bundle $E$ on $X$ is said to be a $\tau$-twisted $G$-equivariant vector bundle if there are vector bundle maps $U_p : E \to E$ which cover the left actions $g : X \to X$ of $g \in G$ and are subject to the relations
\begin{equation*}
U_g(g'{\bm k}) U_{g'}({\bm k}) 
= e^{i\tau_{g, g'}(gg'{\bm k})} U_{gg'}({\bm k}) 
\end{equation*}
on the fiber of $E$ at ${\bm k} \in X$. Since the direct sum of these vector bundles makes sense, the same argument as in \ref{Topological classification and the $K$-group} leads to the formulation of the $\tau$-twisted $G$-equivariant $K$-group $K^{\tau + 0}_G(X)$ by using twisted vector bundles.

\medskip

The odd $K$-group $K^{\tau -1}_G(X)$ can also be formulated in terms of the twisted equivariant vector bundle: For a $\tau$-twisted $G$-equivariant vector bundle $E$, let us consider an automorphism $q : E \to E$ of vector bundles which cover the identity map of the base space $X$ and are subject to the relations
\begin{equation*}
q(g{\bm k}) U_g({\bm k}) = U_g({\bm k}) q({\bm k})
\end{equation*}
on the fiber of $E$ at ${\bm k } \in X$. The equivalence classes of such automorphisms constitute $K^{\tau - 1}_G(X)$. Automorphisms of twisted vector bundles $q : E \to E$ and $q' : E' \to E'$ are equivalent if there is a twisted bundle $F$ such that $E \oplus F$ and $E' \oplus F$ are isomorphic and $q \oplus {\bm 1}_F$ and $q' \oplus {\bm 1}_F$ are homotopic in the way compatible with the symmetries.

\section{Mayer-Vietoris sequence}
\label{app:Mayer-Vietoris sequence}

Given a finite group $G$ acting on a space $X$ and a group two-cocycle $\tau = \{ \tau_{g, g'}({\bm k}) \} \in Z^2(G; C(X, \R/2\pi \Z))$, we have the twisted equivariant $K$-theory $K^{\tau + 0}_G(X)$. (More generally, $\tau$ can be a twist \cite{FHT}, a geometric object classified by the Borel equivariant cohomology $H^3_G(X; \Z)$.) If $Y \subset X$ is a closed subspace, then the relative $K$-group $K^{\tau + 0}_G(X, Y)$ can be defined. In Karoubi's formulation, the equivalence classes of triples $(E, H, H')$ such that $H({\bm k}) = H'({\bm k})$ for ${\bm k} \in Y$ constitute $K^{\tau}_G(X, Y) = K^{\tau + 0}_G(X, Y)$. For $n \ge 0$, we use $n$ chiral symmetries to define $K^{\tau - n}_G(X, Y)$ similarly. These $K$-groups are naturally modules over the representation ring $R(G)$ of $G$, and there is a natural $R(G)$-module isomorphism, called the Bott periodicity:
$$
K^{\tau - n}_G(X, Y) \cong K^{\tau -n - 2}_G(X, Y).
$$
Extending this isomorphism, we define $K^{\tau + n}_G(X, Y)$ for all $n \in \Z$. 

Let $X'$ be another $G$-space, and $Y' \subset X'$ a closed subspace. If $f : X' \to X$ is a $G$-equivariant map such that $f(Y') \subset Y$, then we write $f : (X, Y) \to (X', Y')$. Such a map induces by pull-back an $R(G)$-module homomorphism $f^* : K^{\tau + n}_G(X, Y) \to K^{f^*\tau + n}_G(X', Y')$. For $g : (X'', Y'') \to (X', Y')$, it holds that $(f \circ g)^* = g^* \circ f^*$. The basic behaviour of the groups $\{ K^{\tau + n}_G(X, Y) \}_{n \in \Z}$ and the homomorphisms $f^*$ are summarized as the axioms of generalized equivariant cohomology theory as follows \cite{FHT}:
\begin{itemize}
\item
(The homotopy axiom)
Let $f_0 : X' \to X$ and $f_1 : X' \to X$ be $G$-equivariant maps such that $f_i(Y') \subset Y$. Suppose that $f_0$ and $f_1$ are $G$-equivariantly homotopic, in the sense that there is a $G$-equivariant map $F : X' \times [0, 1] \to X$ such that $F(x', i) = f_i(x')$ for $i = 0, 1$ and $x' \in X$. Here the $G$-action on $[0, 1]$ is trivial. If in addition $F(Y' \times [0, 1]) \subset Y$, then there is an isomorphism of twists $f_0^*\tau \cong f_1^*\tau$ and we have the equality of the $R(G)$-module homomorphisms $f_0^* = f_1^*$.

\item
(The excision axiom)
Let $A, B \subset X$ be closed invariant subspaces. Then the inclusion $j : A \to A \cup B$ induces an isomorphism of $R(G)$-modules
$$
j^* : K^{\tau|_{A \cup B} + n}_G(A \cup B, B) \to
K^{\tau|_A + n}_G(A, A \cap B),
$$
where we put $\tau|_{A \cup B} = i_{A \cup B}^*\tau$ and $\tau|_A = i_A^*\tau$ by using the inclusion maps $i_{A \cup B} : A \cup B \to X$ and $i_A : A \to X$.

\item
(The exactness axiom)
For a pair $(X, Y)$ consisting of a space $X$ with $G$-action and an invariant closed subspace $Y \subset X$, there is a long exact sequence of $R(G)$-modules
$$
\cdots \to
K^{\tau + n}_G(X, Y) \to
K^{\tau + n}_G(X) \overset{i^*}{\to}
K^{\tau|_Y + n}_G(Y) \to
K^{\tau + n + 1}_G(X, Y) \to
\cdots,
$$
where $i^* : K^{\tau + n}_G(X) \to K^{\tau|_Y + n}_G(X)$ is induced form the inclusion $i : Y \to X$.

\item
(The additivity)
Suppose that spaces $X_\lambda$ with $G$-action, their invariant subspaces $Y_\lambda \subset X_\lambda$ and twists $\tau_\lambda$ of $X_\lambda$ are given. Then the inclusions $X_\lambda \to \bigsqcup_\lambda X_\lambda$ induce an isomorphism of $R(G)$-modules
$$
K^{\sqcup_\lambda \tau_\lambda + n}_G(\bigsqcup_\lambda X_\lambda, \bigsqcup_\lambda Y_\lambda) 
\cong 
\prod_\lambda K^{\tau + n}_G(X_\lambda, Y_\lambda).
$$

\end{itemize}

The above axioms are parallel to the Eilenberg-Steenrod axioms of ordinary cohomology theory but the dimension axiom. To state the counterpart of the dimension axiom, we remark that, for the space $pt$ consisting of a single point, the equivariant cohomology $H^3_G(pt; \Z)$ classifies central extensions $G^\omega$ of $G$ by $U(1)$:
$$
1 \to U(1) \to G^\omega \to G \to 1.
$$
Since $G$ is a finite group, we have $H^3_G(pt; \Z) \cong H^3(G; \Z) \cong H^2(G; U(1))$, and a two-cocycle $\omega = \{ \omega_{g, g'} \}$ defines a central extension $G^\omega$ by introducing the multiplication $(g, u) \cdot (g', u') = (gg', uu'e^{i \omega_{g, g'}})$ to the set $G \times U(1)$. 
\begin{itemize}
\item
For $\omega \in Z^2(G; \R/2\pi\Z)$, there are isomorphisms:
\begin{align*}
K^{\omega + 0}_G(pt) &\cong R^\omega(G), &
K^{\omega + 1}_G(pt) &= 0,
\end{align*}
where $R^\omega(G)$ is the free abelian group generated by the equivalence classes of representations of $G^\omega$ such that the central $U(1) \subset G^\omega$ acts by the scalar multiplication, or equivalently $\omega$-projective representations of $G$. 
 \end{itemize}

Some direct consequences of the axioms of cohomology theory are as follows:
\begin{itemize}
\item
If $f : X' \to X$ is a $G$-equivariant homotopy equivalence, then $f^* : K^{\tau + n}_G(X) \to K^{f^*\tau + n}_G(X')$ is an isomorphism.

\item
(Mayer-Vietoris exact sequence)
For closed invariant subspaces $A, B \subset X$, there is an exact sequence of $R(G)$-modules:
$$
\cdots \to
K^{\tau|_{A \cup B} + n}_G(A \cup B) \overset{(i^*_A, i^*_B)}{\to}
K^{\tau|_A + n}_G(A) \oplus K^{\tau|_B + n}_G(B) \overset{\Delta}{\to}
K^{\tau|_{A \cap B} + n}_G(A \cap B) \to
K^{\tau|_{A \cup B} + n + 1}_G(A \cup B) \to
\cdots,
$$
where $i_A : A \to A \cup B$ and $i_B : B \to A \cup B$ are the inclusions, and $\Delta$ is expressed as $\Delta(a, b) = j_A^*(a) - j_B^*(b)$ by using the inclusions $j_A : A \cap B \to A$ and $j_B^* : A \cap B \to B$. 
\end{itemize}

It is often useful to introduce the reduced $K$-theory. This is defined only when the cocycle $\tau$ is a constant function on $X$, that is, $\tau \in Z^2(G; U(1))$. In this case, we choose a point $pt \in X$ to define the reduced $K$-theory as follows:
$$
\widetilde{K}^{\tau + n}_G(X) = K^{\tau + n}_G(X, pt).
$$
It turns out that $\widetilde{K}^{\tau + n}_G(X, pt)$ is isomorphic to the kernel of the homomorphism $i^* : K^{\tau + n}_G(X) \to K^{\tau + n}_G(pt)$ induced from the inclusion $i : pt \to X$. We also have a natural direct sum decomposition
$$
K^{\tau + n}_G(X) \cong
K^{\tau + n}_G(pt) \oplus \widetilde{K}^{\tau + n}_G(X).
$$

So far, the equivariant $K$-theory $K^{\tau + n}_G(X, Y)$ twisted by an ungraded twist $\tau$ is considered. In general, a twist $\tau$ can be graded by an element of $H^1_G(X; \Z_2)$. For example, a homomorphism $c : G \to \Z_2$ defines an element of $H^1_G(X; \Z_2)$, and hence a grading. For the equivariant $K$-theory $K^{(\tau, c) + n}_G(X, Y)$ twisted by the graded twist $(\tau, c)$, the axioms of cohomology theory and their consequences above are valid. In the presence of a homomorphism $\phi : G \to \Z_2$, the same claims hold true for ${}^\phi K^{(\tau, c) + n}_G(X, Y)$, for which the Bott periodicity is ${}^\phi K^{(\tau, c) + n}_G(X, Y) \cong {}^\phi K^{(\tau, c) + n + 8}_G(X, Y)$.

\section{Thom isomorphism}
\label{app:Thom}
We let $\pi: V \to X$ be a $G$-equivariant real vector bundle of real rank $r$. Assuming that $V$ has a $G$-invariant Riemannian metric, we write $\pi : D(V) \to X$ for the unit disk bundle of $V$, and $\pi : S(V) \to X$ for the unit sphere bundle of $V$. These spaces inherit $G$-actions from $V$. We also let $\tau$ be a twist with its $\Z_2$-grading $c$. In this setting, the Thom isomorphism theorem \cite{FHT} for $V$ in twisted $K$-theory states the existence of an $R(G)$-module isomorphism
\begin{align}
K_G^{(\tau, c) + n}(X) \cong 
K_G^{\pi^*((\tau, c) + (\tau_V, c_V)) + n + r}(D(V),S(V)).
\end{align}

The twist $\tau_V$ of $X$ and its grading $c_V$ are associated to $V$. In terms of characteristic classes of $V$, the twist $\tau_V$ is classified by the equivariant third integral Stiefel-Whitney class $W_3^G(V) \in H^3_G(X; \Z)$, which is the obstruction for $V$ to admitting a $G$-equivariant $\mathrm{Pin}^c$-structure. Similarly, the $\Z_2$-grading $c_V$ is classified by the equivariant first Stiefel-Whitney class $w_1^G(V) \in H^1_G(X; \Z_2)$, which is the obstruction for $V$ to being $G$-equivariantly orientable. 

In the special case that $V$ underlies a $G$-equivariant complex vector bundle, we have $w_1^G(V) = 0$ and $W_3^G(V) = 0$. 

To illustrate a non-trivial case, let us assume for a moment that $X = pt$. Under this assumption, a $G$-equivariant real vector bundle on $pt$ is nothing but a real representation $\rho : G \to O(V)$. In this case, we have $H^1_G(pt; \Z_2) \cong \mathrm{Hom}(G, \Z_2)$. Then $w_1^G(V) \in H^1_G(pt; \Z_2)$ is given by the homomorphism $\det \circ \rho : G \to \Z_2$. For an interpretation of $W_3^G(V) \in H^3_G(pt; \Z)$, recall that the $\mathrm{Pin}$-group is a double covering of the orthogonal group
$$
\begin{CD}
1 @>>> \Z_2 @>>> \mathrm{Pin}(r) @>>> O(r) @>>> 1,
\end{CD}
$$
and the group $\mathrm{Pin}^c(r)$ is defined to be the quotient of $\mathrm{Pin}(r) \times U(1)$ under the diagonal $\Z_2$-action. Accordingly, $\mathrm{Pin}^c(r)$ is a central extension of $O(r)$ by $U(1)$. The pull-back under $\rho : G \to O(r)$ gives a central extension of $G$:
$$
\begin{CD}
1 @>>> U(1) @>>> \mathrm{Pin}^c(r) @>>> O(r) @>>> 1 \\
@. @| @AAA @AA{\rho}A @. \\
1 @>>> U(1) @>>> \rho^*\mathrm{Pin}^c(r) @>>> G @>>> 1.
\end{CD}
$$
Recall also that $H^3_G(pt; \Z)$ classifies central extensions of $G$ by $U(1)$. Then the characteristic class $W_3^G(V) \in H^3_G(pt; \Z)$ classifies the central extension $\rho^*\mathrm{Pin}^c(r)$.

Finally, we clarify the meaning of $(\tau, c) + (\tau_V, c_V)$. This is a product of graded twists. If we identify a twist $\tau_i$ with a cohomology class $\tau_i \in H^3_G(X; \Z)$ and its $\Z_2$-grading $c_i$ with $c_i \in H^1_G(X; \Z_2)$ through the classifications, then the graded twist $(\tau_0, c_0) + (\tau_1, c_1)$ is identified with the following cohomology class:
$$
(\tau_0, c_0) + (\tau_1, c_1)
= (\tau_0 + \tau_1 + \beta(c_0 \cup c_1), \
c_0 + c_1) 
\in H^3_G(X; \Z) \times H^1_G(X; \Z_2),
$$
where $c_0 \cup c_1 \in H^2_G(X; \Z_2)$ is the cup product, and $\beta : H^2_G(X; \Z_2) \to H^3_G(X; \Z)$ is the Bockstein homomorphism associated to the exact sequence of coefficients $0 \to \Z \overset{2}{\to} \Z \to \Z_2 \to 0$.

\section{Gysin exact sequence}
\label{Sec:Gysin}

As above, let $\pi : V \to X$ be a $G$-equivariant real vector bundle of rank $r$. From the exact sequence for the pair $(D(V), S(V))$ and the Thom isomorphism, we can derive the Gysin exact sequence for the sphere bundle $\pi : S(V) \to X$, which is the following six-term exact sequence of $R(G)$-modules. 
$$
\begin{CD}
K_G^{\pi^*((\tau, c) + (\tau_V, c_V)) + r + 1}(S(V)) @<{\pi^*}<<
K_G^{(\tau, c) + (\tau_V, c_V) + r + 1}(X) @<<<
K_G^{(\tau, c) + 1}(X) \\
@VVV @. @AAA \\
K_G^{(\tau, c) + 0}(X) @>>>
K_G^{(\tau, c) + (\tau_V, c_V) + r}(X) @>{\pi^*}>>
K_G^{\pi^*((\tau, c) + (\tau_V, c_V)) + r}(S(V)).
\end{CD}
$$

Suppose now that there is a fixed point $pt \in S(V)$. In this case, the equivariant map $s : X \to S(V)$ given by $s(x) = pt$ obeys $\pi \circ s = 1$, so that the Gysin exact sequence splits:
$$
K_G^{\pi^*((\tau, c) + (\tau_V, c_V)) + r + n}(S(V))
\cong 
K_G^{(\tau, c) + (\tau_V, c_V) + r + n}(X) \oplus 
K_G^{(\tau, c) + n - 1}(X).
$$

For topological insulators, the following are useful: 
\begin{itemize}
\item (Index for boundary gapless state)
If $G$ trivially acts on $S^1$, then:
\begin{align}
K_G^{\pi^*(\tau, c) + n}(X \times S^1) 
\cong K_G^{(\tau, c) +n}(X) \oplus K_G^{(\tau, c) +n-1}(X). 
\end{align}
\item (Dimensional reduction for $\Z_2$ symmetry)
If $\Z_2 = \{1,\sigma\}$ acts on $S^1$ as ``reflection" $\sigma : e^{i \theta} \mapsto e^{- i \theta}$, then:
\begin{align}
K_{\Z_2}^{\pi^*(\tau, c)+n}(X \times S^1) 
\cong K_{\Z_2}^{(\tau, c) +n}(X) 
\oplus K_{\Z_2}^{(\tau, c + w) + n-1}(X),
\label{eq:app_Gysin_reflection}
\end{align}
where the ``anti-symmetry'' $w \in H^1_{\Z_2}(X; \Z_2)$ is the pull-back of the identity map $1 \in H^1_{\Z_2}(pt; \Z_2) = \mathrm{Hom}(\Z_2, \Z_2)$ under the collaption map $X \to pt$.

\item (Defect gapless state as a boundary state)
If $G$ acts on $S^r$ through $G \to O(r+1)$ with a point fixed, then:
\begin{equation}\begin{split}
K_G^{\pi^*(\tau, c)+n}(X \times S^r \times S^r) 
&\cong 
K_G^{(\tau, c) + n}(X \times S^r) \oplus 
K_G^{(\tau, c) + (\tau_V, c_V) + n-r}(X \times S^r) \\
&\cong 
K_G^{(\tau, c) + n}(X) \oplus 
K_G^{(\tau, c) + (\tau_V, c_V)+ n-r}(X) \oplus 
K_G^{(\tau, c) + (\tau_V, c_V)+ n-r}(X) \oplus 
K_G^{(\tau, c) + n}(X).
\end{split}\end{equation}
Here the first three direct summands are ``weak" indices. 
\end{itemize}

\section{Ext functor $\mathrm{Ext}^1_R(A, B)$}
\label{app:ext}
Let $A$ and $B$ be modules over a ring $R$. An $R$-module $E$ fitting into the exact sequence of $R$-modules
$$
0 \to B \to E \to A \to 0
$$
is called an extension of $A$ by $B$. Such an extension is generlly not unique, and the isomorphism classes of the extensions are in one to one correcpondence with the elements in the group $\mathrm{Ext}^1_R(A, B)$. To definte $\mathrm{Ext}^1_R(A, B)$, let us choose a free resolution of $A$, which is an exact sequence of $R$-modules
$$
\cdots \overset{\partial}{\to}
F_n \overset{\partial}{\to} 
F_{n-1} \overset{\partial}{\to} \cdots 
F_1 \overset{\partial}{\to} 
F_0 \to 
A \to 0
$$
such that each $F_i$ is a free $R$-module, that is, the direct sum of copies of $R$. Setting $F^n = \mathrm{Hom}_R(F_n, B)$ and defining $\delta : F^n \to F^{n+1}$ to be $\delta (f) = f \circ \partial$ for $f \in F^n$, we have a cochain complex $(F^n, \delta)$. Its $1$st cohomology is $\mathrm{Ext}^1(A, B)$.

\subsection{A proof of (\ref{eq:K0_p4_T2})}
Now, we apply the above classification of extensions to the case where $R = R(\Z_4) = \Z[t]/(1-t^4)$ is the represetation ring of $\Z_4$, $A$ is the ideal $A = (1 - t + t^2 - t^3)$, and $B = R(\Z_4)$. A free resolution of $A$ can be given by taking $F_n = R(\Z_4)$, in which $\partial : F_n \to F_{n-1}$ is the multiplication by $1 + t$ if $n$ is odd and that by $1 - t + t^2 - t^3$ if $n$ even. Any $R$-module homomorphism $f \in F^1 = \mathrm{Hom}_{R(\Z_4)}(R(\Z_4), R(\Z_4))$ is uniquely specified by the value of $1 \in R(\Z_4)$. Let $a_0, \ldots, a_3 \in \Z$ be defined by $f(1) = a_0 + a_1 t + a_2 t^2 + a_3 t^3$. On the one hand, the condition for $f$ to be $\delta (f) = 0$ is $a_0 - a_1 + a_2 - a_3 = 0$. Therefore any $f \in F^1 \cap \mathrm{Ker}(\delta)$ is of the form $f(1) = a_0(1 + t^3) + a_1(t - t^3) + a_2(t^2 + t^3)$. On the other hand, If we define $g \in F^0$ to be the multiplication by $a_1 t + (a_2 - a_1) t^2 + a_0 t^3$, then $\delta (g) = f$ for $f$ as above. This means that the kernel of $\delta : F^1 \to F^2$ agrees with the image of $\delta : F^0 \to F^1$, and hence $\mathrm{Ext}^1_{R(\Z_4)}((1 - t + t^2 - t^3), R(\Z_4)) = 0$. Consequetly, any extension of $(1 - t + t^2 - t^3)$ by $R(\Z_4)$ is isomorphic to the obvious extension $R(\Z_4) \oplus (1 - t + t^2 - t^3)$.

\bibliography{refs}

\end{document}